\documentclass{article}
\usepackage[utf8]{inputenc}

\pdfoutput=1
\usepackage{bbold}
\usepackage{amssymb}
\usepackage{amsmath}
\usepackage[dvips]{graphicx}
\usepackage{setspace}
\usepackage{slashed}
\usepackage{mathtools}
\usepackage{amsfonts}
\usepackage{fancyhdr}
\usepackage{dsfont}
\usepackage{xcolor}
\usepackage{graphicx}
\usepackage{rotating}
\usepackage{comment}
\usepackage{color} 
\usepackage{subcaption}
\usepackage[percent]{overpic}
\usepackage{cite}
\usepackage{braket}
\definecolor{darkgreen}{rgb}{0,0.5,0}
\definecolor{darkblue}{rgb}{0,0,0.6}
\definecolor{purple}{rgb}{0.4,.2,0.7}

\newcommand{\be}{\begin{equation}}
\newcommand{\ee}{\end{equation}}

\usepackage[colorlinks=true,citecolor=darkgreen,linkcolor=black,urlcolor=purple]{hyperref}

\usepackage{pdfsync}

\makeatletter
\newcommand*{\defeq}{\mathrel{\rlap{%
                     \raisebox{0.3ex}{$\m@th\cdot$}}%
                     \raisebox{-0.3ex}{$\m@th\cdot$}}%
                     =} 
\makeatother

\def\be{\begin{eqnarray}}
\def\ee{\end{eqnarray}}

\newcommand{\tr}{\textrm{Tr}\,}

\newcommand{\bea}{\begin{eqnarray}}
\newcommand{\eea}{\end{eqnarray}}
\def\ben{\begin{equation}}
\def\een{\end{equation}}

     \let\r=v

\def\be{\begin{equation}}
\def\ee{\end{equation}}
\def\ba{\begin{eqnarray}}
\def\ea{\end{eqnarray}}

\def\bal#1\eal{\begin{align}#1\end{align}}
\def\bs#1\es{\begin{split}#1\end{split}}

\allowdisplaybreaks  

\numberwithin{equation}{section}

\def\be{\begin{equation}}
\def\ee{\end{equation}}
\def\ba{\begin{eqnarray}}
\def\ea{\end{eqnarray}}
\def\bal#1\eal{\begin{align}#1\end{align}}

\def\r{\rightarrow}

\def\r{\right}

\usepackage{tikz}
\usetikzlibrary{positioning,arrows}
\usetikzlibrary{decorations.pathmorphing}
\usetikzlibrary{decorations.markings}
\tikzset{
particle/.style={postaction={decorate}},
graviton/.style={decorate, decoration={snake, amplitude=0.8 mm, segment length=1.5 mm, pre length=0.8 mm, post length=0.8 mm}},
photon/.style={
        decoration={complete sines, amplitude=0.15cm, segment length=0.2cm},
        decorate    
    },
gluon/.style={
        decoration={coil, aspect=0.75, mirror, segment length=1.5mm},
        decorate
    }
}
 
\def \be {\begin{equation}}
\def \ee {\end{equation}}

\usepackage{framed}

\begin{document}
\onehalfspacing

\begin{center}

~
\vskip5mm

{\LARGE  {
Hartle-Hawking state and its factorization in 3d gravity
\\
\ \\
}}

\vskip10mm

Wan Zhen Chua${}^{1}$ and Yikun Jiang${}^{1,2}$

\vskip5mm

\it{${}^1$ Department of Physics, Cornell University, Ithaca, New York, USA}
\vskip2mm
\it{${}^2$ Department of Physics, Northeastern University, Boston, MA 02115, USA}

\vskip5mm

\end{center}

\vspace{4mm}

\begin{abstract}
\noindent
We study 3d quantum gravity with two asymptotically anti-de Sitter regions, in particular, using its relation with coupled Alekseev-Shatashvili theories and Liouville theory. Expressions for the Hartle-Hawking state, thermal $2n$-point functions, torus wormhole correlators and Wheeler-DeWitt wavefunctions in different bases are obtained using the ZZ boundary states in Liouville theory. Exact results in 2d Jackiw-Teitelboim (JT) gravity are uplifted to 3d gravity, with two copies of Liouville theory in 3d gravity playing a similar role as Schwarzian theory in JT gravity. The connection between 3d gravity and the Liouville ZZ boundary states are manifested by viewing BTZ black holes as Maldacena-Maoz wormholes, with the two wormhole boundaries glued along the ZZ boundaries. In this work, we also study the factorization problem of the Hartle-Hawking state in 3d gravity. With the relevant defect operator that imposes the necessary topological constraint for contractibility, the trace formula in gravity is modified in computing the entanglement entropy. This trace matches with the one from von Neumann algebra considerations, further reproducing the Bekenstein-Hawking area formula from entanglement entropy. Lastly, we propose a calculation for off-shell geometrical quantities that are responsible for the ramp behavior in the late time two-point functions, which follows from the understanding of the Liouville FZZT boundary states in the context of 3d gravity, and the identification between Verlinde loop operators in Liouville theory and ``baby universe'' operators in 3d gravity.

\end{abstract}

\pagebreak
\pagestyle{plain}

\setcounter{tocdepth}{2}
{}
\vfill
\tableofcontents

\newpage

\date{}

\section{Introduction}
To better understand the quantum nature of gravity, we study exactly soluble simple models of gravity. This machinery has been successful in low dimensional theories of pure gravity with negative cosmological constant, where bulk degrees of freedom are absent. In particular, fruitful results have been achieved in 2d Jackiw-Teitelboim (JT) gravity, please refer to \cite{Mertens:2022irh} for a nice review. The next logical step is to study 3d gravity, where only boundary graviton excitations are involved \cite{Brown:1986nw}. At the level of action, the theory in the bulk is topological as classical 3d gravity has a famous Chern-Simons theory formulation\cite{Achucarro:1986uwr, Witten:1988hc}. In the presence of one or two asymptotically AdS boundaries, 3d quantum gravity is related to copies of geometrical Alekseev-Shatashvili theories, which corresponds to quantization for the coadjoint orbits of the Virasoro group \cite{Witten:1987ty, Alekseev:1988ce, Alekseev:1990mp, Maloney:2007ud, Cotler:2018zff, Cotler:2020ugk}. Recently, it has been proposed that exact quantization of 3d gravity around on-shell solutions is equivalent to ``Virasoro TQFT'', based on considerations on quantization of Teichm$\ddot{\text{u}}$ller spaces\cite{Verlinde:1989ua, Krasnov:2005dm, Scarinci:2011np, Collier:2023fwi}. 

In this paper, we mainly focus on 3d quantum gravity with two asymptotic boundaries, where the spatial slice takes the topology of an annulus. It has been understood that in such cases, 3d gravity is related to four copies of Alekseev-Shatashvili theories that are coupled across boundaries \cite{Henneaux:2019sjx, Cotler:2020ugk}. Via a field redefinition, the coupled theories become two copies of Liouville theory with non-trivial mixing between the zero modes\cite{Henneaux:1999ib, Henneaux:2019sjx}, and the path integral measure is reduced to flat Liouville measures from Alekseev-Shatashvili symplectic measures\cite{Mertens:2018fds}. Hence, Liouville theory plays a similar role as Schwarzian theory in JT gravity  \cite{Maldacena:2016upp}. Using this correspondence, we identify the Hartle-Hawking state in 3d gravity as a quantum state in two copies of Liouville theory, further uplifting exact calculations in 2d JT gravity to answers in 3d gravity. We exhaust the list of calculations here: partition function \cite{Maldacena:2016upp, Stanford:2017thb, Saad:2019lba}, correlation functions \cite{Mertens:2017mtv}, Wheeler-DeWitt wavefunctions in different bases \cite{Harlow:2018tqv, Yang:2018gdb}, factorization, modification of trace and entanglement entropy\cite{Jafferis:2019wkd, Penington:2023dql}, and wormholes describing quantum noise \cite{Stanford:2020wkf}. As expected, dimensional reduction of our 3d results reproduces all the 2d JT answers \cite{Mertens:2018fds, Ghosh:2019rcj}. For off-shell configurations, we also reinterpret the calculation for the spectral form factor \cite{Saad:2019lba} in \cite{Cotler:2020ugk} in terms of overlaps between quantum states \cite{Saad:2019pqd}. Further identifying the Verlinde loop operators in Liouville theory as 3d ``baby universe'' operators, we propose a calculation for the late time two-point functions \cite{Saad:2019pqd} that is related to resolving the version of information paradox proposed in\cite{Maldacena:2001kr}. These results have been more or less anticipated in \cite{Mertens:2017mtv, Mertens:2018fds, Blommaert:2018iqz, Mertens:2022ujr}, in particular when Liouville theory is used as an intermediate tool to calculate Schwarzian quantities. In this paper, we use techniques developed in \cite{Collier:2022bqq, Collier:2023fwi} to show directly that such an identification follows from a geometrical reasoning.

3d quantum gravity on two asymptotic boundaries is related to four copies of coupled Alekseev-Shashvili theories summed over orbits\cite{Henneaux:2019sjx, Cotler:2020ugk}, which is further equivalent to two copies of Liouville theory upon field redefinition\cite{Henneaux:1999ib, Henneaux:2019sjx}, provided that we consider the correct path integral measure. In particular, the connection between 3d gravity and Liouville theory includes description of Hilbert spaces, which further allows us to obtain expressions for gravitational quantities, i.e. two-boundary torus wormhole correlators. The wormhole contribution for the product of correlation functions is the variance that describes quantum noise \cite{Stanford:2020wkf}.

We also propose the Hartle-Hawking state in 3d gravity to be given by two copies of the Liouville ZZ boundary states,
\begin{equation}\label{eq:HH_BTZ_ZZ2}
    | \Psi^{\text{HH}}_{\beta/2} \rangle \cong e^{-\beta H/4} | ZZ \rangle  e^{-\beta H/4}| \widetilde{ZZ} \rangle ~,
\end{equation}
where $\cong$ is to distinguish the two Hilbert spaces, and the non-trivial pairing between different chiralities is shown in Figure \ref{fig:ZZ}. $|ZZ \rangle$, known as the ZZ boundary state, is a Liouville conformal boundary state and is constructed from certain superposition of Ishibashi states to ensure conformal invariance \cite{Zamolodchikov:2001ah}. $H$ is the Hamiltonian in Liouville theory and the tilde is to denote a second copy. It is known that Liouville theory is not the holographic dual of gravity. For one thing, Liouville theory has a flat spectrum whereas  holographic CFTs have an exponential Cardy density of states\cite{Cardy:1986ie}. Hence, the Liouville thermofield double is not the dual of Hartle-Hawking state as the information on the Cardy spectrum is encoded in the ZZ boundary state.

The motivation of proposing \eqref{eq:HH_BTZ_ZZ2} stems from having the partition function of non-rotating Euclidean BTZ black holes as two copies of Liouville overlaps \cite{Zamolodchikov:2001ah, Mertens:2017mtv}
\begin{equation}\label{eq:Z_BTZ_ZZ}
\begin{split}
    Z_{\text{BTZ}} &= \langle ZZ|e^{-\beta H/2}|ZZ \rangle \langle \widetilde{ZZ} |e^{-\beta H/2}|\widetilde{ZZ} \rangle~, \\
    &= \chi_{\mathds{1}}\left(i\frac{2\pi}{\beta} \right)\chi_{\mathds{1}}\left(i\frac{2\pi}{\beta} \right)~,
\end{split}
\end{equation}
where $\chi_{\mathds{1}}$ is the Virasoro vacuum character. Slicing the thermal partition function in half in the thermal direction allows us to identify the Hartle-Hawking state in \eqref{eq:HH_BTZ_ZZ2} naturally. \footnote{Similar ideas have been considered in \cite{Mertens:2022ujr}.} To further understand this proposal, we notice that non-rotating BTZ metric takes the following parametrization,
\begin{equation}\label{eq:MM_metric2}
    ds^2 = d\rho^2 + \cosh^2 \rho e^{\Phi(z,\bar{z})}dz d\bar{z}~,
\end{equation}
where from Einstein's equations, $\Phi$ has to satisfy the Liouville equation and the Liouville field takes the following expression
\begin{equation}\label{eq:BTZ wormhole3}
\begin{aligned}
    e^\Phi& = \frac{4\pi^2}{\beta^2}\frac{1}{\sin^2 \left(\frac{2\pi}{\beta} \text{Im}(z)\right)}~.
\end{aligned}
\end{equation}
The singular behavior of the Liouville field near $\text{Im}(z) \rightarrow 0;\beta/2$ corresponds to the ``ZZ boundary condition''\cite{Zamolodchikov:2001ah}.

\begin{figure}[h]
\begin{center}
\begin{minipage}[b]{0.4\linewidth}
\begin{overpic}[scale=0.2]{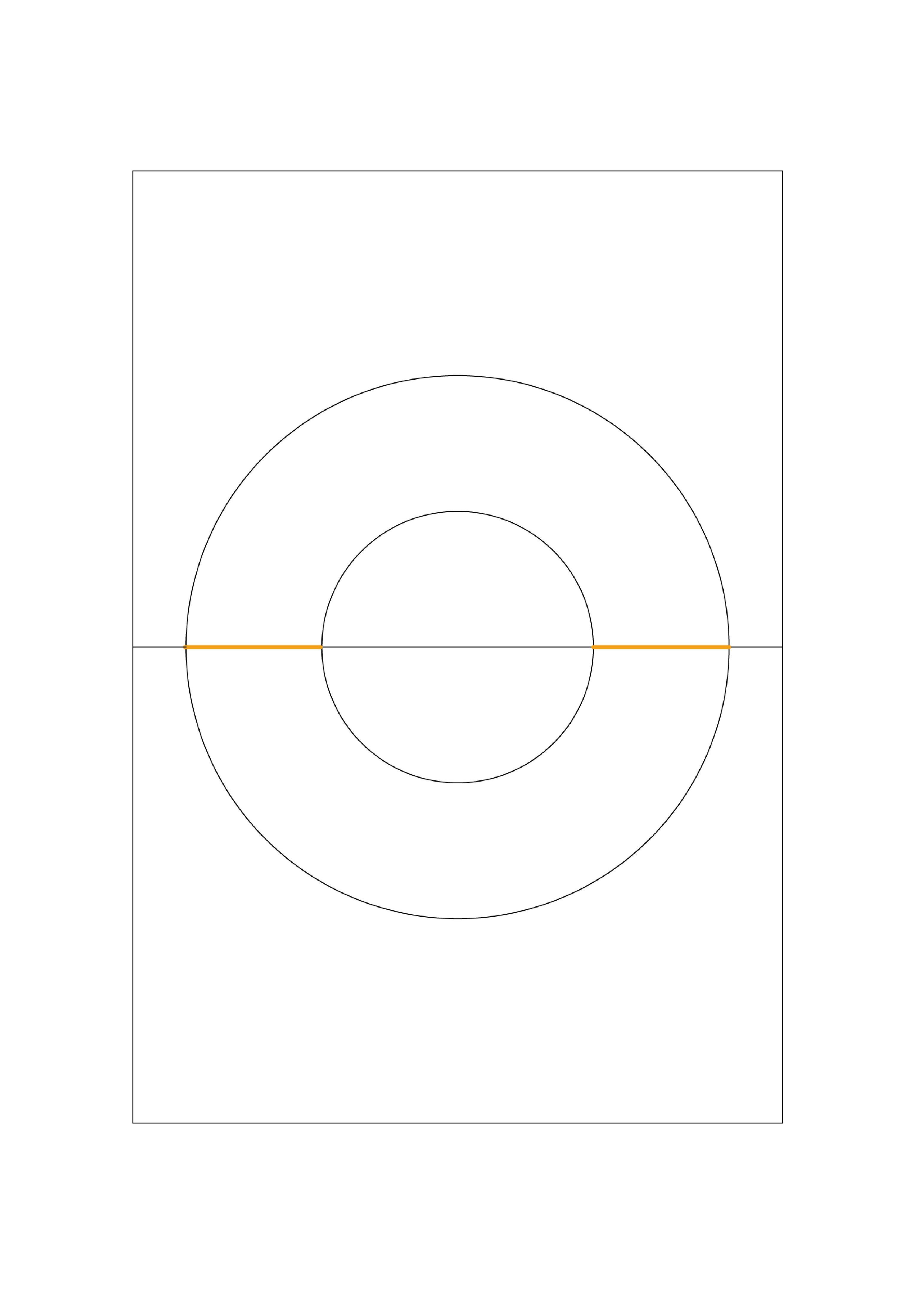}
\end{overpic} 
        \end{minipage}
        \begin{minipage}[b]{0.45\linewidth}
\begin{overpic}[scale=0.22]{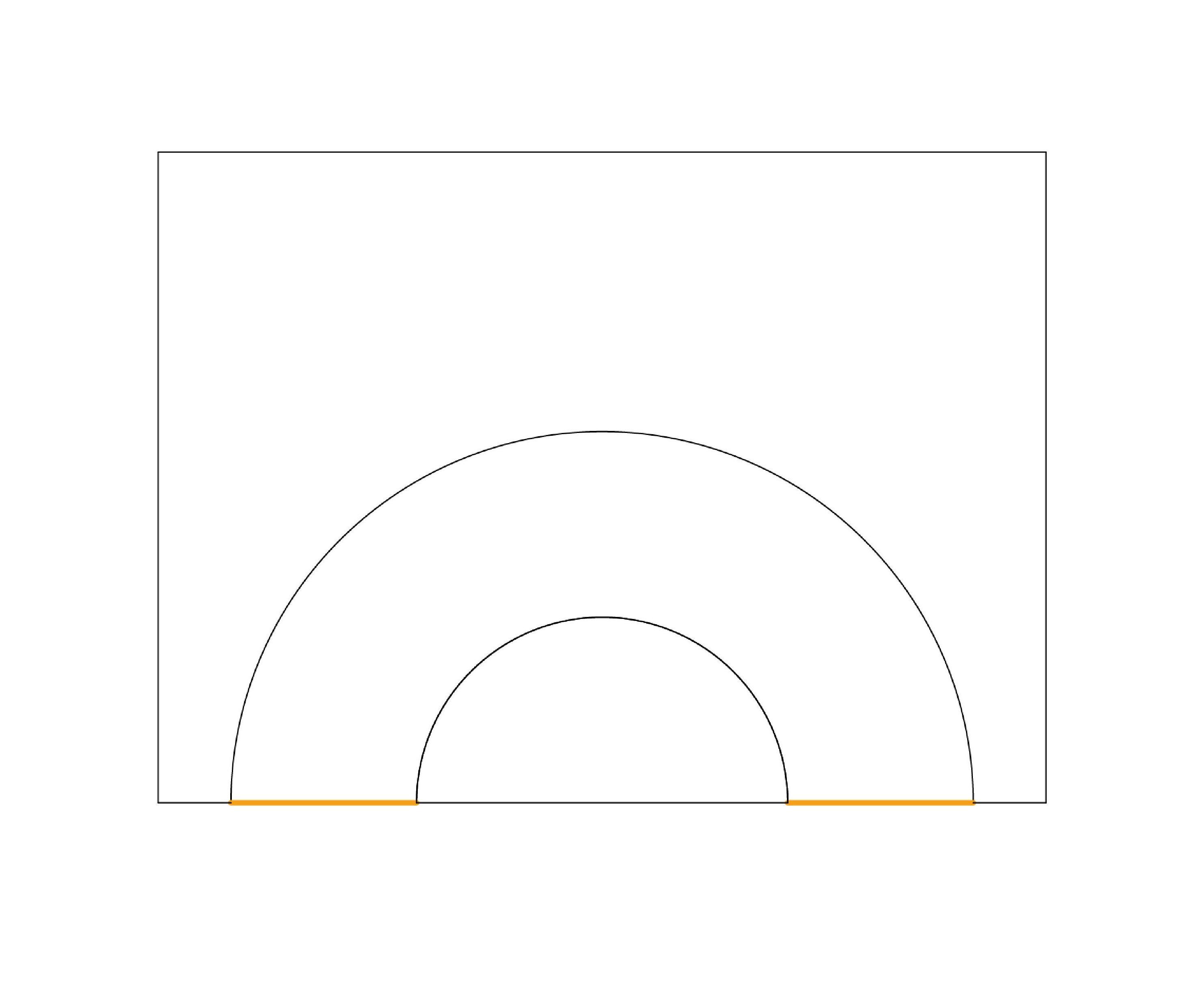}
\end{overpic}  
  \end{minipage}
\end{center}
 \caption{(Left): The asymptotic boundary region of the upper half-space construction has the topology of a torus where the concentric circles are identified with each other. (Right): Using the folding trick \cite{Wong:1994np, Cardy:2004hm}, we can fold the anti-chiral half of the partition function in the lower half plane onto the upper half plane, allowing us to have a non-chiral Liouville transition amplitude (one $SL(2,R)$ factor in gravity) on the upper half plane. We can also use the folding trick to fold the chiral half in the upper half plane onto the lower half plane, getting another copy of non-chiral Liouville theory (another $SL(2,R)$ factor in gravity) in the lower half plane. In the end, there is a full Liouville theory in the upper and lower half plane respectively, which corresponds to the two $SL(2,R)$ factors in gravity.} \label{zz doubling}
\end{figure}

The asymptotically AdS boundaries lie at $\rho \rightarrow \pm \infty$ respectively. In previous works \cite{Maldacena:2004rf, Collier:2022bqq}, similar metric ansatz \eqref{eq:MM_metric2}, known as the hyperbolic slicing, which we now call it the ``wormhole slicing'' as it is used to study two-boundary observables, i.e. Maldacena-Maoz (or Fuchsian) wormholes connecting two asymptotic boundaries, and the semiclassical gravity calculation in 3d matches the large $c$ behavior of Liouville CFT quantities. \cite{Collier:2023fwi} argue that the matching of semiclassical behavior further implies an exact equivalence between the 3d gravity and Liouville results at finite central charge $c$. The wormhole slicing  \eqref{eq:MM_metric2} seems to provide a contradicting geometry for quantities with one asymptotic boundary, i.e., the Hartle-Hawking wavefunction and BTZ partition function. However, the ZZ boundary condition glues the ``two'' halves that are denoted by $\rho \rightarrow \pm \infty$ respectively. In the language of boundary CFT, this is can be viewed as the doubling or folding trick as shown in Figure \ref{zz doubling} \cite{Cardy:1984bb, Wong:1994np, Cardy:2004hm}.

With the wormhole slicing, we can study $2n$-point correlation functions for below black hole threshold probe operators that are inserted symmetrically across $\rho=0$ in BTZ background. In gravity, this corresponds to the study of Einstein action coupled to $2n$ massive probe particles whereas in Liouville theory, this corresponds to the computation of $2n$-point functions with two ZZ boundaries on a finite cylinder. The on-shell renormalized gravitational and Liouville actions match after a careful treatment of counterterms. The exact matching of these two results beyond semiclassical level follows from arguments in \cite{Collier:2023fwi}. We can also understand this exact matching directly using CFT techniques, and we provide a detailed analysis for the case of two-point functions. The ZZ boundary state being a superposition of Ishibashi states\cite{Ishibashi:1988kg, Fateev:2000ik}, encodes all contributions from descendents. As an explicit example, we show that a probe operator insertion between Ishibashi states is exactly equal to the two-point torus conformal block, as expected from the doubling trick\cite{Cardy:1984bb, Cardy:2004hm}. In addition, the ZZ wavefunction and together with the DOZZ structure constant \cite{Zamolodchikov:1995aa} in the Liouville transition amplitude becomes the crossing kernel \cite{Ponsot:1999uf, Ponsot:2000mt, Collier:2019weq}. These facts turn the Liouville overlap into a two-point identity block on a torus. We like to mention that correlation functions in terms of Liouville overlaps also match with the ensemble-averaged result that is proposed in  \cite{Collier:2022bqq}.

The relation in \eqref{eq:HH_BTZ_ZZ2} is further verified in semiclassical limit through the calculation of Wheeler-DeWitt wavefunctions. We first analyze in detail the boundary value problem in gravity, in particular paying attention to possible corner terms. We find two bases: the fixed $(\Phi_0,J)$ basis and fixed $(E,J)$ basis, generalizing results in JT gravity\cite{Harlow:2018tqv, Yang:2018gdb}. The $(E,J)$-basis, including higher-dimensional wavefunctions, take the form of a ``Pacman'' geometry and are studied in great detail in \cite{Chua:2023srl}. It is worth mentioning that the $(E,J)$-states in the bulk is holographically dual to Liouville primary states, further giving us the following identification between Wheeler-DeWitt wavefunctions and Liouville overlaps
\begin{equation}
\begin{split}
    \Psi^{\text{HH}}_{\beta/2}(E,J) &= \langle P | e^{-\beta H/4} | ZZ \rangle \langle \widetilde{P}| e^{-\beta H/4}| \widetilde{ZZ} \rangle~,
\end{split}
\end{equation}
where the matching involves analytic continuation of Liouville momenta $P,\widetilde{P}$ such that $P,\widetilde{P}$ are related to the ADM mass $E$ and imaginary angular momentum $J$ of black holes in Euclidean signature.
In this work, we focus on the $(\Phi_0,J)$-states, which again can be studied using the wormhole slicing and matching with Liouville theory beyond large $c$ involves techniques developed in \cite{Collier:2023fwi}. For $J=0$, $\Phi_0$ is related to the renormalized geodesic length between the ``two'' halves of the asymptotic boundaries at $\rho \rightarrow \pm \infty$ respectively at fixed angular coordinates, making $(\Phi_0,0)$-basis being the uplift of the fixed geodesic length basis in JT gravity\cite{Harlow:2018tqv, Yang:2018gdb}. With one extra dimension, the bulk slice of $(\Phi_0,0)$-states is a hyperbolic cylinder with scalar curvature $R^{(2)} = -2$ instead of a 1d geodesic. With that being said, in addition to determining the height of the hyperbolic cylinder, $\Phi_0$ also parametrizes the waist. In Liouville theory, we define the state $\ket{\Phi_0}$ using Liouville zero mode wavefunction $\langle  \Phi_0  | P \rangle \equiv \psi_{P}(\Phi_0)$\cite{Fateev:2000ik}. The on-shell action is the large $c$ limit of the corresponding Liouville transition amplitude
\begin{equation}\label{eq:HHPhi0_BTZ_ZZ2}
    \Psi^{\text{HH}}_{\beta/2}(\Phi_0) = \langle \Phi_0| e^{-\beta H/4} | ZZ \rangle \langle \Phi_0| e^{-\beta H/4}| \widetilde{ZZ} \rangle ~.
\end{equation}
The connection of our results to 2d JT gravity is straightforward and can be achieved by performing dimensional reduction or taking the near-extremal limit\cite{Mertens:2018fds, Ghosh:2019rcj}, which we show explicit examples respectively. 

The $(\Phi_0,J)$-basis involves mixing of moduli between the ``two'' boundaries, and the spacetime geometry of the Wheeler-DeWit wavefunction corresponds to a quasi-Fuchsian wormhole \cite{Collier:2022bqq}
\begin{equation} 
    \begin{split}
        ds^2 =  d\rho^2 +\cosh^2 \rho e^{\Phi(z,\bar{z})}\big{|}dz + 4 G_N J(1+\tanh \rho) e^{-\Phi(z,\bar{z}) } d\bar{z}\big{|}^2~.
    \end{split}
\end{equation}
The on-shell action is given by
\begin{equation} 
    \begin{split}
        -S_{\text{grav}}(\Phi_0,J) = -\frac{c}{6}S_{\text{Liouv}(z,\bar{w})}(\Phi_0,J)-\frac{c}{6}S_{\text{Liouv}(w,\bar{z})}(\Phi_0,J) ~,
    \end{split}
\end{equation}
where $(z,\bar{z})$ are the coordinates for the flat metric at the ``left'' boundary and $(w,\bar{w})$ are the coordinates for the ``right'' boundary. For the right hand side of the equation, we have one action that corresponds to a Liouville field $\Phi_-(z,\bar{w})$ living on a complex metric $dz d\bar{w}$ and its complex conjugate $\Phi_+(w,\bar{z})$ living on another complex metric $dw d\bar{z}$.

We use our two-sided Hartle-Hawking state to study the factorization problem in 3d gravity. In terms of Chern-Simons theory, we find a local boundary condition that factorizes the state into two single-sided gravity Hilbert spaces. Similar to the case in JT gravity, this ``cutting map'' is not isometric\cite{Jafferis:2019wkd}. We find the relevant defect operator that provides an isometric factorization map: $\mathcal{J}:\mathcal{H} \rightarrow \mathcal{H}_L \otimes \mathcal{H}_R$, which subsequently modifies the definition of trace to $Z_n=\tr_{\mathcal{H}_R}(\mathcal{D} \tilde{\rho}^n)$ in the calculation of $n$-th gravitational R$\acute{\text{e}}$nyi entropy. Interestingly, the extra single-sided defect operator $\mathcal{D}$ that we need for the correct calculation of trace in gravity takes the following form
\be \label{one-sided defect operator0}
\begin{aligned}
\mathcal{D}&=\int_0^\infty dP' d\widetilde{P}' \sum_{N_1,\widetilde{N}_1} S_{\mathds{1}P'} S_{\mathds{1}\widetilde{P}'} \ket{h_{P'},N_1}\ket{\widetilde{h_{P'},N_1}}\bra{h_{P'},N_1}  \bra{\widetilde{h_{P'},N_1}}~,
\end{aligned}
\ee
where $h_{P'}, h_{\widetilde{P}'}$ are the conformal weights for Virasoro primary operators and $N_1, \widetilde{N_1}$ denote the levels for the descendents. $S_{\mathds{1}P'}$ is the modular S-matrix element between identity operators and primary operators. Implicitly, the defect operator projects to identity in the dual cycle in the language of Chern-Simons theory, which further corresponds to zero flux projection. This is exactly the topological contractibility condition in the thermal cycle in gravity. Furthermore, the modified trace due to the defect operator matches with the proposed unique trace formula for gravity in the context of Type II$_\infty$ von Neumann algebra with a trivial center\cite{Chandrasekaran:2022eqq, Penington:2023dql}.

Using these ingredients, we calculate the entanglement entropy and obtain
\begin{equation}
	S_{\text{EE}} = -\partial_n \left(\frac{Z_n}{(Z_1)^n}\right) \Big{|}_{n = 1} = -\text{Tr}_{\mathcal{H}_R} \hat{\rho}\ln \hat{\rho} +\text{Tr}_{\mathcal{H}_R} \hat{\rho} \ln \mathcal{D} ~,
\end{equation}
where $\hat{\rho}$ is the normalized reduced density matrix. The answer takes a similar form as the FLM formula \cite{Faulkner:2013ana}, and we show that the second term is the expectation value of an ``area operator'' in a precise sense. This term reproduces the Bekenstein-Hawking area formula at the saddle point in the large c limit \cite{Bekenstein:1972tm, PhysRevD.7.2333, PhysRevD.9.3292, 1974Natur.248...30H, 1975CMaPh..43..199H}. Since $S_{\mathds{1}P}$ is also the Plancherel measure of the quantum semi-group $SL^+_{q}(2,R)$ \cite{Ponsot:1999uf, Ponsot:2000mt, Teschner:2003em, Teschner:2005bz, Mertens:2022ujr, Wong:2022eiu}, our derivation provides a canonical interpretation on the Bekenstein-Hawking entropy as a topological entanglement entropy\cite{McGough:2013gka}. In particular, our derivation follows from the observation that the topological contractibility condition in gravity can be imposed by an operator in 2d CFT constructed from modular invariance. This is in the spirit of the original derivation of Cardy's formula\cite{Cardy:1986ie}.

Our formalism is also useful for computing off-shell geometrical quantities. We explain the Hilbert space description on the spectral form factor calculation \cite{Cotler:2020ugk} in 3d gravity, generalizing the 2d results\cite{Saad:2019lba, Saad:2019pqd}. In this procedure, we explain the roles played by the FZZT-boundary states\cite{Fateev:2000ik, Teschner:2000md}, and find that Verlinde loop operators in Liouville theory are the holographic dual of ``baby universe'' operators in gravity\cite{Penington:2023dql}. We follow a similar proposal to the 2d JT case \cite{Saad:2019pqd}, and show that ``double-trumpet'' geometries that contribute to late time two-point functions have density of states that exhibit level repulsion in random matrix theory, which further governs the linear growth of correlators at late times, thus providing a potential solution to the version of information paradox in \cite{Maldacena:2001kr} for 3d gravity.

The paper is organized as follows. In Section \ref{sec:3d_Liouville_AS}, we review relations between 3d gravity with two boundaries, Alekseev-Shatashvili theories and Liouville theory. In Section \ref{sec:HH-state}, we explain our identification of the Hartle-Hawking state with the Liouville ZZ boundary states. We provide a geometric explanation for this identification, in particular using the wormhole slicing of BTZ black holes to explain the gluing of ``two'' boundaries that arises from ZZ boundary conditions. In Section \ref{sec:correlation_function}, we further generalize results of partition functions by including operator insertions. Using wormhole slicing, we first show the matching of semiclassical results between gravity and Liouville theory results. We then use the torus two-point functions as an explicit example to show how overlaps of Ishibashi states become the torus conformal blocks, and how the ZZ wavefunctions and DOZZ structure coefficients in Liouville theory combine together to give the crossing kernel. The combination of these gives us the exact formula of the Virasoro identity block. Finally, we match Liouville correlation functions with results obtained from an ensemble average interpretation of CFTs. In Section \ref{sec:WdW}, we show the correct boundary value problem for the study of Wheeler-DeWitt wavefunctions in gravity. We find two different bases for the wavefunctions: the fixed $(\Phi_0,J)$-basis and fixed $(E,J)$-basis. We study the fixed $(\Phi_0,J=0)$-basis using the wormhole slicing, further matching the large $c$ limit of the corresponding transition amplitude in Liouville theory. We then analyze the geometry that corresponds to the fixed $(\Phi_0,J)$-basis using quasi-Fuchsian wormholes. In addition, the dimensional reduction of the fixed $(\Phi_0,J=0)$-basis to the geodesic length basis in JT gravity is demonstrated. In Section \ref{sec:factorization}, we show how to factorize the two-sided Hartle-Hawking state into two single-sided Hilbert spaces and find the relevant modification to the trace formula in gravity through implementing topological contractibility conditions using CFT data. Using these ingredients, we reproduce the Bekenstein-Hawking entropy formula in 3d gravity from a canonical calculation. In Section \ref{sec:late time}, we show how the Hilbert space formalism can be used to study off-shell gravitational quantities. Further understanding the roles played by the FZZT boundary states and identifying Verlinde loop operators as ``baby universe'' operators, we reproduce the ``double trumpet'' spectral form factor in 3d gravity. Finally, we give a proposal on the relevant off-shell wormhole geometry that contributes to the linear ramp behavior in late time two-point functions,  further suggesting a possible candidate in resolving the information paradox in 3d gravity.

\textbf{Note added:} While this work was in preparation, the papers \cite{Mertens:2022ujr, Wong:2022eiu} appeared, which studied properties of the Hartle-Hawking state and the factorization problem from a quantum group perspective.

\section{3d gravity with two asymptotic boundaries as Liouville theory}\label{sec:3d_Liouville_AS}

We first review the canonical quantization of 3d gravity with two asymptotic boundaries, which has been done in previous works \cite{Coussaert:1995zp, Henneaux:1999ib, Henneaux:2019sjx, Cotler:2018zff, Cotler:2020hgz, Cotler:2020ugk}. We focus on the connections between 3d gravity, Alekseev-Shatashvili theory and Liouville theory, in particular on the role played by Liouville zero modes and their connection to holonomies along the non-contractible cycle in gravity. 

\subsection{3d gravity as coupled Alekseev-Shatashvili theories summed over orbits}\label{subsec:brickwall}

The 3d Einstein-Hilbert action on an asymptotically AdS spacetime $\mathcal{M}$ is given by
\begin{equation}\label{eq:3d_action}
	S_0 = \frac{1}{16 \pi G_N}\int_{\mathcal{M}} d^3 x \sqrt{-g}(R+2)~,
\end{equation}
up to boundary terms. We consider a spacetime manifold $\mathcal{M}$ with spatial topology of an annulus $\mathcal{A}$, and parametrize the spacetime using coordinates $(r,t,\phi)$. Taking $L_\alpha$ and $\widetilde{L}_\alpha$ to be the generators in the fundamental representation of $SL(2,R)$, and combine the dreibein $e^\alpha$ and spin connection $\omega^\alpha$ into $A^\alpha$ and $\widetilde{A}^{\alpha}$ as
\begin{equation}
	A^\alpha = \omega^\alpha +e^\alpha ~,~~\widetilde{A}^{\alpha} = \omega^\alpha - e^\alpha~,
\end{equation}
we construct algebra-valued one forms, $A = A^\alpha L_\alpha$ and $\widetilde{A} = \widetilde{A}^\alpha \widetilde{L}_\alpha$. 

\begin{figure}[h]
	\centering
	\includegraphics[scale=0.3]{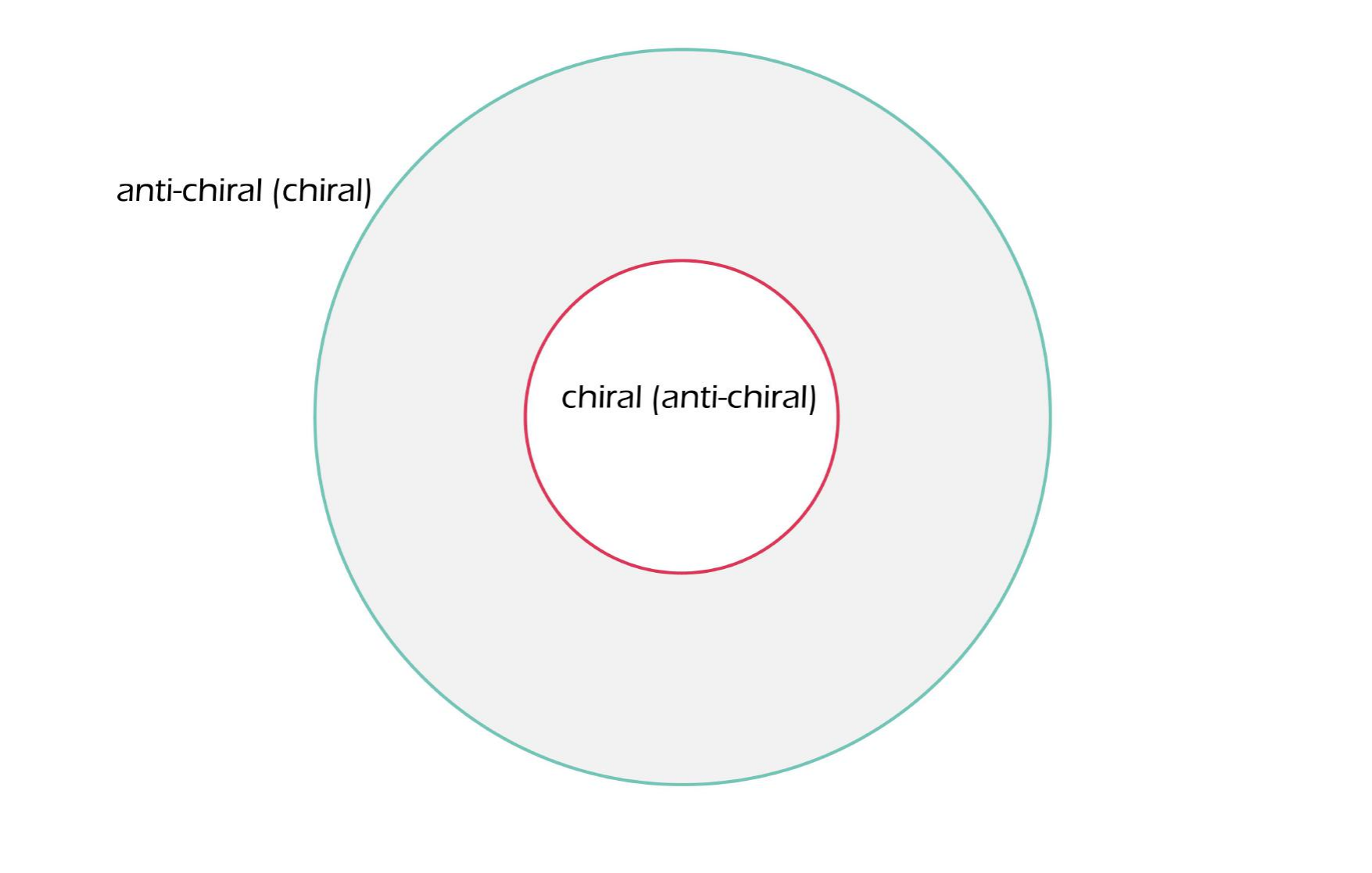}
	\caption{On the annulus, the gravity Hilbert space is equivalent to a superposition of coupled chiral(anti-chiral) and anti-chiral(chiral) Alekseev-Shatashvili theories that lives on the inner and outer circle respectively. } \label{fig:gravity annulus}
\end{figure}

The Einstein-Hilbert action can be expressed as a difference of Chern-Simons actions for $A$ and $\widetilde{A}$ \cite{Cotler:2020ugk,Achucarro:1986uwr, Witten:1988hc} 
\begin{equation} \label{eq:full_action}
S_0[A,\widetilde{A}]  = -S_{\text{cs}}[A] + S_{\text{cs}}[\widetilde{A}]~,~~S_{\text{cs}}[A]= \frac{1}{16 \pi G_N} \int_{\mathcal{M}} \text{Tr}\left(A \wedge dA + \frac{2}{3}A \wedge A \wedge A\right)~,
\end{equation}
up to boundary terms. More explicitly, we have 
\begin{equation}\label{eq:ScsHam}
	S_{\text{cs}}[A] = -\frac{1}{16\pi G_N} \int_{\mathcal{M}} dr dt  d\phi \, \text{Tr} \left( A_{\phi} \dot{A}_{r}-A_{r} \dot{A}_{\phi}  + 2 A_{t} F_{ r\phi } \right) ~,
\end{equation}
where the dot indicates a derivative with respect to $t$ and the field strength is given by
\begin{equation}
	\label{Fphir}
	F_{r\phi } = \partial_r A_\phi -\partial_{\phi} A_r + [A_r,A_\phi]\,.
\end{equation}
The variation of the above bulk term gives\footnote{The relative sign between the two boundary terms comes from having the normal vectors to the boundary to be outward pointing in global coordinates. }
\begin{equation}\label{eq:variation_bulk}
	 -\delta S_{\text{cs}}[A] + \delta S_{\text{cs}}[\widetilde{A}]=\int_{\mathcal{M}} (\text{EOM})+\frac{1}{8 \pi G_N}\int_{\partial \mathcal{M}^+} dt d\phi \, \text{Tr}(A_t \delta A_\phi - \widetilde{A}_t \delta \widetilde{A}_\phi)-\frac{1}{8 \pi G_N}\int_{\partial \mathcal{M}^-} dt d\phi \, \text{Tr}(A_t \delta A_\phi - \widetilde{A}_t \delta \widetilde{A}_\phi)~,
\end{equation}
where $\partial \mathcal{M}$ is composed from two disconnected boundaries, outer circle $\partial \mathcal{M}^+$ and inner circle $\partial \mathcal{M}^-$, and each boundary has spatial topology $S^1$. This is shown in Figure \ref{fig:gravity annulus}. To impose asymptotically AdS$_3$ boundary conditions at the inner and outer circle, we need to add the following boundary terms to the original action in \eqref{eq:full_action}
\begin{equation}
	S_{\text{bdy}}[A,\widetilde{A}] =  -\frac{1}{16 \pi G_N}\int_{\partial \mathcal{M}^+} dt d\phi \, \text{Tr}(A_\phi^2 + \widetilde{A}_\phi^2)-\frac{1}{16 \pi G_N}\int_{\partial \mathcal{M}^-} dt d\phi \, \text{Tr}(A_\phi^2 +\widetilde{A}_\phi^2)~,
\end{equation}
such that $A_{\phi} - A_t =\widetilde{A}_\phi +\widetilde{A}_{t} = 0$ at $\partial \mathcal{M}^+$ and $A_{\phi} + A_t =\widetilde{A}_\phi -\widetilde{A}_{t} = 0$ at $\partial \mathcal{M}^-$.  

In the following, we focus on the gauge field $A$ at $\partial \mathcal{M}^+$ for simplicity and the analysis for the remaining gauge field $\widetilde{A}$ at the outer boundary follows suit. The terms at the inner boundary also follow a similar story. The relevant terms are given by
\begin{equation}\label{eq:Chiral_AS}
\begin{split}
	S_-[A] =  \frac{1}{16\pi G_N} \left(\int_{\mathcal{M}} dr dt  d\phi \, \text{Tr} \left( A_{\phi} \dot{A}_{r}-A_{r} \dot{A}_{\phi}  + 2 A_{t} F_{r \phi } \right) - \int_{\partial \mathcal{M}^+}  dt d\phi \, \text{Tr}(A_\phi^2 )\right)~.
\end{split}
\end{equation}
Integrating out $A_t$, we get $F_{ r\phi }=0$ and the spatial gauge fields are pure gauge
\begin{equation}
\begin{split}
	&A_{r}=G^{-1} \partial_r G~,\\
	&A_{\phi}=G^{-1} (\partial_\phi+K(t)) G~,
\end{split}
\end{equation}
where $G(t,r,\phi+2\pi)=G(t,r,\phi)$ is an element of $SL(2,R)$ and $K(t)$ parametrizes the holonomy. 

With the above parametrization of gauge fields and $F_{r \phi} = 0$, the action \eqref{eq:Chiral_AS} becomes,
\begin{equation}\label{eq:Chiral_AS_2}
\begin{split}
	S_-[G,K(t)] &= \frac{1}{48 \pi G_N}\int_{\mathcal{M}}  \text{Tr}\, (G^{-1} dG)^3-\frac{1}{16 \pi G_N}\int_{\partial \mathcal{M}^+} dt d\phi \text{Tr}\, (G^{-1} \partial_\phi G G^{-1} \partial_- G +2 G^{-1} K(t) \partial_- G + K^2(t))~,
\end{split}
\end{equation}
where $\partial_- = \partial_\phi - \partial_t  $.  The bulk term is the Wess-Zumino term and with it being a total derivative, it depends only on the boundary group element. We pick the ``canonical'' gauge for the hyperbolic holonomy $K(t)$ and the path integral for its conjugate momentum forces it to be independent of time\cite{Henneaux:2019sjx ,Cotler:2020ugk}
\begin{equation}\label{eq:Kt}
	K(t) = \gamma L_0 = \frac{\gamma}{2}\begin{pmatrix}
	    1 & 0 \\
     0 & -1
	\end{pmatrix}~,
\end{equation}
with $\gamma$ being a constant. Next, we use Gauss decomposition to decompose the group element on the outer boundary, $h(t,\phi)=G(t, r=r^{\text{outer}}, \phi)$ \cite{Alekseev:1988ce, Cotler:2018zff},
\begin{equation} \label{eq:Gauss}
h(t,\phi) = e^{Y L_-} e^{ \Psi L_0} e^{X L_+} = \begin{pmatrix}
	1 & 0 \\
	Y & 1
\end{pmatrix}\begin{pmatrix}
\exp(\Psi/2) & 0 \\
0 & \exp(-\Psi/2) 
\end{pmatrix}\begin{pmatrix}
1 & X \\
0 & 1
\end{pmatrix}~,
\end{equation}
where $Y,\Psi$ and $X$ are functions of $(t,\phi)$. 
\eqref{eq:Chiral_AS_2} becomes
\begin{equation}\label{eq:Chiral_AS_3}
	\begin{split}
		S_-[h,K(t)] =-\frac{1}{16 \pi G_N}\int_{\partial \mathcal{M}^+} dt d\phi \left(\frac{1}{2} \partial_- \Psi (\partial_\phi\Psi+2  \gamma) + \frac{1}{2}\gamma^2 +2 e^\Psi \partial_- X(\partial_\phi Y - \gamma Y)\right)~.
	\end{split}
\end{equation}

To make connection with the geometric Alekseev-Shatashvili theories that is related to the quantization of coadjoint orbits for the Virasoro group \cite{Alekseev:1988ce, Alekseev:1990mp, Witten:1987ty,Cotler:2018zff, Cotler:2020ugk}, we introduce the following parametrization for $Y(t,\phi)$
\begin{equation}\label{eq:Ytphi}
	Y(t,\phi) = \exp\left(- \gamma(f_L(t,\phi) - \phi) \right) ~,
\end{equation}
where $f_L(t,\phi)$ is an element of Diff$(S^1)$ with periodicity $f_L(t,\phi+2\pi) = f_L(t,\phi) + 2\pi $. In addition, asymptotically AdS$_3$ boundary conditions gives us the following constraints \cite{Coussaert:1995zp}
\begin{equation}
A_r = 0, \qquad A_\phi= L_- + {\mathcal{L}} (t,\phi) L_+~,
\end{equation}
leading us to
\begin{equation}
	e^\Psi(\partial_\phi Y - \gamma Y) = 1~, \quad		\partial_\phi \Psi+ \gamma = 2 X~, \quad \mathcal{L}(t,\phi)  = \partial_\phi X+X^2~.
\end{equation}
With $Y(t,\phi)$ in \eqref{eq:Ytphi}, we can solve for $X$ and $e^\Psi$
\begin{equation}\label{eq:PsiX}
		e^{-\Psi} = - \gamma f'_L(t,\phi)e^{- \gamma(f_L(t,\phi) - \phi)}~, \quad		X(t,\phi) = \frac{1}{2}  \gamma f'_L - \frac{f''_L}{2 f'_L}~,
\end{equation}
where $'$ indicates the derivative with respect to the angular coordinate $\phi$. With \eqref{eq:Ytphi} and \eqref{eq:PsiX}, we obtain the chiral Alekseev-Shatashvili action\cite{Alekseev:1988ce, Alekseev:1990mp} 
\begin{equation}\label{Sgeom}
	S_-[f_L] =  -\frac{1}{32\pi G_N} \int_{\partial \mathcal{M}^+} dt d\phi \; \left[ \frac{\partial_- f'_L f''_L}{f'_L{}^2} +  \gamma^2 \partial_- f_L f'_L \right]\,.
\end{equation}
The Hilbert space for this theory corresponds to a Virasoro highest weight representation with central charge and conformal weight \cite{Cotler:2018zff}
\be
c=\frac{3}{2 G_N}, \qquad h=\frac{c-1}{24}(1+\gamma^2)~.
\ee
Performing a similar analysis for the antichiral gauge field $\widetilde{A}$, which is parametrized by a Diff($S^1$) element $\widetilde{f}_R$ on the outer boundary gives,
\begin{equation}\label{Sgeom2}
	S_+[\widetilde{f}_R] =  -\frac{1}{32\pi G_N} \int_{\partial \mathcal{M}^+} dt d\phi \; \left[ \frac{\partial_+ \widetilde{f}'_R \widetilde{f}''_R}{\widetilde{f}'{}^2_R} + \widetilde{ \gamma}^2 \partial_+ \widetilde{f}_R \widetilde{f}'_R \right]\,.
\end{equation}
where $\partial_+ =  \partial_\phi +\partial_t $.

The full action \eqref{eq:full_action} is given by a sum of four coupled Alekseev-Shatashvili actions on the boundaries,
\begin{equation}\label{eq:4AS_actions} 
    \begin{split}
        S_0[f_L,f_R;\widetilde{f}_L,\widetilde{f}_R] &= S_-[f_L]+ S_+[\widetilde{f}_R] +S_-[\widetilde{f}_L]+S_+[f_R]~, \\
        &=-\frac{1}{32\pi G_N} \int_{\partial \mathcal{M}^+} dt d\phi \; \left[ \frac{\partial_- f'_L f''_L}{f_L'{}^2} +  \gamma^2 \partial_- f_L f'_L \right]- \frac{1}{32\pi G_N} \int_{\partial \mathcal{M}^+} dt d\phi \; \left[ \frac{\partial_+ \widetilde{f}'_R \widetilde{f}''_R}{\widetilde{f}_R'{}^2} + \widetilde{ \gamma}^2 \partial_+ \widetilde{f}_R \widetilde{f}'_R \right]~, \\
        &- \frac{1}{32\pi G_N} \int_{\partial \mathcal{M}^-} dt d\phi \; \left[ \frac{\partial_- \widetilde{f}'_L \widetilde{f}''_L}{\widetilde{f}_L'{}^2} + \widetilde{ \gamma}^2 \partial_- \widetilde{f}_L \widetilde{f}'_L \right]-\frac{1}{32\pi G_N} \int_{\partial \mathcal{M}^-} dt d\phi \; \left[ \frac{\partial_+ f'_R f''_R}{f_R'{}^2} +  \gamma^2 \partial_+ f_R f'_R \right]~,
    \end{split}
\end{equation}
where $f_L(\widetilde{f}_R)$ and $f_R(\widetilde{f}_L)$ are functions related to the group element $G(\widetilde{G})$ at the outer and inner circle respectively. 

We pay attention to how both the zero modes $\gamma$ and $\widetilde{\gamma}$ couple the outer and inner boundaries. As shown in Figure \ref{fig:gravity annulus}, the pairing of the left and right-moving sectors between two copies of Liouville theory at the two asymptotic boundaries follows a similar permutation.  
\begin{figure}
\begin{center}
\begin{overpic}[scale=0.5]{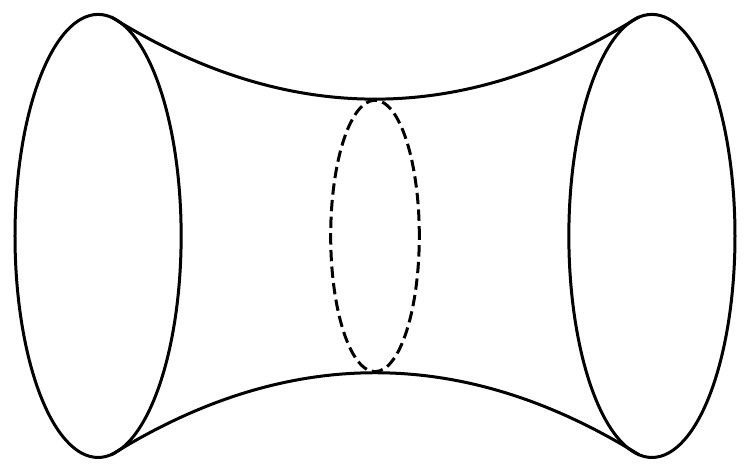}
\put(30,5){\parbox{0.2\linewidth}{
		\begin{equation*}
			\pi(\gamma+\Bar{\gamma})
\end{equation*}}}
\end{overpic}  
\end{center}
\caption{The spatial geometry that corresponds to fixed holonomies $K=\gamma L_0, \tilde{K}=\tilde{\gamma} \widetilde{L_0}$. The length of the waist is $\pi(\gamma+\bar{\gamma})$.}\label{fig:bottleneck}
\end{figure}
We can write down the metric on the spatial slice that corresponds to fixed holonomy wormhole geometries \cite{Cotler:2020ugk}, as shown in Figure \ref{fig:bottleneck},
\be \label{bottleneck}
ds^2=\left(\gamma \tilde{\gamma} \sinh^2(\rho)+\frac{(\gamma+\tilde{\gamma})^2}{4}\right)d\phi^2+d\rho^2~,
\ee
where $\rho \in (-\infty,\infty)$ and $\phi \sim \phi+2\pi$.
The whole theory consists of quantum states that comes from the quantization of these wormhole backgrounds and we will show that the BTZ black hole state can be viewed as a specific superposition of these wormhole states.

The path integral for the quantization of the whole theory is given by \cite{Mertens:2018fds, Blommaert:2018iqz,Cotler:2020ugk}\footnote{We have modded out the $U(1)$ redundancies in the field redefinition for integration space of \eqref{eq:Z_Annulus}, see (6.21) and Appendix D of \cite{Blommaert:2018iqz}.}
\be\label{eq:Z_Annulus}
Z_{\mathcal{A} \times \mathbb{R}} =\int_0^\infty  d\gamma  d\widetilde{\gamma} \int \mathcal{D} f_L \mathcal{D}f_R \mathcal{D} \widetilde{f}_L \mathcal{D}\widetilde{f}_R \frac{\text{Pf}(\omega_{f_L})  \text{Pf}(\omega_{f_R})\text{Pf}(\omega_{\widetilde{f}_L})  \text{Pf}(\omega_{\widetilde{f}_R})}{U(1) \times U(1)}e^{i S_0[f_L,f_R;\widetilde{f}_L,\widetilde{f}_R]}~,
\ee
where $\omega_{f}$ is the symplectic two form for the geometric action of $f$. Subsequently, the Hilbert space of four coupled Alekseev-Shatashvili theories is composed from a sum over two copies of Virasoro modules\footnote{The signs of $\gamma$ and $\tilde{\gamma}$ are chosen to correspond to smooth and nonsingular bottleneck geometries\cite{Cotler:2020ugk}. In Liouville theory, this also eliminates the double counting on degenerate $\gamma$ and -$\gamma$ states.  }
\be
\mathcal{H}=\int_0^\infty d\gamma \int_0^\infty d\widetilde{\gamma} (\mathcal{H}^{\text{Vir}}_{\gamma} \otimes \mathcal{\widetilde{H}}^{\text{Vir}}_{\widetilde{\gamma}}) \otimes (\mathcal{H}^{\text{Vir}}_{\widetilde{\gamma}} \otimes \mathcal{\widetilde{H}}^{\text{Vir}}_{\gamma} )~.
\ee

\subsection{Coupled Alekseev-Shatashvili theories and Liouville theory}

In this section, we show that under a field redefinition, the action and measure can be described by two copies of Liouville theory\cite{Henneaux:1999ib, Henneaux:2019sjx, Cotler:2018zff, Cotler:2020hgz, Cotler:2020ugk, Mertens:2018fds, Blommaert:2018iqz}.

The non-trivial coupling by the zero modes $\gamma$ and $\widetilde{\gamma}$ respectively sheds light on how the four fields $(f_L,f_R,\widetilde{f}_L,\widetilde{f}_R)$ should be paired to give two copies of Liouville theory. Focusing on the pairing of $(f_L,f_R)$ through the superselection sector $\gamma$, we introduce the following field redefinition \cite{Gervais:1981gs, Gervais:1983am,Mertens:2018fds}
\begin{equation}
	\begin{split}
\Phi(t,\phi) &=\ln \left(-2 \gamma^2 \frac{f'_L f'_R}{\sinh^2\left(\frac{\gamma}{2}(f_L-f_R)\right)} \right)~,\\
\Pi_{\Phi}(t,\phi) &=\frac{1}{b^2}\left( \frac{f''_{L}}{f'_{L}}-\frac{f''_{R}}{f'_{R}}-\gamma (f'_{L}+f'_{R}) \coth\left(\frac{\gamma}{2}(f_L-f_R)\right) \right)~,
\end{split}
\end{equation}
Under this field redefinition, the symplectic two form becomes
\begin{equation}
\omega_\Phi =\int d\phi \, \delta \Pi_{\Phi} \wedge \delta \Phi ~.
\end{equation}
which is the correct flat measure for Liouville theory. The second pair labelled by tildes follows a similar non-canonical field redefinition, i.e. $(\widetilde{f}_L,\widetilde{f}_R) \rightarrow(\widetilde{\Phi},\Pi_{\widetilde{\Phi}})$. The action in \eqref{eq:4AS_actions} then becomes
\be
\begin{split}
S_0&=\frac{1}{8 \pi}\int dt d\phi \left(\Pi_{\Phi} \dot{\Phi}-H_\Phi  + \Pi_{\widetilde{\Phi}} \dot{\widetilde{\Phi}}-H_{\widetilde{\Phi}}  \right) \\
&=\frac{1}{8 \pi}\int dt d\phi \left(\Pi_{\Phi} \dot{\Phi}-\left(\frac{1}{2}b^2 \Pi_{\Phi}^2+\frac{1}{2} \frac{\Phi^{\prime 2}}{b^2}+\frac{e^{\Phi}}{b^2}-2\frac{\Phi''}{b^2} \right)  + \Pi_{\widetilde{\Phi}} \dot{\widetilde{\Phi}}- \left(\frac{1}{2}b^2 \Pi_{\widetilde{\Phi}}^2+\frac{1}{2} \frac{\widetilde{\Phi}^{\prime 2}}{b^2}+\frac{e^{\widetilde{\Phi}}}{b^2}-2\frac{\widetilde{\Phi}''}{b^2} \right)\right)~,
\end{split}
\ee
where the path integral of the four coupled Alekseev-Shatashvili theories becomes two copies of the phase space path integral form of Liouville theory, which is equivalently
\begin{equation}
S_0=\frac{1}{8 \pi b^2} \int dt d\phi \left( \frac{1}{2}(\dot{\Phi}^2-\Phi'^2)-e^{\Phi}+2\Phi''+\frac{1}{2}(\dot{\widetilde{\Phi}}^2-\widetilde{\Phi}'^2)-e^{\widetilde{\Phi}}+2\widetilde{\Phi}'' \right)~.
\end{equation}

In conclusion, we have shown that 3d gravity with two asymptotic boundaries is equivalent to two copies of Liouville theory, where the left and right-moving sectors of each Liouville copy are paired across the boundaries as shown in Figure \ref{fig:gravity annulus}. This is similar to the correspondence between Chern-Simons theories and WZW models for compact gauge groups where we need to further impose gauge constraints to get gauge invariant wavefunctions in the bulk theory, which further translates to imposing Ward identity on the boundary \cite{Witten:1988hf, Elitzur:1989nr}. This provides the famous identification of the bulk Hilbert space with boundary conformal blocks, where in particular for gravity, the bulk Hilbert space is given by Liouville conformal blocks \cite{Verlinde:1989ua}. In later parts of the paper, we show that the state in gravity is given by a  superposition of Ishibashi states, which has the same energy for the left and right-movers \cite{Ishibashi:1988kg}.

Before we move on to the study of wavefunctions, we want to point out that the identification of the total bulk Hilbert space as two copies of Liouville theory is useful in computing observables. For example, the exact two-boundary wormhole correlation functions can be obtained by taking trace in the presence of operator insertions in 
Liouville theory \cite{Collier:2022bqq, Collier:2023fwi}. In Appendix \ref{app:Stanford_WH}, we show an example of a two-boundary torus wormhole that is obtained from the product of torus two-point functions, and in the Schwarzian limit $b\rightarrow 0$, we reproduce wormholes in JT gravity that have two insertions on each of its disk boundary. It is mentioned in \cite{Stanford:2020wkf} that wormholes are the gravity dual description of quantum noise.

\section{Hartle-Hawking state in 3d gravity}\label{sec:HH-state}
\subsection{BTZ partition function and the Liouville ZZ boundary states}\label{subsec:Liouville_background}

 Euclidean non-rotating BTZ black hole is topologically a solid torus with a contractible  Euclidean time $\tau_E$ circle, whose periodicity is $\beta$. As shown in previous works \cite{Ghosh:2019rcj,Cotler:2018zff,Maloney:2007ud,Cotler:2020ugk,Giombi:2008vd}, the one-loop exact partition function for non-rotating BTZ can be expressed as a product of vacuum characters
\begin{equation}\label{eq:BTZ_partition}
\begin{split}
Z_{\text{BTZ}} = \chi_{\mathds{1}}\left(-\frac{1}{\tau} \right)\chi_{\mathds{1}}\left(\frac{1}{\bar{\tau}} \right) ~,
\end{split}
\end{equation}
with modular parameter $\tau=\frac{i\beta}{2\pi}$. The Virasoro vacuum character $\chi_{\mathds{1}}$ is the trace over the identity module of the Virasoro algebra:
\begin{equation}
\begin{split}
\chi_{\mathds{1}}(\tau) \equiv \text{Tr}_{\mathcal{H}_{\mathds{1}}}\left(q^{L_0 - \frac{c}{24}}\right)&= \frac{(1-q)q^{- \frac{c-1}{24}}}{\eta(\tau)}~,~~q = e^{2 \pi i \tau}~,
\end{split}
\end{equation} 
and 
\begin{equation}
\eta(\tau) \equiv q^{1/24}\prod_{n=1}^{\infty}(1-q^n)~,
\end{equation}
is the Dedekind eta function. 

We can define the Hartle-Hawking state for non-rotating BTZ by slicing open the path integral for the partition function at the time reflection symmetric position. The spatial surface where the slicing  occurs  has the topology of an annulus. In Section \ref{sec:3d_Liouville_AS}, we reviewed the relation between the bulk Hilbert space of gravity on an annulus to the Hilbert space of two copies of Liouville theory, where a non-trivial matching of chiralities across the boundaries is involved.

The product of vacuum characters takes the following expression in terms of Liouville theory states
\begin{equation}\label{eq:Z_BTZ_ampltiude}
\chi_{\mathds{1}}\left(-\frac{1}{\tau} \right)\chi_{\mathds{1}}\left(\frac{1}{\bar{\tau}} \right) = \langle ZZ | q^{L_0 - \frac{c}{24}} |ZZ \rangle \langle \widetilde{ZZ} | \bar{q}^{{\bar{L}}_0 - \frac{c}{24}} | \widetilde{ZZ} \rangle=\langle ZZ |e^{-\beta H/2}|ZZ \rangle \langle \widetilde{ZZ} | e^{-\beta H/2} | \widetilde{ZZ} \rangle ~,
\end{equation}
where $|ZZ \rangle$ and $| \widetilde{ZZ} \rangle$ are the two copies of ZZ boundary states in Liouville theory \cite{Zamolodchikov:2001ah}, and $H=H_L+H_R=L_0 +{\bar{L}}_0- \frac{c}{12}$ is the total Hamiltonian for Liouville CFT. For non-rotating BTZ, $q=\bar{q}$ is real. We will comment on complex $q$, i.e. the case for rotating BTZ black holes, in the end.

\begin{figure}[h]
	\centering
	\includegraphics[scale=0.45]{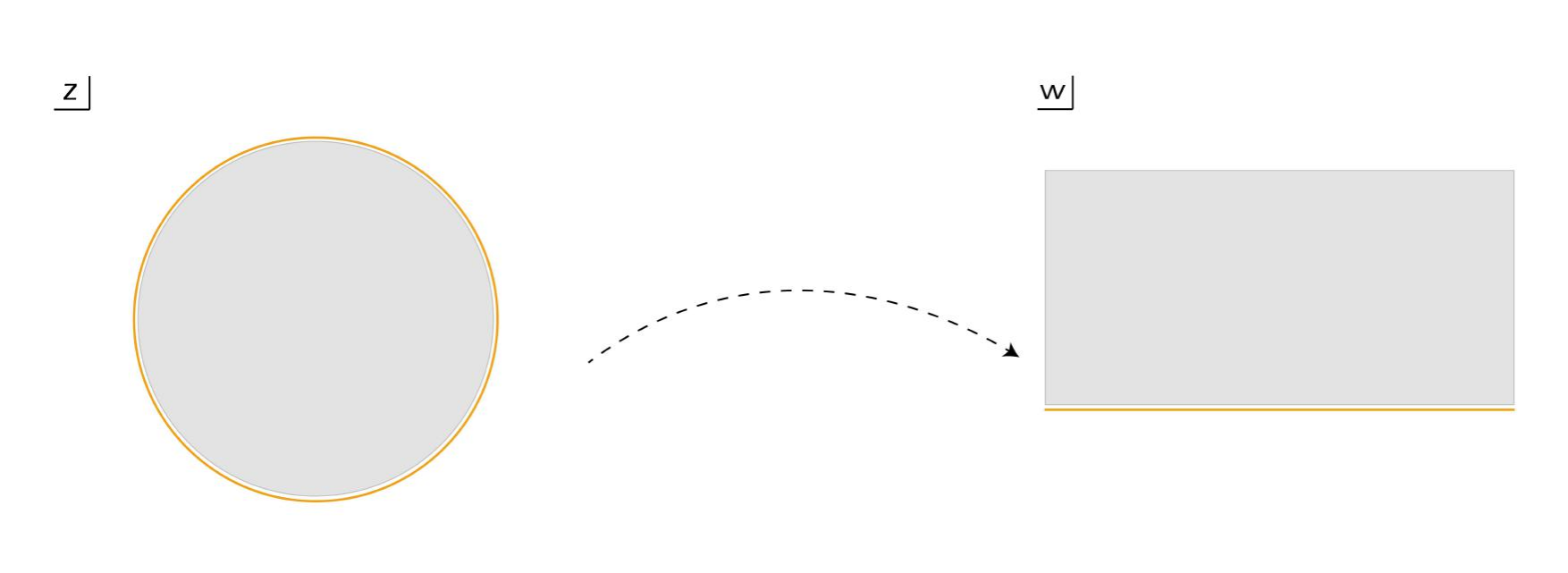}
	\caption{(Left): On the Poincare disk, the ZZ boundary condition (indicated by orange line) is imposed at $|z| \rightarrow 1$. (Right): Alternatively, we can conformally map the disk to the upper half plane and impose the ZZ boundary condition on the real line $w=\bar{w}$.} \label{fig:zz conformal mapping}
\end{figure}

We first briefly review the ZZ boundary states in Liouville theory. Zamolodchikov and Zamolodchikov (ZZ) quantized Liouville theory on a Poincare disk and found a solution for the quantized geometry \cite{Zamolodchikov:2001ah}. The ZZ boundary condition for the classical solution is defined at the location at which the Liouville field blows up. To be more specific, with $|z| = 1$ being the radius of the Poincare disk, the ZZ boundary condition is
\begin{equation}\label{eq:classical_liouville}
e^{\Phi(z,\bar{z})} \rightarrow \frac{4}{(1-z \bar{z})^2}~,
\end{equation}
and the boundary is at $|z| \rightarrow 1$. With the holomorphic coordinate transformation $w = f(z) = -i\frac{z-1}{z+1}$, we map the Poincare disk to the upper half plane, and the ZZ boundary condition can be stated as
\begin{equation}
e^{\Phi(w,\bar{w})} \rightarrow \frac{1}{(\text{Im} w)^2}~,
\end{equation}
which is defined on the real line $w=\bar{w}$, as shown in Figure \ref{fig:zz conformal mapping}. 
\begin{figure}
	\begin{center}
		\begin{overpic}[scale=0.6]{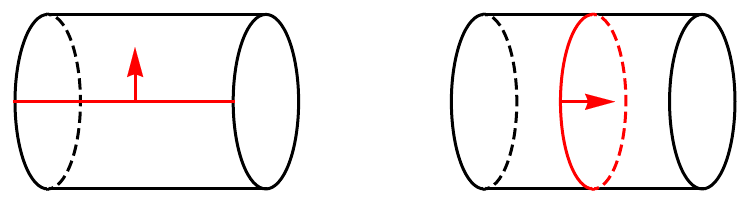}
					\put(30,10){\parbox{0.2\linewidth}{
		\begin{equation*}
\text{\Huge  $=$}
\end{equation*}}}
		\end{overpic}  
	\end{center}
	\caption{From the open-closed duality, we interpret the boundary condition in the open channel as a boundary state in the closed channel\cite{Cardy:2004hm}.} \label{fig:openclose}
\end{figure}  
In quantizing the Liouville field in \eqref{eq:classical_liouville} and as shown in Figure \ref{fig:openclose}, we can change to the closed string channel, and view the ZZ boundary condition as giving us the ZZ conformal boundary state that satisfies\cite{Ishibashi:1988kg, Cardy:2004hm}
\begin{equation}\label{boundary condition}
\begin{split}
(L_n - \Tilde{L}_{-n}) | ZZ \rangle = 0~.
\end{split}
\end{equation} 
A complete set of solutions to \eqref{boundary condition} is given by Ishibashi states $| P \rangle \rangle$,  labeled by the normalizable primary operators in Liouville theory, 
\be
V_{\alpha}=e^{2 \alpha \Phi}=e^{2 \left(\frac{Q}{2}+i P\right)\Phi}~,
\ee
where $P \geq 0$, and the Ishibashi states are normalized as
\be
\langle \langle P | q^{L_0 - \frac{c}{24}} | Q \rangle \rangle = \chi_P(\tau) \delta(P-Q)=\frac{q^{P^2}}{\eta(\tau)}\delta(P-Q)~.
\ee
Using the expression of the bulk one point function $\langle V_\alpha(w) \rangle$ in the presence of the ZZ boundary condition, the ZZ boundary state can be written as a superposition of Ishibashi states \cite{Zamolodchikov:2001ah}
\begin{equation}
    | ZZ \rangle = \int_{0}^{\infty} dP \Psi^{*}_{ZZ}(P) | P \rangle \rangle~,
\end{equation}
where the ZZ wavefunction is given by 
\begin{equation}
\Psi_{ZZ}(P)=2^{9/4}  i \pi P \left(\pi \mu \gamma(b^2) \right)^{-iP/b} \frac{1}{\Gamma(1-2i P b) \Gamma(1-2 iP/b)}~,
\end{equation}
and $\gamma(x) = \Gamma(x)/\Gamma(1-x)$. $\mu$ is the cosmological constant and it scales as $\mu = (4 \pi b^2)^{-1}$ in the semiclassical limit $b \rightarrow 0$. We can express the ZZ wavefunction in a compact form
\begin{equation} \label{ZZ wavefunction}
\begin{split}
\Psi_{ZZ}(P) = \sqrt{S_{\mathds{1}P} S_L(P)}~,
\end{split}
\end{equation} 
where $S_{\mathds{1}P}$ is the modular S-matrix
\begin{equation}
S_{\mathds{1}P} = 4 \sqrt{2}\sinh(2 \pi P b)\sinh \left(\frac{2 \pi P}{b}\right)~.
\end{equation}
Besides encoding the Cardy density of states in holographic 2d CFTs, $S_{\mathds{1}P}$ is also the Plancherel measure on the quantum semi-group $SL^+_{q}(2,R)$ \cite{McGough:2013gka, Mertens:2022ujr}. $S_L(P)$ is a pure phase known as  the Liouville reflection amplitude
\begin{equation}
S_L(P) = -\left(\pi \mu \gamma(b^2)\right)^{-2 i P/b} \frac{\Gamma\left(1+\frac{2 i P}{b}\right)\Gamma(1+2 i P b)}{\Gamma\left(1-\frac{2 i P}{b}\right)\Gamma(1-2 i P b)}~.
\end{equation}

In this paper, we consider another convention in normalizing the Liouville primary operators \cite{Collier:2022bqq}
\be\label{eq:Pp_Liouville}
\hat{V}_\alpha=\frac{e^{2 \alpha \Phi}}{\sqrt{S_L(P')}} =\frac{e^{2 \left(\frac{Q}{2}+i P'\right)\Phi}}{\sqrt{S_L(P')}}~.
\ee
In addition to putting the formulas in a much compact form, $\hat{V}_\alpha$ is Hermitian due to the reflection property of Liouville primaries $V_\alpha$ under $P \rightarrow -P$\cite{Zamolodchikov:2001ah}.
Hence, they are more natural to be considered in the context of AdS/CFT. In terms of the normalized operators $\hat{V}_\alpha$, the ZZ boundary state is given by\footnote{We will use $\ket{P'}$ to denote the states related to the normalized primary operators $\hat{V}_{\alpha}$.}
\begin{equation}
    | ZZ \rangle = \int_{0}^{\infty} dP' \Psi^{*\prime}_{ZZ}(P') | P' \rangle \rangle=\int_{0}^{\infty} dP' \sqrt{S_{\mathds{1}P'}} | P' \rangle \rangle~,
\end{equation}
which takes the same form of the Cardy state for the identity operator in rational CFTs\cite{Cardy:1989ir}.

\begin{figure}[h]
	\centering
	\includegraphics[scale=0.4]{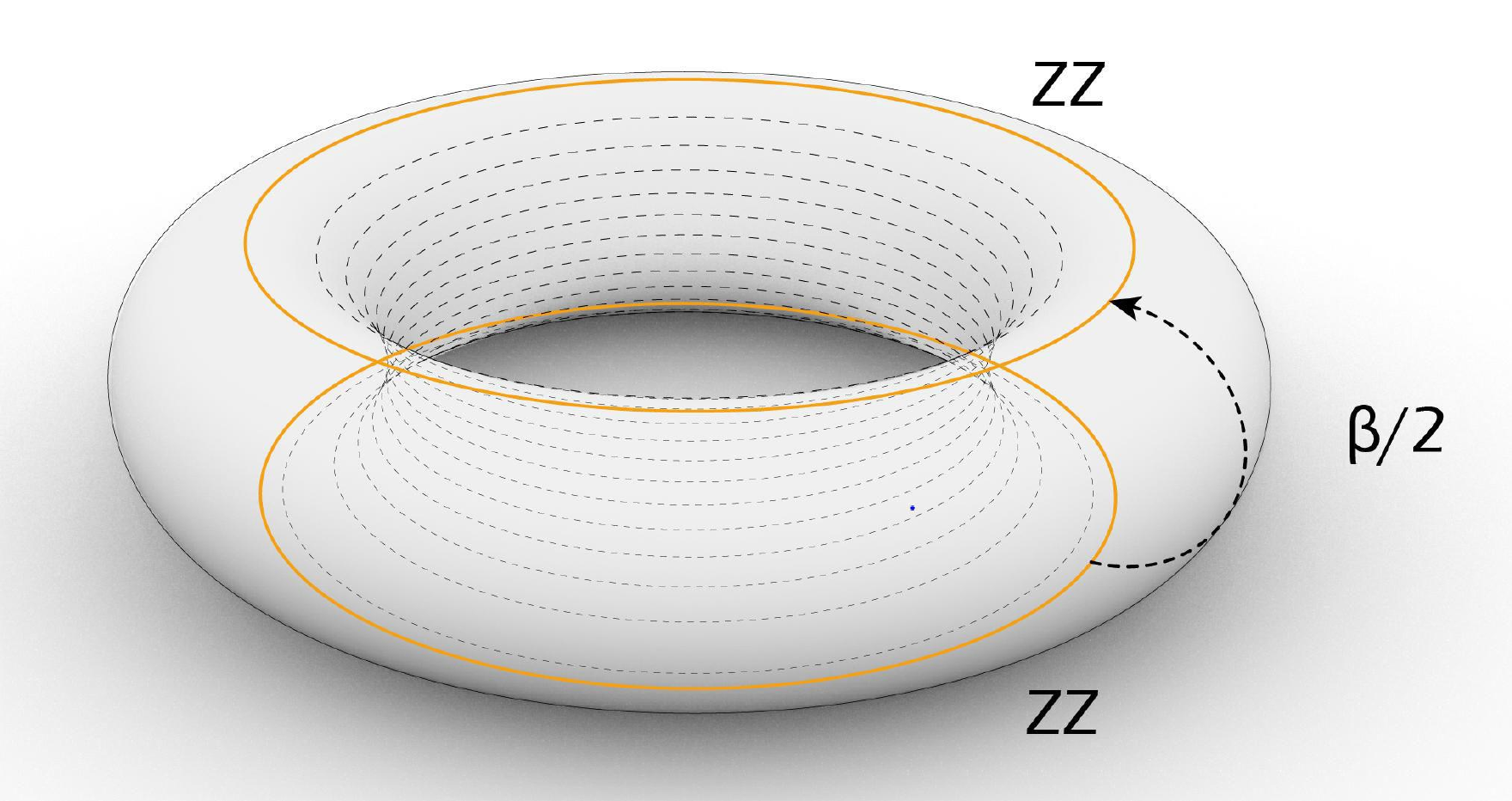}
	\caption{The transition amplitude of the ZZ boundary states, as shown in \eqref{eq:Z_BTZ_ampltiude},  has the topology of a cylinder and is equal to the identity character defined on the torus via the doubling trick.}\label{fig:doubling partition function}
\end{figure}

We can easily check that
\begin{equation}
\langle ZZ | q^{L_0 - \frac{c}{24}} |ZZ \rangle = \int_0^\infty dP dQ \Psi_{ZZ}(P) \Psi_{ZZ}^*(Q) \langle \langle P | q^{L_0 - \frac{c}{24}} | Q \rangle \rangle = \int_0^\infty dP S_{\mathds{1}P} \chi_P(\tau) = \chi_{\mathds{1}}\left(-\frac{1}{\tau} \right)~,
\end{equation}  
where the above result can also be anticipated from the doubling trick, as shown in Figure \ref{fig:doubling partition function} \cite{Cardy:1984bb, Cardy:2004hm, Mertens:2018fds}. 
In Section \ref{sec:correlation_function}, we demonstrate the doubling trick works similarly with operator insertions, thus relating operators between the ZZ boundary states to Liouville conformal blocks. 

\subsection{Hartle-Hawking state from the Liouville  ZZ boundary states}

\begin{figure}[h]
	\centering
	\includegraphics[scale=0.9]{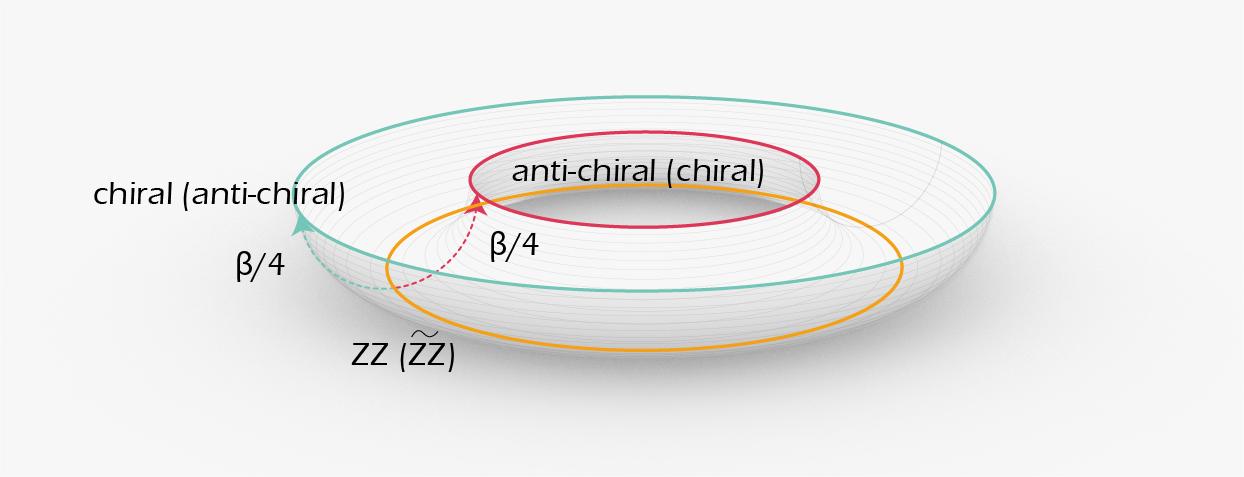}
	\caption{As shown, the chiral(anti-chiral) and anti-chiral(chiral) sectors of $e^{-\beta H/4}|ZZ \rangle(| \widetilde{ZZ} \rangle)$ live on the outer and inner circle of the Hartle-Hawking state respectively. The ZZ boundaries are indicated by the orange $S^1$ circle in the angular direction. The two chiralities move in opposite thermal directions, hence, we see a mixing of chiralities between the two boundaries on a constant time slice. }\label{fig:ZZ}
\end{figure}

With the partition function of non-rotating BTZ written as a transition amplitude in \eqref{eq:Z_BTZ_ampltiude}, it is natural to slice open the path integral and expect an identification of the Hartle-Hawking state with the evolved Liouville ZZ boundary states
\be \label{HH ZZ state}
\begin{aligned}
 | \Psi^{\text{HH}}_{\beta/2} \rangle &\cong q^{\frac{L_0}{2} - \frac{c}{48}} |ZZ \rangle \bar{q}^{\frac{\bar{L}_0}{2} - \frac{c}{48}} |\widetilde{ZZ} \rangle \\
 &=e^{-\beta H/4}|ZZ \rangle e^{-\beta H/4}|\widetilde{ZZ} \rangle~,
 \end{aligned}
\ee
where $\cong$ is used as the two states live in different Hilbert spaces, i.e. one being a gravitational bulk state and the other being a boundary Liouville state. With $|ZZ \rangle$ as one of the copies of the boundary states that lives on the spatial circle $S^1$, $e^{-\beta H/4}$ evolves the ZZ boundary state such that the chiral and anti-chiral sectors of $|ZZ \rangle$ live on the outer and inner circle respectively, as shown in Figure \ref{fig:ZZ}. This also applies to $|\widetilde{ZZ} \rangle$, but $e^{-\beta H/4}$ evolves the moving sectors with opposite chirality. Hence, there is a mixing of left and right-moving sectors between the two copies of Liouville theories on the two asymptotic boundaries of the torus. 

The relation between gravity and Liouville theory has been anticipated in \cite{Mertens:2017mtv, Callebaut:2018nlq}, where Liouville theory was argued to be living on the kinematic space of Schwarzian theory, and observables in JT gravity take the form of certain limits of Liouville quantities. In fact, the Hartle-Hawking state in JT gravity follows a similar logic, and can be derived from considering the near extremal limit of our result in 3d gravity\cite{Lin:2022rzw, Lin:2022zxd}.\footnote{We thank Henry Lin for mentioning this to us.} In this paper, we show why this duality is naturally manifested from a geometrical point of view in 3d gravity,  provided we view Euclidean BTZ as a ``two-boundary wormhole'' and use techniques developed in \cite{Collier:2022bqq} to match the large c behavior of Liouville quantities to results of semiclassical 3d gravity. The further exact identification between 3d gravity and Liouville at finite c will be a consequence of formulating 3d quantum gravity as quantization of the cotangent bundle of Teichm$\ddot{\text{u}}$ller space, i.e.  ``Virasoro TQFT'' \cite{Collier:2023fwi}.

Before we move on, we want to make an important comment. The Hartle-Hawking state was famously shown to be holographically dual to the thermofield double state of the holographic boundary CFT \cite{Maldacena:2001kr}:
\begin{equation}\label{eq:CFT_tfd}
    | \Psi^{\text{HH}}_{\beta/2} \rangle = \sum_n e^{-\beta E_n/2}|E_n\rangle_1 |E_n \rangle_2~.
\end{equation} 
We clearly see a similarity in expression between \eqref{HH ZZ state} and \eqref{eq:CFT_tfd}, as $|ZZ \rangle$, being constructed from a superposition of Ishibashi states, have the same left and right-moving energies, both for the primaries and all of the descendents. However, Liouville theory is not the holographic CFT dual of 3d gravity, and \eqref{HH ZZ state} is not the thermofield double state for 3d gravity. For one thing, Liouville theory has a flat spectrum and gravity exhibits a Cardy spectrum\cite{Cardy:1986ie}. In addition, the identity operator is absent in the spectrum of normalizable operators in Liouville theory \cite{Seiberg:1990eb}. However, we later will see that the ZZ boundary state in Liouville encodes the Cardy behavior in addition to projecting onto the identity module in the dual channel.

\subsection{BTZ as a Maldacena-Maoz wormhole }\label{subsec:ZZ-cylinder-WH}

\begin{figure}[!tbp]
  \centering
{\includegraphics[width=0.6\textwidth]{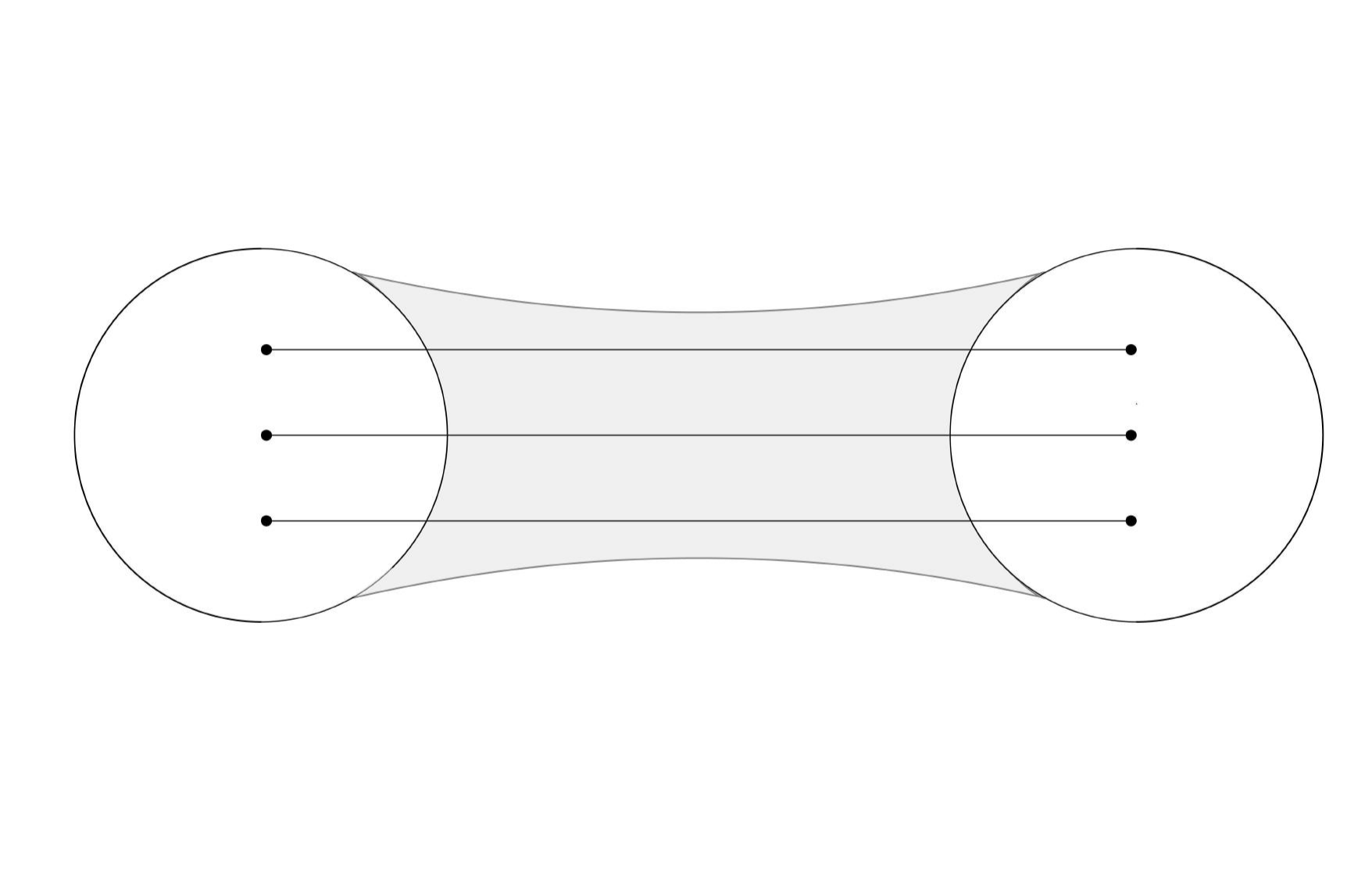}\label{3pointworm}}
\hfill
{\includegraphics[width=0.3\textwidth]{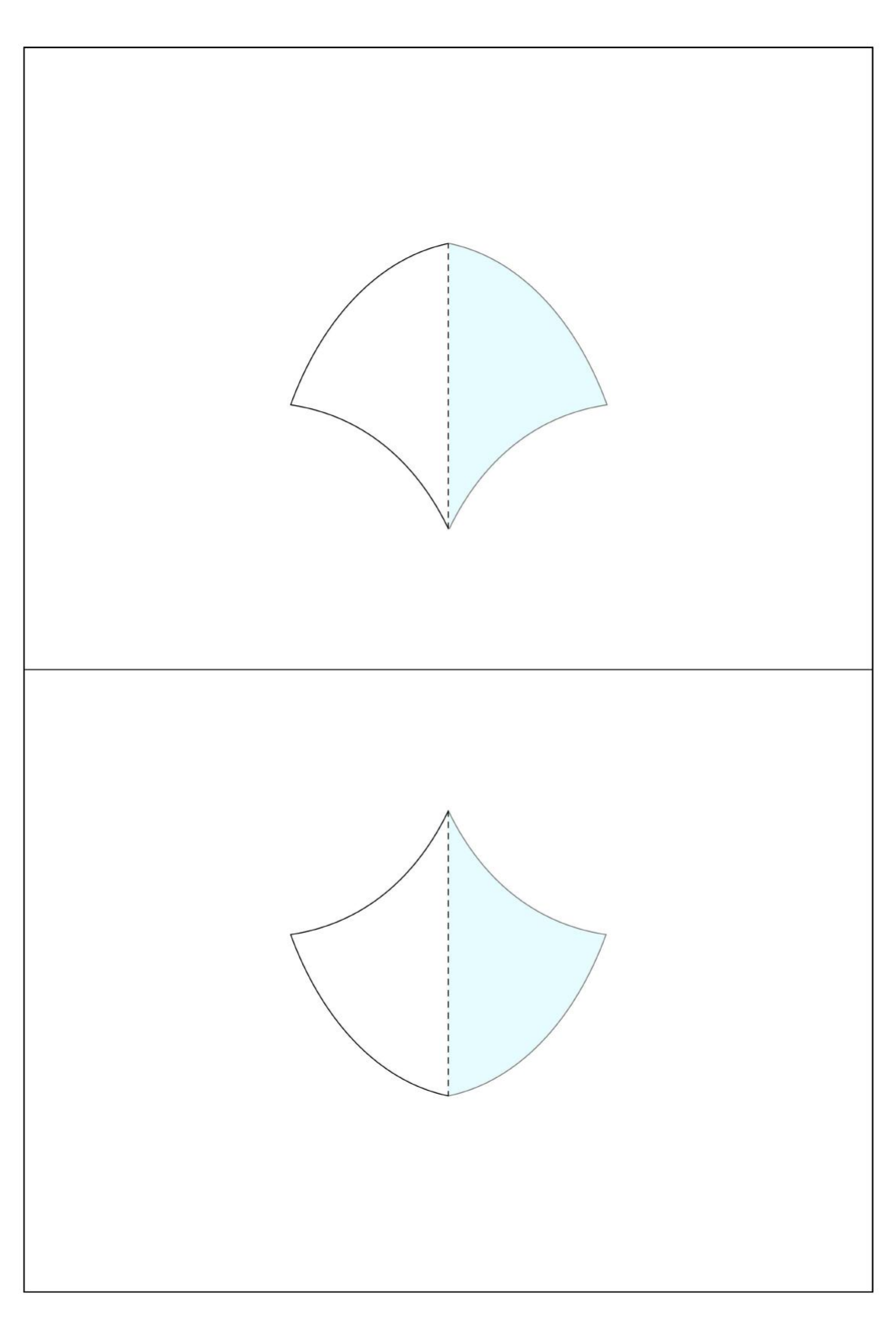}\label{worm hole top}}
  \hfill
{\includegraphics[width=1\textwidth]{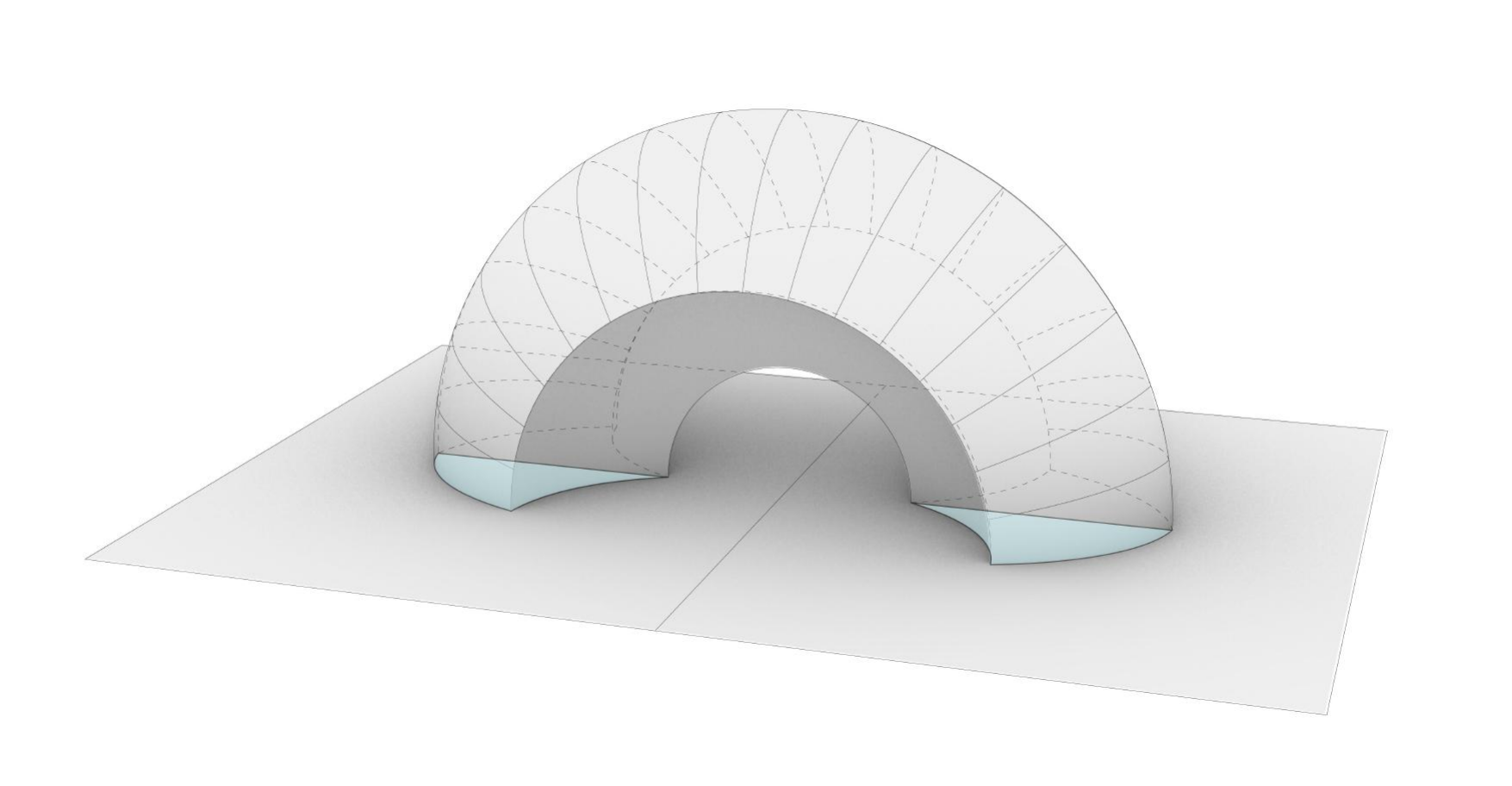}\label{worm hole}}
 \caption{The wormholes in \cite{Maldacena:2004rf,Collier:2022bqq} have two asymptotic boundaries. The  Maldacena-Maoz (Fuchsian) wormhole with three defects on the two spherical boundaries calculates the variance
of the three-point functions in the CFT ensemble. }\label{fig:3d wormholes}
\end{figure}

In the absence of matter sources, Einstein equations in 3d gravity can be satisfied if we use the ``wormhole slicing'' ansatz
\begin{equation}\label{eq:WH_metric}
    ds^2 = d\rho^2 + \cosh^2 \rho e^{\Phi(z,\bar{z})} dz d\bar{z}~,
\end{equation}
where $\Phi(z,\bar{z})$ satisfies the Liouville equation
\begin{equation}\label{eq:Liouville_background}
     \partial \bar{\partial} \Phi = \frac{e^{\Phi}}{2}~.
\end{equation}
In  \cite{Maldacena:2004rf, Collier:2022bqq} and illustrated in Figure \ref{fig:3d wormholes}, the above ``wormhole slicing'' allows us to get an identification between the semiclassical two-boundary wormhole gravitational path integral and Liouville correlators. By solving 3d gravity with ``Virasoro TQFT'', the exact matching between the two theories beyond large c follows naturally \cite{Collier:2023fwi}.

Here, in matching Liouville quantities with results from 3d gravity with boundaries, we make one further identification between the ZZ boundary condition in Liouville theory and the Hartle-Hawking ``no boundary'' condition in gravity\cite{Hartle:1976tp}. 

\begin{figure}[h]
	\centering
	\includegraphics[scale=0.5]{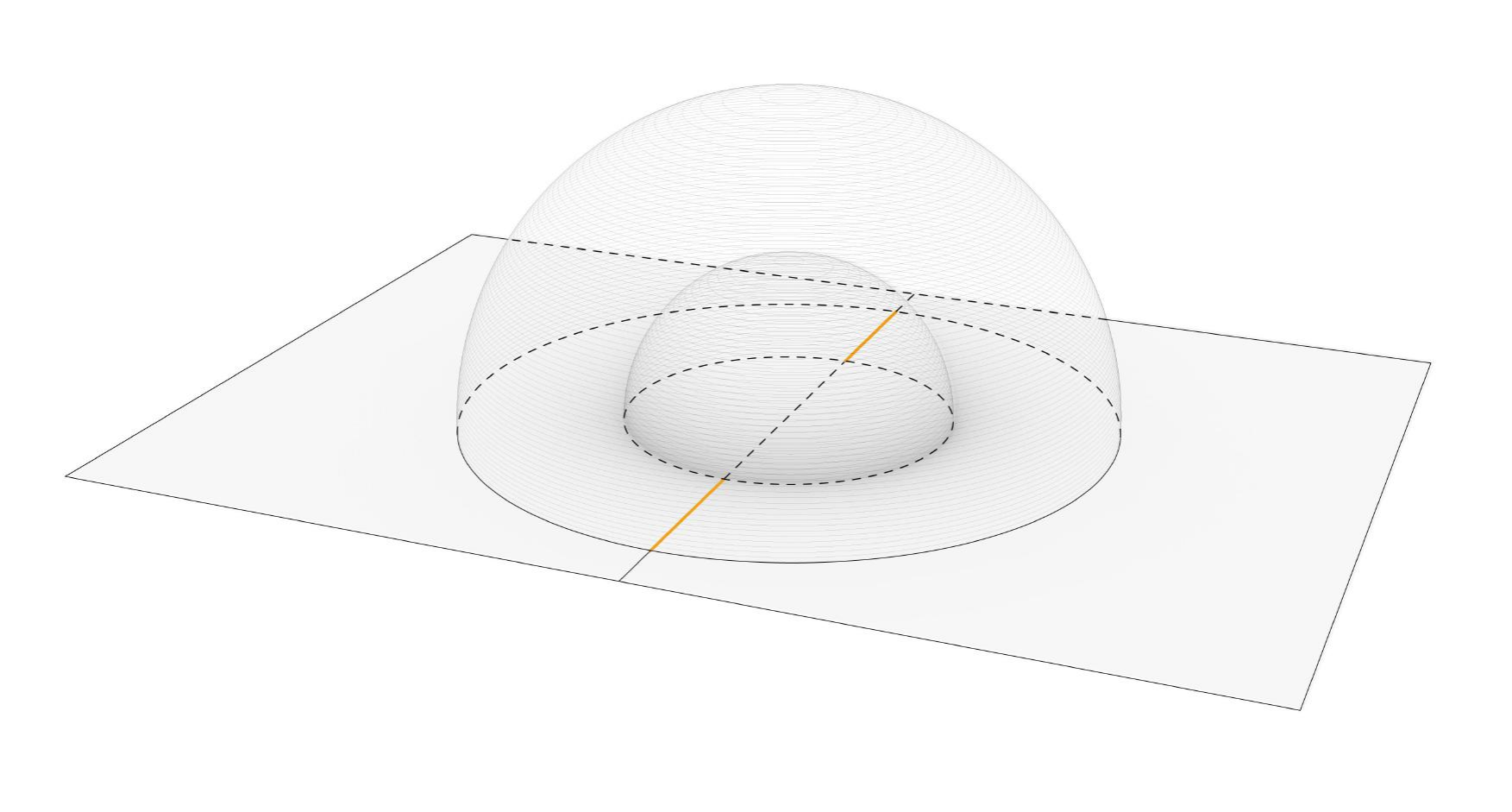}
	\caption{Although BTZ black holes have only one asymptotic boundary, they have a  ``wormhole'' description as the ZZ boundaries (indicated by the disconnected orange lines) are glued together.}\label{fig:btz dome}
\end{figure}  
\begin{figure}
\begin{center}
\begin{overpic}[scale=1]{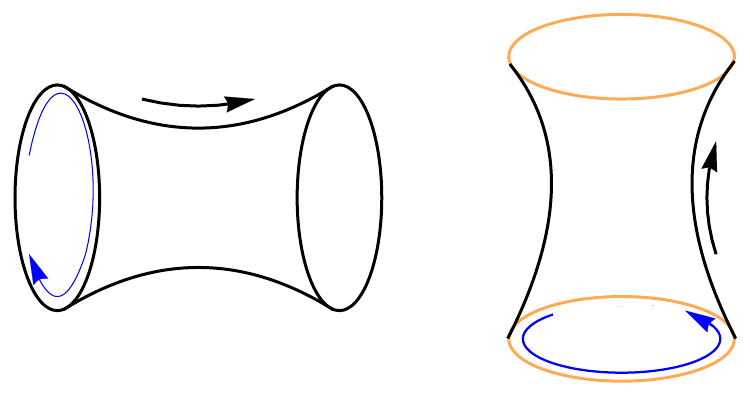}
\put(-5,25){\parbox{0.2\linewidth}{
		\begin{equation*}
			\color{blue}{2 \pi}
\end{equation*}}}
\put(70,5){\parbox{0.2\linewidth}{
		\begin{equation*}
			\color{blue}{2 \pi}
\end{equation*}}}
\put(88,25){\parbox{0.2\linewidth}{
		\begin{equation*}
			\text{Im}(z)
\end{equation*}}}
\put(13,40){\parbox{0.2\linewidth}{
		\begin{equation*}
			\rho
\end{equation*}}}
\put(95,5){\parbox{0.2\linewidth}{
		\begin{equation*}
			\text{Im}(z) = 0
\end{equation*}}}
\put(95,45){\parbox{0.2\linewidth}{
		\begin{equation*}
			\text{Im}(z) = \beta/2
\end{equation*}}}
\end{overpic}  
\end{center}
\caption{(Left): Each Im$(z)=$ constant leaf is a hyperbolic cylinder, with its two ends living on the ``two'' asymptoticaly AdS boundaries respectively. (Right): Each $\rho=$ constant leaf is also a hyperbolic cylinder, with the ZZ boundaries (orange $S^1$ circles) lying at both ends at Im$(z)=0$ and Im$(z)=\beta/2$  respectively.}\label{fig:cylinders}
\end{figure}

We use the BTZ partition function as a concrete example. The upper half-space construction for non-rotating Euclidean BTZ was studied in \cite{Carlip:1994gc, Carlip:1995qv}. The metric is given by
\begin{equation}\label{eq:non_BTZ_metric}
    ds^2 = (r^2 - r_+^2) d\tau_E^2 + \frac{dr^2}{r^2 - r_+^2}+r^2 d\phi^2~,
\end{equation}
where $\phi \sim \phi+2 \pi$ is the periodic identification of the spatial circle, and the Euclidean time is also identified, i.e. $\tau_E \sim \tau_E+\beta$, such that we have a smooth metric near the horizon. $\beta$ is the inverse temperature of the black hole and is related to the horizon radius $r_+$ through $\beta = 2\pi/r_+$. The spacetime metric is locally isometric to the hyperbolic space $\mathbb{H}^3$ and hence, we can perform a coordinate transformation to bring the Euclidean BTZ metric to the upper half-space metric of $\mathbb{H}^3$ in Poincare coordinates,
\begin{equation}
\begin{split}
x &= \sqrt{\frac{r^2 - r_+^2}{r^2 }}e^{r_+ \phi}\cos r_+ \tau_E ~, \\
y &=\sqrt{\frac{r^2 - r_+^2}{r^2 }} e^{r_+ \phi}\sin r_+ \tau_E~, \\
w &= \frac{r_+}{r }e^{r_+ \phi}~.
\end{split}
\end{equation}
We further introduce spherical coordinates $(R,\psi,\chi)$ for the upper half-space
\begin{equation}
    (x,y,w) = (R \cos \psi \cos \chi, R \sin \psi \cos \chi, R \sin \chi)~,
\end{equation}
where the Euclidean BTZ metric becomes
\begin{equation}
    ds^2 = \frac{1}{w^2}(dx^2 + dy^2 + dw^2) = \frac{1}{R^2 \sin^2 \chi}(R^2 d\chi^2 + R^2 \cos^2 \chi d\psi^2+dR^2 ) ~,~(w>0)~.
\end{equation}
The periodic identification of $\phi$ requires $R \sim Re^{2 \pi r_+}$, whereas the periodic identification of $\tau_E$ requires $\psi \sim \psi+2 \pi$. Hence, as shown in Figure \ref{fig:btz dome}, the BTZ black hole is the region between the two domes in the upper half plane. To see the connection with the ``wormhole slicing'' in \eqref{eq:WH_metric}, we introduce the following coordinate transformation
\begin{equation}\label{eq:coord_trans_WH}
    \begin{split}
        \tan \psi &= \tan\left(\frac{2\pi}{\beta} \text{Im}(z)\right) \tanh \rho~, \\
        R &= e^{\frac{2\pi}{\beta} \text{Re}(z)}~, \\
        \sin \chi &= \text{sech} \rho \sin \left(\frac{2\pi}{\beta} \text{Im}(z)\right) ~,
    \end{split}
\end{equation}
where $\text{Re}(z) \sim \text{Re}(z)+2 \pi$ and Im$(z)\in [0,\frac{\beta}{2}]$. 

The BTZ metric takes the following form, 
\begin{equation}\label{eq:BTZ wormhole metric}
\begin{aligned}
    ds^2 &= d\rho^2 + \cosh^2 \rho e^{\Phi(z,\bar{z})} dz d\bar{z}~,
\end{aligned}
\end{equation}
with the Liouville field being
\begin{equation}\label{eq:BTZ wormhole}
\begin{aligned}
    e^\Phi& = \frac{4\pi^2}{\beta^2}\frac{1}{\sin^2 \left(\frac{2\pi}{\beta} \text{Im}(z)\right)}~.
\end{aligned}
\end{equation}
We also like to point out the connection between the solid torus picture in Figure \ref{fig:doubling partition function} and the upper half-space construction in Figure \ref{fig:btz dome}. The radial direction $R$ in spherical coordinates is the Schwarzschild angular coordinate $\phi$ of non-rotating BTZ. The polar angle $\chi$ is the radial direction $r$, where $\chi = \pi/2$ is the location of the horizon $r=r_+$ and $\psi$ is the azimuthal angle that is responsible for the thermal direction. The ZZ boundaries are located at $\psi = 0;\pi$, as indicated by the disconnected orange lines (They are $S^1$ circles because of the identification in the spatial direction.) in the figures. 

As shown in Figure \ref{fig:cylinders}, the ZZ boundaries are located at Im$(z) =0$ and Im$(z)=\frac{\beta}{2}$ on the cylindrical surface respectively and they glue the ``two'' copies of cylinder along their $S^1$ at the asymptotic boundary. Furthermore, the periodicity of the thermal circle is restored through the gluing. As shown in Figure \ref{fig:cylinders}, with each constant $\rho$ slice as a hyperbolic cylinder on $\mathbb{H}^2$, we can think of the upper half-space construction of the bulk being composed from constant $\rho$ slices where $\rho$ acts as an angle that rotates the hyperbolic cylinder $e^\Phi dz d\bar{z}$ around the ZZ boundaries.
 
    \begin{figure}[h]
\includegraphics[width=1\textwidth]{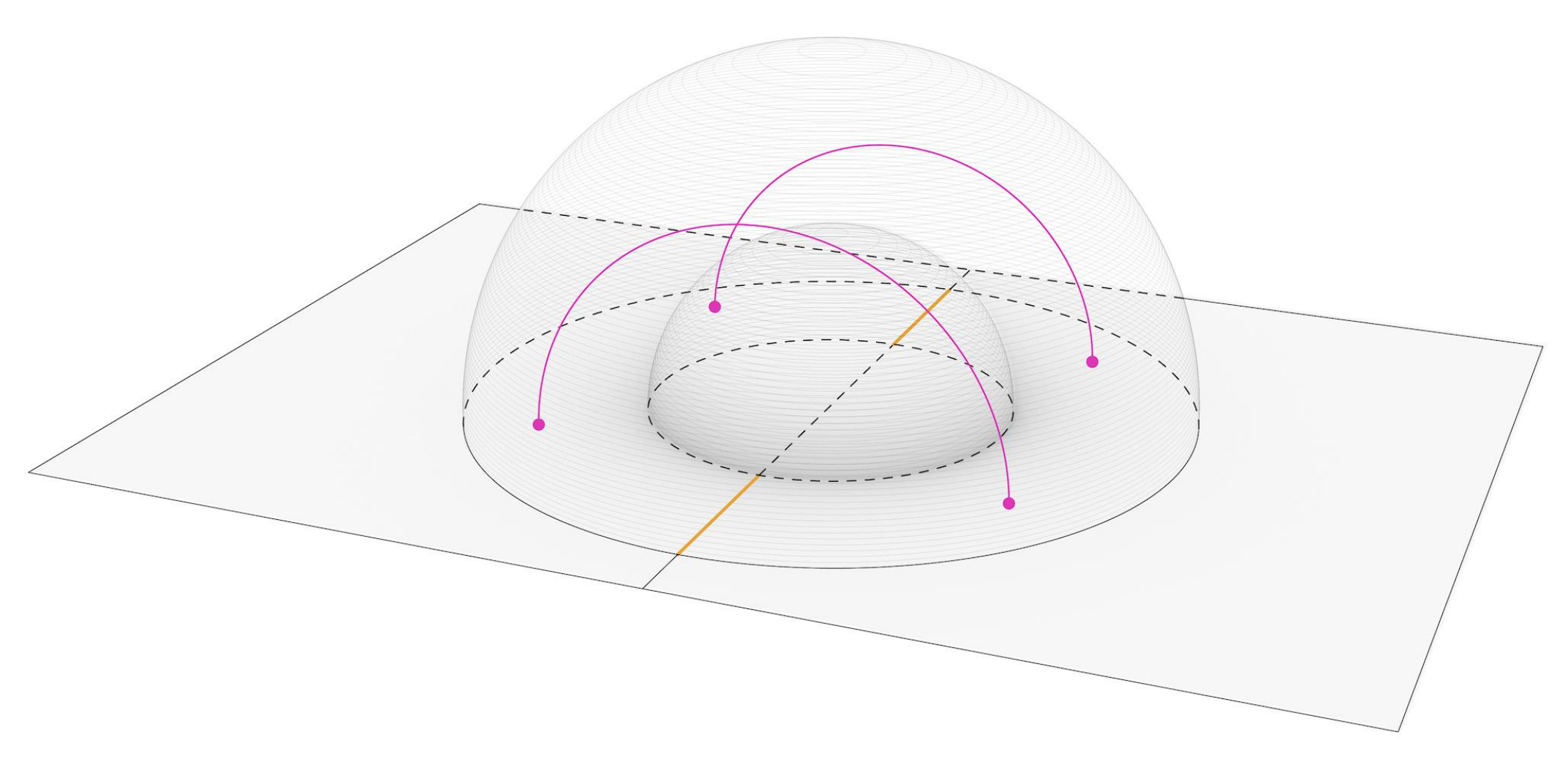}
    \caption{Geometric description of holographic thermal $2n$-point functions in Section \ref{sec:correlation_function} from wormhole slicing ($n=2$ in the diagram). }\label{dome defect}
    \end{figure}
   \begin{figure}[h]
\begin{center}
\begin{minipage}[b]{0.4\linewidth}
\begin{overpic}[scale=0.2]{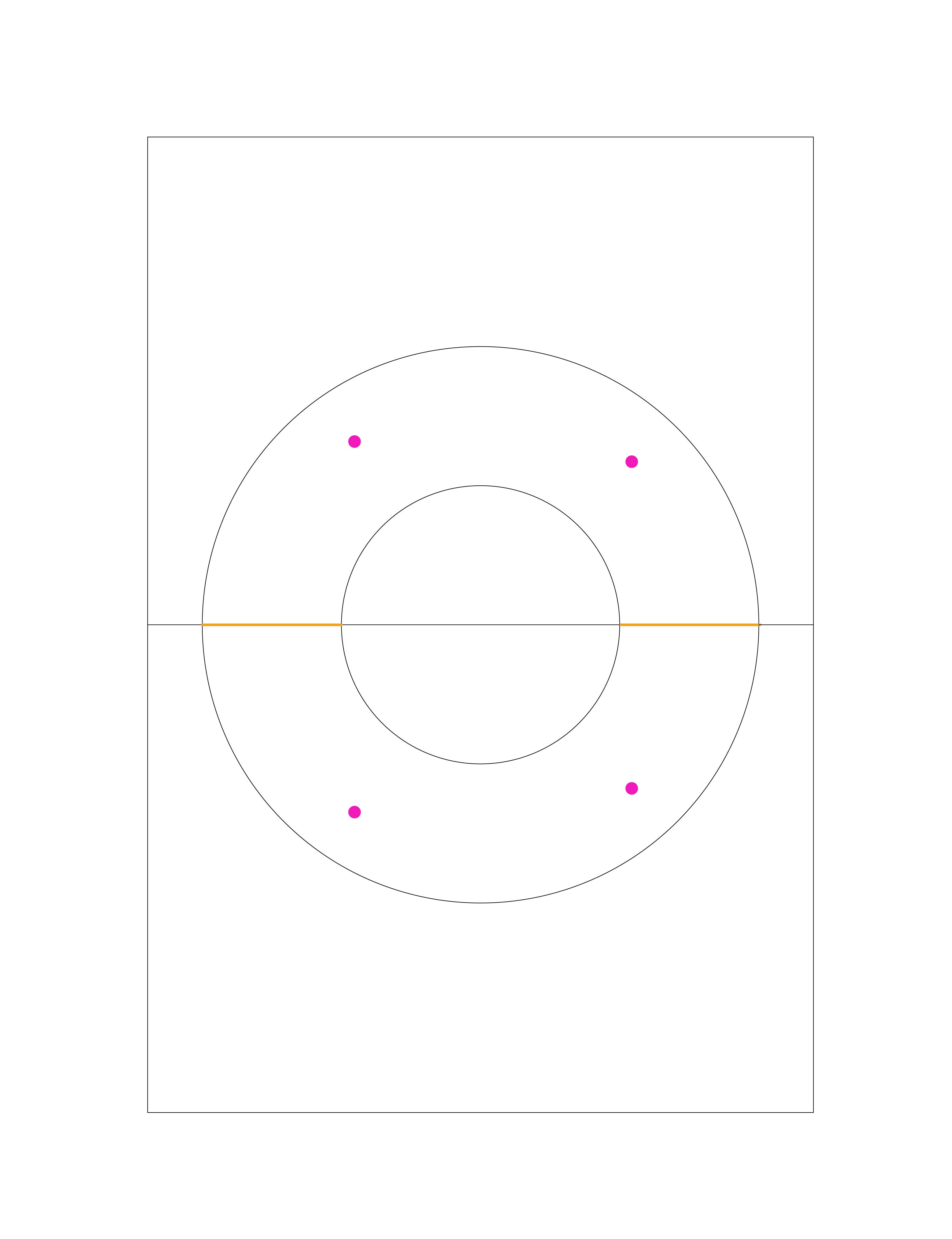}
\end{overpic} 
        \end{minipage}
        \begin{minipage}[b]{0.45\linewidth}
\begin{overpic}[scale=0.2]{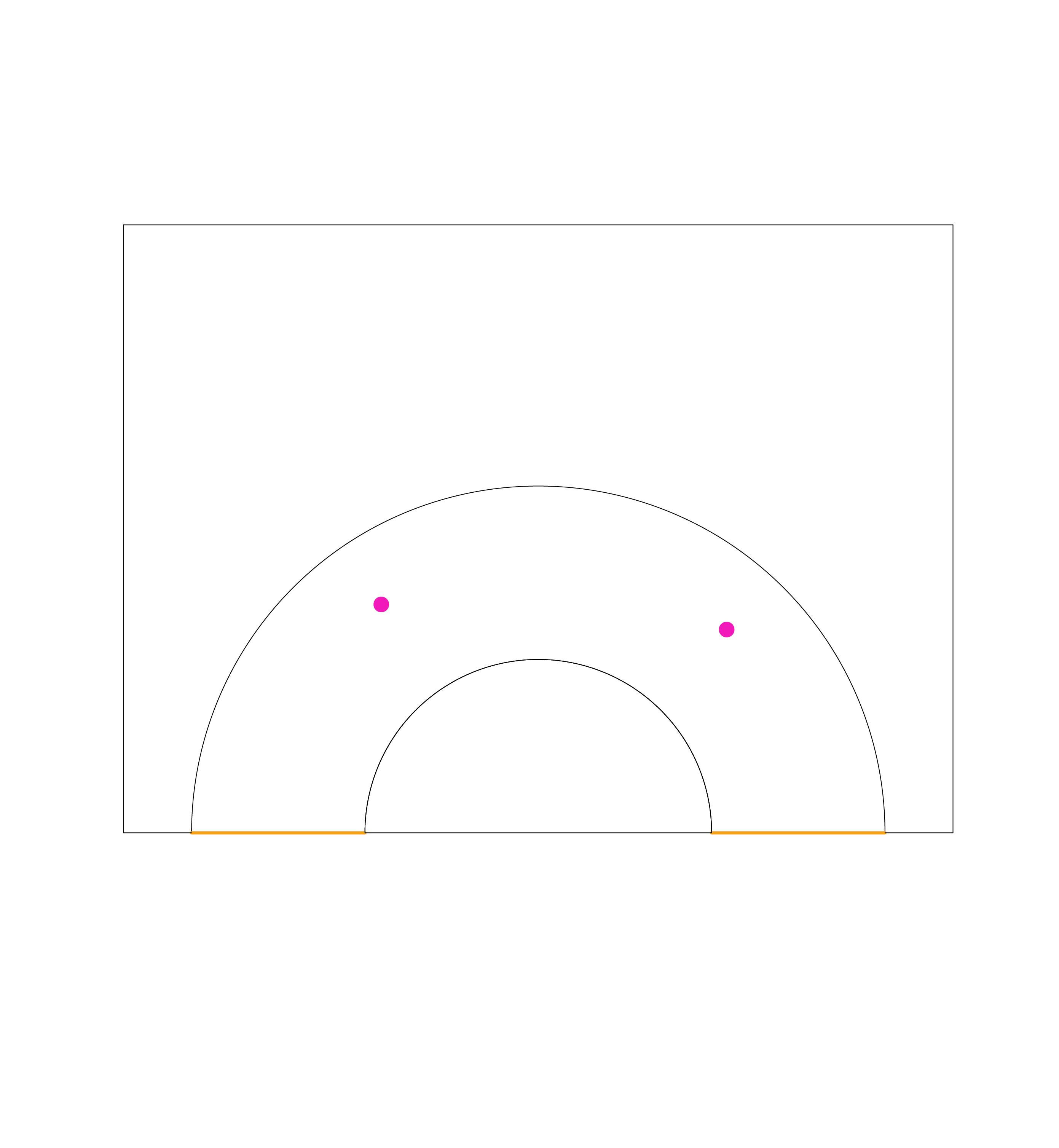}
\end{overpic}  
  \end{minipage}
\end{center}
    \caption{(Left): $n=2$ operators are inserted on each cylindrical region of the asymptotic boundary. With the ZZ boundaries (indicated by orange line) being glued together, we have a total of four operator insertions on the asymptotic torus. (Right): With $n=2$ operators between the ZZ boundary states on the upper half plane, we view the correlation function as non-chiral Liouville transition amplitudes. }\label{region dome defect}
\end{figure}
\begin{figure}[h]
    \centering
\includegraphics[width=0.9\textwidth]{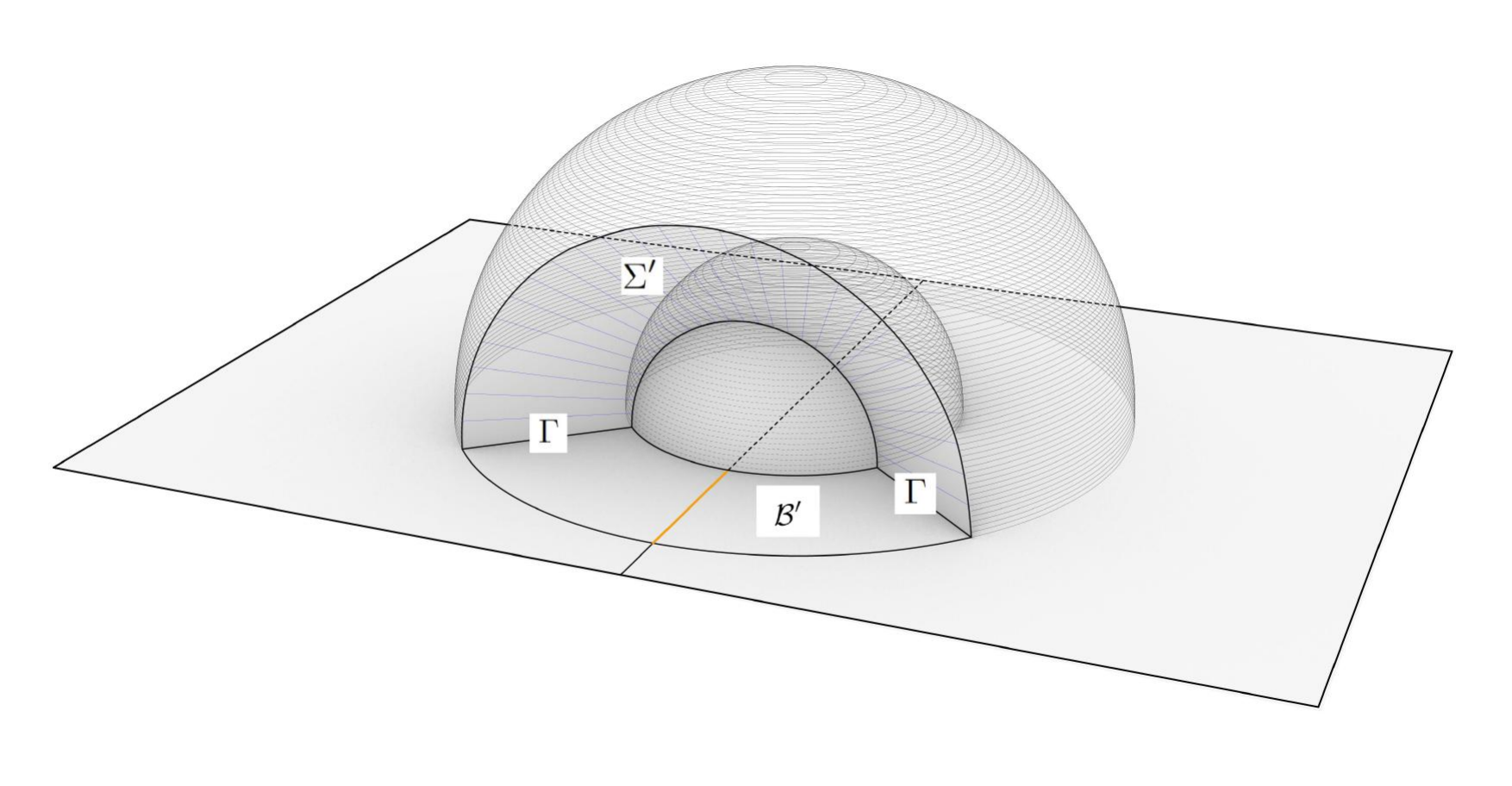}
\caption{Geometric description of Wheeler-DeWitt wavefunction in the $\Phi_0$ basis in Section \ref{sec:WdW} from wormhole slicing. }\label{dome wavefunction}
    \end{figure}
    \begin{figure}[h]
\begin{center}
\begin{minipage}[b]{0.45\linewidth}
\begin{overpic}[scale=0.2]{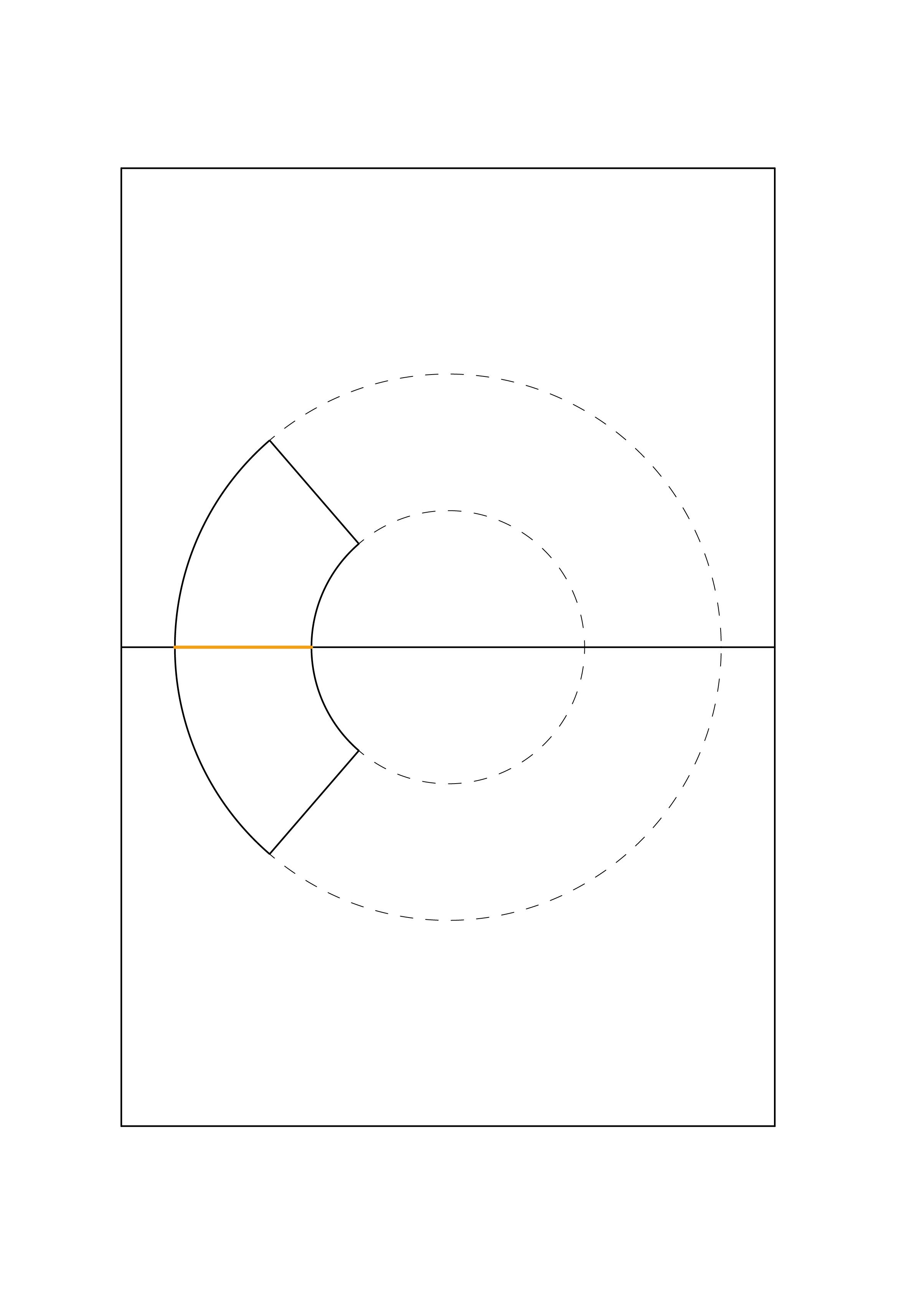}
\end{overpic} 
        \end{minipage}
        \begin{minipage}[b]{0.5\linewidth}
\begin{overpic}[scale=0.25]{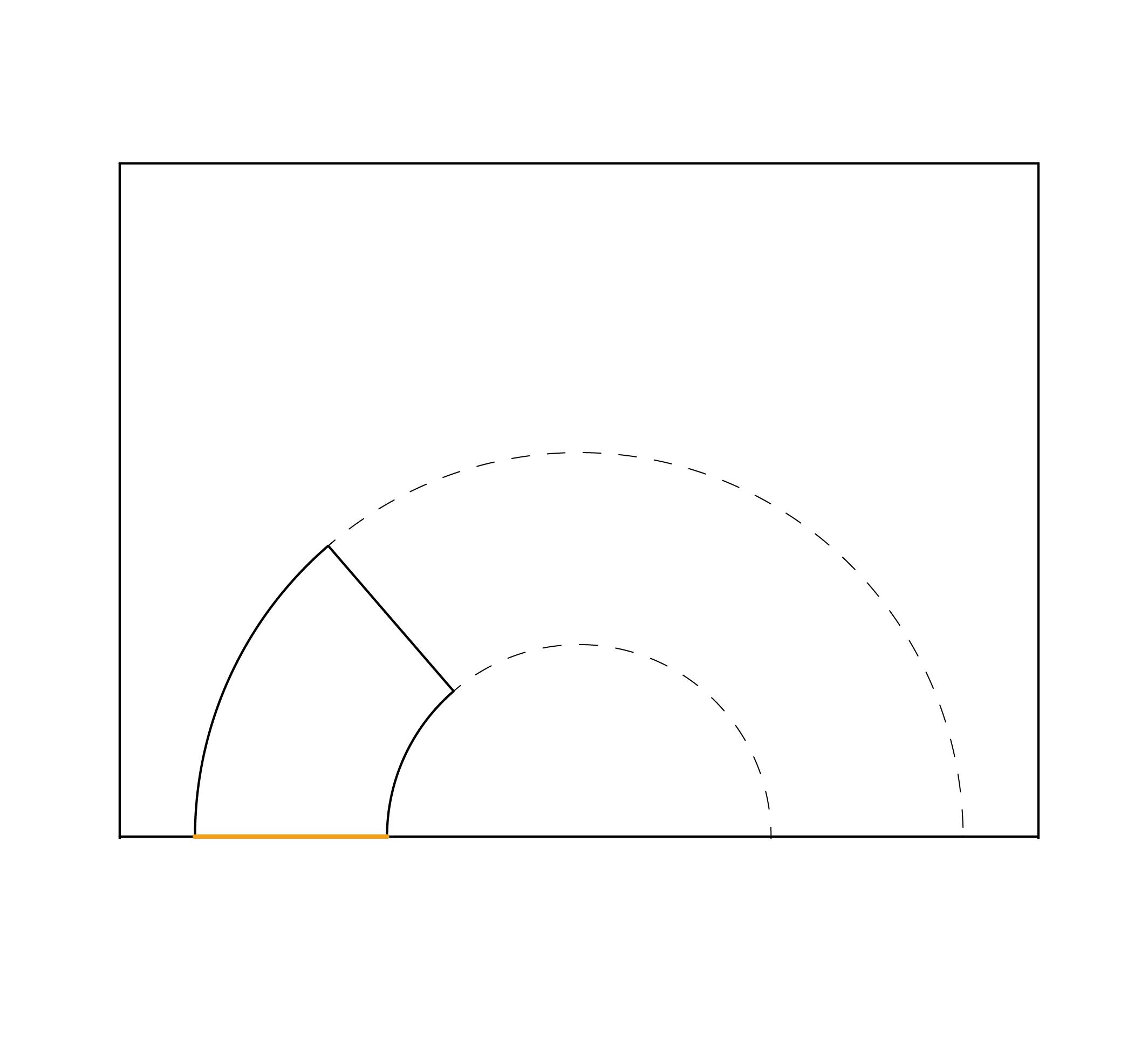}
\end{overpic}  
  \end{minipage}
\end{center}
\caption{(Left): The corresponding domain of the wavefunction at the asymptotic boundary. (Right): The corresponding cylindrical region of the wavefunction on the upper half plane.}\label{fig: domain dome wavefunction}
\end{figure}

For the computation of BTZ $2n$-point functions in Section \ref{sec:correlation_function}, we show that there are two ZZ boundaries residing on the ends of each $\rho \rightarrow \pm \infty$ cylindrical surface, in addition to $n$ operators inserted in each cylindrical region. This is illustrated in Figure \ref{dome defect}. Similar to the BTZ partition function, the dual description of $2n$-point function is just Liouville correlators between two copies of ZZ boundaries. 

For the study of Wheeler-DeWitt wavefunctions, we just need to impose the ZZ boundary condition on one end and another boundary condition that corresponds to some bulk quantum states on the other end of the cylindrical surface. We show an example of the $\Phi_0$-states in Section \ref{subsec:bvp_WdW}, where through a direct calculation, the gravitational path integral matches the transition amplitude between the ZZ boundary state and another state in Liouville. This wormhole description for the wavefunction is illustrated in Figure \ref{dome wavefunction} and \ref{fig: domain dome wavefunction}.
\subsection{Gluing of AdS boundaries by ZZ boundary condition}\label{gluing}
In the above discussion for BTZ black hole, we like to avoid a naive geometrical contradiction on viewing BTZ as a ``wormhole''. Namely, the ``wormhole slicing'' is used to describe geometries that have two asymptotic boundaries at $\rho \rightarrow \pm \infty$ respectively (Figure \ref{fig:3d wormholes}) \cite{Collier:2022bqq, Maldacena:2004rf}, 
whereas the BTZ black holes have only one asymptotic boundary (Figure \ref{fig:btz dome}).\footnote{A similar problem has been addressed in \cite{Maldacena:2004rf} for the hyperbolic slicing of global AdS.} The reason for this discrepancy is that when we impose the ZZ boundary conditions at the two ends of the asymptotic cylindrical surface, the ZZ boundaries at $\rho \rightarrow \pm \infty$ respectively are automatically glued together. This is Cardy's doubling trick\cite{Cardy:1984bb} in the language of CFT, where we can view correlation functions on the upper half plane with conformal boundary conditions on the real axis as correlation functions for a chiral theory on the whole plane, which also comes naturally from the Ward identity. In gravity, this corresponds to the ``transparent boundary condition''\cite{Maldacena:2004rf}, and the gluing of the ZZ boundaries can be seen through the shrinking of the codimension one cap formed by the ZZ cutoff $\epsilon_y$.\footnote{Geometrically, the gluing is seen through the extrinsic curvature contribution on the codimension-one cap formed in the bulk by the ZZ boundary conditions. The contribution is equivalent to the bulk+AdS boundary terms of the cylindrical region that we carve out in regulating divergences near the ZZ boundaries. As a consequence, there seem to be ``two'' asymptotic regions but the ``two'' halves are glued together by the zero-size  cap formed by ZZ. This is demonstrated in Appendix \ref{app:ZZ_gluing}. } As a result, the distance between the ``two'' halves goes to zero and the ZZ state being a ``no boundary'' state is justified. 

The ``wormhole'' construction of BTZ has the topology annulus times interval $S^1_{\phi}\times I_{\rho} \times I_{\tau_E} =A_{\phi,\rho} \times I_{\tau_E}$, which is what we expect from canonical quantization in Section \ref{sec:3d_Liouville_AS}. The spatial $S^1$ circles of constant $\rho$ slices are identified at $\tau_E=0$ and $\tau_E=\frac{\beta}{2}$ respectively, resulting in the geometry formed by foliation of constant $\rho$ slices to be a solid torus. From the CFT point of view, the ZZ boundary condition ensures only the identity module runs in the dual open channel and with the doubling trick, we get the vacuum conformal block in the $\tau_E$ channel of the torus. Using the language of  Chern-Simons theory, this ensures only the trivial Wilson loop exists in the $\tau_E$ cycle, which further corresponds to contractibility in the bulk of Euclidean black holes. In \cite{Harlow:2021lpu}, it is stated that holographic duality has been taken into account when we require a contractible  Euclidean time cycle in ensuring a smooth near horizon metric. Contractibility in the bulk plays an important role in quantum gravity, in particular in the Euclidean path integral derivation of the Bekenstein-Hawking entropy \cite{Gibbons:1976ue}. In Section \ref{sec:factorization}, we see how this condition is related to the definition of trace in gravity and its interplay with modular invariance.

\section{Correlation functions and the identity block}\label{sec:correlation_function}
We have shown in Section \ref{sec:HH-state} that the overlap between the ZZ boundary states on a finite cylinder gives the chiral identity character in the dual channel, agreeing with what we expect from holography duality and doubling trick. In this section, we further show that the ZZ boundary states provide an easy way to capture the $2n$-point functions for below black hole threshold scalar operators (we will also call them scalar probe operators) in holographic CFTs. From the doubling trick, we expect the expression to be given by the corresponding Virasoro identity block in certain channel, just as what we obtain from modular transformation on vacuum characters in one-loop partition function.

We propose the equivalence of the following three equations
\begin{equation}\label{eq:Liouville_correlators}
\begin{aligned}
    \mathcal{G}^{(2 n)}_{\alpha_1, \cdots, \alpha_n}(z_1,\cdots,z_n) &=\langle \tilde{O}_{h_{\alpha_1}, h_{\alpha_1}}(z_1,\bar{z}_1)  \tilde{O}_{h_{\alpha_1}, h_{\alpha_1}}(\bar{z}_1,z_1)...\tilde{O}_{h_{\alpha_n}, h_{\alpha_n}}(z_n,\bar{z}_n)  \tilde{O}_{h_{\alpha_n}, h_{\alpha_n}}(\bar{z}_n,z_n)\rangle_{\beta,\text{BTZ}}\\
    &= \langle  \hat{V}_{\alpha_1}(z_1,\bar{z}_1)   \cdots \hat{V}_{\alpha_n}(z_n,\bar{z}_n)  \rangle_{\beta/2,ZZ}\langle  \hat{V}_{\alpha_1}(\bar{z}_1,z_1) \cdots   \hat{V}_{\alpha_n}(\bar{z}_n,z_n) \rangle_{ \beta/2, \widetilde{ZZ}}~,\\
    &= c_b^{-2 n} \overline{\langle O_{h_{\alpha_1}, h_{\alpha_1}}(z_1,\bar{z}_1)  O_{h_{\alpha_1}, h_{\alpha_1}}(\bar{z}_1,z_1)...O_{h_{\alpha_n}, h_{\alpha_n}}(z_n,\bar{z}_n)  O_{h_{\alpha_n}, h_{\alpha_n}}(\bar{z}_n,z_n)\rangle}_{\beta,\text{BTZ}}~.
\end{aligned}
\end{equation}
In the first line, we have the expectation value of scalar fields $\tilde{O}_{h_{\alpha_1}, h_{\alpha_1}}(z_1,\bar{z}_1)$ in gravity, located at boundary points $z_i = \phi_i+i \tau_i$ in BTZ background where  $\langle \cdot \rangle_{\beta,\text{BTZ}}$ indicates the thermal expectation value in BTZ with inverse temperature $\beta$. In this paper, we will consider below black hole threshold operators, with large conformal dimensions, such that $1 \ll h_{\alpha_i}<c/6$ and $m_i \approx 2 h_{\alpha_i}$. These fields in the bulk correspond to
massive particles that travel on geodesics. When they are heavy enough, they backreact on the geometry and produce conical deficits, hence, we will also call them ``defects'' in such cases. We introduce the defect strength $\eta \in [0,1/2]$ as in \eqref{defect strength} following \cite{Collier:2022bqq}. In the second line, we have the rescaled Liouville operators
\begin{equation}\label{eq:rescaled_Liouville}
    \hat{V}_{\alpha}= \begin{cases}
    \frac{V_{\alpha}}{\sqrt{S_L(P) \rho_0(P)}}~&,~~\eta \in[0,\frac{1}{2}]~, \\
        \frac{V_{\alpha}}{\sqrt{S_L(P))}}~&,~~P\text{  real}~.
    \end{cases}
\end{equation}
where we use $\rho_0(P)$ to denote the density of states, i.e. 
\begin{equation}
    \rho_0(P) = 4 \sqrt{2} \sinh(2 \pi P b)\sinh \left(\frac{2 \pi P}{b}\right)~,
\end{equation}
in this section. Notice in \eqref{eq:rescaled_Liouville} that there is a further rescaling of the probe operators by $\frac{1}{\sqrt{\rho_0(P)}}$ compared to operators with real $P$, i.e. operators above the black hole threshold, and the reason will be clear later. The second line in \eqref{eq:Liouville_correlators} calculates the insertion of rescaled Liouville operators between the Hartle-Hawking states, where more precisely $\langle \cdot \rangle_{\beta/2,ZZ}$ means the expectation value obtained from the ZZ overlaps defined on a cylinder with height $\beta/2$. In other words, we propose to represent each pair of scalar probe operator insertions by a product of  two Liouville operators
\be
\tilde{O}_{h_{\alpha_1}, h_{\alpha_1}}(z_1,\bar{z}_1)  \tilde{O}_{h_{\alpha_1}, h_{\alpha_1}}(\bar{z}_1,z_1) \to \hat{V}_{\alpha_1}(z_1,\bar{z}_1) \hat{V}_{\alpha_1}(\bar{z}_1, z_1)~.
\ee
In the third line of \eqref{eq:Liouville_correlators}, we have the ensemble-averaged expression in large c CFTs where $O_{h_{\alpha_1}, h_{\alpha_1}}(z_1,\bar{z}_1)$ is a scalar operator with conformal dimension $\Delta_1 = 2 h_{\alpha_1}$. $c_b$ is an overall constant that is independent of the conformal weights and its expression is given in Appendix \ref{app:Liouville}. In this work, we focus on operator insertions that are symmetric in time reflection, which is the analogue of the two-boundary Fuchsian wormholes in \cite{Collier:2022bqq}. The situation for more general operator insertions and spinning operators are discussed in the end, and will be left as future work.

To verify the relations in \eqref{eq:Liouville_correlators}, we first compute the on-shell action that has $2n$ operator insertions at the AdS boundary of BTZ. We adopt the ``wormhole slicing'' metric in \eqref{eq:BTZ wormhole metric} and with the associated Liouville solution from Einstein's equations, we reproduce Liouville correlators in the semiclassical limit, both for heavy and light probe operators. In doing so, we have carefully taken into account the renormalization of operators with the associated counterterms, generalizing results in \cite{Collier:2022bqq} by considering regions with  boundaries. To understand the exact equivalence at finite c\cite{Collier:2023fwi}, we study the two-point function as an example in operator formalism in detail. We show the insertion of Liouville primary operators between Hartle-Hawking states gives the corresponding CFT identity block in the dual channel. From the doubling trick, the scalar one-point function of $V_\alpha$ between Ishibashi states is equal to the chiral two-point Virasoro block on the torus. The product of the ZZ wavefunctions gives the modular S-matrix and combining with the DOZZ structure constant, we get exactly the conformal crossing kernel\cite{Ponsot:1999uf, Ponsot:2000mt, Collier:2019weq}. Hence, two copies of one-point function $V_\alpha$ between the ZZ boundary states on a finite cylinder gives us the corresponding two-point identity block on the torus in the dual channel. Finally, we also match Liouville correlators obtained from ZZ overlaps with the ensemble averaged result of holographic CFTs.

\subsection{Heavy probe operators}\label{subsec:heavy_operators}

We first consider heavy probes\footnote{The conformal weight for probe operators are all below the black hole threshold $c/6$, the ``heavy'' refers to operators with $h/c$  held fixed in the semiclassical limit.} where $\alpha = \eta/b$ is of order $\mathcal{O}(b^{-1})$ in the semiclassical limit. Similar to \cite{Collier:2022bqq}, we use the wormhole slicing ansatz \eqref{eq:BTZ wormhole metric}, and the probe fields are inserted at $z=z_i$ for $i=1,\cdots,n$. In the presence of these fields, Einstein equations for a hyperbolic metric require the Liouville field $\Phi$ to satisfy the Liouville equation with sources
\begin{equation}
    \partial \bar{\partial} \Phi = \frac{e^{\Phi}}{2}-4 \pi G_N \sum_i m_i \delta^{(2)}(z-z_i)~,
\end{equation}
where $m_i = \eta_i/2 G_N$ is the local mass parameter that shows up in the worldline action of the particle $i$. The ADM mass of the particle is related to the conformal dimension of the dual operator through \cite{Collier:2022bqq}
\be\label{eq:mass_heavy}
\Delta_i = m_i(1-2 G_N m_i)~.
\ee
The geometry for the solution is shown in Figure \ref{dome defect}, and we expect the answer to be given by Liouville theory defined on the boundary as Figure \ref{region dome defect}. 

The properly normalized gravitational action is given by
\begin{equation}\label{eq:S_heavydefects}
   -S_{2n} =  -S_{\text{bulk}} - S_{\text{bdry}} -S_{ZZ} - S_{\text{defect}} -  S_{\text{ct}}(\eta_i)~,
\end{equation}
where the terms are defined as
\begin{equation}\label{eq:npoint}
\begin{split}
 -S_{\text{bulk}} &= \frac{1}{16 \pi G_N}\int_{\mathbb{B} \times \mathbb{R}}  \sqrt{g}(R+2)~, \\
  -S_{\text{bdry}} &= \frac{1}{8 \pi G_N}\left(\int_{\mathbb{B}}  \sqrt{\gamma}(\Theta-1)+   \sum_i \int_{D_i}  \sqrt{\gamma}\Theta\right)~, \\
  -S_{ZZ} &= \frac{1}{8 \pi G_N}\int_{\mathbb{\Sigma}_{ZZ}}  \sqrt{h}K ~, \\
  - S_{\text{defect}} &=  \sum_i\left(\frac{1}{16 \pi G_N}\int_{D_i \times \mathbb{R}}\sqrt{g}R -  m_i \int dl_i\right)~, \\
   -  S_{\text{ct}}(\eta_i) &= \frac{1}{2 G_N}\sum_i \left((1-2 \eta_i)\ln(1-2 \eta_i) - 2 \eta_i \ln \epsilon + 2 \eta_i^2 \ln \epsilon_i +\ln 2-1\right) ~.
\end{split}
\end{equation}
We define the regions $\mathbb{B} = \{|z-z_i| > \epsilon_i; X=\text{Re}(z) \in [0, 2 \pi); Y=\text{Im}(z) \in [ \epsilon_y , \beta/2 -\epsilon_y]\}$ and $D_i = \{|z-z_i| < \epsilon_i\}$. $\mathbb{\Sigma}_{ZZ}$ is the surface that describes the ZZ boundary conditions that glue the ``two'' asymptotic halves together and is located both at Im$(z) = \epsilon_y;\beta/2-\epsilon_y$. $\epsilon_y$ is introduced as a cutoff for the carved out cylindrical region to account for the divergences from the ZZ boundary conditions and $\mathbb{\Sigma}_{ZZ}$ glues the familiy of constant $\rho$ leaves at the cutoff surface. As argued in Section \ref{sec:HH-state}, this is equivalent to studying the full BTZ background without the carved out cylindrical region $\mathbb{\Sigma}_{ZZ}$. We provide a careful analysis in showing the equivalence of these two perspectives in gravity in Section \ref{sec:WdW}.

We follow a similar technique to \cite{Collier:2022bqq} in determining $S_{\text{ct}}(\eta_i) $ to renormalize the defect operators and the steps are shown in Appendix \ref{app:counterterm}. The Liouville field has the following behavior as we approach the defects at $z_i$ and the ZZ boundary condition respectively,
\begin{equation}
    \Phi(z,\bar{z}) \sim \begin{cases}
    -4 \eta_i \ln |z-z_i| &\quad z \rightarrow z_i \\
    -2 \ln (\epsilon_y) &\quad \text{Im}(z) \rightarrow \epsilon_y;~~\beta/2 - \epsilon_y~.
    \end{cases}
\end{equation}
As mentioned, we want to relate the $2n$-point function to correlators in the CFT. We choose an AdS cutoff $\rho_0(z,\bar{z})$ that is dependent on the behavior of the Liouville field $\Phi$,
\begin{equation}
    \rho_0(z,\bar{z}) =  
    \begin{cases}
   \ln \frac{2}{\epsilon} - \frac{\Phi}{2} &\quad |z-z_i|>\epsilon_i \\
    \ln \frac{2}{\epsilon} + 2 \eta_i \ln \epsilon_i -\frac{C_i}{2} &\quad |z-z_i|<\epsilon_i~,
    \end{cases}
\end{equation}
where $C_i$ are the $\mathcal{O}(1)$ terms of the expansion of the Liouville field around the defect. Away from the defects, the induced metric at the asymptotic boundary $\rho = \pm \rho_0(z,\bar{z})$ is flat,
\begin{equation}
ds^2_{\text{bdry}} \approx \left(\frac{1}{\epsilon^2} + \frac{e^\Phi}{2}\right)dz d\bar{z} + \frac{1}{4}\left(\partial \Phi dz + \bar{\partial} \Phi d\bar{z}\right)^2~.
\end{equation}
The behavior of the particle in the bulk are given by geodesics. It is shown in \cite{Collier:2022bqq} that the worldline action of each particle $i$ cancels the delta function concentrated at $z_i$ from the scalar curvature due to defect and hence,
\begin{equation}
    - S_{\text{defect}} = 0~.
\end{equation}

The sum of the bulk and AdS boundary terms is given by
\begin{equation}\label{eq:BTZ_correlation}
\begin{split}
    -S_{\text{bulk}}-S_{\text{bdy}} &= -\frac{1}{4 \pi G_N}\int_{\mathbb{B}}  dz d\bar{z} \left(\frac{1}{4}(\partial \Phi \bar{\partial} \Phi + e^\Phi) - \frac{1}{2}\bar{\partial}(\Phi \partial \Phi) -\frac{e^\Phi}{2} \left(1+ \ln \frac{\epsilon}{2}\right)\right)~,  \\
    &=-\frac{1}{4 \pi G_N}\int_{\mathbb{B}}  dz d\bar{z} \left(\frac{1}{4}(\partial \Phi \bar{\partial} \Phi + e^\Phi) - \frac{1}{2}\bar{\partial}(\Phi \partial \Phi) \right) +\frac{1}{ G_N} \left(1+ \ln \frac{\epsilon}{2}\right)\left(\frac{1}{\epsilon_y}+\sum_i \eta_i\right)~,
\end{split}
\end{equation}
where we have used the fact that $\int_{\Gamma_1} dz d\bar{z} e^\Phi$ is just the hyperbolic area of $\Gamma_1$ in the second line of \eqref{eq:BTZ_correlation}. 
The contribution by the ZZ boundary condition is given by 
\begin{equation}
    -S_{ZZ} =-\frac{i}{2 \pi G_N} \oint_{\text{Im}(z) = \epsilon_y} dz \left(\ln \frac{2}{\epsilon}-\frac{\Phi}{2}\right)\partial_{z}\Phi ~,
\end{equation}
which cancels some of the boundary terms in \eqref{eq:BTZ_correlation}. The total gravitational action, including the counterterms gives us
\begin{equation}
    \begin{split}
        -S_{2n}  &= -\frac{c}{3}S_{\text{Liouville}} +\frac{c}{6}\sum_i s(\eta_i)~,
    \end{split}
\end{equation}
where the renormalized Liouville action $S_{\text{Liouville}}$ and $s(\eta_i)$ are given by
\begin{equation}
    \begin{split}
        -S_{\text{Liouville}} &= -\frac{1}{2 \pi}\int_{\mathbb{B}}  dz d\bar{z} \left(\frac{1}{4}(\partial \Phi \bar{\partial} \Phi + e^\Phi) \right) +\frac{1}{4 \pi \epsilon_y}\oint_{\text{Im}(z) = \beta/2 -\epsilon_y} dz \Phi +\frac{1}{4 \pi \epsilon_y}\oint_{\text{Im}(z) = \epsilon_y} dz \Phi + \frac{2}{  \epsilon_y} + \frac{2\ln \epsilon_y}{ \epsilon_y} \\
        &+\sum_i \left[ \frac{i}{4 \pi }\oint_{|z-z_i| = \epsilon_i}dz \Phi \partial \Phi+2 \eta_i^2 \ln \epsilon_i  \right]~, \\
        s(\eta_i) &= 2(1-2 \eta_i)(\ln(1-2 \eta_i) + \ln 2 -1)~,
    \end{split}
\end{equation}
respectively. $s(\eta_i)$ is related to the semiclassical limit of the Liouville reflection amplitude
\begin{equation}
    e^{-\frac{c}{6}s(\eta)} \approx S_L(P)~,
\end{equation}
matching the normalization of heavy probe operators that we introduced earlier and  $\eta \sim \frac{1}{2}+i P \sqrt{\frac{6}{c}}$ \cite{Collier:2022bqq}. 
Hence, we have matched the semiclassical gravity calculation for heavy probe operator insertions in BTZ to Liouville.

\subsection{Light probe operators}

The calculation in Section \ref{subsec:heavy_operators} parallels the two-boundary Fuchsian wormhole calculation with heavy defects in \cite{Collier:2022bqq}. In our current situration, we can also consider light probe operator insertions as in the absence of operator insertions, the BTZ black hole is a solution, unlike the case for two-boundary sphere wormholes\cite{Collier:2022bqq}. If we consider light probe operator insertions, the $2n$-point function of these operators is fully determined by the renormalized geodesic lengths in BTZ background. 

To be more specific, let us consider the defect strength to be small with $\alpha =b h \sim \mathcal{O}(b)$ in the small $b$ limit \cite{Zamolodchikov:1995aa}. As a result, there is no backreaction to the background metric. From the Liouville overlap in \eqref{eq:Liouville_correlators}, we obtain
\begin{equation} \label{eq:Liouville_light_classical}
     \mathcal{G}^{(2 n)}_{\alpha_1, \cdots, \alpha_n}(z_1,\cdots,z_n) \approx Z_{\text{BTZ}} e^{\frac{n c}{3}(\ln 2 -1)}e^{ h_{\alpha_1} \Phi(z_1,\bar{z}_1)} e^{ h_{\alpha_1} \Phi(\bar{z}_1,z_1)}\cdots e^{ h_{\alpha_n} \Phi(z_n,\bar{z}_n)}e^{ h_{\alpha_n}  \Phi(\bar{z}_n,z_n)} ~,
\end{equation}
where $\Phi(z_i,\bar{z}_i)$ is now the background solution of the Liouville equation in \eqref{eq:Liouville_background} on a finite cylindrical region and $h_{\alpha_i}$ is the associated conformal weights of the exponential Liouville operator. In addition, we have taken into account the semiclassical behavior of the reflection coefficient and density of states in the light defect regime, i.e. $P = -i b h+\frac{i}{2}\left(\frac{1}{b}+b\right)$,
\begin{equation}
    S_{L}(P)\rho_0(P) \approx e^{\frac{c}{3}(1- \ln 2)}~.
\end{equation}

We can reproduce \eqref{eq:Liouville_light_classical} from the gravitational action in \eqref{eq:S_heavydefects} by shrinking the disks $D_i$ such that the radius $\epsilon_i$ vanishes. The action becomes
\begin{equation}\label{eq:S_lightdefects}
    \begin{split}
        -S_{2n} = \frac{1}{16 \pi G_N}\int_{\mathbb{B}\times \mathbb{R}}  \sqrt{g}(R+2) + \frac{1}{8 \pi G_N}\int_{\mathbb{B}}  \sqrt{\gamma}(\Theta-1) + \frac{1}{8 \pi G_N}\int_{\mathbb{\Sigma}_{ZZ}}  \sqrt{h}K - \sum_i \left(m_i \int dl_i +S_{\text{ct}}(0)\right)~.
    \end{split}
\end{equation}
These light fields are dual to operators on the boundary CFT, with the usual mass-dimension formula $\Delta_i=2h_{\alpha_i}=1+\sqrt{1+m_i^2}$ in AdS/CFT\cite{Witten:1998qj}. Note that the mass-dimension formula for light proble operators is different from \eqref{eq:mass_heavy} due to absence of backreaction. We immediately see that the first three terms give us the contribution to the thermal partition function in \eqref{eq:Liouville_light_classical} and the counterterm  is exactly the normalization of Liouville operators of the same defect strength
\begin{equation}
    -S_{\text{ct}}(0) = \frac{1}{2 G_N}(\ln 2 -1) \approx \frac{1}{S_{L}\left(-i b h+\frac{i}{2}\left(\frac{1}{b}+b\right)\right)\rho_0\left(-i b h+\frac{i}{2}\left(\frac{1}{b}+b\right)\right)}~.
\end{equation}
It remains for us to determine the renormalized geodesic length between the two operators of the same local mass $m_i$, which is given by the value of the Liouville field at $z_i$
\begin{equation}
    \int dl_i = \int_{-\rho_0(z_i,\bar{z}_i)}^{\rho_0(z_i,\bar{z}_i)} d\rho  -2 \ln \frac{2}{\epsilon}= -\Phi(z_i,\bar{z}_i)~.
\end{equation}
We therefore match the Liouville CFT calculation in \eqref{eq:Liouville_light_classical} with the gravitational action in \eqref{eq:S_lightdefects} as we have $m_i = 2 h_{\alpha_i}$.

If we set the defect strength to be zero, it is similar to inserting $2n$ identity operators where the semiclassical limit of the correlation function is given by
\begin{equation}
     \mathcal{G}^{(2 n)}_{0, \cdots,0}(z_1,\cdots,z_n) \approx Z_{\text{BTZ}} e^{\frac{n c}{3}(\ln 2 -1)}~,
\end{equation}
and the $( S_{L}(0)\rho_0(0) )^{-n} \approx e^{\frac{n c}{3}(\ln 2 -1)}$ normalization factor matches the total counterterm contribution from the gravity side. In the absence of defects, the gravitational action reproduces the thermal partition function for Euclidean BTZ with the following on-shell Liouville action
\begin{equation}
    -S_L = \frac{\pi^2}{\beta}~. 
\end{equation}

As an explicit example, we further show that the above calculation matches with earlier holographic computations \cite{Keski-Vakkuri:1998gmz, Louko:2000tp} on the thermal two-point function of scalar operators $O_{h_{\alpha_1}, h_{\alpha_1}}$ with dimension $\Delta_1 =2h_{\alpha_1}$. More explicitly, from the AdS/CFT dictionary, the thermal two-point function on $S^1 \times \mathbb{R}$ in CFT is obtained by evaluating the regularized bulk geodesic length $L_{\text{reg}}$ between the two operators
\be
\begin{split} \label{eq:2pt_cylinder}
\langle O_{h_{\alpha_1}, h_{\alpha_1}}(z_1,\bar{z}_1) O_{h_{\alpha_1}, h_{\alpha_1}}(\bar{z}_1,z_1) \rangle_{\beta,\text{BTZ}}&=Z_{\text{BTZ}} e^{-\Delta_1  L_{\text{reg}}}~, \\
&= Z_{\text{BTZ}}\left( - \frac{\left(\frac{2\pi}{\beta}\right)^2}{\sinh^2\left(\frac{\pi}{\beta}(z_1 -\bar{z}_1) \right)} \right)^{h_{\alpha_1}}\left(  -\frac{\left(\frac{2\pi}{\beta}\right)^2}{\sinh^2\left(\frac{\pi}{\beta}(\bar{z}_1 -z_1) \right)} \right)^{h_{\alpha_1}}~.
\end{split}
\ee
Up to a normalization factor $e^{\frac{1}{3}(\ln 2-1)}$, we match the correlation functions in \eqref{eq:Liouville_light_classical} and \eqref{eq:2pt_cylinder}
\begin{equation}
    \mathcal{G}_{\alpha_1}^{(2)}(z_1) =e^{\frac{1}{3}(\ln 2-1)}\langle O_{h_{\alpha_1}, h_{\alpha_1}}(z_1,\bar{z}_1) O_{h_{\alpha_1}, h_{\alpha_1}}(\bar{z}_1,z_1)\rangle_{\beta,\text{BTZ}}~.
\end{equation}

\subsection{Torus identity block from crossing kernel and ensemble average}
From \cite{Collier:2023fwi}, the matching between 3d gravity and Liouville is beyond the semiclassical limit. Here, we use the simplest two-point function as an explicit example to illustrate this fact, i.e. 
\begin{equation}\label{eq:Liouville_correlators3}
\begin{aligned}
    \mathcal{G}^{(2)}_{\alpha_1}(z_1)& = \langle  \hat{V}_{\alpha_1}(z_1,\bar{z}_1)   \rangle_{\beta/2,ZZ}  \langle   \hat{V}_{\alpha_1}(\bar{z}_1,z_1)  \rangle_{ \beta/2, \widetilde{ZZ}}~.
\end{aligned}
\end{equation}
From the doubling trick and Cardy behavior of the ZZ wavefunctions, we show that the final expression takes the form of the two-point identity block on a torus.

Focusing on the chiral half, we have the following expression
\begin{equation}\label{eq:Liouville_correlators1}
\begin{aligned}
    \langle  \hat{V}_{\alpha_1}(z_1,\bar{z}_1) \rangle_{\beta/2,ZZ} 
    &=\int_0^\infty dP' dQ' \Psi_{ZZ}'(P') \Psi_{ZZ}^{* \prime}(Q') \langle \bra{P'} e^{-(\beta/2 - \tau_1)H }\hat{V}_{\alpha_1} e^{- \tau_1 H} \ket{Q'} \rangle\\
   &=\int_0^\infty dP' dQ'  \hat{C}_{DOZZ}(P',P_{\alpha_1},Q') \Psi_{ZZ}'(P') \Psi_{ZZ}^{* \prime}(Q') \mathcal{F}_{h_{P'},h_{Q'}}(h_{\alpha_1}, h_{\alpha_1},z_1 - \bar{z}_1,\beta)~,
\end{aligned}
\end{equation}
where $\tau_1 = (z_1 - \bar{z}_1)/2 i$ is the location of the operator $V_{\alpha_1}(z_1,\bar{z}_1)$ on the thermal circle and $P_{\alpha_1} = -i(\alpha_1-Q/2)$. $\hat{C}_{DOZZ}$ is the normalized DOZZ structure constant defined in Appendix \ref{app:Liouville}. From the doubling trick, the insertion of $V_{\alpha_1}(z_1,\bar{z}_1)$ between overlaps of Ishibashi states gives a torus conformal block, more explicitly,\footnote{The phase in front of the sum is due to the phase difference of $\pi/2$ between the $z$ coordinate and Euclidean time evolution. \cite{Ghosh:2019rcj}}
\be
\begin{aligned}
&\frac{ \langle \bra{P'} e^{-(\beta/2 - \tau_1)H }\hat{V}_{\alpha_1} e^{- \tau_1 H} \ket{Q'} \rangle}{\hat{C}_{DOZZ}(P',P_{\alpha_1},Q')}\\
&=e^{i\pi h_{\alpha_1}}\sum_{N_1,N_2} \frac{\bra{P',N_1,N_1}\hat{V}_{\alpha_1} \ket{Q',N_2,N_2}}{\bra{P'}\hat{V}_{\alpha_1} \ket{{Q'}}}\exp\left(-\left(\frac{\beta}{2} - \tau_1\right)\left(h_{P'}+N_1-\frac{c}{24}\right)-\tau_1 \left(h_{Q'}+N_2-\frac{c}{24}\right)\right)\\
&=e^{i\pi h_{\alpha_1}}\sum_{N_1,N_2} \bra{h_{P'},N_1} O_{h_{\alpha_1}} \ket{h_{Q'},N_2} \bra{h_{Q'},N_2} O_{h_{\alpha_1}} \ket{h_{P'},N_1}  \\
&\times\exp\left(-\left(\frac{\beta}{2} - \tau_1\right)\left(h_{P'}+N_1-\frac{c}{24}\right)-\tau_1 \left(h_{Q'}+N_2-\frac{c}{24}\right)\right)\\
&=\mathcal{F}_{h_{P'},h_{Q'}}(h_{\alpha_1}, h_{\alpha_1},z_1 - \bar{z}_1,\beta)~.
\end{aligned}
\ee
where $O_{h_{\alpha_1}}$ is a formal chiral operator with conformal weight $h_{\alpha_1}$ at the origin. The states $\ket{h_{P'},N_1}$ contain a whole orthonormal Virasoro module with conformal weight $h_{P'}$. In expressing the conformal block, we have chosen the normalization of the three point function $\bra{h_{P'}} O_{h_{\alpha_1}} \ket{h_{Q'}} $ to be unity.

Using the expression of the ZZ wavefunctions of rescaled Liouville operators labelled by $P'$, we obtain
\be
\begin{aligned}
 \langle  \hat{V}_{\alpha_1}(z_1,\bar{z}_1) \rangle_{\beta/2,ZZ} 
 &=\int_0^\infty dP' dQ' \sqrt{\rho_0(P')} \sqrt{\rho_0(Q')}  \hat{C}_{DOZZ}(P',P_{\alpha_1},Q') \mathcal{F}_{h_{P'},h_{Q'}}(h_{\alpha_1}, h_{\alpha_1},z_1 - \bar{z}_1,\beta)~,\\
 &=\int_0^\infty dP' dQ' \sqrt{\rho_0(P')} \sqrt{\rho_0(Q')} \hat{C}_{DOZZ}(P',P_{\alpha_1},Q') \vcenter{\hbox{
	\begin{tikzpicture}[scale=0.75]
	\draw[thick] (0,0) circle (1);
	\draw[thick] (-1,0) -- (-2,0);
	\node[above] at (-2,0) {$O_{h_{\alpha_1}}$};
	\node[above] at (0,1) {$P'$};
	\node[below] at (0,-1) {$Q'$};
	\draw[thick] (1,0) -- (2,0);
	\node[above] at (2,0) {$O_{h_{\alpha_1}}$};
	\node[scale=0.75] at (0,0) {$\beta$};
	\end{tikzpicture}
	}}~.
\end{aligned}
\ee
The prefactor in front of the torus conformal block is exactly the crossing kernel\cite{Ponsot:1999uf, Ponsot:2000mt, Collier:2019weq}, which subsequently gives us the torus identity block in the dual channel, 
\begin{equation}\label{eq:torus2ptIdentityCrossing}
\int_0^\infty dP' dQ' \sqrt{\rho_0(P')} \sqrt{\rho_0(Q')} \hat{C}_{DOZZ}(P',P_{\alpha_1},Q') \vcenter{\hbox{
	\begin{tikzpicture}[scale=0.75]
	\draw[thick] (0,0) circle (1);
	\draw[thick] (-1,0) -- (-2,0);
	\node[above] at (-2,0) {$O_{h_{\alpha_1}}$};
	\node[above] at (0,1) {$P'$};
	\node[below] at (0,-1) {$Q'$};
	\draw[thick] (1,0) -- (2,0);
	\node[above] at (2,0) {$O_{h_{\alpha_1}}$};
	\node[scale=0.75] at (0,0) {$\beta$};
	\end{tikzpicture}
	}}
= 
\vcenter{\hbox{
	\begin{tikzpicture}[scale=0.75]
	\draw[thick] (0,0) circle (1);
	\draw[thick] (0,1) -- (0,2);
	\draw[thick] (0,2) -- (0.866,2+1/2);
	\draw[thick] (0,2) -- (-0.866,2+1/2);
	\node[left] at (0,3/2) {$\mathbb{1}$};
	\node[left] at (-1.2,0) {$\mathbb{1}$};
	\node[left] at (-0.866,2+1/2) {$O_{h_{\alpha_1}}$};
	\node[right] at (0.766,2+1/2) {$O_{h_{\alpha_1}}$};
	\node[scale=1] at (0,0) {$\frac{4\pi^2}{\beta}$};
	\end{tikzpicture}
	}}  ~.
\end{equation}

This is indeed what we expect from the doubling trick.

It is known that gravitational observables have an ensemble-averaged interpretation in large c CFTs\cite{Collier:2022bqq, Collier:2023fwi}. We like to match the result obtained in \eqref{eq:torus2ptIdentityCrossing} with the ensemble averaged result. In particular, the thermal two-point function has the conformal block expansion
\be\label{eq:thermal_2-point}
\langle O_{h_{\alpha_1}, h_{\alpha_1}}(z,\bar{z}) O_{h_{\alpha_1}, h_{\alpha_1}}(\bar{z},z) \rangle_{\beta,\text{BTZ}} =\sum_{p,q} |c_{p1q}|^2\mathcal{F}_{h_p,h_q}(h_{\alpha_1},h_{\alpha_1},z_1-\bar{z}_1,\beta)\mathcal{F}_{h_{\bar{p}},h_{\bar{q}}}(h_{\alpha_1},h_{\alpha_1},\bar{z}_1-z_1,\beta)~,
\ee
where $h_p,h_q$ are the conformal weights of intermediate states. If we average \eqref{eq:thermal_2-point} over the large $c$ CFT ensemble introduced in \cite{Collier:2022bqq}, we can make the following replacement when performing the average
\be
\begin{aligned}
\sum_{p,q} &\rightarrow \int_{0}^\infty dP' d\Bar{P}' dQ' d\Bar{Q}' \, \rho_0(P') \rho_0(\Bar{P}')  \rho_0(Q')\rho_0(\Bar{Q}') \\
|c_{p1q}|^2 &\rightarrow C_0(h_{P'},h_{\alpha_1},h_{Q'}) \bar{C}_0(h_{\Bar{P}'},h_{\alpha_1},h_{\Bar{Q}'})\\
&=c_b^2 \frac{\hat{C}_{DOZZ}(P',P_{\alpha_1},Q') \hat{C}_{DOZZ}(\Bar{P}',P_{\alpha_1},\Bar{Q}') }{\sqrt{\rho_0(P') \rho_0(\Bar{P}')  \rho_0(Q')\rho_0(\Bar{Q}')}}~, \\
\mathcal{F}_{h_p,h_q}(h_{\alpha_1},h_{\alpha_1},z_1-\bar{z}_1,\beta) &\rightarrow \mathcal{F}_{h_{P'},h_{Q'}}(h_{\alpha_1},h_{\alpha_1},z_1-\bar{z}_1,\beta)~, 
\end{aligned}
\ee
where the density of states of the ensemble coincides with the modular S-matrix of Liouville and $C_0$ is the universal OPE function. We finally arrive at the expression of the averaged thermal two-point function
\begin{equation}
    \begin{split}
        &c_b^{-2}\overline{\langle O_{h_{\alpha_1}, h_{\alpha_1}}(z,\bar{z}) O_{h_{\alpha_1}, h_{\alpha_1}}(\bar{z},z) \rangle }_{\beta,\text{BTZ}} \\
        &\approx \Big{|}\int_{0}^\infty dP' dQ' \sqrt{\rho_0(P')  }\sqrt{\rho_0(Q')} \hat{C}_{DOZZ}(P',P_{\alpha_1},Q')\mathcal{F}_{h_{P'},h_{Q'}}(h_{\alpha_1},h_{\alpha_1},z_1-\bar{z}_1,\beta)\Big{|}^2~, \\
        &=  \langle  \hat{V}_{\alpha_1}(z_1,\bar{z}_1) \rangle_{\beta/2,ZZ}  \langle  \hat{V}_{\alpha_1}(\bar{z}_1,z_1) \rangle_{\beta/2,\widetilde{ZZ}} ~,
    \end{split}
\end{equation}
matching the result that we obtain from two copies of scalar one-point function inserted between two ZZ overlaps on a finite cylinder.\footnote{Similar results have also been obtained in \cite{Mertens:2022ujr}.} In contrast to the two-boundary wormhole calculation in \cite{Collier:2022bqq}, the ensemble average calculation for BTZ correlation functions doesn't involve delta functions as we are performing the trace on a finite cylinder, and hence, the intermediate states automatically have the same conformal weights. 

\subsection{ZZ boundary states for Virasoro conformal blocks on a sphere}
We have been focusing on obtaining observables and Hartle-Hawking state at finite temperature from the Liouville ZZ boundary states. In this subsection, we briefly illustrate on obtaining Virasoro conformal blocks on the sphere from the ZZ boundary states.\footnote{We thank Tom Hartman for suggesting this to us.} From Section \ref{subsec:Liouville_background} and Figure \ref{fig:zz conformal mapping}, the ZZ boundary condition on $|z| = 1 $ of the Poincare disk can be mapped to the real axis of the upper half plane using the holomorphic conformal transformation $w = f(z) = -i \frac{z-1}{z+1}$. In particular, we can compute the normalized two point function on the upper half plane with only one ZZ boundary condition on the real axis (which can also be interpreted as the overlap between the ZZ boundary state and vacuum) and obtain \cite{Zamolodchikov:2001ah}
\be \label{eq:sphere_4pt}
G_{\alpha_1,\alpha_2}(x) = \frac{\left\langle \hat{V}_{\alpha_{1}}(z_{1},\bar{z}_1)\hat{V}_{\alpha_{2}}(z_{2},\bar{z}_2)\right\rangle _{ZZ}}{\left\langle \hat{V}_{\alpha_{1}%
}(z_{1},\bar{z}_1)\right\rangle _{ZZ}\left\langle \hat{V}_{\alpha_{2}}(z_{2},\bar{z}_2)\right\rangle _{ZZ}}~,
\ee
where $x = \frac{(z_1-z_2)(\bar{z}_1-\bar{z}_2)}{(z_1-\bar{z}_2)(\bar{z}_1-z_2)}$ is the cross ratio.  This indeed matches with the doubling trick, as $G_{\alpha_1,\alpha_2}(x)$ is the chiral identity Virasoro block on the sphere in the dual channel
\begin{equation} \label{ZZ BLOCK}
    G_{\alpha_1,\alpha_2}(x)  = (1-x)^{2\Delta_{1}} \mathcal{F}\left(
\begin{array}
[c]{cc}%
h_{\alpha_1} & h_{\alpha_2}\\
h_{\alpha_1}  & h_{\alpha_2}%
\end{array},
\mathbb{1},1-x\right) ~.
\end{equation}

It's interesting to point out that although Liouville theory is not chaotic and has a flat spectrum, the information about gravitational scattering and quantum chaos is included in Liouville data. For example, we can study the out-of-time-ordered correlation function (OTOC) using the exact formula \eqref{ZZ BLOCK} in the context of quantum chaos in 2d CFTs \cite{Roberts:2014ifa}. As an exact formula, it doesn't have the ``forbidden singularity'' that appears in approximations\cite{Fitzpatrick:2016ive, Collier:2018exn}. More explicitly, the information that describes gravitational scattering and scrambling behind the horizon is encoded in operator ordering and the monodromy properties of CFT conformal blocks are all encoded in the R-matrix of Liouville theory \cite{Jackson:2014nla, Mertens:2017mtv, Teschner:2012em}.

\section{Wheeler-DeWitt Wavefunctions}\label{sec:WdW}
For 2d JT gravity, the Wheeler-DeWitt wavefunctions have been studied in \cite{Harlow:2018tqv,Yang:2018gdb}, and the authors found two bases that correspond to fixing the renormalized geodesic length $L$ and fixing energy $E$ respectively. Following a similar spirit as previous works, we propose the Wheeler-DeWitt wavefunctions to be represented by two bases: the $(\Phi_0,J)$-basis and $(E,J)$-basis where $\Phi_0$ is the analogue of renormalized geodesic length. In addition, $\Phi_0$ is related to the height and waist of a hyperbolic cylinder at a spatial slice in 3d, as we demonstrate soon. $E$ is the ADM energy and $J$ is the angular momentum of the black hole.

\subsection{Boundary value problem for the wavefunctions}\label{subsec:bvp_WdW}
In this subsection, we describe the boundary value problem for the study of Wheeler-DeWitt wavefunctions where the geometry of the wavefunction $(\mathcal{M},g_{\mu\nu})$ has two hypersurfaces   as its boundary:  $\mathcal{B}$ and $\Sigma$. We generalize the machinery developed in \cite{Brown:1992bq, Brown:1992br, Brown:2000dz} to Euclidean signature to study the boundary value problem.

The gravitational action is given by
\begin{equation}\label{eq:Sgrav_Phistates1}
    \begin{split}
        -S_{\text{grav}} &= \frac{1}{16 \pi G_N}\int_{\mathcal{M}} d^3 x\sqrt{g}\left(R+2\right)+ \frac{1}{8 \pi G_N}\int_{\mathcal{B}} d^2 x\sqrt{\gamma}\left(\Theta-1\right)\\
        &+ \frac{1}{8 \pi G_N}\int_{\Sigma} d^2 x\sqrt{h} K -\frac{1}{8 \pi G_N}\int_{\Gamma} dx \sqrt{\sigma_\Gamma} \frac{\pi}{2}~,
    \end{split}
\end{equation}
where $\Gamma = \mathcal{B} \cap \Sigma $ is composed of two disconnected codimension two joints and $(\sigma_\Gamma)_{AB}$ is the induced metric of $\Gamma$. With $\gamma_{ab}$ as the induced metric of $\mathcal{B}$ and $r^\mu$ as the outward pointing normal vector, the extrinsic curvature of $\mathcal{B}$ is given by $\Theta_{\mu\nu} = \gamma^\alpha_\mu \nabla_\alpha r_\nu$. With $h_{ij}$ as the induced metric of $\Sigma$ and $u^\mu$ as the outward pointing normal vector, the extrinsic curvature is given by $K_{\mu\nu} = h^\alpha_\mu \nabla_\alpha u_\nu$. We soon clarify the last term of \eqref{eq:Sgrav_Phistates1}, which is added by hand in defining the gravitational action for studying the wavefunction \cite{Takayanagi:2019tvn, Harlow:2018tqv}.

\begin{figure}[h]
	\centering
	\includegraphics[scale=0.4]{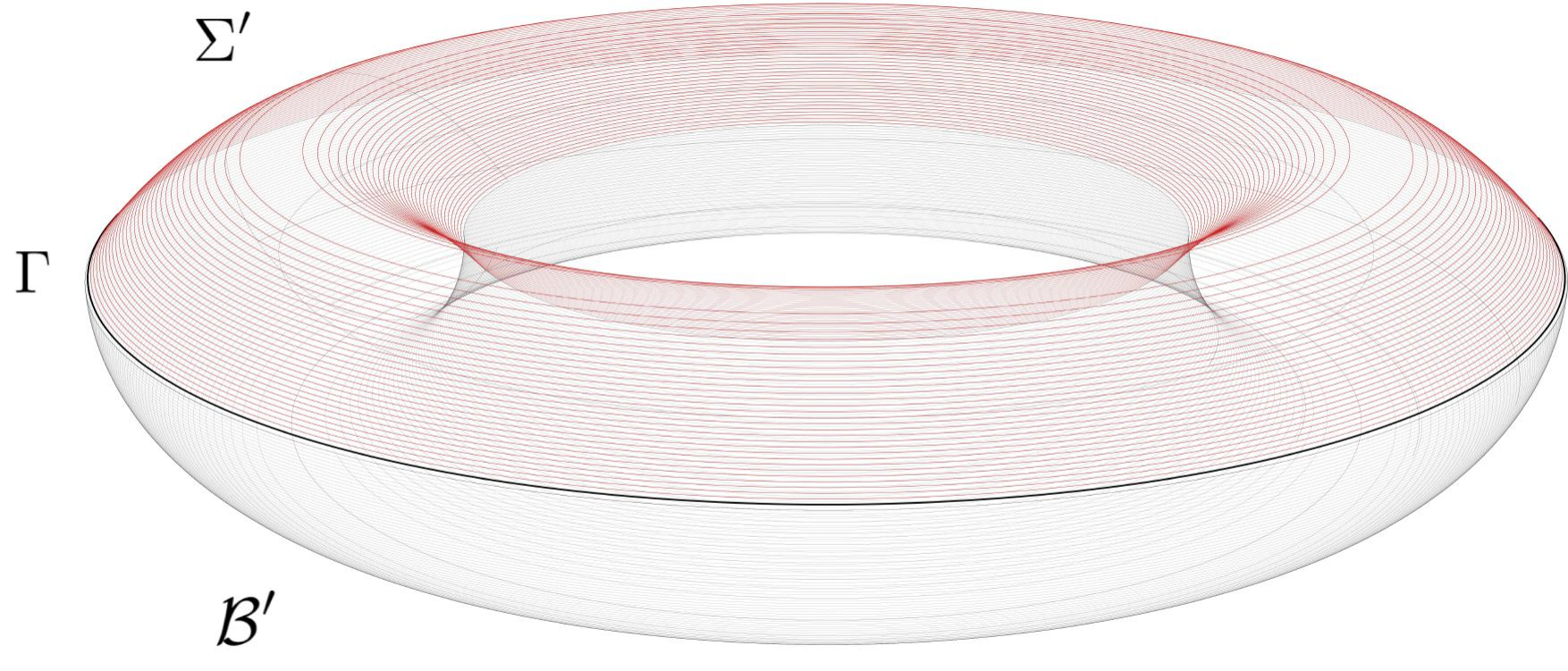}
	\caption{The geometry corresponding to the calculation of the 3d Wheeler-DeWitt wavefunctions. $\mathcal{B}'$ is the smooth part of the asymptotically AdS boundary, $\Sigma'$ is where we impose the boundary condition corresponding to the argument of the wavefunction and $\Gamma$ is the codimension two intersection between $\mathcal{B}$ and $\Sigma$.} \label{fig:HH_Wavefunction}
\end{figure}

Since we have the codimension two joints $\Gamma$ from the intersection of $\mathcal{B}$ and $\Sigma$, the Hayward corner terms arise in terms of the local angle $\theta$ at the joints when we smoothen out the region near $\Gamma$ \cite{Brown:2000dz,Hayward:1993my,Takayanagi:2019tvn},
 \begin{equation}\label{eq:Sgrav_Hayward}
 \begin{split}
     -S_{\text{Hayward}} &= \frac{1}{8 \pi G_N}\int_{\Gamma} dx \sqrt{\sigma_\Gamma}\theta = \frac{1}{8 \pi G_N}\int_{\Gamma} dx \sqrt{\sigma_\Gamma}\cos^{-1}(r \cdot u)~.
\end{split}     
 \end{equation}
 When the local angle at the two disconnected joints $\Gamma$ are right angles respectively, the contribution from the Hayward term is given by $\theta = \pi/2$, thus cancelling the last term in  \eqref{eq:Sgrav_Phistates1}. 
After the smoothening procedure, the two GHY boundary terms can be expressed as a sum of three terms, as shown in Figure \ref{fig:HH_Wavefunction},
\begin{equation}
    \frac{1}{8 \pi G_N}\left[\int_{\mathcal{B}} d^2 x\sqrt{\gamma}\left(\Theta-1\right)+\int_{\Sigma} d^2 x\sqrt{h} K \right] \rightarrow  \frac{1}{8 \pi G_N}\left[\int_{\mathcal{B}'} d^2 x\sqrt{\gamma}\left(\Theta-1\right)+\int_{\Sigma'} d^2 x\sqrt{h} K+ \int_{\Gamma} dx \sqrt{\sigma_\Gamma} \theta\right] ~,
\end{equation}
where $\mathcal{B}'$ is the part of region $\mathcal{B}$ excluding $\Gamma$ at the AdS boundary. This applies too for $\Sigma'$. 

We rewrite \eqref{eq:Sgrav_Phistates1} in terms of smooth extrinsic curvature components at the corresponding boundary regions\footnote{\eqref{eq:Sgrav_Phistates_smooth1} is similar to Trace $K$ action defined in (1.2) of \cite{Brown:2000dz}.}
\begin{equation}\label{eq:Sgrav_Phistates_smooth1}
\begin{split}
    -S_{\text{grav}} &= \frac{1}{16 \pi G_N}\int_{\mathcal{M}} d^3 x\sqrt{g}\left(R+2\right)+ \frac{1}{8 \pi G_N}\int_{\mathcal{B}'} d^2 x\sqrt{\gamma}\left(\Theta-1\right)\\
        &+ \frac{1}{8 \pi G_N}\int_{\Sigma'} d^2 x\sqrt{h} K +\frac{1}{8 \pi G_N}\int_{\Gamma} dx \sqrt{\sigma_\Gamma} \left(\theta-\frac{\pi}{2}\right)~.
\end{split}
\end{equation}
The variation of the action $\delta S_{\text{grav}}$ with respect to a metric variation $\delta g_{\mu\nu}$ is given by \cite{Brown:2000dz}
\begin{equation}\label{eq:deltaS_grav1}
	\begin{split}
		-\delta S_{\text{grav}} &= -\frac{1}{16 \pi G_N}\int_{\mathcal{M}} d^{3} x \sqrt{g}\left(G^{\mu\nu}- g^{\mu\nu}\right)\delta g_{\mu\nu} -\frac{1}{2}\int_{\mathcal{B}'} d^{2} x \sqrt{\gamma} T^{ab}\delta \gamma_{ab} \\
  &-  \int_{\Sigma'} d^{2} x P^{ij} \delta h_{ij}+\frac{1}{8 \pi G_N}\int_{\Gamma} dx\left(\theta-\frac{\pi}{2}\right) \delta \sqrt{\sigma_\Gamma}~,
	\end{split}
\end{equation}
where $G^{\mu\nu}$ is the Einstein tensor, $T^{ab}$ is the Brown-York stress tensor \cite{Brown:1992br}
\begin{equation}
	\begin{split}
		T^{ab} &= \frac{1}{8 \pi G_N}\left(\Theta^{ab} - \gamma^{ab} \Theta+ \gamma^{ab}\right)~,
	\end{split}
\end{equation}
and $P^{ij}$ is the momentum conjugate to $h_{ij}$
\begin{equation} \label{Pij}
    P^{ij} = \frac{\sqrt{h}}{16 \pi G_N}(K^{ij} - h^{ij} K)~.
\end{equation}
For the variational terms on the $\Sigma'$ surface, we like to further perform an ADM splitting for the induced metric $h_{ij}$
\begin{equation}\label{eq:ADM_hij}
    h_{ij}dx^i dx^j = M^2 d\rho^2 +\sigma_{AB}(d\zeta^A +W^A d\rho)(d\zeta^B +W^B d\rho)~,
\end{equation}
where $M$ is the  ``radial'' lapse function, $W^A$ is the radial shift and $\sigma_{AB}$ is the metric of a codimension two surface that foliates $\Sigma'$. The normal vector to this foliation is $n_i = M D_i \rho$, where $D_i$ is the covariant derivative compatible with $h_{ij}$. The variation of the induced metric $\delta h_{ij}$ follows a similar decomposition \cite{Brown:2000dz}
\begin{equation}
    	\delta h_{ij} = \frac{2}{M}n_i n_j \delta M + \frac{2}{M}\sigma_{A(i}n_{j)} \delta W^A +\sigma^A_{(i}\sigma^B_{j)}\delta \sigma_{AB}~,
\end{equation}
and we obtain the following equation
\begin{equation}\label{eq:deltaS_Sigma}
     -  \int_{\Sigma'}  d^{2} x P^{ij} \delta h_{ij} = \int_{\Sigma'}d^2 x \sqrt{\sigma}\left( l\delta M+j_A \delta W^A- \frac{1}{\sqrt{\sigma}}P^{ij}\sigma^A_{i}\sigma^B_{j}\delta \sigma_{AB}\right)~,
\end{equation}
where $l$ is the normal momentum density and $j_A$ is the tangential (angular) momentum density of the codimension two surface $\sigma_{AB}$
\begin{equation}
    \begin{split}
        l &=-\frac{2 P_{ij} n^i n^j }{\sqrt{h}}  ~, \\
j_A &= -\frac{2 P_{ij}\sigma_A^i n^j }{\sqrt{h}}  ~.
    \end{split}
\end{equation}

With the variational terms written down explicitly in \eqref{eq:deltaS_grav1}, we are ready to specify the boundary conditions imposed for the study of wavefunctions. 

$\mathcal{B}'$ is the surface that describes the asymptotically AdS boundary and we impose the usual Dirichlet boundary condition 
\begin{equation}\label{eq:AdS_bdy_cond}
    ds^2|_{\mathcal{B}'} = \frac{1}{\epsilon^2}(d\tau_E^2 + d\phi^2)+\mathcal{O}(1)~,
\end{equation}
where $\epsilon$ is the AdS cutoff. The topology of $\mathcal{B}'$ on the AdS boundary is a cylinder, where the range of $\tau_E$ is $\beta/2$ and the spatial circle has periodicity $\phi \sim \phi+2 \pi$.  

On $\Sigma'$, we first discuss boundary conditions corresponding to wavefunctions with zero angular momentum. The boundary conditions on $\Sigma'$ that corresponds to the fixed $\Phi_0$-basis is given by 
\begin{equation}\label{eq:bdy_Phi_HH1}
	\begin{split}
		M = 1~,~~j_A = 0,~~P^{ij}\sigma^A_i \sigma^B_j = 0~,
	\end{split}
\end{equation} 
where the last equation implies $n^\mu$ is a tangent vector for a geodesic curve in $\mathcal{M}$, i.e. $n^\mu \nabla_\mu n^\nu = 0$.\footnote{To show $n^\mu$ being the tangent of a geodesic curve in $\mathcal{M}$, we first note that from \eqref{Pij} and \eqref{eq:bdy_Phi_HH1},
\begin{equation}\label{eq:Sigma_Pdd}
\begin{split}
    \sigma_\mu^A \sigma_\nu^B K^{\mu\nu} - \sigma^{AB}K &= 0 \\
    K &= \sigma^{\mu\nu}\nabla_\mu u_\nu~, 
    \end{split}
\end{equation}
where the second line of \eqref{eq:Sigma_Pdd} comes from $\sigma_{AB}$ having only one component in 3d. From the original expression of $K_{\mu\nu}$, we have
\begin{equation}\label{eq:geodesic}
    \begin{split}
        K_{\mu\nu} &= h^\alpha_\mu \nabla_\alpha u_\nu \\
        &= (\sigma^\alpha_\mu +n^\alpha n_\mu )\nabla_\alpha u_\nu~, \\
        K &=  \sigma^{\mu\nu}\nabla_\mu u_\nu - u_\mu n^\alpha  \nabla_\alpha n^\mu ~,
    \end{split}
\end{equation}
and immediately from the last line of \eqref{eq:geodesic}, we see $n^\alpha  \nabla_\alpha n^\mu = 0$ to satisfy boundary conditions on $\Sigma'$. In deriving these relations, we have used the completeness relation, i.e. $\sigma^{\mu\nu} = \sigma^\mu_A \sigma^\nu_B \sigma^{AB}, n^\mu = h^\mu_i n^i,\cdots$ and the Leibniz rule, i.e. $n_\mu n^\alpha  \nabla_\alpha u^\mu=- u_\mu n^\alpha  \nabla_\alpha n^\mu$. } The range of $\rho$ on $\Sigma'$ is chosen to be $2 \ln \frac{2}{\epsilon} - \Phi_0$ where $-\Phi_0$ plays the role of a renormalized geodesic length in the $\rho$ direction.

The fixed $(E,J=0)$-states have the following boundary conditions on $\Sigma'$,\footnote{In writing down the boundary conditions, a gauge has been chosen. This is similar to the 2d case \cite{Harlow:2018tqv}. We like to clarify that $l_{\text{here}} = q_{\text{there}}$  in \cite{Chua:2023srl} as we use $q$ for something else in this paper.} 
\begin{equation} \label{E basis}
    l,j_A = 0~,~~\sigma_{AB} = 8 G_N E \cosh^2 \rho \, \delta_{A,\phi} \delta_{B,\phi}~,
\end{equation}
where the horizon with $r_+ = \sqrt{8 G_N E}$ is located at $\rho = 0$ and $\rho =\mp \ln \left(\frac{1}{\sqrt{2 G_N E} \epsilon}\right)\rightarrow \pm \infty$ is where the AdS boundary resides. $l = 0$ implies that the codimension two surface $\Gamma$ that foliates $\Sigma'$ is extremal in the direction parallel to $u^\mu$ and $j_A= 0$ implies the absence of angular momentum.\footnote{This differs from the $E$-states studied in \cite{Chua:2023srl}, where $W^A = 0$ is chosen instead of $j_A = 0$ here. $W^A = 0$ allows $j_A$ to take any value in the path integral, and at the saddle-point, $j_A = 0$. On the other hand, we choose to study the state that corresponds to $j_A=0$ and subsequently, match it with Liouville.}
 
On $\Gamma$, we impose Dirichlet boundary condition such that the induced metric is
\begin{equation}\label{eq:Gamma_Phi_bc}
    ds^2|_{\Gamma} = \frac{1}{\epsilon^2}d\phi^2+\mathcal{O}(1)~,
\end{equation}
where $\phi \in [0,2 \pi)$ parameterizes the spatial circle on $\Gamma$.

We can also study wavefunctions with fixed non-zero angular momentum. We need to further add the following boundary term to the action in  \eqref{eq:Sgrav_Phistates_smooth1} \cite{Brown:1992bq,Chua:2023srl},
\be
-\int_{\Sigma'}d^2 x \sqrt{\sigma}j_A  W^A~,
\ee
such that the following variational term on $\Sigma'$ in \eqref{eq:deltaS_Sigma} becomes
\be
 \int_{\Sigma'}d^2 x \sqrt{\sigma}j_A  \delta W^A \rightarrow -\int_{\Sigma'}d^2 x W^A \delta (\sqrt{\sigma}j_A)~.
\ee
We can thus impose the following boundary conditions for the $(\Phi_0,J)$-basis 
\begin{equation}\label{eq:bdy_Phi_HH2}
	\begin{split}
		M = 1~,~~\sqrt{\sigma}j_A = \frac{-i J}{2 \pi} \delta_{A,\phi},~~P^{ij}\sigma^A_i \sigma^B_j = 0~,
	\end{split}
\end{equation} 
where $J$ is imaginary for Euclidean solutions. For the $(E,J)$-basis in \cite{Chua:2023srl}, the following boundary conditions are imposed on $\Sigma'$,
\begin{equation} \label{E,J basis}
    l=0~,~~\sqrt{\sigma}j_A = \frac{-i J}{2 \pi} \delta_{A,\phi}~,~~\sigma_{AB} = r_+^2(E,J) \cosh^2 \rho \, \delta_{A,\phi} \delta_{B,\phi}~,
\end{equation}
where $r_+^2(E,J) = 4 G_N(E+\sqrt{E^2 - J^2})$.

\subsection{Geometry for the $\Phi_0$-basis}\label{subsec:Phi0_geometry}
We first show the solution for the $\Phi_0$-basis, and it takes the following metric parametrization
\begin{equation}\label{eq:metric_HH_Phi0_states}
    ds^2 = d\rho^2 + \cosh^2 \rho e^\Phi dz d\bar{z}~,
\end{equation}
where 
\begin{equation}\label{eq:HH_Liouville_sol1}
    e^\Phi = \left(\frac{\beta r_+}{2 \pi}\right)^2\frac{1}{\sin^2 \left(\frac{\beta r_+}{2 \pi} \text{Im}(z)\right)}~,
\end{equation}
and the parameter $r_+$ is determined by the following implicit equation
\begin{equation}\label{eq:Phi0_rp1}
    e^{\Phi_0} = \left(\frac{\beta r_+}{2 \pi}\right)^2\frac{1}{\sin^2 \left(\frac{\beta r_+}{4 }\right)}~.
\end{equation}

We like to emphasize that $\beta \neq 2 \pi/r_+$ in general, but the relation $\beta = 2 \pi/r_+$ only holds at the peak of the wavefunction, or at the saddle-point when we use overlaps of the wavefunction to compute the partition function. The range of $z$ is given by $\text{Re}(z) \in [0, 4 \pi^2/\beta); \text{Im}(z) \in [0 ,\pi/2],$\footnote{In principle, we have to further rescale the AdS cutoff $\epsilon$ to match with the cylindrical region defined in \eqref{eq:AdS_bdy_cond}. However, with the conformal ratio of the boundary cylinder being fixed to $\beta/4 \pi$, we choose not to keep track on the overall rescaling of the cylindrical region.} whereas $\rho$ has the range of  $\rho \in [-\rho_0(\Phi),\rho_0(\Phi)]$ with $\rho_0(\Phi) = \ln \frac{2}{\epsilon} - \frac{\Phi}{2}$. In fact, this geometry is a portion of a BTZ black hole solution with inverse temperature $2 \pi/r_+$ that has thermal length $\beta/2$ at the asymptotic boundary.  We remind ourselves that there are ``two'' halves of the AdS boundary slices defined by $\rho = \pm \rho_0(\Phi)$ respectively when we use the wormhole slicing to describe the spacetime geometry of non-rotating BTZ. We demonstrate the gluing of the ``two'' halves in gravity in a moment.

The $\mathcal{B}'$ surface, which is the asymptotically AdS boundary, is given by the union of the $\rho=\pm \rho_0(\Phi) $ surfaces with
\begin{equation}
ds^2|_{\mathcal{B}'} \approx \left(\frac{1}{\epsilon^2} + \frac{e^\Phi}{2}\right)dz d\bar{z} + \frac{1}{4}\left(\partial \Phi dz + \bar{\partial} \Phi d\bar{z}\right)^2~.
\end{equation}
The $\Sigma'$ slice is given by the $\text{Im}(z) =\frac{\pi}{2}$ surface, and the induced metric is
\begin{equation}\label{eq:metric_Sigma}
    ds^2|_{\Sigma'} = d\rho^2 + \cosh^2 \rho e^{\Phi_0} dX^2~,
\end{equation}
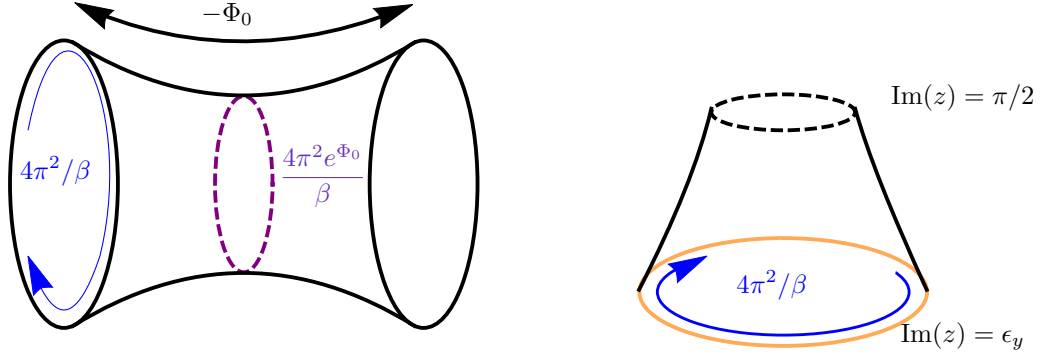
\begin{figure}
\begin{center}
\begin{overpic}[scale=0.8]{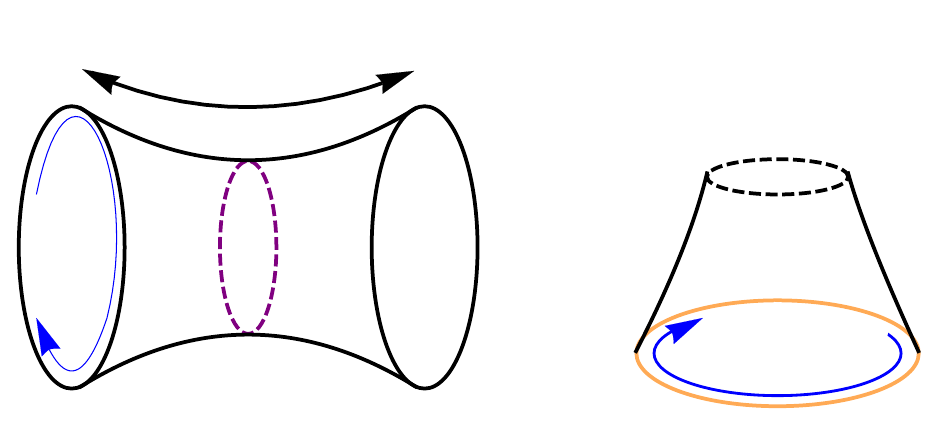}
\put(90,30){\parbox{0.2\linewidth}{
		\begin{equation*}
			\text{Im}(z) = \pi/2
\end{equation*}}}
\put(90,5){\parbox{0.2\linewidth}{
		\begin{equation*}
			\text{Im}(z) = \epsilon_y
\end{equation*}}}
\put(70,8){\parbox{0.2\linewidth}{
		\begin{equation*}
			\color{blue}{4 \pi^2/\beta}
\end{equation*}}}
\put(-5,20){\parbox{0.2\linewidth}{
		\begin{equation*}
			\color{blue}{4 \pi^2/\beta}
\end{equation*}}}
\put(13,38){\parbox{0.2\linewidth}{
		\begin{equation*}
			-\Phi_0
\end{equation*}}}
\put(23,20){\parbox{0.2\linewidth}{
		\begin{equation*}
		\color{purple}{	\frac{4 \pi^2 e^{\Phi_0}}{\beta}}
\end{equation*}}}
\end{overpic}  
\end{center}
\caption{(Left): $\Sigma'$ is a hyperbolic cylinder with renormalized height $-\Phi_0$ and waist $4 \pi^2 e^{\Phi_0}/\beta$. (Right): Each constant $\rho$ slice of the Hartle-Hawking wavefunction is a hyperbolic cylinder with ZZ-boundary conditions (indicated by orange line) at one end and on the other end at Im$(z) = \pi/2$, the Liouville field takes a constant value $\Phi_0$. The spatial circle is identified as $X \sim X+4 \pi^2/\beta$. }\label{fig:cylinderHH2}
\end{figure}
such that the boundary value problem in \eqref{eq:bdy_Phi_HH1} is satisfied and $X = \text{Re}(z)$. The Liouville field $\Phi$
takes a constant value of $\Phi_0$ on the $\Sigma'$ surface. Geometrically and as shown in Figure \ref{fig:cylinderHH2}, $\Sigma'$ is a hyperbolic cylinder with waist $4 \pi^2 e^{\Phi_0}/\beta$ and renormalized length $-\Phi_0$.\footnote{We thank Tom Hartman for useful discussions.}   

The geometry of the solution is given in Figure \ref{dome wavefunction}. Similar to the geometric picture we had in Section \ref{subsec:ZZ-cylinder-WH} in using the wormhole slicing to describe the BTZ partition function, the $\Phi_0$-basis Hartle-Hawking wavefunction can also be thought as being composed from constant $\rho$ slices, where each $\rho=$constant leaf is a part of a hyperbolic cylinder, as shown in Figure \ref{fig:cylinderHH2}. $\rho$ acts as an angle that rotates the $\rho=$constant leaf around the ZZ boundaries.\footnote{We thank Tom Hartman for useful discussions.} 

\begin{figure}[h]
	\centering
	\includegraphics[scale=0.5]{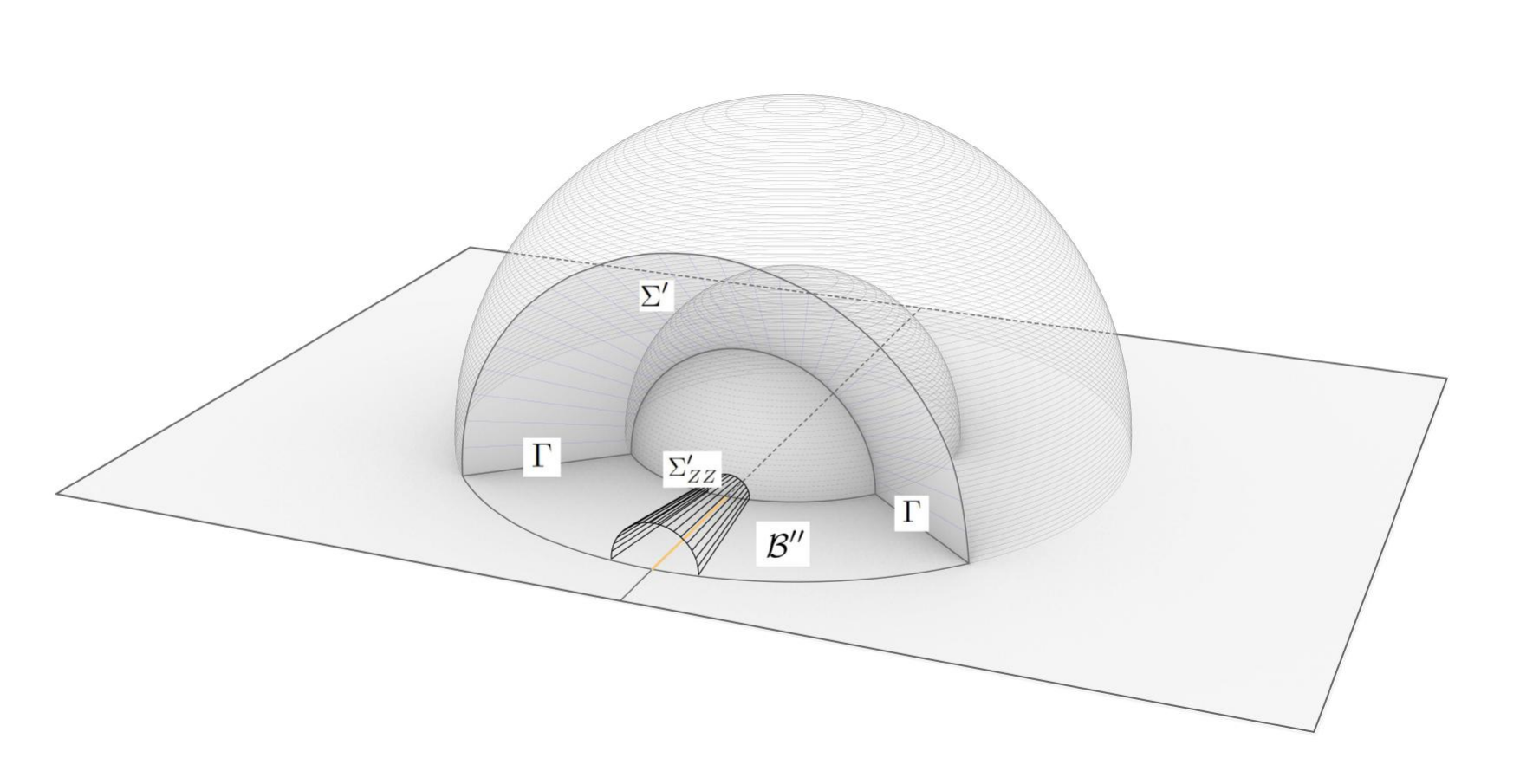}
	\caption{We illustrate the Wheeler-DeWitt wavefunction in the $\Phi_0$-basis as a portion of non-rotating BTZ. With the cutoff surface formed from the ZZ boundary conditions, we see that the ``two'' disconnected AdS boundaries $\mathcal{B}''$ are reduced to a connected $\mathcal{B}'$ surface when we shrink the ZZ surface to zero size. This justifies the fact of the ZZ boundary condition in Liouville being a ``no-boundary'' condition in gravity. } \label{fig:HH_Wavefunctioncutoff}
\end{figure}  

\begin{figure}[h]
	\centering
\includegraphics[scale=0.4]{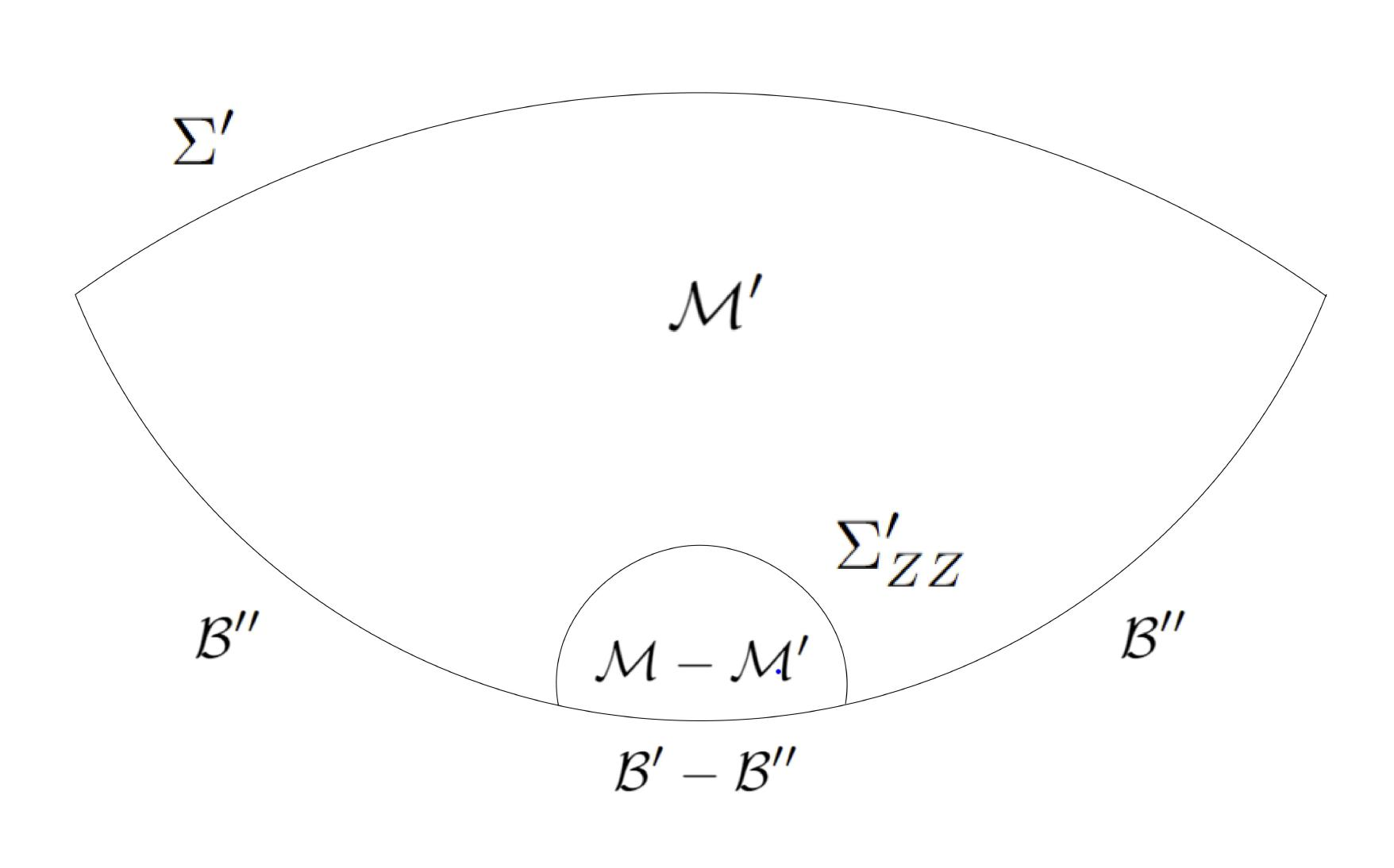}
	\caption{The cross section of the carved out cylindrical region $\mathcal{M}-\mathcal{M}'$ is shown. The bulk+AdS boundary terms of the cylindrical region is equivalent to the GHY contribution at the ZZ cutoff surface $\Sigma'_{ZZ}$. } \label{fig:ZZ_term}
\end{figure}

We are now ready to evaluate the on-shell action. First of all, we like to explain the problem related to the gluing of the ZZ boundaries in gravity. In using the wormhole slicing, we see that our solution for the metric diverges near $\text{Im} (z)=0$. This is an artifact in the choice of parametrizing the solution, which follows from our discussion on the BTZ partition function in Section \ref{subsec:ZZ-cylinder-WH}. We impose a cutoff at Im($z)=\epsilon_y$ and study the geometry in Figure \ref{fig:HH_Wavefunctioncutoff} in calculating the action. This corresponds to carving out a cylindrical region $\mathcal{M}-\mathcal{M}'$ in the bulk and we soon show the contribution of the bulk+AdS boundary terms of $\mathcal{M}-\mathcal{M}'$ to be given by the boundary term at the Im($z)=\epsilon_y$ cutoff surface. As shown in Figure \ref{fig:ZZ_term}, with the Im($z)=\epsilon_y$ surface formed from ZZ boundary conditions, we see the gluing of the ``two'' halves being manifested by shrinking $\mathcal{M}-\mathcal{M}'$ to zero size. The on-shell action in \eqref{eq:Sgrav_Phistates_smooth1} is then simplified to\footnote{The extrinsic curvature contribution at the surface formed from ZZ boundary conditions excludes the Hayward term contribution at the spacelike joint $\mathcal{B} \cap \Sigma_{ZZ}$. This is indicated through the prime label in $\Sigma'_{ZZ}$. We later will see with the gluing of the ZZ boundaries, there is no Hayward contribution at $\mathcal{B}'' \cap \Sigma_{ZZ}$. }
\begin{equation}\label{eq:Sgrav_Phistates3}
    \begin{split}
        -S_{\text{grav}}(\Phi_0) &= \frac{1}{16 \pi G_N}\int_{\mathcal{M'}} d^3 x\sqrt{g}\left(R+2\right)+ \frac{1}{8 \pi G_N}\int_{\mathcal{B''}} d^2 x\sqrt{\gamma}\left(\Theta-1\right)\\
        &+ \frac{1}{8 \pi G_N}\int_{\Sigma'_{ZZ}}^{\Sigma'} d^2 x\sqrt{h} K +\frac{1}{8 \pi G_N}\int_{\Gamma} dx \sqrt{\sigma_\Gamma}\left(\theta-\frac{\pi}{2}\right)~,
    \end{split}
\end{equation}
where $\mathcal{B''}$ is the part of $\mathcal{B'}$ with Im$(z) \geq \epsilon_y$ at $\rho=\pm \rho_0(\Phi)$.  We introduce the shorthand notation $\int_{\Sigma'_{ZZ}}^{\Sigma'}$ to indicate  $\int_{\Sigma'} - \int_{\Sigma'_{ZZ}}$ where $\Sigma'_{ZZ}$ is the surface at Im$(z) = \epsilon_y$. $\mathcal{M'}$ is the bulk region enclosed by the boundaries $\Sigma$, $\mathcal{B''}$ and $\Sigma_{ZZ}$.
In Appendix \ref{app:ZZ_gluing}, we show that 
\be
 \frac{1}{16 \pi G_N} \left(\int_{\mathcal{M} - \mathcal{M}'}d^3 x\sqrt{g}(R+2)+ 2\int_{\mathcal{B}' - \mathcal{B}''}d^2 x \sqrt{\gamma}(\Theta-1) \right)= - \frac{1}{8 \pi G_N}\int_{\Sigma'_{ZZ}} d^2 x \sqrt{h} K
\ee
such that the on-shell actions \eqref{eq:Sgrav_Phistates_smooth1} and \eqref{eq:Sgrav_Phistates3} are indeed the same. In fact, when we compute the local angle $\theta$ at the joint $\mathcal{B}'' \cap \Sigma_{ZZ}$ in the $\epsilon_y \rightarrow 0$ limit, we see that it vanishes when we shrink the carved out cylindrical region to zero size. Hence, in the $\epsilon_y \rightarrow 0$ limit, the ``two'' boundaries at $\pm \rho_0(\Phi)$ are glued together, and they combine into the smooth surface $B'$. If we start with the action in \eqref{eq:Sgrav_Phistates3} in defining the gravitational action for the study of wavefunction, we would have imposed the Dirichlet boundary condition for a disconnected $\mathcal{B}''$ and this is equivalent to the connected ``no-boundary'' AdS boundary condition, as argued in Section \ref{subsec:ZZ-cylinder-WH}.

Performing a similar calculation to Section \ref{subsec:heavy_operators}, we obtain the following expression from the bulk and AdS boundary terms
\begin{equation}\label{eq:AdS_Phi0}
\begin{split}
    &\frac{1}{16 \pi G_N}\int_{\mathcal{M'}} d^3 x\sqrt{g}\left(R+2\right)+ \frac{1}{8 \pi G_N}\int_{\mathcal{B}''} d^2 x\sqrt{\gamma}\left(\Theta-1\right) \\
    &= -\frac{1}{4 \pi G_N}\int_{B} dz d\bar{z} \left(\frac{1}{4}(\partial \Phi \bar{\partial} \Phi + e^\Phi) - \frac{1}{2}\bar{\partial}(\Phi \partial \Phi) - \partial \bar{\partial} \Phi \left(1+ \ln \frac{\epsilon}{2}\right)\right)~.
\end{split}
\end{equation}
where $B$ is the cylindrical domain: Re$(z) \in [0,4\pi^2/\beta]$,  Im$(z)  \in [\epsilon_y,\pi/2]$. For the contribution from the smooth components of $K$ on the $\Sigma$ surfaces, we obtain
\begin{equation}\label{eq:Sigma_Phi0}
    \frac{1}{8 \pi G_N}\int_{\Sigma'_{ZZ}}^{\Sigma'} d^2 x\sqrt{h} K =\frac{i}{4 \pi G_N}\left(\oint_{\text{Im}(z) = \pi/2} dz \left(\ln \frac{2}{\epsilon}-\frac{\Phi_0}{2}\right)\partial_{z}\Phi - \oint_{\text{Im}(z) = \epsilon_y} dz  \left(\ln \frac{2}{\epsilon}-\frac{\Phi_{ZZ}}{2}\right)\partial_{z}\Phi  \right)~.
\end{equation}

Given that $\Gamma$ is composed from two disconnected joints, the corresponding $\theta$ is given by the sum of the two contributions. With  $r_\mu^{\pm}$ being the outward unit normal to the $\rho = \pm\rho_0(\Phi)$ surfaces respectively and $u_\mu$ to be the forward pointing normal vector to the Im$(z)=$ constant surface, we obtain the Hayward term

\begin{equation}\label{eq:Hayward_Phi01}
    \begin{split}
        \frac{1}{8 \pi G_N}\int_{\Gamma} dx \sqrt{\sigma_\Gamma} \left(\theta-\frac{\pi}{2}\right)&= \frac{1}{8 \pi G_N \epsilon}\oint_{\text{Im}(z) = \pi/2} dz    \left( \cos^{-1} (r^+ \cdot u) +\cos^{-1} (r^- \cdot u)  - \pi\right)~, \\
         &= -\frac{i}{4 \pi G_N }\oint_{\text{Im}(z) = \pi/2} dz  \partial_z \Phi ~.
    \end{split}
\end{equation}
One can check that the local angle $\theta$ between $\mathcal{B}''$ and $\Sigma_{ZZ}$ vanishes in the $\epsilon_y \rightarrow 0$ limit as $ -r^\pm \cdot u |_{\mathcal{B}'' \cap \Sigma_{ZZ}}= 1$ when $\mathcal{M} - \mathcal{M}'$ is of zero size, hence, Hayward term is absent at this joint. 

Summing the terms above, we obtain the expression for the on-shell action in the $\Phi_0$-basis as
\begin{equation}\label{eq:Sgrav_Phi0}
    \begin{split}
        -S_{\text{grav}}(\Phi_0)  &= -\frac{1}{4 \pi G_N}\int_{B} dz d\bar{z} \left(\frac{1}{4}(\partial \Phi \bar{\partial} \Phi + e^\Phi)\right) + \frac{\pi}{G_N \beta \epsilon_y}~, \\
        &= \frac{\beta r_+^2(\Phi_0)}{16 G_N} + \frac{r_+(\Phi_0)}{2G_N }\cot \frac{\beta r_+(\Phi_0)}{4}~,
    \end{split}
\end{equation}
where $r_+(\Phi_0)$ is given by the implicit function in \eqref{eq:Phi0_rp1}.
Hence, the semiclassical Hartle-Hawking wavefunction in the $\Phi_0$-basis is given by
\begin{equation}
    \Psi_{\beta/2}^{\text{HH}}(\Phi_0) \approx e^{ \frac{\beta r_+^2(\Phi_0)}{16 G_N} + \frac{r_+(\Phi_0)}{2G_N }\cot \frac{\beta r_+(\Phi_0)}{4}}~,
\end{equation}
Similar to JT gravity, we see that the peak of the wavefunction occurs when the thermal length  $\beta/2$ takes half the value of the inverse temperature of the black hole, i.e. $\beta =  2 \pi/r_+$.  
\subsection{$\Phi_0$-basis wavefunction from Liouville theory}\label{subsec_Phi0}
\begin{figure}[h]
\begin{center}
\begin{minipage}[b]{0.4\linewidth}
\begin{overpic}[scale=0.2]{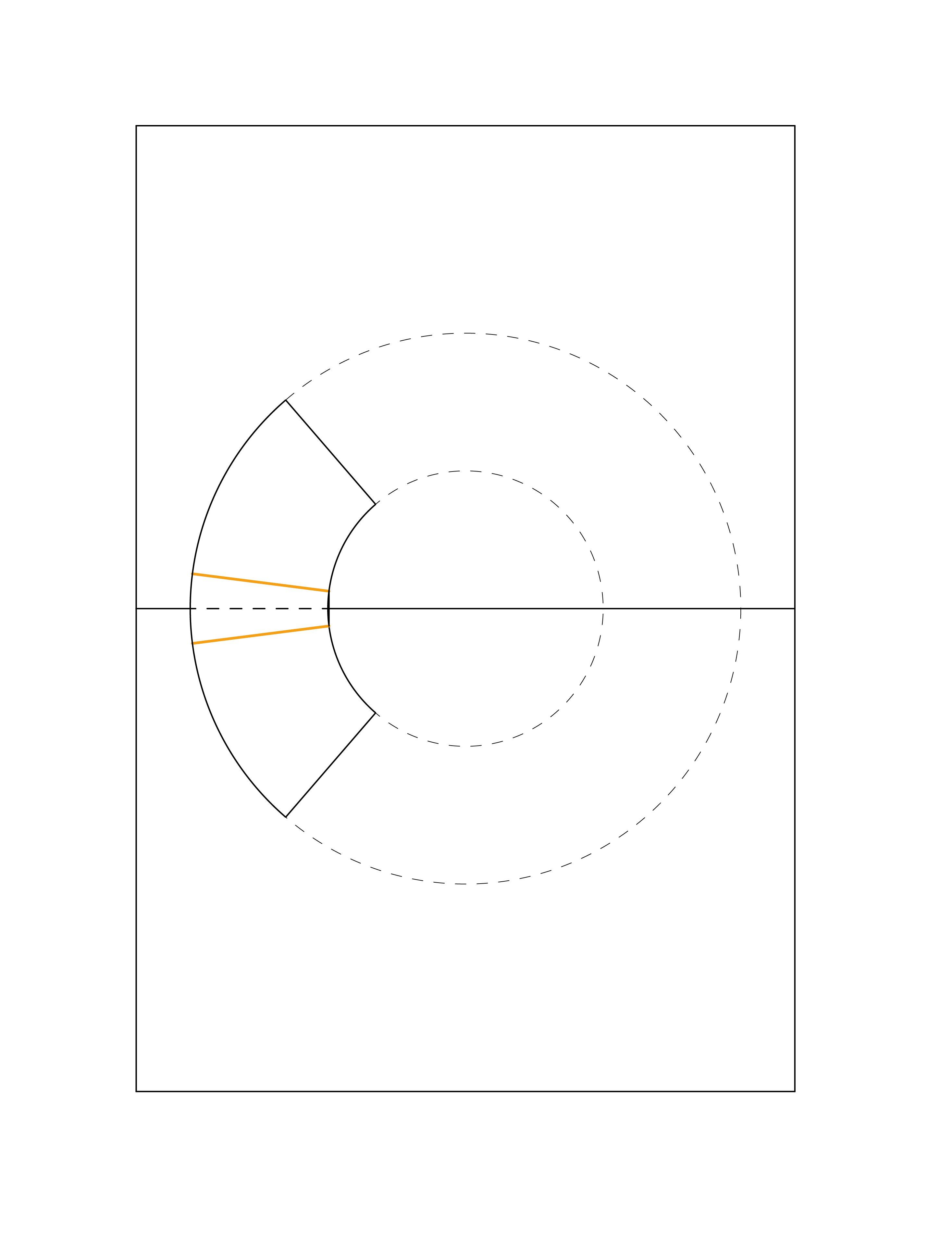}
\put(6,55){\parbox{0.2\linewidth}{
		\begin{equation*}
			\epsilon_y
\end{equation*}}}
\put(10,70){\parbox{0.2\linewidth}{
		\begin{equation*}
			\pi/2
\end{equation*}}}
\end{overpic} 
        \end{minipage}
        \begin{minipage}[b]{0.45\linewidth}
\begin{overpic}[scale=0.22]{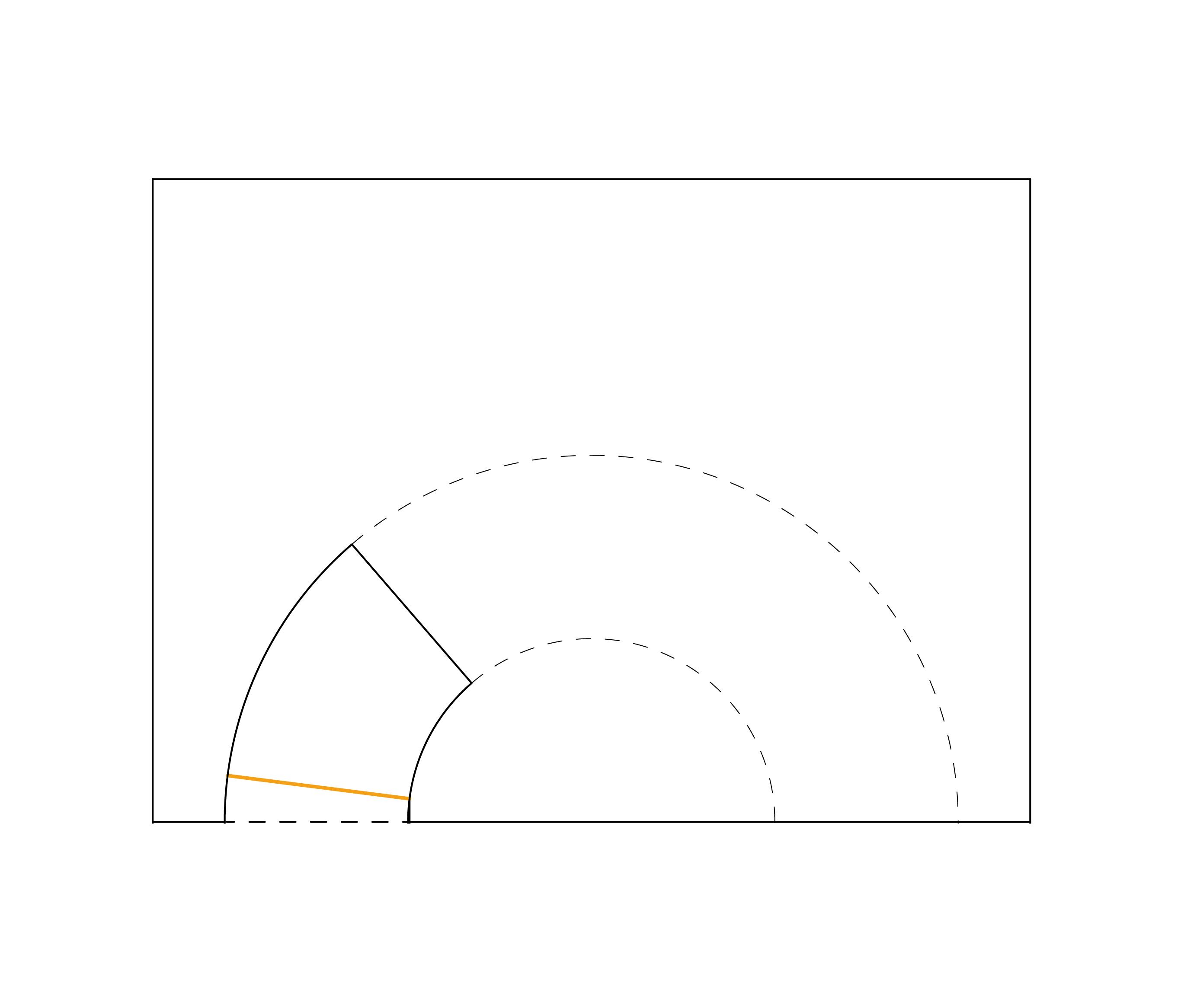}
\put(6,20){\parbox{0.2\linewidth}{
		\begin{equation*}
			\epsilon_y
\end{equation*}}}
\put(15,45){\parbox{0.2\linewidth}{
		\begin{equation*}
			\pi/2
\end{equation*}}}
\end{overpic}  
  \end{minipage}
\end{center}
\caption{(Left): The asymptotic boundary region of the wavefunction wormhole with the cutoff $\epsilon_y$. (Right): Using the folding trick, the anti-chiral (chiral) half of the Wheeler-DeWitt wavefunction is folded onto the upper half plane. Slicing the partition function open in half of the thermal direction is equivalent of having the height of the finite cylinder to be $\pi/2$, in addition to having the conformal ratio to be $\beta/4 \pi$. }\label{fig:Liouville_transition}
\end{figure}
As our gravitational solution takes the form of the wormhole slicing, we expect the gravity calculation to be directly connected with Liouville theory, similar to correlation functions in Section \ref{sec:correlation_function}. At $\rho \rightarrow \pm \rho_0(\Phi)$, we see from Figure \ref{fig:Liouville_transition} that the boundary geometry can be viewed as a transition amplitude on a cylinder in Liouville theory, where on one end of the cylinder, we have the ZZ boundary condition and on the other one, we have a constant $\Phi = \Phi_0$ slice.

More explicitly,  we consider the Liouville action
\begin{equation}\label{eq:Liouville_action_phi}
    \begin{split}
        -S_{\text{Liouville}}(\Phi_0) &= -\frac{1}{2 \pi} \int_B dz d\bar{z} \left(\frac{1}{4}(\partial \Phi \bar{\partial} \Phi + e^\Phi) \right)  + \frac{1}{4 \pi \epsilon_y }\oint_{\text{Im}(z) = \epsilon_y} dz  \Phi  + \frac{2 \pi \ln \epsilon_y}{\beta  \epsilon_y}+\frac{2 \pi}{\beta\epsilon_y}~.
    \end{split}
\end{equation}
With the expression of Liouville field obtained from Einstein's equations in \eqref{eq:HH_Liouville_sol1}, the on-shell Liouville action \eqref{eq:Liouville_action_phi} takes the following finite value
\begin{equation}\label{eq:Liouville_Phi0}
    -S_{\text{Liouville}}(\Phi_0) = \frac{\beta r_+^2(\Phi_0)}{8} + r_+(\Phi_0)\cot \frac{\beta r_+(\Phi_0)}{4}~,
\end{equation}
and the gravitational Wheeler-DeWitt wavefunction is related to the on-shell Liouville action by
\begin{equation} \label{grav and liouville on shell}
      e^{-S_{\text{grav}}(\Phi_0) }\approx e^{-\frac{1}{2 G_N}S_{\text{Liouville}}(\Phi_0)}~.
\end{equation}

Besides the on-shell Liouville action, we like to obtain the gravitational Wheeler-DeWitt wavefunction from Liouville overlaps. We consider the following state,
\be
\ket{\Phi_0}=\int_0^\infty dP  \langle   P|\Phi_0   \rangle \ket{P}\rangle
\ee
where $\langle  \Phi_0  | P \rangle $ is the Liouville zero mode wavefunction, which is obtained by solving the Liouville Hamiltonian for the eigenfunctions in the mini-superspace approximation where only the zero mode is kept, and we get \cite{Fateev:2000ik},\footnote{There is a rescaling of $2 \pi/\beta$ to $e^{\Phi_0/2}$ comparing to $\psi_{P}(\Phi_0)$ in \cite{Fateev:2000ik}. This comes from how the classical Liouville field varies when we perform a holomorphic coordinate transformation in changing the circumference of the spatial circle from $2 \pi$ to $4 \pi^2/\beta$.}
\begin{equation}
   \langle  \Phi_0  | P \rangle \equiv \psi_{P}(\Phi_0) = \frac{2\left(\frac{\pi \mu}{b^2} \right)^{-i P/b}}{\Gamma\left(-\frac{2 i P}{b} \right)}K_{\frac{2 i P}{b}}\left( \frac{2 \pi}{b^2 \beta}e^{\frac{\Phi_{0}}{2}}\right)~.
\end{equation}

The semiclassical Wheeler-DeWitt wavefunction in the $\Phi_0$-basis can be obtained by  considering the semiclassical limit of the overlap
\begin{equation}\label{eq:HH_Phi0}
\begin{split}
 \langle \Phi_0  |e^{-\beta H/4} | ZZ \rangle \langle \Phi_0   |e^{-\beta H/4}| \widetilde{ZZ} \rangle= \frac{1}{|\eta(\tau)|^2}\int_0^\infty dP \Psi^*_{ZZ}(P) \psi_{P }(\Phi_0)e^{-\beta P^2/2}\int_0^\infty d\widetilde{P}  \Psi^*_{\widetilde{ZZ}}( \widetilde{P})  \psi_{\widetilde{P}}(\Phi_{0})e^{-\beta \widetilde{P}^2/2}~,
\end{split}
\end{equation}
To get the small $b$ limit of the above expression, we first consider the integral formula for $K_{i \mu}\left(x \right)$ 
\begin{equation}
\begin{split}
K_{i \mu}\left(x \right) = \frac{1}{2}\int_{-\infty}^\infty d\xi e^{-x  \cosh \xi-i \mu \xi}~,
\end{split}
\end{equation}
and the semiclassical limit is contributed from primary fields,
\begin{equation}\label{eq:Liouville_overlap}
\begin{split}
\langle \Phi_0  |e^{-\beta H/4} | ZZ \rangle 
&\approx -2^{5/4}\int_{-\infty}^\infty d\xi \int_0^\infty dP P e^{\frac{2 \pi P}{b}-\frac{2 \pi}{b^2 \beta}e^{\frac{\Phi_0}{2}} \cosh \xi -\frac{\beta P^2}{2}-\frac{2 i P \xi}{b}}~.
\end{split}
\end{equation}
Performing a similar semiclassical analysis to the other copy, the saddle point result of the two copies of Liouville overlap is given by
\begin{equation}\label{eq:HH_saddle_Liouville}
    \begin{split}
         \langle \Phi_0  |e^{-\beta H/4} | ZZ \rangle \langle \Phi_0   |e^{-\beta H/4}| \widetilde{ZZ} \rangle &\approx \exp\left(\frac{\beta P^2_*}{2}+\frac{2  P_*}{b}\cot \left(\frac{b P_* \beta}{2}\right) +\frac{\beta \widetilde{P}^2_*}{2}+\frac{2 \widetilde{P}_*}{b}\cot \left(\frac{b \widetilde{P}_* \beta}{2}\right)\right)~,
    \end{split}
\end{equation}
where $P_*,\widetilde{P}_*$ satisfy the following relations at the saddle point
\begin{equation}\label{eq:P_saddle}
     P_*  =\frac{\pi}{ b \beta}e^{\Phi_0/2}\sin  \left(\frac{b P_* \beta}{2} \right)~,~~ \widetilde{P}_*  =\frac{\pi}{ b \beta}e^{\Phi_0/2}\sin  \left(\frac{b \widetilde{P}_* \beta}{2} \right)~.
\end{equation}

To match with the gravity and Liouville on-shell actions in \eqref{grav and liouville on shell}, it remains for us to identify the saddle point value of the Liouville momenta with the horizon radius via $P_*(\Phi_0) = r_+(\Phi_0)/2b$, which is similar to the relation found in \cite{Hadasz:2005gk}. We subsequently have
\begin{equation} \label{Liouvlle wavefunction as amplitude}
      \langle \Phi_0  |e^{-\beta H/4} | ZZ \rangle \langle \Phi_0   |e^{-\beta H/4}| \widetilde{ZZ} \rangle\approx e^{-\frac{c}{3}S_{\text{Liouville}}(\Phi_0)} \approx  e^{-S_{\text{grav}}(\Phi_0) }~,
\end{equation}
where $c=\frac{6}{b^2} \rightarrow+ \infty$ in the semiclassical limit. The last equation is obtained through the Brown-Henneaux central charge formula, $c =\frac{3}{2G_N}$.

In short, we have shown that the $\Phi_0$-state gravitational wavefunction can be viewed as two copies of ZZ-$\Phi_0$ Liouville transition amplitudes, which is similar to having two copies of Liouville observables as BTZ correlation functions in Section \ref{sec:correlation_function}.
\subsection{$(\Phi_0, J)$-basis Wheeler-DeWitt wavefunction}\label{subsec:Quasi_Fuchsian}
In this section, we study the Wheeler-DeWitt wavefunction in the $(\Phi_0, J)$ basis, which corresponds to the boundary conditions in \eqref{eq:bdy_Phi_HH2}.

The solution takes the form of a quasi-Fuchsian wormhole metric \cite{Collier:2022bqq}\footnote{We thank Tom Hartman for helpful discussions on the quasi-Fuchsian wavefunction.}
\begin{equation} \label{quasi solution}
    \begin{split}
        ds^2 =  d\rho^2 +\cosh^2 \rho e^{\Phi(z,\bar{z})}\big{|}dz + \frac{1}{2}(1+\tanh \rho)\bar{t}(\bar{z}) e^{-\Phi(z,\bar{z}) } d\bar{z}\big{|}^2~,
    \end{split}
\end{equation}
where $\Phi(z,\bar{z})$ is again the Liouville solution \eqref{eq:HH_Liouville_sol1} on a finite cylinder and $\Phi_0$ is given by \eqref{eq:Phi0_rp1}. 

In general, with the Beltrami coefficient $\mu = \bar{t}(\bar{z})  e^{-\Phi}$, the complex structure on the ``right'' boundary is deformed, as
\begin{equation}
    \begin{split}
        ds^2 \approx \begin{cases}
            d\rho^2 + \frac{1}{4}e^{-2 \rho+\Phi} |dz|^2 ~,~~&\rho \rightarrow -\infty \\
            d\rho^2 + \frac{1}{4}e^{2 \rho+\Phi} |dz +\mu d\bar{z}|^2~,~~&\rho \rightarrow +\infty~.
        \end{cases}
    \end{split}
\end{equation}
If we choose a similar $z$-dependent AdS cutoff $\rho = -\ln \frac{2}{\epsilon}+\frac{\Phi}{2}$ for the ``left'' boundary, we have a flat induced metric at leading order in $\epsilon$ 
\begin{equation}
    ds^2_{\text{left}} = \frac{1}{\epsilon^2}|dz|^2~.
\end{equation}
We put quotes for ``left'' and ``right'' since the Hartle-Hawking wavefunction has only one asymptotic boundary, and the ``two'' boundaries are glued together, as shown in Section \ref{gluing} and Section \ref{subsec:Phi0_geometry}. 

For the ``right'' boundary, we first let $w(z,\bar{z})$ be a solution to the Beltrami equation,
\begin{equation}
    \mu = \frac{\bar{\partial} w}{\partial w}~,
\end{equation}
such that
\begin{equation}
    |dz +\mu d\bar{z}|^2 = \Big{|}\frac{\partial w}{\partial z}\Big{|}^{-2} |dw|^2~.
\end{equation}
If we choose a different cutoff $\rho = \ln\frac{2}{\epsilon}-\frac{\Phi}{2} +\ln |\frac{\partial w}{\partial z}|$ for the ``right'' boundary, the metric is again flat,
\begin{equation}
    ds^2_{\text{right}} = \frac{1}{\epsilon^2}|dw|^2~.
\end{equation}
The solution to our boundary value problem has $\bar{t}(\bar{z}) = it=8 G_N J$ and the solution to the Beltrami equation is given by
\begin{equation}\label{eq:Beltrami}
    w(z,\bar{z}) = \bar{z}+\frac{2 i \cot^{-1}\left(\frac{ \frac{\beta r_+}{2 \pi} \cot \left(  \frac{\beta r_+}{2 \pi} \text{Im}(z)\right)}{\sqrt{\left(\frac{\beta r_+}{2 \pi}\right)^2 + i t}}\right)}{\sqrt{\left(\frac{\beta r_+}{2 \pi}\right)^2 + i t}}~.
\end{equation}
Close to $\Sigma_{ZZ}$, i.e. $\text{Im}(z) = \epsilon_y$, the Beltrami coefficient vanishes and we obtain
\begin{equation}
    w(z,\bar{z})|_{\text{Im}(z) = \epsilon_y}= z~.
\end{equation}

We are ready to compute the gravitational action with the quasi-Fuchsian metric as our solution for the Wheeler-DeWitt wavefunction. The on-shell action takes a similar form as \eqref{eq:Sgrav_Phistates3}
\begin{equation}\label{eq:Sgrav_Phistates4}
\begin{split}
    -S_{\text{grav}}(\Phi_0,J) &= \frac{1}{16 \pi G_N}\int_{\mathcal{M}'} d^3 x\sqrt{g}\left(R+2\right)+ \frac{1}{8 \pi G_N}\int_{\mathcal{B}''} d^2 x\sqrt{\gamma}\left(\Theta-1\right)\\
        &+ \frac{1}{8 \pi G_N}\int_{\Sigma'_{ZZ}}^{\Sigma'} d^2 x\sqrt{h} K -\int_{\Sigma'} d^2 x \sqrt{\sigma} j_A W^A  +\frac{1}{8 \pi G_N}\int_{\Gamma} dx \sqrt{\sigma_\Gamma} \left(\theta-\frac{\pi}{2}\right)~.
\end{split}
\end{equation}
As mentioned, a $z$ dependent cutoff for $\rho$ is chosen such that the induced metric at the asymptotic boundary $\mathcal{B}'$ is flat. The cutoff surfaces are defined as $\rho = - \ln \frac{2}{\epsilon}+\frac{\Phi}{2}$ for the ``left'' boundary and  $\rho = \ln \frac{2}{\epsilon}-\frac{\Phi}{2}+\ln |\frac{\partial w}{\partial z}|$ for the ``right'' boundary. 

For $\Sigma'$, the induced metric is given by
\begin{equation}
    ds^2|_{\Sigma'} = d\rho^2 + \cosh^2 \rho e^\Phi\left(1+ \frac{1}{4}(1+\tanh \rho)^2 t^2 e^{-2 \Phi}\right)dX^2~,
\end{equation}
such that the boundary conditions in \eqref{eq:bdy_Phi_HH2} are satisfied. 

We can simplify the computation using the decomposition of the Ricci scalar in \eqref{eq:R_decomp} and extrinsic curvature splittings in \eqref{eq:K_splitting}, and the action in \eqref{eq:Sgrav_Phistates4} becomes \eqref{eq:ADM_Fuchsian}
\begin{equation}\label{eq:ADM_Quasi_Fuchsian}
\begin{split}
        -S_{\text{grav}}(\Phi_0,J) &= \frac{1}{16 \pi G_N}\Big{(}\int_{\mathcal{M}'} d^3 x\sqrt{g}\left(R^{(2)}+K^2 - K_{\mu\nu}K^{\mu\nu}+2\right)+2\int_{\mathcal{B}''} d^2 x\sqrt{\gamma}\left(\gamma_v k-1 -  t^\mu \nabla_\mu \alpha\right) \\
     &+2 \int_{\Gamma} dx \sqrt{\sigma_\Gamma} \left(\theta-\frac{\pi}{2}\right)\Big{)}~,
\end{split}
\end{equation}
where the boundary terms on $\Sigma'$ and $\Sigma'_{ZZ}$ cancel with the total derivative contribution from the bulk and the added boundary term vanishes as $W^A = 0$. 

Using the quasi-Fuchsian metric in \eqref{quasi solution}, we obtain
\begin{equation}\label{eq:qF_HH_wavefunction}
\begin{split}
        -S_{\text{grav}}(\Phi_0,J)&=  -\frac{1}{16 \pi G_N}\int_{B}dz d\bar{z}\left(\frac{1}{1+t^2 e^{-2\Phi}}\partial \Phi \bar{\partial} \Phi+e^{ \Phi}\right)+\frac{\pi}{ G_N \beta \epsilon_y}~, \\
        &= \frac{r_+(\Phi_0)}{4 G_N}\text{Re}\left[\sqrt{1+\frac{4 i \pi^2 t}{\beta^2 r_+^2(\Phi_0)}}\tan^{-1} \left(\sqrt{1+\frac{4 i \pi^2 t}{\beta^2 r_+^2(\Phi_0)}} \tan \frac{\beta r_+(\Phi_0)}{4}\right)\right]+\frac{r_+(\Phi_0)}{2 G_N}\cot \frac{\beta r_+(\Phi_0)}{4}~,
\end{split}
\end{equation}
Next, we want to reproduce this gravity result from Liouville theory. With different moduli between the ``two'' asymptotic boundaries, we remind ourselves that  we must take into account a mixing of moduli between the ``two'' asymptotic boundaries \cite{Collier:2022bqq, Cotler:2018zff}. This suggests us to propose the following matching between 3d gravity and Liouville theory if the two asymptotic boundaries have different moduli
\begin{equation} \label{quasi sum}
    \begin{split}
        -S_{\text{grav}}(\Phi_0,J) = -\frac{c}{6}S_{\text{Liouv}(z,\bar{w})}(\Phi_0,J)-\frac{c}{6}S_{\text{Liouv}(w,\bar{z})}(\Phi_0,J) ~,
    \end{split}
\end{equation}
where $S_{\text{Liouv}(z,\bar{w})}(\Phi_0,J)$ is the Liouville action that governs a Liouville field $\Phi_-$ in a flat complex metric of $dz d\bar{w}$. $S_{\text{Liouv}(w,\bar{z})}(\Phi_0,J)$ is the complex conjugate of $S_{\text{Liouv}(z,\bar{w})}(\Phi_0,J)$.

The actions for the two Liouville fields $\Phi_\pm$ are given by
\begin{equation}
    \begin{split}
        -S_{\text{Liouv}(z,\bar{w})}(\Phi_0,J) &= -\frac{1}{2 \pi}\int_{B} dz d\bar{w} \left(\frac{1}{4}\left(\partial_z \Phi_- \partial_{\bar{w}}\Phi_- +e^{\Phi_-}\right)\right)+\frac{1}{4 \pi \epsilon_y}\oint_{\text{Im}(z) = \epsilon_y} dz \Phi_-+\frac{2 \pi}{ \beta \epsilon_y}+\frac{2 \pi\ln \epsilon_y}{\beta \epsilon_y}~, \\
        -S_{\text{Liouv}(w,\bar{z})}(\Phi_0,J) &= -\frac{1}{2 \pi}\int_{B} dw d\bar{z} \left(\frac{1}{4}\left(\partial_w \Phi_+ \partial_{\bar{z}}\Phi_+ +e^{\Phi_+}\right)\right)+\frac{1}{4 \pi \epsilon_y}\oint_{\text{Im}(w) = \epsilon_y} dw \Phi_++\frac{2 \pi}{ \beta \epsilon_y}+\frac{2 \pi\ln \epsilon_y}{\beta \epsilon_y}~, 
    \end{split}
\end{equation}
where the two Liouville fields satisfy the following classical equations
\begin{equation}\label{eq:Liouville_eom_QF}
    \begin{split}
        \partial_z \partial_{\bar{w}}\Phi_- &= \frac{e^{\Phi_-}}{2}~, \\
        \partial_w \partial_{\bar{z}}\Phi_+ &= \frac{e^{\Phi_+}}{2}~.
    \end{split}
\end{equation}
If we use the solution to the Beltrami equation in \eqref{eq:Beltrami}, we obtain
\begin{equation}\label{eq:Liouville_action_QF}
    \begin{split}
        -S_{\text{Liouv}(z,\bar{w})}(\Phi_0,J)
        &= -\frac{1}{8 \pi}\int_{B} dz d\bar{z} \left(\frac{\partial \Phi \bar{\partial} \Phi}{1-i t e^{-\Phi}}+e^\Phi\right)+\frac{1}{4 \pi \epsilon_y}\oint_{\text{Im}(z) = \epsilon_y} dz \Phi+\frac{2 \pi }{ \beta \epsilon_y}+\frac{2 \pi\ln \epsilon_y}{\beta \epsilon_y}~, \\
        -S_{\text{Liouv}(w,\bar{z})}(\Phi_0,J) 
        &= -\frac{1}{8 \pi}\int_{B} dz d\bar{z} \left(\frac{\partial \Phi \bar{\partial} \Phi}{1+i t e^{-\Phi}}+e^\Phi\right)+\frac{1}{4 \pi \epsilon_y}\oint_{\text{Im}(z) = \epsilon_y} dz \Phi+\frac{2 \pi}{ \beta \epsilon_y}+\frac{2 \pi \ln \epsilon_y}{\beta \epsilon_y}~,
    \end{split}
\end{equation}
where we have used the solutions to the Liouville equation in \eqref{eq:Liouville_eom_QF}
\begin{equation}
\begin{split}
    \Phi_-(z,\bar{w}(z,\bar{z})) &= \Phi+\ln(1 -i t e^{-\Phi})~, \\
    \Phi_+(w(z,\bar{z}),\bar{z})&= \Phi+\ln(1 +i t e^{-\Phi})~.
\end{split}
\end{equation}
In addition, we use $\partial \Phi = -\bar{\partial} \Phi$ to simplify some relations. Combining the two Liouville actions with permuted moduli in \eqref{eq:Liouville_action_QF}, we match the gravitational action as in \eqref{quasi sum}.

\subsection{$(E,J)$-basis Wheeler-DeWitt wavefunction}
\begin{figure}[h]
	\centering
	\includegraphics[scale=0.4]{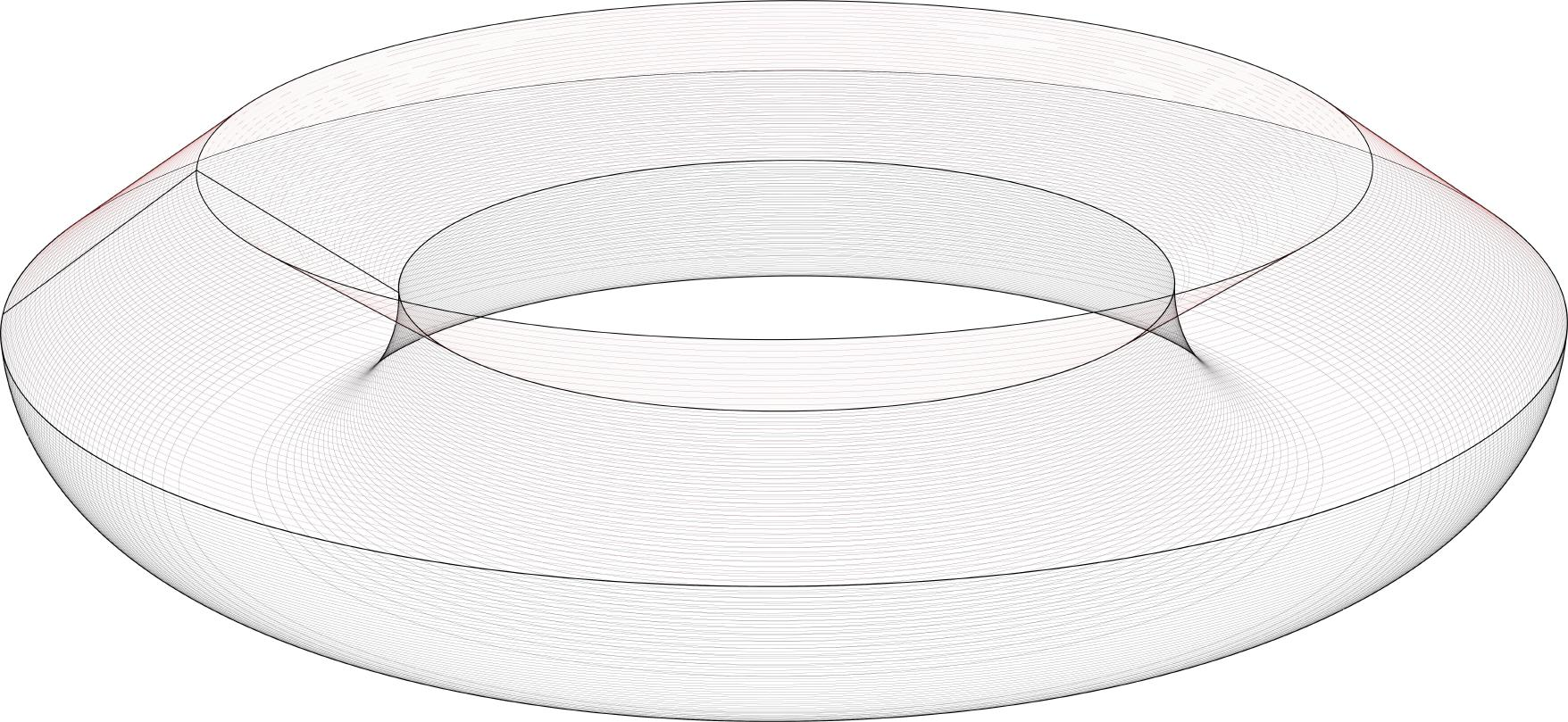}
	\caption{The ``Pacman'' geometry that corresponds to fixed $(E,J)$-basis Wheeler-DeWitt wavefunctions in 3d gravity.} \label{pacman}
\end{figure}  

The boundary value problem in \eqref{E,J basis} and its higher dimensional analogue is studied in detail in \cite{Chua:2023srl}. As shown in Figure \ref{pacman}, the solutions are given by the ``Pacman'' geometry, which is a wedge of rotating BTZ black holes,\footnote{Although our Hartle-Hawking state corresponds to a ``ket state'' of non-rotating black holes, the geometry for wavefunctions in general $(E,J)$ basis is a portion of rotating black holes, as the ``bra state'' has non-zero angular momentum. } labeled by ADM mass $E$ and angular momentum $J$, generalizing the 2d result in \cite{Harlow:2018tqv}. If we choose $J = 0$, it will be the ``Pacman'' geometry of non-rotating BTZ, corresponding to the study of \eqref{E basis}. 

The semi-classical wavefunction in the $(E,J)$-basis is given by,
\begin{equation}
    \Psi_{\beta/2}^{\text{HH}}(E,J) \approx e^{S(E,J)/2 -\beta E/2}~.
\end{equation}
where $S(E,J)$ and $E$ are the entropy and ADM energy of the corresponding black hole respectively. The physical meaning of the boundary condition in \eqref{E,J basis} corresponds to fixing energy and angular momentum of a black hole and hence, we expect to reproduce the gravity wavefunction from the overlap of the Hartle-Hawking state with Liouville primary states $|P'\rangle$.\footnote{In principle, by considering general Banados solutions, we can also try to get geometries that correspond to the overlap of the Hartle-Hawking state with descendent states \cite{Banados:1998gg, Compere:2015knw, Sheikh-Jabbari:2016unm}. Banados geometries are related to rotating BTZ black hole solutions through large diffeomorphisms, which can be interpreted as boundary graviton excitations. The large diffeomorphisms will also provide  the uniformization coordinate to match the gravity calculation with Liouville overlaps on a finite cylinder. We leave a more detailed analysis on this problem to the future.} Indeed, the on-shell gravitational result can be reproduced from the large c limit of 
\begin{equation} 
\begin{split}
\langle P' |e^{-\beta H/4} | ZZ \rangle \langle \widetilde{P}' | e^{-\beta H/4}| \widetilde{ZZ} \rangle &= \Psi_{ZZ}(P')e^{-\beta P'^2/2} \Psi_{\widetilde{ZZ}}(\widetilde{P}')e^{-\beta \widetilde{P}'^2/2}~,
\end{split}
\end{equation}
where we analytically continue the Liouville momenta\cite{Harlow:2011ny}, with $P^{\prime *} = \widetilde{P}'$, and $P',\widetilde{P}'$ are related to the energy $E$ and angular momentum $J$ through
\begin{equation}
    E = P'^2 + \widetilde{P}'^2~,~~J = P'^2 - \widetilde{P}'^2~,~~S(E,J) = \frac{2 \pi(P'+\widetilde{P}')}{b}~.
\end{equation}
In fact, the $\Sigma$ slice is composed of two constant time slices at $\tau_E=0$ and $\tau_E=\beta/2$ in rotating BTZ, and the two slices are glued at the horizon \cite{Chua:2023srl}. The rotating BTZ metric is given by \cite{Banados:1998gg}
\begin{equation} \label{banados btz}
\begin{split}
ds^2 =\left(-\frac{(\gamma+\bar{ \gamma })^2}{4}+\gamma \bar{ \gamma } \cosh^2\rho \right) d\tau_E^2+\frac{i(\gamma^2 - \bar{ \gamma }^2)}{2}d\tau_E d\phi+d\rho^2 + \left(\frac{(\gamma+\bar{ \gamma })^2}{4}+ \gamma \bar{ \gamma } \sinh^2 \rho\right)d\phi^2~,
\end{split}
\end{equation}
where $r^2  =\left(\frac{(\gamma+\bar{ \gamma })^2}{4}+ \gamma \bar{ \gamma } \sinh^2 \rho\right) $ is the radial direction of the black hole and $r \rightarrow + \infty$ denotes the location of the asymptotic boundary. On each fixed time  $\tau_E$ slice, the spatial metric is exactly \eqref{bottleneck} if we identify $\gamma^2 = \frac{24}{c}P^{\prime 2}, \widetilde{\gamma}^2 = \frac{24}{c}\widetilde{P}^{\prime 2}$ where $\gamma,\widetilde{\gamma}$ are finite in the semiclassical limit. This aligns with our claim in identifying quantum states using spatial geometries in Section \ref{subsec:brickwall} as the overlap with $\bra{P'} \bra{\tilde{P}'}$  extracts individual contribution from the superposition of geometries in the Hartle-Hawking state. 


\subsection{Reduction to 2d JT gravity}

\begin{figure}
\begin{center}
\begin{minipage}[b]{0.45\linewidth}
\begin{overpic}[scale=0.38]{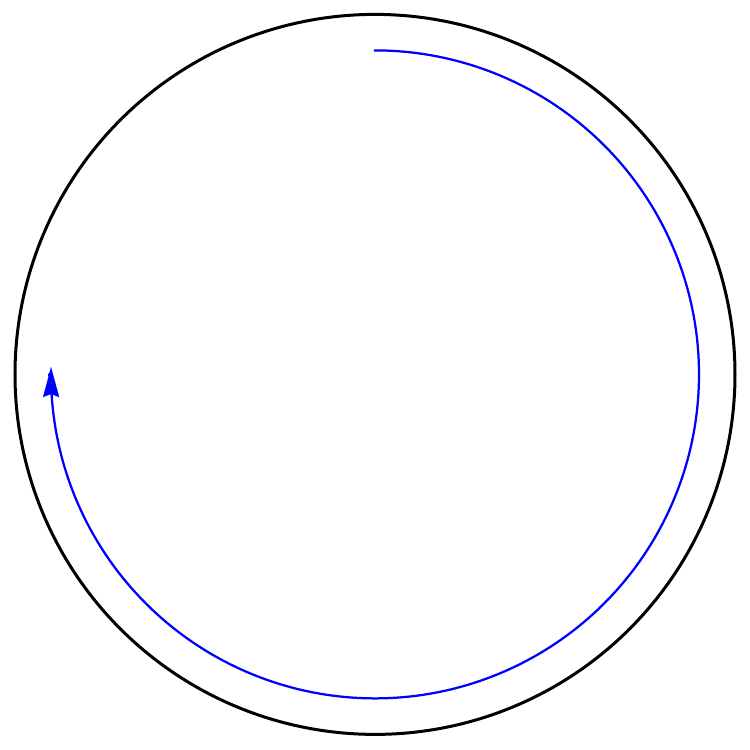}
\put(35,50){\parbox{0.2\linewidth}{
		\begin{equation*}
			\color{blue}{\widetilde{\beta}}
\end{equation*}}}
        \end{overpic}
        \end{minipage}
        \begin{minipage}[b]{0.45\linewidth}
\begin{overpic}[scale=0.4]{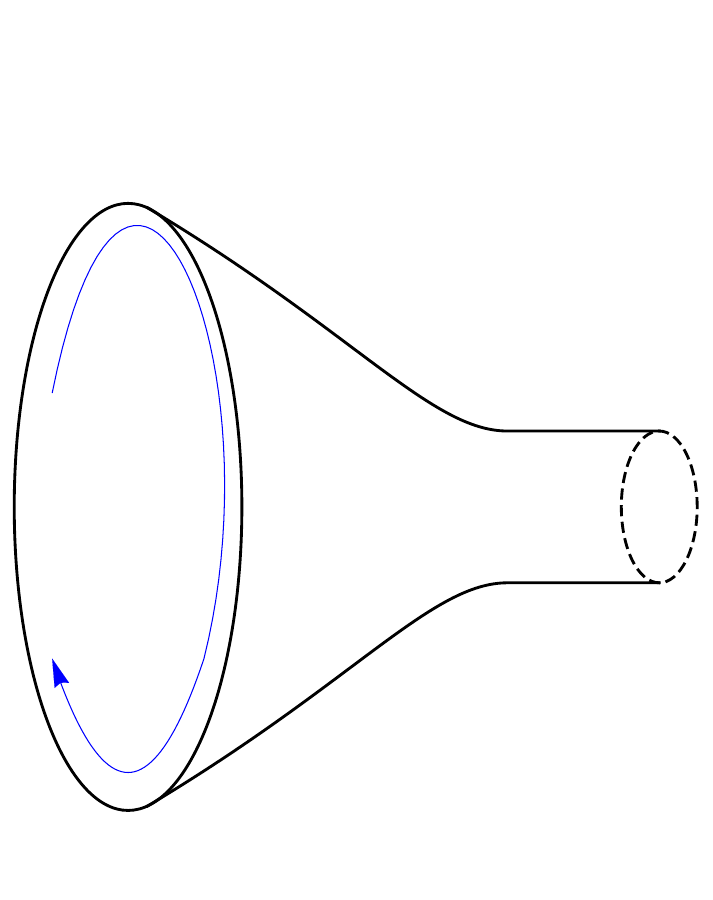}
\put(5,40){\parbox{0.2\linewidth}{
		\begin{equation*}
			\color{blue}{\widetilde{\beta}}
\end{equation*}}}
\put(55,45){\parbox{0.2\linewidth}{
		\begin{equation*}
			\lambda
\end{equation*}}}
\end{overpic}  
  \end{minipage}
\end{center}
\caption{(Left): Disk partition function (Right): Trumpet partition function with geodesic boundary that has length $\lambda$, and the geodesic is indicated by the dashed line.}\label{fig:Ztrumpet}
\end{figure}
It has been shown that we can reproduce the corresponding results in Section \ref{sec:HH-state} and Section \ref{sec:correlation_function} for JT gravity by considering the Schwarzian limit of a chiral half of 3d gravity \cite{Mertens:2017mtv, Mertens:2018fds, Ghosh:2019rcj}. This can be achieved by performing an S-wave dimensional reduction \cite{Mertens:2018fds, Gross:2019ach} or taking the near extremal limit \cite{Ghosh:2019rcj}. For the partition function, the vacuum character $\chi_{\mathbb{1}}(-1/\tau)$ becomes the Schwarzian disk partition function \cite{Ghosh:2019rcj}, and as shown in Section \ref{subsec:3d_doubletrumpet}, the character of nondegenerate representations of weight $h_\lambda$, $\chi_{\lambda}(-1/\tau) $, 
becomes the trumpet partition function with geodesic length $\lambda$ (Figure \ref{fig:Ztrumpet}). This is because the modular S-matrices,  $S_{\mathds{1}P}$ and $S_{\lambda P}$, are reduced to the density of states of the disk and trumpet partition functions in JT gravity respectively. For correlation functions, the Liouville correlators also reduce to Schwarzian correlators in the Schwarzian limit. In particular, the DOZZ formula is reduced to matrix elements of bilocal operator insertions in JT gravity, as shown in \eqref{eq:JT_DOZZ}. Further using the Schwarzian limit of the 6j symbol, we can reproduce the out-of time ordered correlation functions in JT gravity \cite{Mertens:2017mtv}.

In this subsection, we like to provide the recipe for recovering the Wheeler-DeWitt wavefunctions in 2d JT gravity. Through dimensional reduction, we show that the gravitational action, boundary value problem and the classical solution for the fixed $\Phi_0$-basis correspond to the geodesic length basis wavefunction in 2d\cite{Harlow:2018tqv}. In \cite{Chua:2023srl}, the dimensionally reduced $E$-states of 3d gravity is shown to be the energy basis wavefunction in JT gravity. 

Starting from the gravitational action in \eqref{eq:Sgrav_Phistates_smooth1} where the dimension of the quantities are specified for clarification purposes, we have
\begin{equation}\label{eq:Sgrav_Phistates_smooth_2}
\begin{split}
    -S_{\text{grav}} &= \frac{1}{16 \pi G_N^{(3)}}\int_{\mathcal{M}} d^3 x\sqrt{g^{(3)}}\left(R^{(3)}+2\right)+ \frac{1}{8 \pi G_N^{(3)}}\int_{\mathcal{B}'} d^2 x\sqrt{\gamma^{(2)}}\left(\Theta^{(2)}-1\right)\\
        &+ \frac{1}{8 \pi G_N^{(3)}}\int_{\Sigma'} d^2 x\sqrt{h^{(2)}} K^{(2)} +\frac{1}{8 \pi G_N^{(3)}}\int_{\Gamma} dx \sqrt{\sigma_\Gamma} \left(\theta-\frac{\pi}{2}\right)~.
\end{split}
\end{equation}

To dimensionally reduce the theory, we take the following ansatz for the 3d metric \cite{Mertens:2018fds, Gross:2019ach}
\begin{equation}\label{eq:3dtoJT}
	ds^2 = g^{(2)}_{\alpha\beta}dx^\alpha dx^\beta + \Phi_{\text{2d}}^2(x^\alpha)d\phi^2~,
\end{equation}
where $\phi \sim \phi+2 \pi$ is the spatial circle. Non-rotating Euclidean BTZ, i.e. \eqref{eq:non_BTZ_metric}, automatically satisfies this ansatz. With the metric parametrization in \eqref{eq:3dtoJT}, we have the following relations for the Ricci scalar and extrinsic curvature
\begin{equation}\label{eq:3dand2d}
	\begin{split}
		R^{(3)} = R^{(2)} -2 \Phi_{\text{2d}}^{-1}\nabla^2 \Phi_{\text{2d}}~,~~\Theta^{(2)} = \Theta^{(1)} + \Phi_{\text{2d}}^{-1}  r^\alpha  \partial_\alpha \Phi_{\text{2d}}~,~~K^{(2)} = K^{(1)} + \Phi_{\text{2d}}^{-1}  u^\alpha \partial_\alpha \Phi_{\text{2d}}~.
	\end{split}
\end{equation}
and we arrive at the 2d JT action from \eqref{eq:Sgrav_Phistates_smooth_2}
\begin{equation}
	\begin{split}
		-S_{\text{JT}}  &= \frac{1}{16 \pi G_N^{(2)}}\int_{\mathcal{M}} d^2 x \sqrt{g^{(2)}} \Phi_{\text{2d}}\left(R^{(2)}+2\right)+ \frac{1}{8 \pi G_N^{(2)}}\int_{\mathcal{B}'} d x \sqrt{\gamma^{(1)}} \Phi_{\text{2d}} \left(\Theta^{(1)} - 1\right) \\
  &+ \frac{1}{8 \pi G_N^{(2)}} \int_{\Sigma'} d x \sqrt{h^{(1)}} \Phi_{\text{2d}} K^{(1)} +\frac{\Phi_{\text{2d}}^0}{8 \pi G_N^{(2)}}  (\theta_+ + \theta_- -\pi) ~,
	\end{split}
\end{equation}
where the spacetime manifold $\mathcal{M}$ and boundary surfaces are in 2d and 1d respectively.  $\theta_\pm$ are the local angles at the two disconnected joints respectively. The Newton's constant are related through $G_N^{(3)} = 2 \pi G_N^{(2)}$. $\Phi_{\text{2d}}^0$ is the value of the dilaton and $\theta$ is the local angle at $\mathcal{B} \cap \Sigma$ respectively. 

The variation of the above action $-\delta S_{\text{JT}}$ with respect to the metric and dilaton is given by \cite{Harlow:2018tqv}
\begin{equation}
    \begin{split}
        -\delta S_{\text{JT}}  &= \frac{1}{16 \pi G_N }\int_{\mathcal{M}} d^2 x \sqrt{g} \left[\left(\frac{1}{2}(R+2)\Phi_{\text{2d}} g^{\mu\nu} -R^{\mu\nu} \Phi_{\text{2d}}+\nabla^\mu \nabla^\nu \Phi_{\text{2d}} -g^{\mu\nu}\nabla^2 \Phi_{\text{2d}}\right)\delta g_{\mu\nu} +(R+2)\delta \Phi_{\text{2d}}\right] \\
        &+\frac{1}{8 \pi G_N}\int_{\mathcal{B}'}dx \sqrt{\gamma}\left[(\Theta-1)\delta \Phi_{\text{2d}}+\frac{1}{2}(r^\mu \nabla_\mu \Phi_{\text{2d}} - \Phi_{\text{2d}})\gamma^{\alpha\beta} \delta \gamma_{\alpha\beta} \right]\\
        &+ \frac{1}{8 \pi G_N}\int_{\Sigma'} d x \sqrt{h}\left[K \delta \Phi_{\text{2d}} + \frac{1}{2}u^\mu \nabla_\mu \Phi_{\text{2d}}h^{\alpha\beta} \delta h_{\alpha\beta} \right]+ \frac{\delta \Phi_{\text{2d}}^0}{8 \pi G_N}  (\theta_+ + \theta_--\pi)~,
    \end{split}
\end{equation}
where we have dropped the indices labelling the dimensions. We now review the boundary conditions for the study of Hartle-Hawking wavefunction in the length basis in JT gravity. At $\mathcal{B}'$, the AdS boundary condition is imposed by introducing the AdS cutoff $\epsilon$ such that the induced metric and dilaton are equal to
\begin{equation}
    \begin{split}
        ds^2|_{\mathcal{B}'} &= \frac{1}{\epsilon^2}d\tau_E^2+\mathcal{O}(1)~, \\
        \Phi_{\text{2d}}|_{\mathcal{B}'} &= \frac{1}{\epsilon}~,
    \end{split}
\end{equation}
where the range of $\tau_E$ is $\beta/2$. On the $\Sigma'$ surface, the following boundary conditions are imposed
\begin{equation}\label{eq:JT_SigmaPhi}
    \begin{split}
        K|_{\Sigma'} &= 0~, \\
        ds^2|_{\Sigma'} &= d\rho^2~,
    \end{split}
\end{equation}
where the range of $\rho$ is taken to be $2 \ln \frac{2}{\epsilon} + L_{\text{reg}}$ and $L_{\text{reg}}$ is the regularized geodesic length. At the point where $\mathcal{B}$ and $\Sigma$ coincides, we fix the value of the dilaton
\begin{equation}
    \Phi_{\text{2d}}|_{\mathcal{B} \cap \Sigma}=\Phi_{\text{2d}}^0 = \frac{1}{\epsilon}~,
\end{equation}
such that $\mathcal{B}$ and $\Sigma$ are connected at the AdS boundary.

The equations of motion are satisfied by the following parametrization of the $AdS_2$ black hole metric and dilaton
\begin{equation}\label{eq:JT_HH_metric}
\begin{split}
    ds^2_{\text{2d}} &= d\rho^2 + \frac{r_+^2 \cosh^2 \rho}{\sin^2 r_+ T} dT^2~, \\
    \Phi_{\text{2d}} &= \frac{r_+ \cosh \rho}{\sin r_+ T}~,
\end{split}
\end{equation}
where $r_+$ is the horizon radius of the black hole located at $\rho = 0$. Introducing the following coordinate transformation
\begin{equation}
\begin{split}
    r &= \frac{r_+ \cosh \rho}{\sin r_+ T}~, \\
    \tan (r_+ \tau_E) &= \tan(r_+ T)\tanh \rho~,
\end{split}
\end{equation}
the $AdS_2$ black hole metric in Schwarzschild coordinates is given by
\begin{equation}
    ds^2_{\text{2d}} = (r^2 - r_+^2)d\tau_E^2 + \frac{dr^2}{r^2 - r_+^2}~,
\end{equation}
where the thermal circle is given by $\tau_E \sim \tau_E +2 \pi/r_+$. For the solution to the L-basis Hartle-Hawking wavefunction, we adopt the metric parametrization and dilaton in \eqref{eq:JT_HH_metric}. The AdS boundary condition is achieved by introducing a $T$-dependent cutoff $\rho_0(T) = \ln \left(\frac{2 \sin(r_+ T)}{\epsilon r_+}\right)$ such that the induced metric on $\mathcal{B}'$ is given by
\begin{equation}
    ds^2|_{\mathcal{B}'} \approx \left(\frac{1}{\epsilon^2} + r_+^2 \cot^2 (r_+ T)+\frac{r_+^2}{2 \sin^2 (r_+ T)}\right)dT^2~.
\end{equation}
$T \in [0,\beta/4]$ as we have the ``two'' halves of the AdS boundary defined from $\rho = \pm \rho_0(T)$ respectively. $L_{\text{reg}}$ is related to the horizon radius $r_+$ by
\begin{equation}\label{eq:JT_Lreg}
    e^{-L_{\text{reg}}} = \frac{r_+^2}{\sin^2 \frac{\beta r_+}{4}}~.
\end{equation}
The boundary conditions on $\Sigma'$ in \eqref{eq:JT_SigmaPhi} can be achieved by considering $T = $constant slices, where in particular on $\Sigma'$, $T = \beta/4$ such that the end points of $\Sigma$ hits the AdS boundary. 

Before moving on to compute the on-shell action, we like to show the boundary conditions on the $\Sigma'$ surface in 3d gravity reduces to the 2d version. From \eqref{eq:bdy_Phi_HH2}, fixing the radial lapse function $M = 1$ is aligned with $ds^2 = d\rho^2$ in JT gravity and $\rho$ is the parameter that parametrizes the geodesic. $j_A = 0$ is satisfied automatically through the metric parametrization that we consider in \eqref{eq:3dtoJT}.
For $P^{ij}\sigma^A_i \sigma^B_j = 0$, we 
first relate the extrinsic curvature components on the 2-dimensional $\Sigma'$ slice to the dilaton field
\begin{equation}
	K^{ij} \sigma^\phi_i \sigma^\phi_j  =  \Phi_{\text{2d}}^{-3} u^\alpha \partial_\alpha \Phi_{\text{2d}}~,
\end{equation}
and using the decompositon of the trace of extrinsic curvature in \eqref{eq:3dand2d}, we further simplify the last relation in \eqref{eq:bdy_Phi_HH2} to
\begin{equation}
	\begin{split}
		P^{ij}\sigma^\phi_i \sigma^\phi_j  = -\frac{ \Phi_{\text{2d}}^{-1} K^{(1)}}{32 \pi^2 G_N^{(2)}} = 0~,
	\end{split}
\end{equation}
reproducing the boundary conditions on the bulk slice \eqref{eq:JT_SigmaPhi} in JT gravity. 
In short, after dimensional reduction, the boundary value problem for the $\Phi_0$-basis in 3d gravity is related to the one for the $L$ basis in JT gravity\cite{Harlow:2018tqv} through
\begin{equation}
	\begin{split}
		&M = 1 \Rightarrow ds^2  = d\rho^2~,\\
		&j_A = 0~, \\
		&P^{ij}\sigma^\phi_i \sigma^\phi_j = 0 \Rightarrow K^{(1)} = 0~,
	\end{split}
\end{equation}

The semiclassical Hartle-Hawking wavefunction of JT in the $L$-basis can be obtained from \eqref{eq:Sgrav_Phi0}
\begin{equation}
\begin{split}
    \Psi_{\beta/2}^{\text{HH}}(\Phi_0) &= \Psi_{\beta/2}^{\text{JT}}(L_{\text{reg}})~, \\
    &\approx \exp\left(\frac{\beta r_+^2(L_{\text{reg}})}{32 \pi G_N^{(2)}}+ \frac{r_+(L_{\text{reg}})}{4 \pi G_N^{(2)}}\cot\left(\frac{\beta r_+(L_{\text{reg}})}{4}\right)\right)~,
\end{split}
\end{equation}
where the relation $\frac{2 \pi }{\beta}e^{\Phi_0/2} = e^{-L_{\text{reg}}/2}$  from \eqref{eq:Phi0_rp1} and \eqref{eq:JT_Lreg} respectively is a consequence of dimensional reduction \cite{Harlow:2018tqv}.

\section{Factorization in 3d gravity and defect operator from the ZZ boundary states}\label{sec:factorization}

In this section, we consider the factorization of the two-sided Hartle-Hawking state into two single-sided Hilbert spaces, as required from the calculation of the entanglement entropy. Following a similar spirit to  \cite{Jafferis:2019wkd}, where the factorization in JT gravity was studied, we study the case for 3d gravity. In 3d gravity, a local boundary condition factorizes each Liouville state into a direct sum of entangled left and right-moving Alekseev-Shatashvili states. The superselection sectors in the direct sum are labeled by the holonomies around the horizon. The local boundary condition at the horizon allows us to define a Liouville thermofield double state, however, we like to emphasize that it is not the holographic CFT thermofield double state dual to eternal BTZ \cite{Maldacena:2001kr}. To get an isometric factorization map, we need to insert a ``defect operator'' to account for the contractibility condition in the bulk gravitational theory. This operator insertion also modifies the definition of trace and matches with trace in gravity in the language of algebraic quantum field theory considerations \cite{Chandrasekaran:2022eqq, Penington:2023dql}. We subsequently reproduce the Bekenstein-Hawking entropy formula from the left-right entanglement entropy.

\begin{figure}[h]
	\centering
	\includegraphics[scale=0.5]{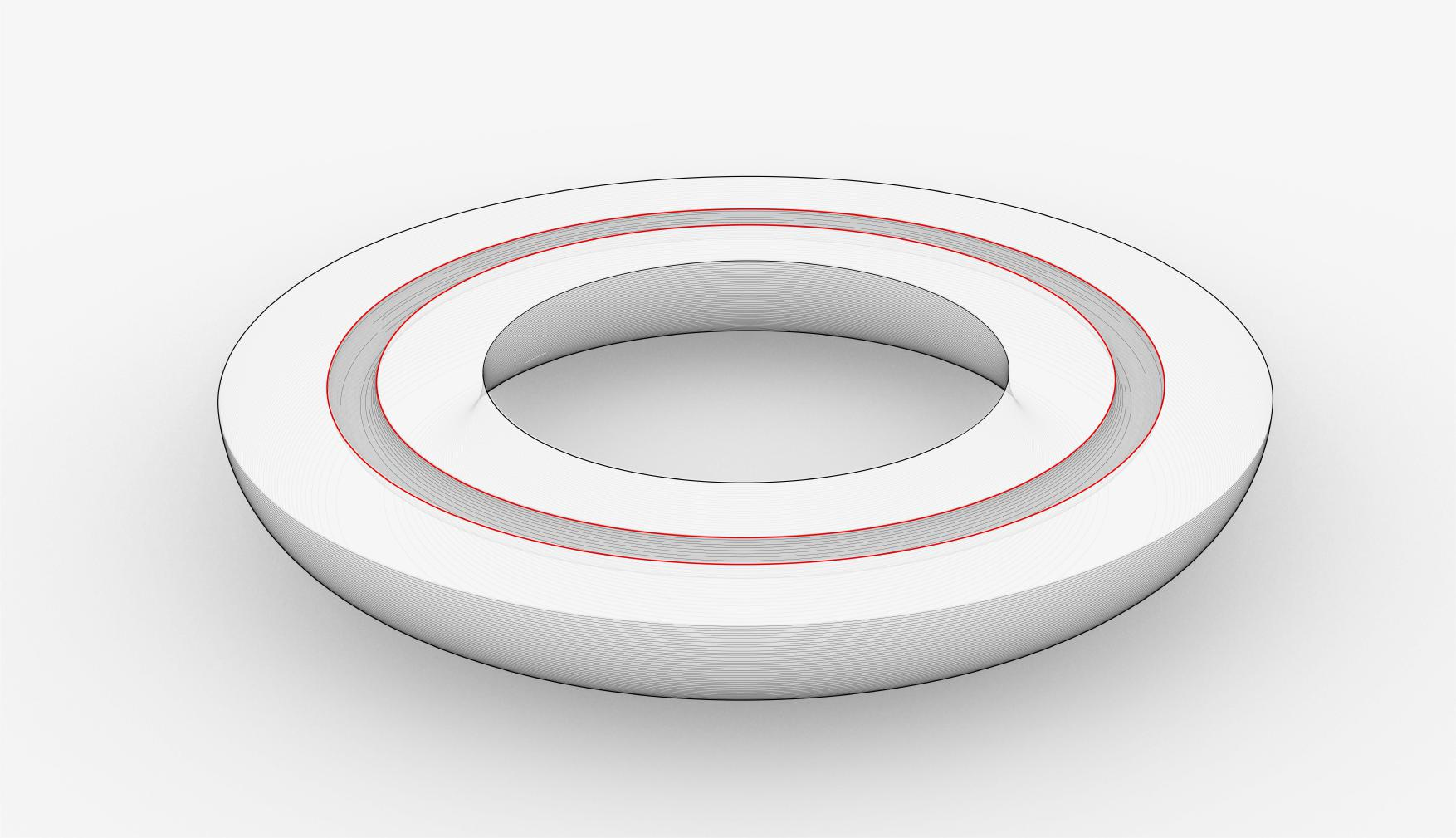}
 	\caption{The spatial annulus is split into two annuli through a trivial topological interface. With the Hilbert space on the inner and outer annulus as the two single-sided Hilbert spaces, we compute the left-right entanglement entropy. } \label{fig:factorization}
\end{figure}  

\subsection{Factorization map in 3d gravity}
As shown in Figure \ref{fig:factorization}, we factorize the bulk Hartle-Hawking state into two copies of single-sided states by allowing the states on the inner and outer annulus to share a common ``horizon boundary''. From the Hilbert space identification in Section \ref{sec:3d_Liouville_AS}, we expect to get the thermofield double state in terms of Alekseev-Shatashvili theories \cite{Maldacena:2001kr, Jafferis:2019wkd}. We obtain this factorization by imposing the following boundary conditions on the horizon boundary
\be \label{eq:local_bc_4AS}
A_t=\widetilde{A}_t=0~.
\ee
This is parallel to the local boundary conditions imposed in the 2d case where the BF formulation of JT gravity is studied \cite{Jafferis:2019wkd}. We further impose 
\be
\partial_t A_{\phi}=\partial_t \widetilde{A_{\phi}}=0~,
\ee
to freeze the graviton degrees of freedom in the bulk such that only boundary graviton excitations are present, similar to the scenario in Section \ref{sec:3d_Liouville_AS}. 
With these boundary conditions, we again obtain four copies of coupled Alekseev-Shatashvili theories, where the chiral action of the outer annulus is coupled with the antichiral action of the inner annulus and vice versa. 

We subsequently obtain the thermofield double state\cite{Maldacena:2001kr} for Alekseev-Shatashvili theories from the ``cutting map'' $\mathcal{I}: \mathcal{H} \rightarrow \mathcal{H}_L \times \mathcal{H}_R$
 \begin{equation}
 \begin{split}
\mathcal{I} |\Psi_{\beta/2}^{\text{HH}} \rangle 
 &=\int_{0}^{\infty} dP'  d\widetilde{P}' \sum_{N_1,\widetilde{N}_1}e^{-\beta E(h_{P'},N_1)/2}e^{-\beta E(h_{\widetilde{P}'},\widetilde{N}_1)/2} |h_{P'},N_1\rangle  |\widetilde{h_{P'},N_1}\rangle \otimes |\widetilde{h_{P'},N_1}\rangle |h_{P'},N_1\rangle ~,
\end{split}
\end{equation}
where $E(h_{P'},N_1) = h_{P'}+N_1-\frac{c}{24}$, and $N_1,\widetilde{N}_1$ are labels for the descendents with unit normalization
\begin{equation}
    \langle P',N_1| Q',N_2 \rangle = \delta(P'-Q')\delta_{N_1,N_2}~.
\end{equation}
However, this local boundary condition is not isometric, which further implies that the new boundary is not contractible after the trace is performed, which is required for a smooth geometry in gravity. 
To get an isometric factorization map $\mathcal{J}$, we need to insert the square root of the two-sided defect operator $D$ through $\mathcal{J}=\sqrt{D} \mathcal{I}: \mathcal{H} \rightarrow \mathcal{H}_L \times \mathcal{H}_R$ after applying $\mathcal{I}$. 

Applying the isometric factorization map $\mathcal{J}$ to the Hartle-Hawking state gives us
\begin{equation}
\begin{split}
\mathcal{J} |\Psi_{\beta/2}^{\text{HH}} \rangle 
&= 
\int_{0}^{\infty} dP'  d\widetilde{P}' \sum_{N_1,\widetilde{N}_1}e^{-\beta E(h_{P'},N_1)/2}e^{-\beta E(h_{\widetilde{P}'},\widetilde{N}_1)/2} \sqrt{S_{\mathds{1}P'}S_{\mathds{1}\widetilde{P}'}}  |h_{P'},N_1\rangle  |\widetilde{h_{P'},N_1}\rangle \otimes |\widetilde{h_{P'},N_1}\rangle |h_{P'},N_1\rangle\\
&=\int_{0}^{\infty} dP'  d\widetilde{P}' \sum_{N_1,\widetilde{N}_1}e^{-\beta E(h_{P'},N_1)/2}e^{-\beta E(h_{\widetilde{P}'},\widetilde{N}_1)/2} \sqrt{\rho_0(P')\rho_0(\widetilde{P}')}  |h_{P'},N_1\rangle  |\widetilde{h_{P'},N_1}\rangle \otimes |\widetilde{h_{P'},N_1}\rangle |h_{P'},N_1\rangle~,
\end{split}
\end{equation}
where the explicit expression of the two-sided defect operator $D$ is given by
\begin{equation}
    D = \int_0^\infty dP' d\tilde{P}'\sum_{N_1,\widetilde{N}_1}S_{\mathds{1}P'}S_{\mathds{1} \widetilde{P}'}|h_{P'},N_1\rangle  |\widetilde{h_{P'},N_1}\rangle |\widetilde{h_{P'},N_1}\rangle |h_{P'},N_1\rangle \otimes \langle h_{P'},N_1|  \langle\widetilde{h_P,N_1}| \langle \widetilde{h_{P'},N_1}|  \langle h_{P'},N_1|~.
\end{equation}

Tracing out the degrees of freedom of one side of the factorized Hilbert space, we naturally have the reduced density matrix $\rho$ produced by $\mathcal{J}$,
\begin{equation}
\begin{split}
	\rho &\equiv \text{Tr}_{\mathcal{H}_L} \mathcal{J}	|\Psi_{\beta/2}^{\text{HH}} \rangle \langle 	\Psi_{\beta/2}^{\text{HH}}| \mathcal{J}^\dagger ~, \\
 &= \int_{0}^{\infty} dP'  d\widetilde{P}' \sum_{N_1,\widetilde{N}_1}e^{-\beta E(h_{P'},N_1)}e^{-\beta E(h_{\widetilde{P}'},\widetilde{N}_1)} 
 S_{\mathds{1}P'}S_{\mathds{1}\widetilde{P}'}|h_{P'},N_1\rangle  |\widetilde{h_{P'},N_1}\rangle \otimes \langle h_{P'},N_1|\langle \widetilde{h_{P'},N_1}| \\
 &= \int_{0}^{\infty} dP'  d\widetilde{P}' \sum_{N_1,\widetilde{N}_1}e^{-\beta E(h_{P'},N_1)}e^{-\beta E(h_{\widetilde{P}'},\widetilde{N}_1)} 
 \rho_0(P')  \rho_0(\tilde{P}') |h_{P'},N_1\rangle  |\widetilde{h_{P'},N_1}\rangle \otimes \langle h_{P'},N_1|\langle \widetilde{h_{P'},N_1}|  ~.
\end{split}
\end{equation}
Furthermore, the trace over Alekseev-Shatashvili states of $\mathcal{H}_R$ of the reduced density matrix $\rho$ readily gives us the partition function of Euclidean BTZ, which is also given by the norm of the Hartle-Hawking state,\footnote{We choose not to keep track of the infinite volume factor coming from $\langle h_{P'},N_1| h_{P'}, N_1 \rangle = \delta(0)$, and only focus on the part that depends on temperature. We can also reabsorb the constants into the definition of entropy as in \cite{Jafferis:2019wkd, Blommaert:2018iqz, Lin:2018xkj} for JT gravity. }
\begin{equation}\label{eq:partition_function}
\begin{split}
	Z(\beta) &= \langle \Psi_{\beta/2}^{\text{HH}}| \Psi_{\beta/2}^{\text{HH}} \rangle ~, \\
 &= \text{Tr}_{\mathcal{H}_R} \rho~, \\
 &= \int_{0}^{\infty} dP'  
 S_{\mathds{1}P'}\chi_{P'}(\tau)\int_0^\infty d\widetilde{P}' S_{\mathds{1}\widetilde{P}'}\chi_{\widetilde{P}'}(-\bar{\tau})\\
 &=\int_{0}^{\infty} dP'  
 \rho_0(P')\chi_{P'}(\tau)\int_0^\infty d\widetilde{P}' \rho_0(\widetilde{P}')\chi_{\widetilde{P}'}(-\bar{\tau})~,
\end{split}
\end{equation}
where $\tau = -\bar{\tau} = \frac{i \beta}{2 \pi}$ is the modular parameter of the asymptotic torus and we have used the relation for the non-degenerate Virasoro character
\begin{equation}
    \sum_{N_1}\langle h_{P'},N_1| e^{-\beta E(h_{P'},N_1)}| h_{Q'},N_1 \rangle = \chi_{h_{P'}}(\tau)\delta(P'-Q')~.
\end{equation}

On the other hand, we have the reduced density matrix $\tilde{\rho}$  associated to the local boundary condition, which is produced by the cutting map $\mathcal{I}$,
\begin{equation}
\begin{split}
	\tilde{\rho} &\equiv \text{Tr}_{\mathcal{H}_L} \mathcal{I}	|\Psi_{\beta/2}^{\text{HH}} \rangle \langle 	\Psi_{\beta/2}^{\text{HH}}| \mathcal{I}^\dagger ~, \\
 &= \int_{0}^{\infty} dP'  d\widetilde{P}' \sum_{N_1,\widetilde{N}_1}e^{-\beta E(h_{P'},N_1)}e^{-\beta E(h_{\widetilde{P}'},\widetilde{N}_1)} 
 |h_{P'},N_1\rangle  |\widetilde{h_{P'},N_1}\rangle \otimes \langle h_{P'},N_1|\langle \widetilde{h_{P'},N_1}| ~.
\end{split}
\end{equation}

\subsection{Modification of trace, area operator and entanglement entropy}

\begin{figure}[h]
	\centering
	\includegraphics[scale=0.5]{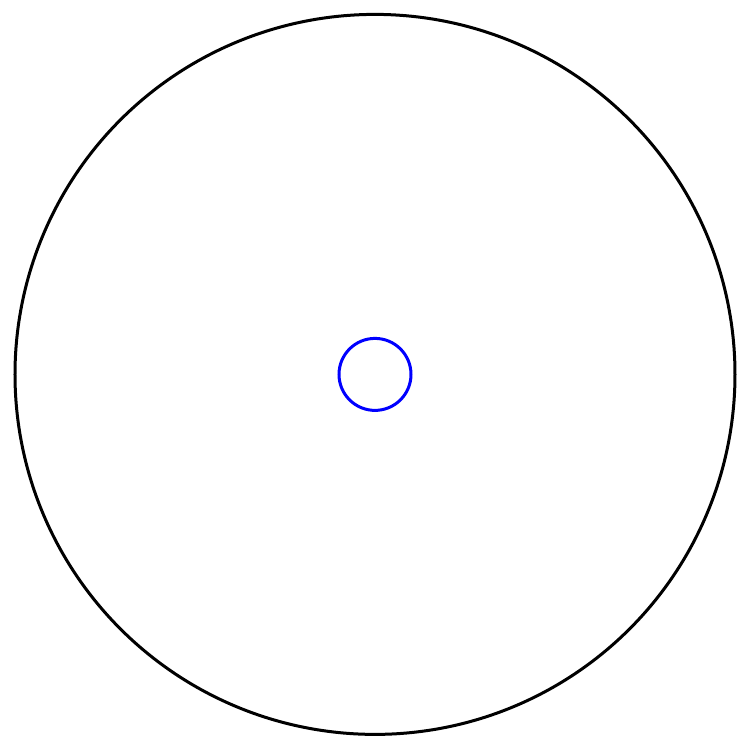}
 	\caption{Figure shows the cross section of the constant $\phi$ slice after performing the trace. In the absence of the defect operator insertion, the blue hole, which is formed by the local boundary condition at the horizon, is not contractible.} \label{fig:hole}
\end{figure}  

To compute the entanglement entropy of the reduced density matrix, we need to modify the notion of trace by inserting the one-sided defect operator $\mathcal{D} = \text{Tr}_{\mathcal{H}_L}D$ on the one-sided Hilbert space, and it fills the gap formed by the local boundary condition at the horizon, as shown in Figure \ref{fig:hole}. The one-sided defect operator $\mathcal{D}$ takes the following form
\be \label{one-sided defect operator}
\begin{aligned}
\mathcal{D}=\text{Tr}_{\mathcal{H}_L}D=\int_0^\infty dP' d\widetilde{P}' \sum_{N_1,\widetilde{N}_1} S_{\mathds{1}P'} S_{\mathds{1}\widetilde{P}'} \ket{h_{P'},N_1}\ket{\widetilde{h_{P'},N_1}}\bra{h_{P'},N_1}  \bra{\widetilde{h_{P'},N_1}}~,
\end{aligned}
\ee
which is a projector onto  trivial flux as in compact Chern-Simons theories\cite{Burnell_2011}. In gravity, this projection means that we only have vacuum Wilson loops\footnote{The holonomy is actually in the center of $SL(2,R)$, and trivial in $PSL(2,R)=SL(2,R)/Z_2$ \cite{Cotler:2018zff}.} in the dual $\tau_E$ cycle, implying contractibility in the bulk. In other words, the modular S-matrix $S_{\mathds{1}P}$ imposes a topological constraint such that we only have vacuum contribution in the dual channel. This is the higher-dimensional analogue of the one-sided defect operator in JT gravity \cite{Jafferis:2019wkd}, and in topological string theory \cite{Donnelly:2020teo,Jiang:2020cqo}.\footnote{In topological string theory, the insertion of the defect operator changes the Euler characteristic such that the Calabi-Yau condition is preserved. } In short, modular invariance in 2d CFT gives us a natural way to capture the topological constraint in the dual gravity theory, which is of the similar spirit as the original derivation of the Cardy formula\cite{Cardy:1986ie}.

In particular, the $n$th Renyi partition function is given by
\begin{equation}\label{eq:Renyi_Zn}
	Z_n = \text{Tr}_{\mathcal{H}_R} ~\mathcal{D} \tilde{\rho}^n = \text{Tr}_{\mathcal{H}_R}~\mathcal{D}^{1-n} \rho^n ~.
\end{equation}
We have seen throughout this paper that the current 3d gravity setup is very similar to what has been done in 2d JT gravity \cite{Jafferis:2019wkd}. Hence, it is also natural to expect that we also obtain a Type II$_{\infty}$ algebra with a trivial center \cite{Penington:2023dql} for 3d gravity coupled to the probe matter fields. For such algebras, there is a unique trace up to multiplicative constants, and in fact, the modification of trace we have matches with the unique trace from algebraic quantum field theory considerations \cite{Chandrasekaran:2022eqq, Penington:2023dql}. To see this, we can consider a one-sided Hilbert space operator $O_R$, and the unique trace is given by
\be
\begin{aligned}
\lim_{\beta\rightarrow 0} \langle 	\Psi_{\beta/2}^{\text{HH}}|O_R|\Psi_{\beta/2}^{\text{HH}} \rangle&={\tr}{}_{\mathcal{H}_L \times \mathcal{H}_R}(\mathcal{J}\ket{ZZ} \ket{\widetilde{ZZ}} \otimes \bra{ZZ}\bra{\widetilde{ZZ}}\mathcal{J}^\dagger O_R)\\
&=\tr_{\mathcal{H}_R} \left( \left( \text{Tr}_{\mathcal{H}_L}(\mathcal{J}\ket{ZZ} \ket{\widetilde{ZZ}}\otimes \bra{ZZ}\bra{\widetilde{ZZ}}\mathcal{J}^\dagger )\right) O_R \right) \\
&=\tr_{\mathcal{H}_R}(\mathcal{D} O_R)~.
\end{aligned}
\ee
For our Hartle-Hawking state, the $n$th Renyi partition function is 
\begin{equation}
	\begin{split}
	Z_n &= \text{Tr}_{\mathcal{H}_R} ~\mathcal{D} \tilde{\rho}^n~,\\
	&= \int_0^\infty dP' S_{\mathds{1}P'} \chi_{h_{P'}}(n\tau)\int_0^\infty d\widetilde{P}' S_{\mathds{1} \widetilde{P}'}  \chi_{h_{\widetilde{P}'}}(-n\bar{\tau})~, \\
	&=  \chi_{\mathds{1}}\left(-\frac{1}{n\tau}\right)  \chi_{\mathds{1}}\left(\frac{1}{n\bar{\tau}}\right)~,
	\end{split}
\end{equation}
where the thermal length is changed from $\beta$ to $n \beta$, matching exactly the expectation from Euclidean path integral in gravity. The calculation of entanglement entropy $S_{\text{EE}}$ follows suit
\begin{equation}
\begin{aligned}
	S_{\text{EE}} = -\partial_n \left(\frac{Z_n}{(Z_1)^n}\right) \Big{|}_{n = 1}  &=\frac{\text{Tr}_{\mathcal{H}_R} \left[(\ln \mathcal{D}) \rho\right]- \text{Tr}_{\mathcal{H}_R} ~\rho \ln \rho}{Z_1}+ \ln Z_1\\
&=-\text{Tr}_{\mathcal{H}_R} \hat{\rho}\ln \hat{\rho} +\text{Tr}_{\mathcal{H}_R} \hat{\rho} \ln \mathcal{D}~.
\end{aligned}
\end{equation}
where $\hat{\rho}=\frac{\rho}{Z_1}$ is the normalized density matrix and the expression takes the form of Faulkner-Lewkowycz-Maldacena (FLM) relation \cite{Faulkner:2013ana}. The first term of the entanglement entropy formula is just the usual von Neumann entropy for the normalized density matrix, and the second term gives the expectation value of the defect operator. It's actually tempting to interpret the defect operator as a ``quantum area operator'', as
\be \label{area operator}
\begin{aligned}
\ln(\mathcal{D})=\int_0^\infty dP' d\widetilde{P}' \sum_{N_1,\widetilde{N}_1} \ln(S_{\mathds{1}P'} S_{\mathds{1}\widetilde{P}'}) \ket{h_{P'},N_1}\ket{\widetilde{h_{P'},N_1}}\bra{h_{P'},N_1}  \bra{\widetilde{h_{P'},N_1}}~,
\end{aligned}
\ee
where 
\be
\begin{aligned}
\ln(S_{\mathds{1}P'} S_{\mathds{1}\widetilde{P}'})& =\ln \left(32 \sinh(2 \pi P' b)\sinh \left(\frac{2 \pi P'}{b}\right) \sinh(2 \pi \tilde{P}' b)\sinh \left(\frac{2 \pi \tilde{P}'}{b}\right) \right)\\
& \underset{b\to 0}{\rightarrow} 2\pi \frac{P'+\tilde{P'}}{b}~.
\end{aligned}
\ee
$2 \pi \frac{P'+\tilde{P'}}{b}$ is proportional to the minimal geodesic length (``area'' in 3d) of the spatial wormhole geometry \eqref{bottleneck}. As shown, the gravitational state is a superposition of all the wormhole geometries, and hence, the operator $\ln(\mathcal{D})$ precisely measures the area in all of these microscopic configurations.

The expectation value of this operator is
\begin{equation}\label{eq:SEE_defect}
\begin{split}
		S_{\text{EE,defect}} &=  \int_0^\infty dP' p(P') \ln S_{\mathds{1}P'}  +\int_0^\infty  d\widetilde{P}' p(\widetilde{P}') \ln S_{\mathds{1}\widetilde{P}'}\\
  &\approx  \ln(S_{\mathds{1}P'_*} S_{\mathds{1}\widetilde{P}'_*}) \\
  &= \frac{2\pi r_+}{b^2}\\
  &=\frac{\pi r_+}{2 G_N}~,
  \end{split}
  \end{equation}
where $p(P')=\frac{S_{\mathds{1}P'} e^{- \beta P^{\prime 2}}}{\int_0^\infty dP' S_{\mathds{1}P'} e^{-\beta P^{\prime 2}}}$ and $P'_*=\widetilde{P}'_* =\frac{\pi}{b \beta}= \frac{r_+}{2 b}$ are the saddle point values of the Liouville momenta.

The notion of trace defined above can be interpreted as a ``quantum trace'', with $S_{\mathds{1}P}$ playing the role of the Plancherel measure of the quantum semi-group $SL^+_{q}(2,R)$. This is aligned with the one-to-one mapping between Virasoro module and representations in $SL^+_{q}(2,R)$ found by Ponsot and Teschner \cite{Ponsot:1999uf, Ponsot:2000mt, Teschner:2003em, Teschner:2005bz, Mertens:2022ujr, Wong:2022eiu}.\footnote{We like to mention the difference between our Hilbert space description and \cite{Mertens:2022ujr, Wong:2022eiu}. Our Hilbert space is based on Alekseev-Shatashvili theories or Liouville theory and is not the same as the formal Hilbert space constructed from representation theory on quantum groups. The local boundary condition that is imposed has insufficient degeneracy of ``edge modes'' and hence, we must insert a defect operator by hand in the computation of trace to account for the degeneracy of states and non-local contractibility condition in the bulk.} This perspective also provides a canonical interpretation for the observation made in \cite{McGough:2013gka} in relating the Bekenstein-Hawking formula to topological entanglement entropy\cite{PhysRevLett.96.110404, PhysRevLett.96.110405}. 

We want to comment on some structural differences between 3d and 2d gravity, although the two computations in factorizing Hilbert space follow a similar spirit. The single-sided Hilbert space is well-defined in 3d but not 2d \cite{Harlow:2018tqv} as  the Hilbert space of two boundary JT gravity describes single-particle quantum mechanics. This can be understood as JT gravity being related to the chiral half of 3d gravity. In 3d gravity, we can directly perform analytic continuation to study the kinematics in Lorentzian signature, whereas for JT gravity, the kinematics between Euclidean and Lorentzian signature is very different. Hence, we can use the advantage of analytic continuation in 3d gravity to understanding some of other puzzles in quantum gravity, which is typically a hard approach in JT gravity.

Finally, we like to mention that the above definition of trace can't be reproduced purely from an edge mode calculation, which is also the case in gauge theories \cite{Donnelly:2014gva, Lin:2017uzr, Wong:2017pdm}, although summing of superselection sectors is involved in both calculations of the trace. In edge mode language, each superselection sector is specified by constant holonomies $K=\gamma L_0, \widetilde{K} = \widetilde{\gamma} \widetilde{L_0}$ as above. However, in the  computation of trace, we are missing the ``modes'' that account for the huge degeneracy $\rho_0(P)$ in each superselection sector. From a topological consideration, we add the degeneracy to each sector by hand using the defect operator \eqref{one-sided defect operator}. To get an edge mode interpretation for the area term in gravity as suggested in \cite{Harlow:2016vwg, Donnelly:2016auv, Lin:2017uzr, Lin:2018xkj, Blommaert:2018iqz, Mertens:2022ujr}, we need to have a UV complete theory of quantum gravity that has a vast amount of states near the edge. The inclusion of these states also provide a local contractible boundary condition \cite{Harlow:2015lma, Harlow:2021lpu}. In such cases, the low energy approximation for the trace over high energy degrees of freedom produces \eqref{one-sided defect operator}. Hence, in the gravitational low energy effective field theory, we can modify the definition of trace through the insertion of the defect operator.

\section{Late time two-point function and ``baby universe'' operators}\label{sec:late time}

All the discussion above involves only on-shell geometries. In this section, we like to mention briefly on off-shell computations in 3d gravity as these geometries are relevant in resolving Maldacena's version of information paradox \cite{Cotler:2016fpe,Saad:2019lba, Saad:2019pqd, Maldacena:2001kr}. As an explicit example, the forever decaying behavior of the semiclassical spectral form factor and holographic thermal two-point function fails to reproduce the ``ramp'' and ``plateau'' features at late times. In particular, the ``ramp'' feature of the late time two-point function in JT gravity was argued to be governed by considering off-shell wormhole contributions  \cite{Saad:2019pqd}.

We first show that the off-shell 3d torus spectral form factor\cite{Cotler:2020ugk} can be calculated using our Hilbert space formalism for gravity with two asymptotic boundaries, generalizing the results in JT gravity\cite{Saad:2019pqd}. The off-shell wormhole, also known as the ``double-trumpet'', is obtained by modifying the ZZ boundary state to FZZT boundary state through insertion of Verlinde loop operators\cite{Verlinde:1988sn, Alday:2009fs, Drukker:2009id, Drukker:2010jp, Dijkgraaf:1988tf, LeFloch:2017lbt, Mertens:2019tcm}. Similar to the 2d case, we propose the Verlinde loop operators to be the 3d ``baby universe'' operators\cite{Saad:2019lba, Penington:2023dql} and hence, we have provided another ingredient in relating gravity observables with Liouville CFT. We like to mention that we decide to leave a careful analysis on the off-shell path integral measure for the future.

\subsection{3d double-trumpet from FZZT boundary states and Verlinde loop operators}\label{subsec:3d_doubletrumpet}
\begin{figure}[h]
\begin{center}
\begin{minipage}[b]{0.45\linewidth}
\begin{overpic}[scale=0.32]{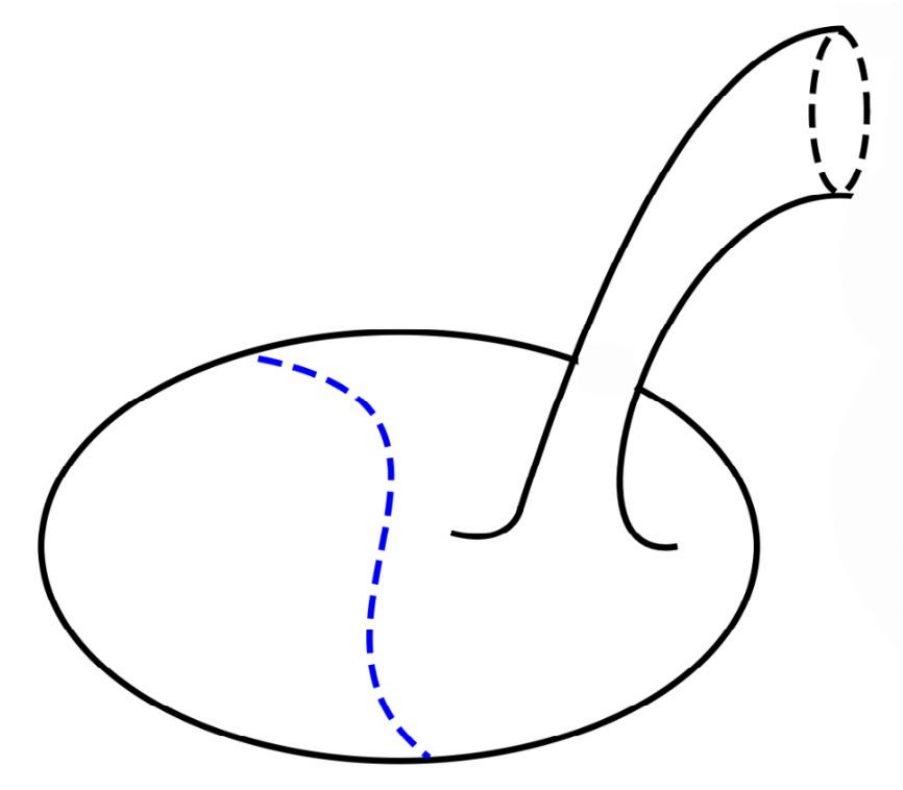}
\put(72,75){\parbox{0.2\linewidth}{
		\begin{equation*}
			\lambda
\end{equation*}}}
\put(-17,20){\parbox{0.2\linewidth}{
		\begin{equation*}
			\widetilde{\beta}/2
\end{equation*}}}
\put(80,20){\parbox{0.2\linewidth}{
		\begin{equation*}
			\widetilde{\beta}/2
\end{equation*}}}
        \end{overpic}
        \end{minipage}
        \begin{minipage}[b]{0.5\linewidth}
\begin{overpic}[scale=0.25]{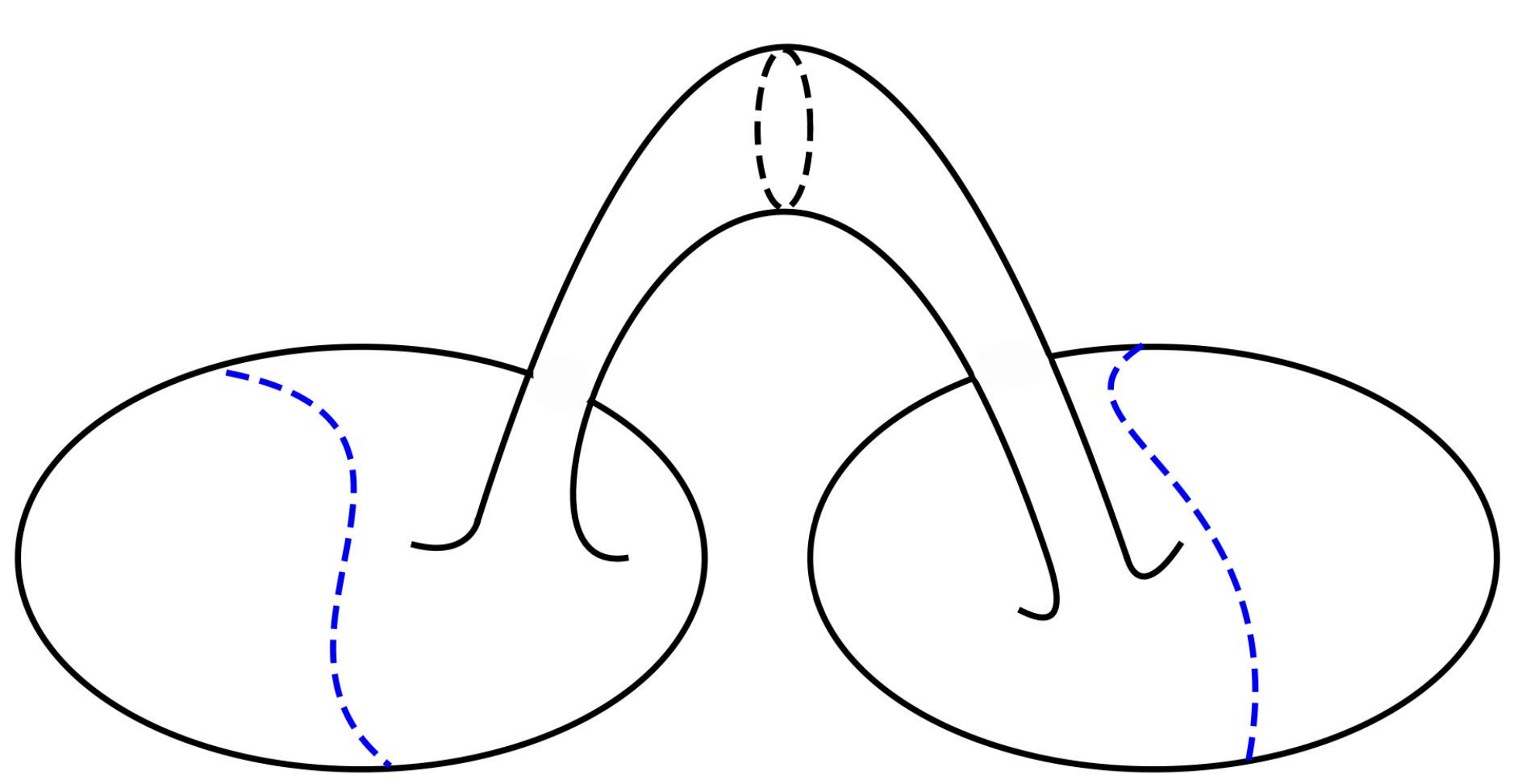}
\put(38,43){\parbox{0.2\linewidth}{
		\begin{equation*}
			\lambda
\end{equation*}}}
\end{overpic}  
  \end{minipage}
\end{center}
\caption{(Left): The trumpet partition function in JT gravity has a circular geodesic boundary of length $\lambda$ in addition to an asymptotic boundary of thermal length $\widetilde{\beta}$. (Right): The ``double trumpet'' spectral form factor in JT gravity is obtained by gluing two trumpet partition functions along the geodesic boundaries.}\label{fig:latetime}
\end{figure}

In JT gravity, the ramp contribution to the spectral form factor comes from trading a baby universe between two Hartle-Hawking states before returning to their original states \cite{Saad:2019pqd,Penington:2023dql}. As shown in Figure \ref{fig:latetime}, the baby universe takes the topology of a cylinder and together with the two AdS boundaries of the Hartle-Hawking states, the Euclidean geometry of the whole system is viewed as a ``double-trumpet''. The trumpet partition function with geodesic boundary whose circumference is $\lambda$ is given by the overlap\footnote{We have used a different notation to \cite{Penington:2023dql,Saad:2019pqd} to prevent possible confusion with other parts of the paper. More explicitly, $\lambda_{\text{here}} = b_{\text{there}}$ is the geodesic length of the trumpet and $\ket{\psi_{\beta/2}^{\text{HH}}}$ denotes the Hartle-Hawking state in JT gravity. }

\be\label{eq:JT_baby}
Z_{\text{trumpet}}(\beta,\lambda)= \bra{\psi_{\beta/2}^{\text{HH}}} \psi_{\beta/2}^{\text{HH}},\lambda \rangle =\bra{\psi_{\beta/2}^{\text{HH}}} \hat{\mathcal{O}}_{\lambda}\ket{\psi_{\beta/2}^{\text{HH}}}~,
\ee
where $\hat{\mathcal{O}}_{\lambda}$ can be interpreted as an operator that initiates the emission of the baby universe with geodesic length $\lambda$\cite{Saad:2019pqd, Penington:2023dql}.  The double-trumpet can be obtained by gluing two copies of trumpet partition function along the geodesic boundary that takes all possible length values
\be
Z_{\text{double trumpet}}(\beta_1,\beta_2)=\int_0^\infty \lambda d\lambda Z_{\text{trumpet}}(\beta_1,\lambda) Z_{\text{trumpet}}(\beta_2,\lambda)~.
\ee
\begin{figure}[h]
\begin{center}
\begin{minipage}[b]{0.45\linewidth}
\begin{overpic}[scale=0.32]{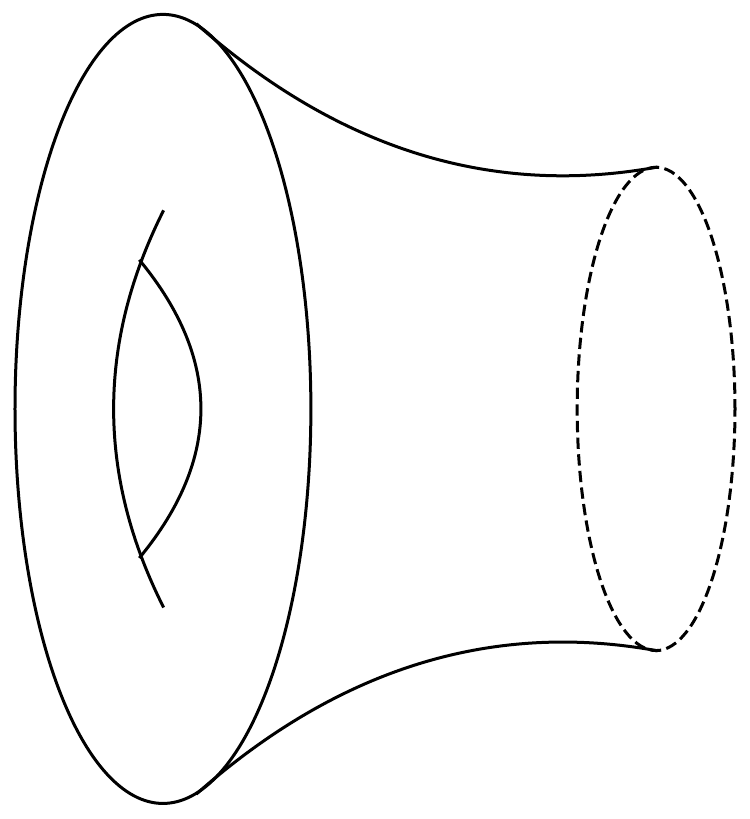}
\put(15,50){\parbox{0.2\linewidth}{
		\begin{equation*}
			\tau
\end{equation*}}}
        \end{overpic}
        \end{minipage}
        \begin{minipage}[b]{0.45\linewidth}
\begin{overpic}[scale=0.5]{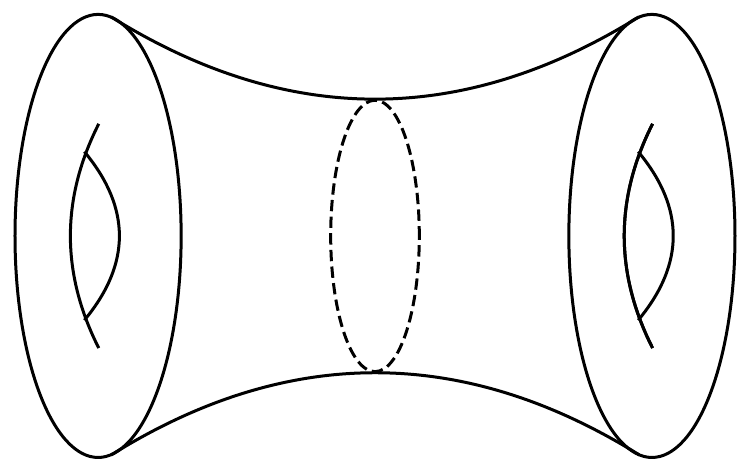}
\put(10,30){\parbox{0.2\linewidth}{
		\begin{equation*}
			\tau_1
\end{equation*}}}
\put(83,30){\parbox{0.2\linewidth}{
		\begin{equation*}
			\tau_2
\end{equation*}}}
\end{overpic}  
  \end{minipage}
\end{center}
\caption{(Left): The trumpet partition function  in 3d gravity is given by the product of characters of nondegenerate representations $\chi_\lambda(-1/\tau)$ with conformal weight $h_\lambda$ (Right): The double trumpet partition function in 3d gravity where the asymptotic tori have moduli $\tau_1$ and $\tau_2$ respectively.}\label{fig:Ztrumpet_3d}
\end{figure}

Similar to JT, we like to apply the Hilbert space formalism in obtaining the double-trumpet partition function in 3d gravity. As an early spoiler for the 3d version of \eqref{eq:JT_baby}, the Verlinde loop operators in Liouville are the ``baby universe'' operators that modify the ZZ boundary states to FZZT-boundary states.

\begin{figure}[h]
	\centering
	\includegraphics[scale=0.5]{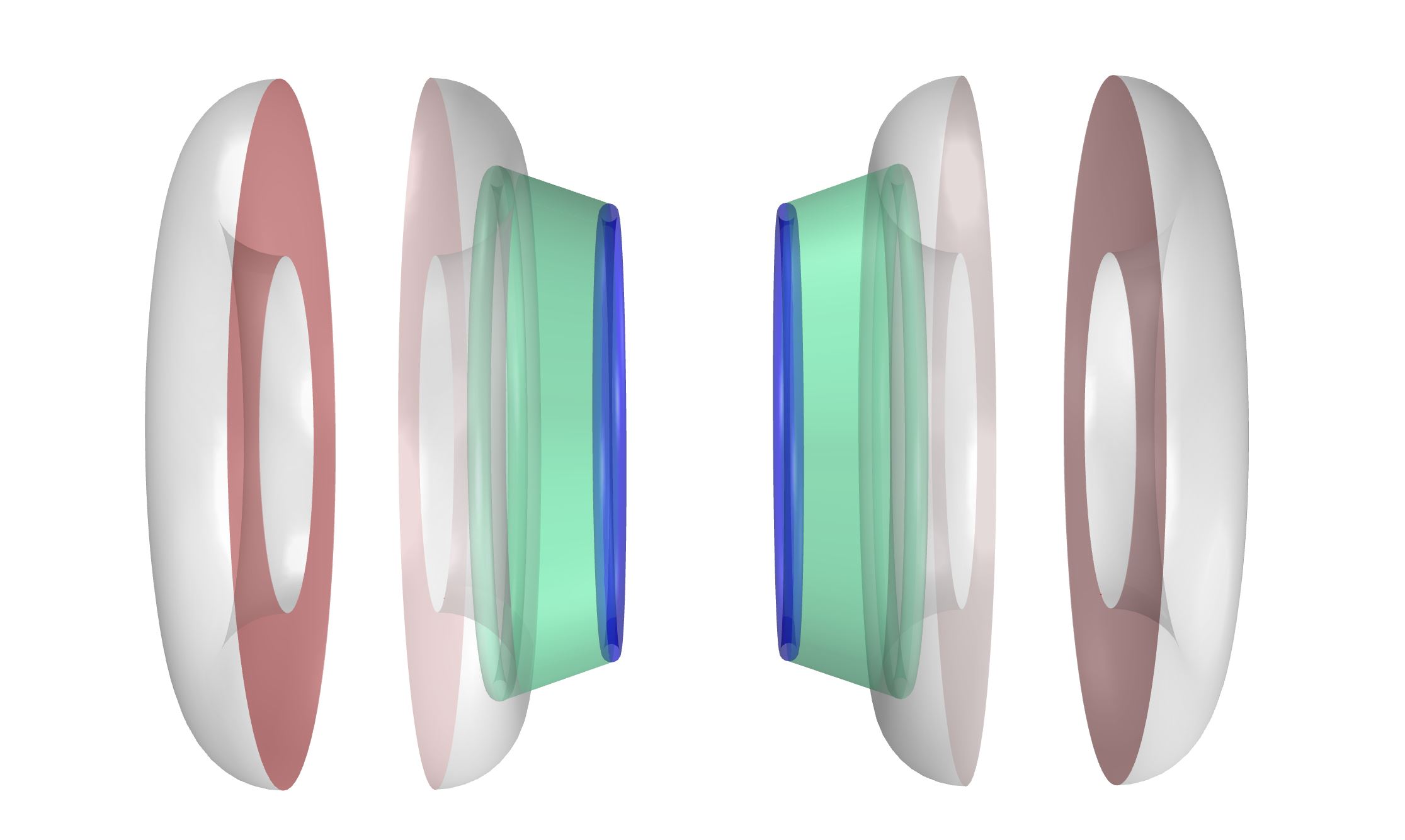}
	\caption{The preamplitude can be obtained by gluing two trumpet partition functions along the baby universes (indicated by the blue tori). Each trumpet partition function can be calculated using the overlap between the Hartle Hawking state $\ket{\psi_{\beta/2}^{\text{HH}}}$ and the Hartle Hawking state with a baby universe emission $\hat{\mathcal{O}}_{\lambda} \hat{\mathcal{O}}_{\widetilde{\lambda}} \ket{\psi_{\beta/2}^{\text{HH}}}$}. \label{double torus}
\end{figure}

As shown in Figure \ref{fig:Ztrumpet_3d}, the 3d double-trumpet that has a topology of torus$\times$interval, takes the following summation \cite{Cotler:2020ugk,Cotler:2020hgz}
\be \label{modular sum}
Z_{T_2 \times I}(\tau_1,\tau_2)=\sum_{g \in PSL(2,Z)} \tilde{Z}(\tau_1, g \tau_2)~,
\ee
where the sum is over preamplitudes $\tilde{Z}(\tau_1, g \tau_2)$ with relative modular transformation on the two asymptotic tori. $\tilde{Z}(\tau_1, \tau_2)$ is given by 
\be
\begin{aligned}
\tilde{Z}(\tau_1, \tau_2)&=\int \mathcal{DM}\,Z_{\text{trumpet}}(\tau_1,\bar{\tau}_1,\lambda,\widetilde{\lambda}) Z_{\text{trumpet}}(\tau_2,\bar{\tau}_2,\lambda,\widetilde{\lambda}) 
\end{aligned}
\ee
where the integration measure is\footnote{Compared to \cite{Cotler:2020ugk}, we have chosen a different preamplitude for the modular sum, hence, obtaining a different measure. The final result is still \eqref{modular sum} due to modular invariance. This choice is made to parallel the calculation done in JT gravity\cite{Saad:2019pqd, Saad:2019lba}.}
\be \label{torus measure}
\int \mathcal{DM} = 8\frac{\sqrt{\text{Im}(\tau_1) \text{Im}(\tau_2)}}{|\tau_1 \tau_2|} \int_0^\infty \lambda d\lambda\int_0^\infty \widetilde{\lambda} d\widetilde{\lambda} ~.
\ee
Up to an overall $8\frac{\sqrt{\text{Im}(\tau_1) \text{Im}(\tau_2)}}{|\tau_1 \tau_2|} $ factor, the measure is just two copies of the Weil-Petersson measure.
\begin{figure}
\begin{center}
\begin{overpic}[scale=0.2]{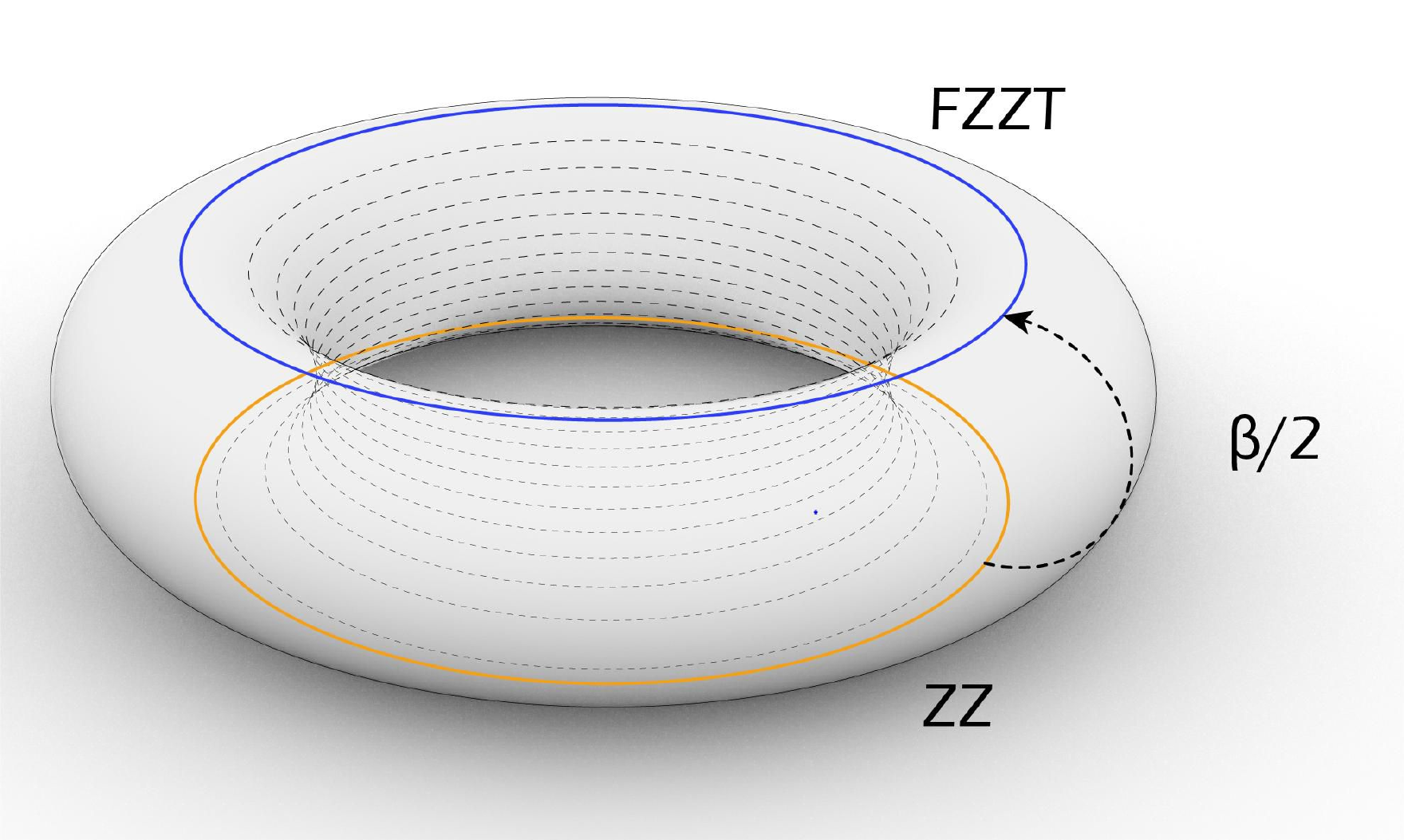}
        \end{overpic}
\begin{overpic}[scale=0.22]{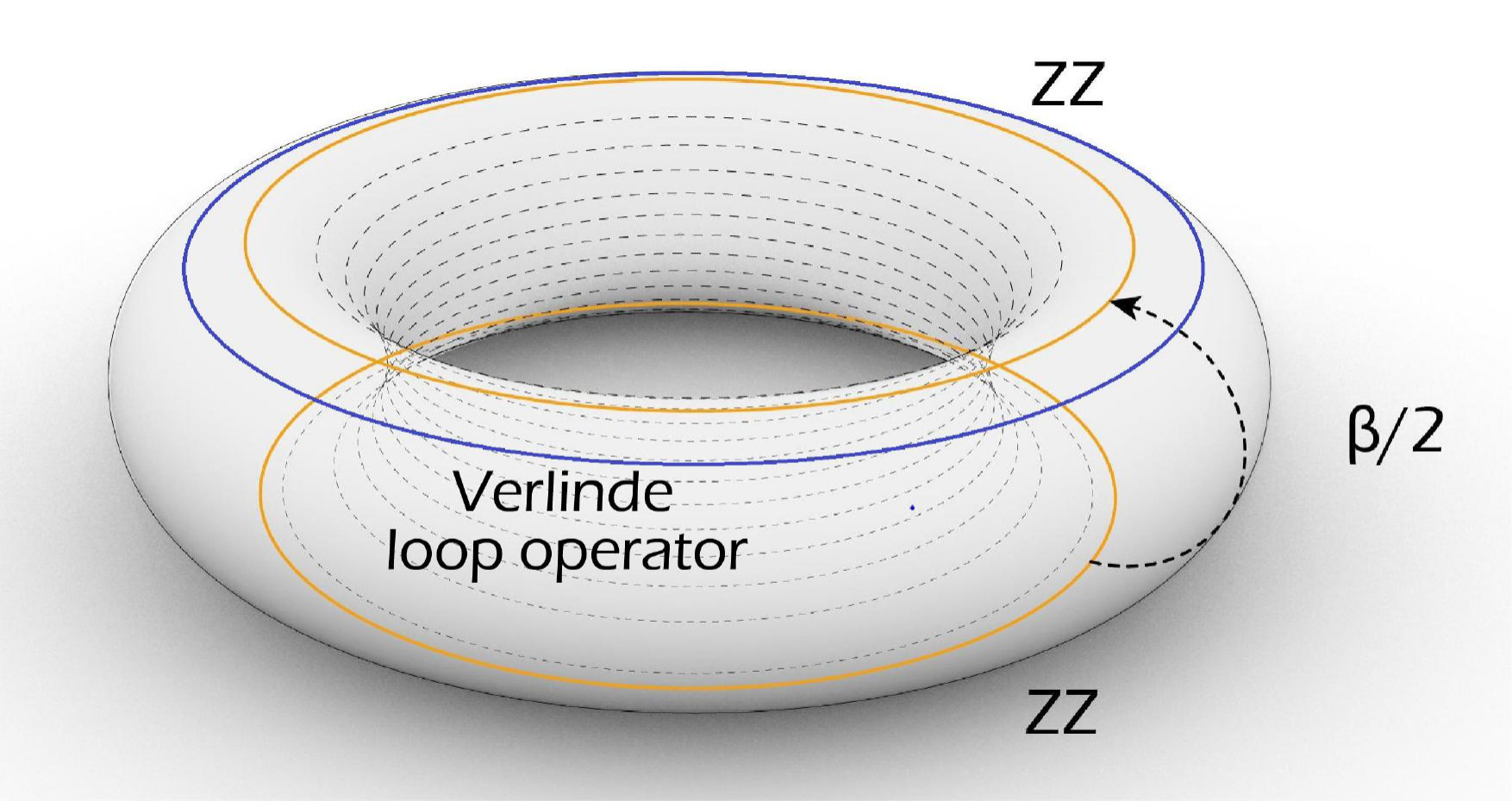}
\end{overpic}  
\end{center}
\caption{(Left): The trumpet partition function can be calculated as a transition amplitude between ZZ boundary states and FZZT boundary states. (Right): The FZZT boundary states can be obtained through acting Verlinde loop operators on the ZZ boundary states}\label{fig:fzzt}
\end{figure}
As illustrated in Figure \ref{fig:fzzt} and \ref{double torus}, the trumpet partition function is given by
\be
\begin{aligned}\label{eq:Z_trumpet_3d}
Z_{\text{trumpet}} (\tau,\bar{\tau},\lambda,\widetilde{\lambda})
&=\bra{\Psi^{\text{HH}}_{\beta/2}} \hat{\mathcal{O}}_{\lambda} \hat{\mathcal{O}}_{\widetilde{\lambda}} | \Psi^{\text{HH}}_{\beta/2} \rangle~, \\
&= \bra{ZZ} \hat{\mathcal{O}}_{\lambda} q^{L_0-c/24} \ket{ZZ}
\bra{\widetilde{ZZ}} \hat{\mathcal{O}}_{\widetilde{\lambda}} \bar{q}^{\bar{L}_0-c/24} \ket{\widetilde{ZZ}}~, \\
&=\chi_{\lambda}\left(-\frac{1}{\tau}\right) \chi_{\widetilde{\lambda}}\left( \frac{1}{\bar{\tau}}\right)~,
\end{aligned}
\ee
where $\hat{\mathcal{O}}_{\lambda}$ is the Verlinde loop operator that carries a similar label to primaries in Liouville theory\cite{Verlinde:1988sn, Alday:2009fs, Drukker:2009id, Drukker:2010jp, Dijkgraaf:1988tf, LeFloch:2017lbt, Mertens:2019tcm}. On the other hand, the product of characters $\chi_{\lambda}\left(-\frac{1}{\tau}\right) \chi_{\widetilde{\lambda}}\left( \frac{1}{\bar{\tau}}\right)$ can be obtained by performing a trace on half of the geometry that is obtained from cutting along the minimal geodesic of the waist on a similar spatial slice to \eqref{bottleneck} \cite{Cotler:2020ugk}.\footnote{We like to clarify what we meant by similar spatial slice. The $\phi$ direction in \eqref{bottleneck} is actually the $\tau_E$ direction of the spatial geometry of the trumpet partition function where $\tau_E \sim \tau_E+\beta$. Hence,  the constant spatial slice here is a constant $\phi$ slice, and  the trace is performed in the $\phi$ direction where $\phi \sim \phi+2 \pi$.} We will make a comment on this at the end of this subsection.

The holomorphic Virasoro character $\chi_{\lambda}\left(-\frac{1}{\tau}\right)$ with conformal weight $h_\lambda = \frac{c-1}{24}+\lambda^2$ is defined by the following overlap

\be
\chi_{\lambda}\left(-\frac{1}{\tau}\right)=\bra{B_{2\lambda}} q^{L_0-c/24} \ket{ZZ}~,
\ee
where
\begin{equation}
    \chi_\lambda(\tau) = \frac{q^{h_\lambda-\frac{c-1}{24}}}{\eta(\tau)}~,
\end{equation}
and $\ket{B_{2\lambda}}$ is the FZZT-boundary state \cite{Fateev:2000ik,Teschner:2000md}
\be \label{fzzt}
\ket{B_{2\lambda}}=\hat{\mathcal{O}}_{\lambda} |ZZ \rangle =\int_0^\infty dP \Psi^{*}_{2\lambda}(P) \ket{P}\rangle~.
\ee
The FZZT-wavefunction is related to the ZZ wavefunction through

\be
\Psi_{\lambda}(P)=\frac{\Psi_{ZZ}(P) \cos(2\pi P \lambda)}{2\sinh(2\pi P b) \sinh(\frac{2\pi P}{b})}~.
\ee
In addition, we have used the relation $\eta(-1/\tau) = \eta(\tau) \sqrt{-i \tau}$ after performing modular transform. 

In terms of normalized operators in \eqref{eq:Pp_Liouville}, the FZZT-boundary state can be expressed as
\be \label{FZZT CARDY}
\ket{B_{2\lambda}}=\int_0^\infty dP' \Psi^{*\prime }_{2\lambda}(P')\ket{P'}\rangle=\int_0^\infty dP' \frac{S_{\lambda P'}}{\sqrt{S_{\mathds{1} P'}}} \ket{P'}\rangle~,
\ee
where
\be
S_{\lambda P'}=2\sqrt{2} \cos (4\pi P' \lambda)~,
\ee
is the modular S-matrix relating two non-degenerate representations in Liouville theory and \eqref{FZZT CARDY} takes a similar form as the Cardy states in rational CFTs \cite{Cardy:1989ir, Cardy:2004hm}. The Schwarzian limit of the chiral half of \eqref{eq:Z_trumpet_3d} reproduces the trumpet partition function in JT gravity \cite{Penington:2023dql,Saad:2019pqd}.

Similar to 2d JT gravity, we again interpret the holographic dual of the operator that initiates the emission of the baby universe as creating a wormhole in the bulk. In hindsight, the ``baby universe operators'' introduced in JT gravity gets uplifted to natural loop observables in 2d CFT, which are the Verlinde loop operators \cite{Verlinde:1988sn, Dijkgraaf:1988tf, Alday:2009fs, Drukker:2009id, Drukker:2010jp, LeFloch:2017lbt, Mertens:2019tcm}. We can understand this duality using the fact that the Verlinde loop operators are shown to be equivalent to topological defects in Liouville theory\cite{Drukker:2009id, Drukker:2010jp, Alday:2009fs}, which in turn are equivalent to Wilson loops in the bulk, similar to the situation in rational CFTs\cite{Witten:1988hf}. The appearance of Verlinde loop operators in our currrent situation stems from the connection between Liouville theory, 3d gravity and quantum Teichm$\ddot{\text{u}}$ller theory\cite{Verlinde:1989ua, Teschner:2003em, Teschner:2005bz, Alday:2009fs, Drukker:2009id}.

Finally, we like to make a comment  regarding the differences between on-shell and off-shell calculations. Naively, we expect the double-trumpet partition function to be obtained through a similar computation as the calculation for the two-boundary torus wormholes in Appendix \ref{app:Stanford_WH}, i.e. inserting a complete basis of Alekseev-Shatashvili states on the boundaries of the annulus and taking trace of the whole Hilbert space. This is a reasonable approach as the two situations have the same topology of torus$\times$interval which is equivalent to annulus$\times$circle. The difference is that the two-boundary torus wormholes are sourced by operator insertions, making it an on-shell geometry whereas the double-trumpet is completely off-shell. If we directly perform a similar trace to obtain the double-trumpet partition function, we end up with the wrong measure, i.e. 
\be
\begin{aligned}
\tilde{Z}'(\tau_1,  \tau_2)&=\int_0^\infty d\lambda  d\widetilde{\lambda} \,Z_{\text{trumpet}}(\tau_1,\bar{\tau}_1,\lambda,\widetilde{\lambda}) Z_{\text{trumpet}}(\tau_2,\bar{\tau}_2,\lambda,\widetilde{\lambda}) \neq \tilde{Z}(\tau_1,  \tau_2)~.
\end{aligned}
\ee
As shown, each trumpet partition function computes the trace for half of the spatial geometry in \eqref{bottleneck}. For on-shell geometrical observables, we can simply glue the two halves together with a flat measure $\int_0^\infty d\lambda d\tilde{\lambda}$. For off-shell configurations, we have to use the measure in \eqref{torus measure}, which comes from two copies of the Weil-Petersson measure that is proportional to $\int_0^\infty d\lambda d\tilde{\lambda} \, \lambda \tilde{\lambda} $, in performing the gluing. This discrepancy exists too in JT gravity where the identity operator limit of the two-point function in JT gravity is divergent and doesn't reproduce the double-trumpet partition function.\footnote{Arguments have been made in \cite{Yan:2023rjh} to attribute the discrepancy on the sum over windings in JT gravity.} We provide a possible resolution to this puzzle. In the off-shell calculation, the extra $\lambda$ and $\bar{\lambda}$ in the gluing measure can be viewed as coming from an integration over relative twists on the two trumpets, similar to JT gravity\cite{Saad:2019lba, Saad:2019pqd}, as there is no gravitational saddles to fix their values. For on-shell calculations, there are extra operator insertions (as shown in Appendix \ref{app:Stanford_WH}) or the ZZ boundaries that act as sources in stabilizing the geometry through a specific twist angle. Hence, the integral over twist angles will be trivial and the gluing measure is flat. We leave a more detailed understanding on the discrepancies between on-shell and off-shell calculations for the future.

\subsection{Late time thermal two-point function}\label{late time two point}
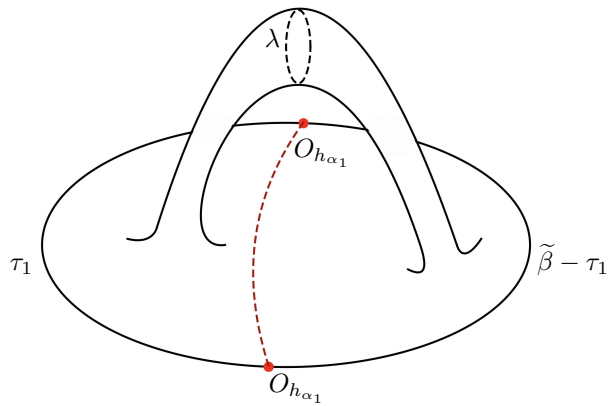
\begin{figure}[h]
\begin{center}
\begin{overpic}[scale=0.2]{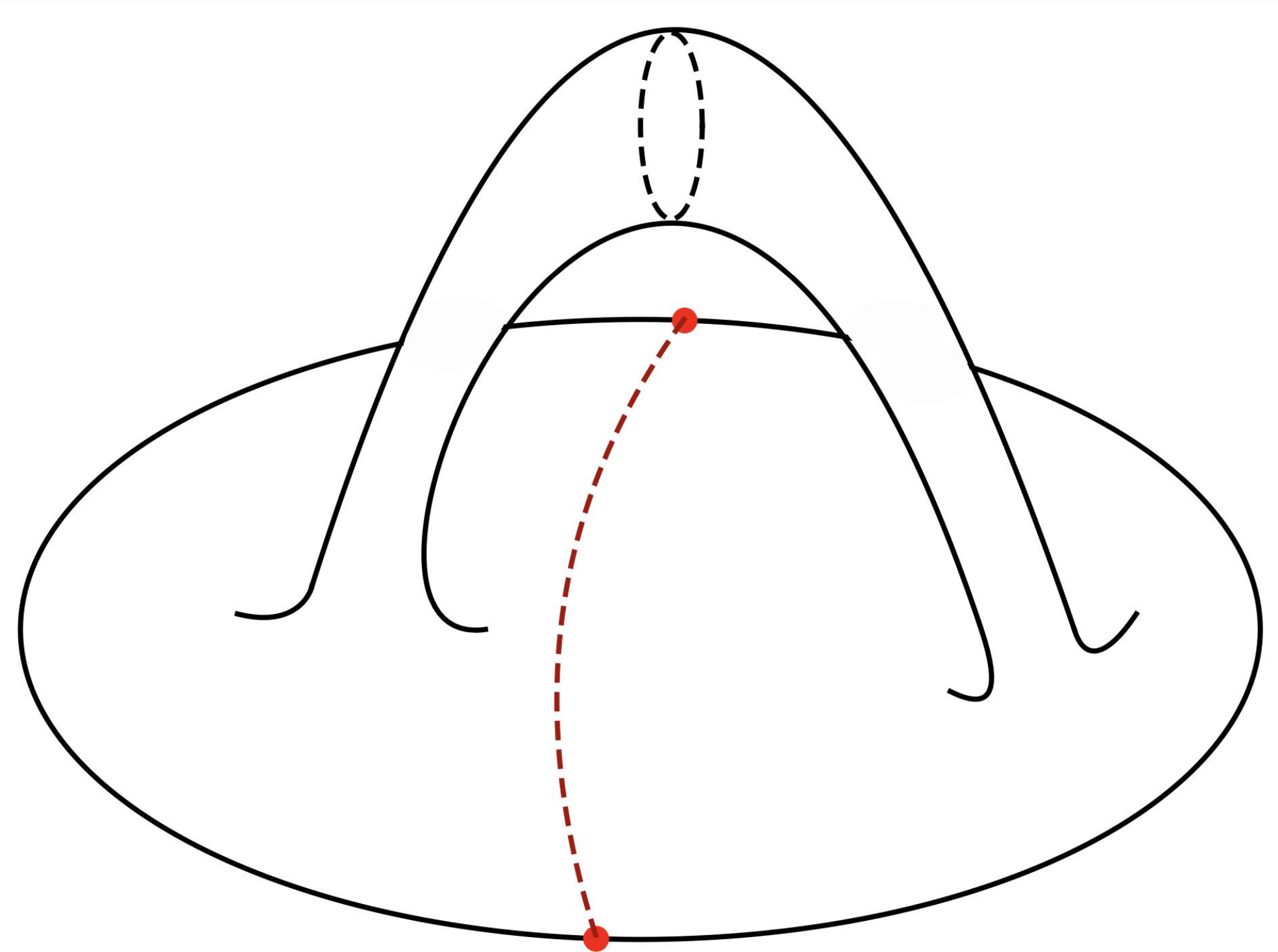}
\put(25,65){\parbox{0.2\linewidth}{
		\begin{equation*}
  \lambda
\end{equation*}}}
\put(-25,20){\parbox{0.2\linewidth}{
		\begin{equation*}
  \tau_1
\end{equation*}}}
\put(85,20){\parbox{0.2\linewidth}{
		\begin{equation*}
  \widetilde{\beta} - \tau_1
\end{equation*}}}
\put(30,-5){\parbox{0.2\linewidth}{
		\begin{equation*}
  O_{h_{\alpha_1}}
\end{equation*}}}
\put(35,42){\parbox{0.2\linewidth}{
		\begin{equation*}
  O_{h_{\alpha_1}}
\end{equation*}}}
\end{overpic}  
\end{center}
\caption{By cutting along the geodesic (red dashed line) that connects the two operators $O_{h_{\alpha_1}}$, the late time two-point function in JT gravity is similar to the ``double trumpet'' geometry. }\label{fig:late2pt}
\end{figure}
If we analytically continue the thermal two-point function to Lorentzian time, \eqref{eq:Liouville_correlators3} has the forever decaying behavior. However, this decaying behavior can't persist as it is in conflict with the AdS/CFT correspondence, and this is the version of information paradox proposed in \cite{Maldacena:2001kr}. We expect to see a ramp and plataeu following the decaying behavior, similar to what is observed in the spectral form factor\cite{Cotler:2016fpe,Saad:2019lba, Saad:2019pqd}, as the dual theory is highly chaotic.

Following a similar spirit to what has been done in JT gravity in reproducing the linear ramp\cite{Saad:2019pqd}, as shown in Figure \ref{fig:late2pt}, we propose the following computation for the late time thermal two-point function
\be \label{modular sum for late}
\mathcal{G}^{(2),\text{late}}_{\alpha_1}(z_1; \tau)=\sum_{g \in PSL(2,Z)} \widetilde{\mathcal{G}^{(2),\text{late}}_{\alpha_1}(g z_1; g \tau)}~,
\ee
\begin{figure}[h]
	\centering
	\includegraphics[scale=0.4]{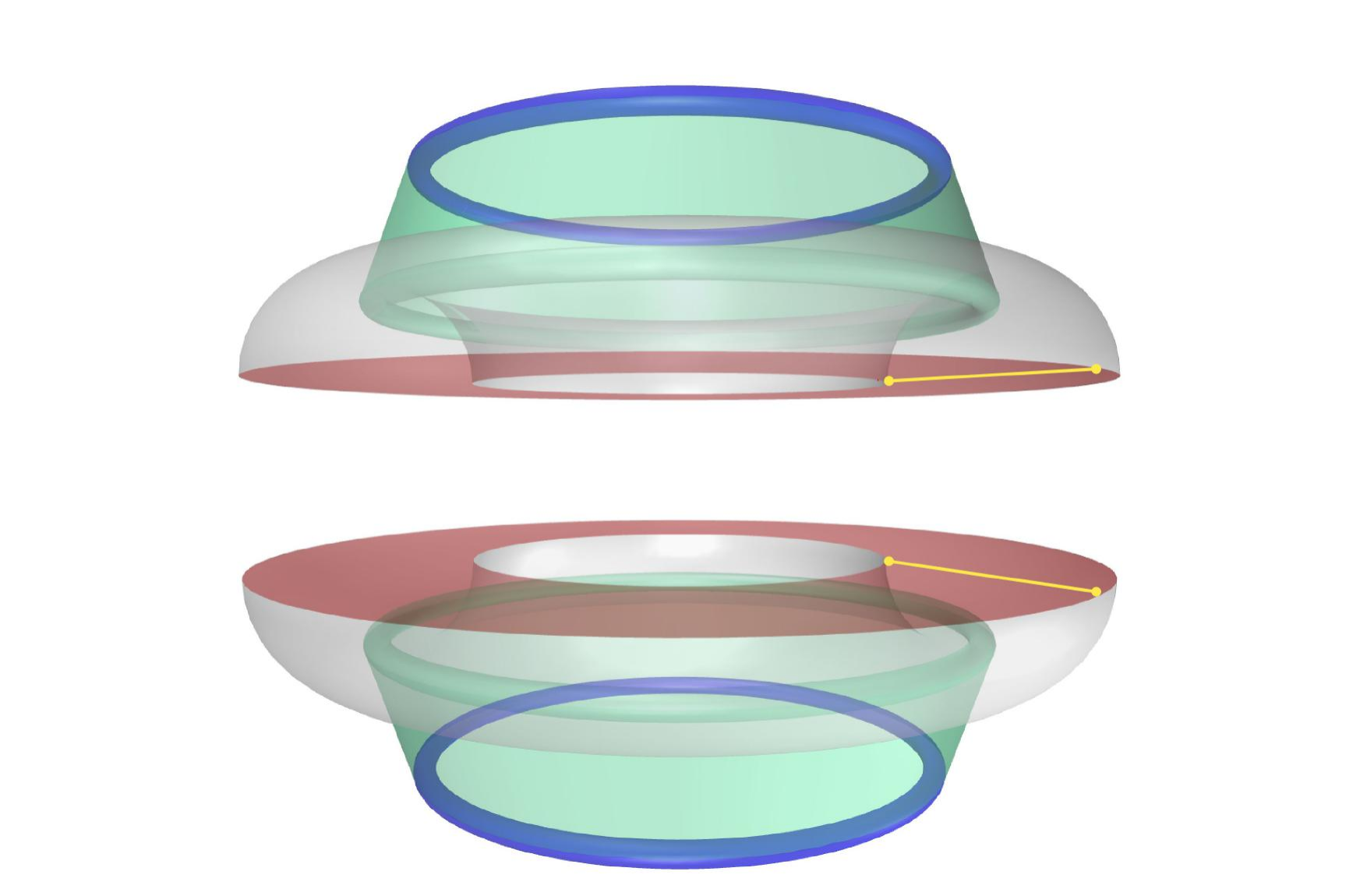}
	\caption{The off-shell wormhole contribution to the preamplitude of the late time two-point function. The preamplitude can be viewed as coming from two copies of Liouville operator insertions, indicated by the yellow line, sandwiched between FZZT boundary states that are responsible for the emission of baby universes. The baby universes, indicated by the blue tori, are glued together, forming a ``double trumpet'' geometry.} \label{late time two point1}
\end{figure}  
where the preamplitude, as shown in Figure \ref{late time two point1}, is given by, 
\be
\begin{aligned}
\widetilde{\mathcal{G}^{(2),\text{late}}_{\alpha_1}( z_1; \tau)}&=\mathcal{N}(z_1,\tau)\int_0^\infty  d\lambda d\widetilde{\lambda} \, \lambda \, \widetilde{\lambda} \, \langle \hat{V}_{\alpha_1}(z_1,\bar{z}_1) \rangle_{\beta/2,\text{FZZT}}^\lambda  \langle \hat{V}_{\alpha_1}(\bar{z}_1,z_1) \rangle_{\beta/2,\widetilde{\text{FZZT}}}^{\widetilde{\lambda}}\\
&=\mathcal{N}(z_1,\tau) \int_0^\infty  d\lambda  \, \lambda \bra{B_{2\lambda}}e^{-(\beta/2-\tau_1) H} \hat{V}_{\alpha_1} e^{-\tau_1 H} \ket{B_{2\lambda}} \int_0^\infty d\widetilde{\lambda} \,\widetilde{\lambda} \bra{B_{2\widetilde{\lambda}}}e^{-(\beta/2+\tau_1) H} \hat{V}_{\alpha_1} e^{\tau_1 H}\ket{B_{2\widetilde{\lambda}}}~,
\end{aligned}
\ee
where $\tau_1 = (z_1 - \bar{z}_1)/2 i$ is the location of the operator $V_{\alpha_1}(z_1,\bar{z}_1)$ on the thermal circle. $\langle \cdot \rangle_{\beta/2,\text{FZZT}}^\lambda$ is the expectation value between two FZZT-boundary states $|B_{2 \lambda} \rangle$ on a finte cylinder of length $\beta/2$ and $\mathcal{N}(z_1,\tau)$ is an overall constant from the integration measure that is related to the moduli of the asymptotic torus. We obtain the following expression for the chiral half of the late time two-point function
\be
\begin{aligned}
 &\int_0^\infty  d\lambda  \, \lambda \bra{B_{2\lambda}}e^{-(\beta/2-\tau_1) H} \hat{V}_{\alpha_1} e^{-\tau_1 H} \ket{B_{2\lambda}} \\
&=\int_0^\infty  d\lambda  \, \lambda \int_0^\infty dP' dQ'  \hat{C}_{DOZZ}(P',P_{\alpha_1},Q') \Psi_{2\lambda}'(P') \Psi_{2\lambda}^{* \prime}(Q') \mathcal{F}_{h_{P'},h_{Q'}}(h_{\alpha_1}, h_{\alpha_1},z_1 - \bar{z}_1,\beta)\\
&=8c_b^{-1}\int_0^\infty  d\lambda  \, \lambda \int_0^\infty dP' dQ'  C_{0}(h_{P'},h_{\alpha_1},h_{Q'})\cos(4 \pi P'\lambda) \cos(4 \pi Q'\lambda) \mathcal{F}_{h_{P'},h_{Q'}}(h_{\alpha_1}, h_{\alpha_1},z_1 - \bar{z}_1,\beta)~.
\end{aligned}
\ee
The $\lambda$ integral doesn't converge, but we immediately see that  the integrand takes a similar form to the density of states for the ramp in random matrix theory. Following \cite{Saad:2019lba}, we perform the $\lambda$ integral using analytic continuation for the correlator of resolvents
\be \label{ramp dos}
\rho_{\text{ramp}}(P',Q')=\int_0^\infty d\lambda \, \lambda {\cos(4\pi P' \lambda) \cos(4\pi Q' \lambda)}=-\frac{P'^2+Q'^2}{16 \pi ^2 \left(P'^2-Q'^2\right)^2}~,
\ee
where $\rho_{\text{ramp}}(P',Q')$ encodes the level repulsion in random matrix theory and details on this calculation can be found in Appendix \ref{app:ramp}.

To summarize, the preamplitude for the late time thermal two-point function is given by
\be\label{eq:late2pt_3d}
\begin{aligned}
\widetilde{\mathcal{G}^{(2),\text{late}}_{\alpha_1}(  z_1; \tau)} &=64 \mathcal{N}(z_1,\tau) c_b^{-2}\Bigg{|}\int_0^\infty dP' dQ' \rho_{\text{ramp}}(P',Q') C_{0}(h_{P'},h_{\alpha_1},h_{Q'})\mathcal{F}_{h_{P'},h_{Q'}}(h_{\alpha_1}, h_{\alpha_1},z_1 - \bar{z}_1,\beta)\Bigg{|}^2~.
 \end{aligned}
\ee

Provided that we have the correct measure, we expect to get the ramp behavior from \eqref{modular sum for late} in the low temperature, late-time limit, similar to \cite{Cotler:2020ugk, Saad:2019pqd}.
In contrast to on-shell gravitational quantities that was argued to capture statistics of OPE coefficients \cite{Collier:2022bqq}, the off-shell quantities capture statistics of density of states and are responsible for the chaotic random matrix behavior\cite{Cotler:2020ugk}, as seen from \eqref{ramp dos}. Similar to on-shell gravitational quantities, the Schwarzian limit of the chiral half of the late time thermal two-point function in \eqref{eq:late2pt_3d} reproduces the contribution from ``double-trumpet'' geometries in JT gravity \cite{Saad:2019pqd}.

\section{Future directions}
\subsection{Rotating BTZ and spinning operators}
In this paper, we focused on Hartle-Hawking state related to non-rotating BTZ. It's natural to expect that for rotating BTZ, we have the following Hartle-Hawking state
\be\label{eq:HH_theta}
|\Psi^{\text{BTZ}}_{\beta/2,\theta/2} \rangle=q^{\frac{L_0}{2} - \frac{c}{48}} |ZZ \rangle \bar{q}^{\frac{\bar{L}_0}{2} - \frac{c}{48}} |\widetilde{ZZ} \rangle=e^{-\beta_L H/4}|ZZ \rangle e^{-\beta_R H/4}|\widetilde{ZZ} \rangle ~,
\ee
where $q=e^{2\pi i \tau}, \tau=\frac{\theta+i\beta}{2\pi}=\frac{i \beta_L}{2\pi}, \bar{\tau}=\frac{\theta-i\beta}{2\pi}=-\frac{i \beta_R}{2\pi}$, $\theta$ is the twist angle and $H$ is the total Hamiltonian in Liouville theory.

For correlation functions in terms of Liouville overlaps, it is natural to consider spinning probe operators at asymmetric insertion points

\be
\mathcal{G}_{\alpha_1,\bar{\alpha}_1,\cdots,\alpha_n,\bar{\alpha}_n}^{(2n)}(z_1,z_1'\cdots,z_n,z_n') = \langle  \hat{V}_{\alpha_1}(z_1,z'_1)   \cdots \hat{V}_{\alpha_n}(z_n,z'_{n})  \rangle_{\beta/2; \theta/2,ZZ}\langle  \hat{V}_{\bar{\alpha}_1}(\bar{z}'_1,\bar{z}_1) \cdots   \hat{V}_{\bar{\alpha}_n}(\bar{z}'_{n},\bar{z}_{n}) \rangle_{ \beta/2;\theta/2, \widetilde{ZZ}}~.
\ee
where $\langle \cdot \rangle_{\beta/2; \theta/2,ZZ}$ represents ZZ overlaps on a finite cylinder $\beta/2$ and via the doubling trick, one of the ends of the cylinder at the ZZ boundaries is glued with a twist angle $\theta$. We use Liouville scalar operator insertions to represent spinning operators in holographic CFTs through 
\be
\tilde{O}_{h_{\alpha_1}, h_{\bar{\alpha}_1}}(z_1,\bar{z}_1)  \tilde{O}_{h_{\bar{\alpha}_1}, h_{\alpha_1}}(z_1',\bar{z}_1') \to \hat{V}_{\alpha_1}(z_1,z_1') \hat{V}_{\bar{\alpha}_1}(\bar{z}_1', \bar{z}_1)~.
\ee
To justify these identifications, we need to consider the quasi-Fuchsian wormhole slicing \eqref{quasi solution}\cite{Krasnov:2005dm} as our metric ansatz for computing gravitational actions with spinning probe operator insertions. This is along the lines of obtaining rotating BTZ partition function using \eqref{quasi solution} \cite{Garbarz:2020fky}.

We leave a more detailed analysis on these problems for the future.

\subsection{Universal Type II$_{\infty}$ algebra in 2d large c CFTs}
As argued in Section \ref{sec:factorization} and  following similar arguments in \cite{Penington:2023dql, Kolchmeyer:2023gwa}, similarity in obtaining the trace formula between 3d gravity and JT gravity strongly hints that we get a Type II$_{\infty}$ algebra when 3d gravity is coupled to probe matter fields. It is definitely worthwhile to understand this algebra better.

Moreover, we can understand this universal algebra from large c chaotic CFTs. Motivated by the eigenstate thermalization hypothesis and CFT ensemble proposal in \cite{Srednicki:1994mfb, PhysRevA.43.2046, Collier:2022bqq}, it seems reasonable to make the following conjecture: In any large c chaotic CFTs, if we perform a coarse graining procedure, i.e. averaging heavy operators over a small energy window above the black hole threshold, we get a universal Type II$_{\infty}$ algebra that's related to the universal CFT ensemble in general cases, and in particular, Liouville theory in the presence of two boundaries. Partial support of this conjecture comes from taking into account universal dynamics of heavy operators in 2d CFTs
\cite{Collier:2018exn, Collier:2019weq, Numasawa:2022cni}, arguments similar to \cite{Hartman:2014oaa} to extend the regime of validity of these formulas in large c CFTs, and Tauberian theorem to control the size of the averaging energy window \cite{Qiao:2017xif, Mukhametzhanov:2018zja, Mukhametzhanov:2019pzy, Ganguly:2019ksp, Pal:2019yhz, Das:2020uax, Pal:2023cgk}. It is the universal sector of algebra for heavy operators that provides emergence of universal dynamics in 3d gravity. The emergence of geometry from algebra is similar to what has been observed in higher dimensions where the algebra is given by the cross product of algebra generated by generalized free fields and the modular automorphism group\cite{Leutheusser:2021frk, Witten:2021unn}. 

We also like to point out one more approach in seeing the emergence of geometry from algebra, especially from the perspective of open CFTs. It has been known that we can use the modular tensor
category of Moore-Seiberg data\cite{Moore:1988qv, Moore:1988uz} to construct TQFTs that characterizes rational CFT observables\cite{Felder:1999mq, Fuchs:2002cm, Fuchs:2003id, Fuchs:2004dz, Fuchs:2004xi, Fjelstad:2005ua}. More explicitly, \cite{Chen:2022wvy, Cheng:2023kxh, Chen:2024unp} generalize insights made in \cite{PhysRevLett.121.177203, Aasen:2016dop, Aasen:2020jwb} through constructing an explicit real space renormalization group (RG) operator using the 6j symbols that are related to the open CFT three point functions. Upon applying the RG procedure, a TQFT in one higher dimension and its associated geometry will be emergent via an exact MERA-like tensor network\cite{Vidal:2007hda, Vidal:2008zz, Swingle:2009bg, 2011JSP...145..891E}. Using this method, we hope that we can see the emergence of 3d gravity from the universal algebraic data in 2d large c CFTs \cite{Chen:2024unp}.

\subsection{3d topological recursion, 3d gravity from topological M-theory and random matrices from conformal bootstrap}

As mentioned in Section \ref{sec:late time}, we need a better understanding in  obtaining the integration measure for the gluing of off-shell gravitational quantities in 3d gravity. If we have a more systematic way of obtaining the integration measure, we can use the building blocks in this paper to construct path integrals that compute genus $g$ partition function with $n$ boundaries, similar to what has been done in JT gravity, and probably show the corresponding ``topological recursion relation'' in 3d gravity \cite{Saad:2019lba}, thus, obtain a better understanding on the averaged dual theory. 

We want to point out another potential approach to address this problem. It is understood that we can understand JT gravity from the worldsheet theory of topological B-model on certain fibration over the spectral curve of the matrix model\cite{Dijkgraaf:2002fc, McNamara:2020uza, Post:2022dfi}. It is  natural to expect that 3d gravity is related to the membrane theory of ``topological M theory''\cite{Dijkgraaf:2004te}. 

Finally, regarding the CFT ensemble, we also like to point out that the ensemble for OPE coefficients is proposed based on bootstrap considerations (identity block dominance in certain channels)\cite{Collier:2019weq}. However, we expect the random matrix behavior, i.e. level repulsion in the density of states \eqref{ramp dos} and its associated ramp behavior, to be universal in chaotic theories. In this paper, it comes from the gluing measure related to the Weil-Petersson form in gravity. It would be interesting to understand this universality from a bootstrap point of view.\footnote{Papers along this direction appeared\cite{DiUbaldo:2023qli, Belin:2023efa} when we are on the final stage of preparing this paper.}

\bigskip

\nocite{Turiaci:2017zwd, Mertens:2020hbs, Mertens:2022aou, Engelsoy:2016xyb, Lam:2018pvp, Lou:2019heg, Shen:2019rck, Kim:2015qoa, Jafferis:2015del, Blommaert:2018oro, Iliesiu:2019xuh, Teschner:2001rv, Teschner:2003at, Nidaiev:2013bda, Carlip:2005tz}

\noindent \textbf{Acknowledgments} We thank Thomas Hartman for helpful discussions on the wormhole slicing in 3d gravity, and collaboration in the early stages of this project. We thank Hao Geng, Janet Hung, Kristan Jensen, Henry Lin, Don Marolf, Steve Shenker and Gabriel Wong for helpful discussions. WZC thanks Ho Tat Lam, Grégoire Mathys, David Meltzer, Baur Mukhametzhanov and Jeevan Chandra Namburi for useful discussions. YJ thanks Simon Caron-Huot, Venkatesa Chandrasekaran, David Kolchmeyer, Chen-Te Ma, Alex Maloney, Thomas Mertens, Eric Perlmutter, Amir Tajdini, Joaquin Turiaci and Yixu Wang for helpful discussions, and his friends for encouraging him to continue pursuing research in physics. We thank Zhenhao Zhou for providing help in plotting many of the figures in this paper. This work is supported by NSF grant PHY-2014071. The work of YJ is also supported by the ASCR EXPRESS grant, Novel Quantum Algorithms from Fast Classical Transforms, and Northeastern University.

\appendix
\section{Liouville basics}\label{app:Liouville}

In this appendix, we list out the basics for Liouville theory that we used in this paper. The conformal weights of primary fields are related to Liouville momentum $P$ through
\begin{equation}
    h_P = \frac{c-1}{24}+P^2~.
\end{equation}
The central charge $c$ is related to the Liouville background charge $Q$ or Liouville parameter $b$ through
\begin{equation}
    Q = \sqrt{\frac{c-1}{6}} = b+b^{-1}~,
\end{equation}
where the semiclassical behavior of CFT quantities are studied by taking the large $c$ limit or small $b$ limit. In studying the semiclassical limit, we consider heavy probe operators with large conformal dimension, and for these operators, it is convenient to adopt the following parametrization of conformal weights 
\begin{equation} \label{defect strength}
    h = \frac{c}{6}\eta(1-\eta) ~,
\end{equation}
where we call $\eta$ the defect strength as in \cite{Collier:2022bqq}, and $\eta \in [0,1/2]$ for below black hole threshold operators. 

In the main text, we compute correlation functions by inserting vertex operators $V_\alpha(z) = e^{2 \alpha \Phi(z)}$ between the ZZ boundary states. The conformal weight of $V_\alpha(z)$ is given by
\begin{equation}
    h_\alpha = \alpha(Q-\alpha)~,
\end{equation}
where $\alpha$ is related to the Liouville momentum through $\alpha = \frac{Q}{2}+i P_{\alpha}$. The 2-point function determines the normalization of vertex operators
\begin{equation}
	\begin{split}
	\langle V_{\alpha_1}(0) V_{\alpha_2}(1) \rangle_L = \delta(P_{\alpha_1} +P_{\alpha_2}) + S_L(P_{\alpha_1}) \delta(P_{\alpha_1} - P_{\alpha_2})~,
	\end{split}
\end{equation} 
where $S_L(P)$ is the Liouville reflection amplitude given by
\begin{equation}
S_L(P) = -\left(\pi \mu \gamma(b^2)\right)^{-2 i P/b} \frac{\Gamma\left(1+\frac{2 i P}{b}\right)\Gamma(1+2 i P b)}{\Gamma\left(1-\frac{2 i P}{b}\right)\Gamma(1-2 i P b)}~.
\end{equation}

The DOZZ structure constant is given by \cite{Dorn:1994xn,Zamolodchikov:1995aa}
\begin{equation}
    C_{\text{DOZZ}}(P_{\alpha_1}, P_{\alpha_2}, P_{\alpha_3}) = \langle V_{\alpha_1}(0) V_{\alpha_2}(1) V_{\alpha_3}(\infty) \rangle~,
\end{equation}
and is related to the universal OPE coefficient $C_0(P_{\alpha_1}, P_{\alpha_2}, P_{\alpha_3})$ through \cite{Collier:2019weq}
\begin{equation}
    \begin{split}
        C_0(P_{\alpha_1}, P_{\alpha_2}, P_{\alpha_3}) &= \frac{\left(\pi \mu \gamma(b^2)b^{2-2b^2}\right)^{\frac{Q}{2b}}}{2^{\frac{3}{4}}\pi}\frac{\Gamma_b(2 Q)}{\Gamma_b( Q)}\frac{ C_{\text{DOZZ}}(P_{\alpha_1}, P_{\alpha_2}, P_{\alpha_3})}{\sqrt{\prod_{k=1}^3 S_L(P_{\alpha_k}) \rho_0(P_{\alpha_k})}}~, \\
        &\equiv c_b \frac{ C_{\text{DOZZ}}(P_{\alpha_1}, P_{\alpha_2}, P_{\alpha_3})}{\sqrt{\prod_{k=1}^3 S_L(P_{\alpha_k}) \rho_0(P_{\alpha_k})}}~,
    \end{split}
\end{equation}
where $c_b$ is a constant that is independent of the primary conformal weights and cancels the dependence on the cosmological constant $\mu$. More explicitly, the universal OPE function $C_0$ takes the following expression \cite{Collier:2019weq}
\begin{equation}
    C_0(P_{\alpha_1}, P_{\alpha_2}, P_{\alpha_3}) = \frac{\Gamma_b(2 Q)}{\sqrt{2} \Gamma_b(Q)^3}\frac{\prod_{\pm1,2,3} \Gamma_b\left(\frac{Q}{2}\pm_1 i P_{\alpha_1} \pm_2 i P_{\alpha_2} \pm_1 i P_{\alpha_1}\right)}{\prod_{k=1}^3 \Gamma_b(Q+2 i P_{\alpha_k}) \Gamma_b(Q-2 i P_{\alpha_k})}~,
\end{equation}
where $\prod_{\pm1,2,3}$ represents all possible sign permutations in the product of the eight terms. $\Gamma_b(x)$ is the ``double'' gamma function with poles at $x = -m b -n b^{-1}$ where $m,n$ being non-negative integers. In the main text, we use $\hat{V}$ to denote a different normalization for Liouville operators that are below and above the black hole threshold respectively, 
\begin{equation}
    \hat{V}_{\alpha}= \begin{cases}
    \frac{V_{\alpha}}{\sqrt{S_L(P) \rho_0(P)}}~&,~~\eta \in[0,\frac{1}{2}]~, \\
        \frac{V_{\alpha}}{\sqrt{S_L(P)}}~&,~~P \text{  real}~.
    \end{cases}
\end{equation}
In the main text, $\ket{P'}$ is used to denote primary states correspond to normalizable primary operators $\hat{V}_{\alpha}$ such that
\begin{equation}
    \hat{C}_{\text{DOZZ}}(P_{\alpha_1}, P_{\alpha_2}, P_{\alpha_3}) = \langle \hat{V}_{\alpha_1}(0) \hat{V}_{\alpha_2}(1) \hat{V}_{\alpha_3}(\infty) \rangle~,
\end{equation}
where the normalized DOZZ structure constant $\hat{C}_{\text{DOZZ}}$ is given by
\begin{equation}
     \hat{C}_{\text{DOZZ}}(P_{\alpha_1}, P_{\alpha_2}, P_{\alpha_3})  = \frac{ C_{\text{DOZZ}}(P_{\alpha_1}, P_{\alpha_2}, P_{\alpha_3})}{\sqrt{\prod_{j \in \text{light}} S_L(P_{\alpha_j}) \rho_0(P_{\alpha_j})\prod_{k \in \text{heavy}} S_L(P_{\alpha_k}) }}~.
\end{equation}

\section{Two-boundary torus wormhole and its Schwarzian limit}\label{app:Stanford_WH}
In this appendix, we show that the two-boundary Hilbert space defined on an annulus in 3d gravity is useful for calculating certain exact two-boundary torus wormhole correlators. In the Schwarzian limit,  we reproduce the two-boundary disk wormholes with heavy probe operator insertions in JT gravity that describes quantum noise\cite{Stanford:2020wkf}.

\begin{figure}
\begin{center}
\begin{overpic}[scale=0.5]{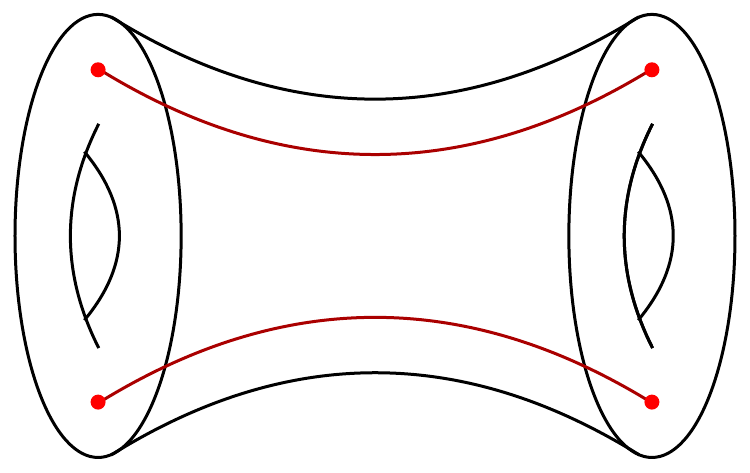}
\put(-3,30){\parbox{0.2\linewidth}{
		\begin{equation*}
			\tau
\end{equation*}}}
\put(71,30){\parbox{0.2\linewidth}{
		\begin{equation*}
			\tau'
\end{equation*}}}
\end{overpic}  
\end{center}
\caption{The two-boundary torus wormhole that is constructed from product of torus two-point functions. }\label{fig:torusWH}
\end{figure}

As an explicit example, we compute the preamplitude of the wormhole contribution for the product of torus two-point functions (Figure \ref{fig:torusWH}), i.e. 
\be
\begin{split}
&G_{\alpha_1,\alpha_2}(\tau,\bar{\tau}) G_{\alpha_1,\alpha_2}(\tau',\bar{\tau}')=\braket{O_{h_{\alpha_1}, h_{\alpha_1}} O_{h_{\alpha_2}, h_{\alpha_2}}}_{T_2(\tau,\bar{\tau})} \braket{O_{h_{\alpha_1}, h_{\alpha_1}}  O_{h_{\alpha_2}, h_{\alpha_2}} }_{T_2(\tau',\bar{\tau}')}~, \\
&= \sum_{p,p',q,q'} c_{1 p q} c_{2 p q} c_{1 p' q'} c_{2 p' q'}\vcenter{\hbox{
	\begin{tikzpicture}[scale=0.5]
	\draw[thick] (0,0) circle (1);
	\draw[thick] (-1,0) -- (-2,0);
	\node[above] at (-2,0) {$O_{h_{\alpha_1}}$};
	\node[above] at (0,1) {$p$};
	\node[below] at (0,-1) {$q$};
	\draw[thick] (1,0) -- (2,0);
	\node[above] at (2,0) {$O_{h_{\alpha_2}}$};
	\node[scale=0.75] at (0,0) {$\tau$};
	\end{tikzpicture}
	}}\overline{\vcenter{\hbox{
	\begin{tikzpicture}[scale=0.5]
	\draw[thick] (0,0) circle (1);
	\draw[thick] (-1,0) -- (-2,0);
	\node[above] at (-2,0) {$O_{h_{\alpha_1}}$};
	\node[above] at (0,1) {$p$};
	\node[below] at (0,-1) {$q$};
	\draw[thick] (1,0) -- (2,0);
	\node[above] at (2,0) {$O_{h_{\alpha_2}}$};
	\node[scale=0.75] at (0,0) {$\tau$};
	\end{tikzpicture}
	}}}\vcenter{\hbox{
	\begin{tikzpicture}[scale=0.5]
	\draw[thick] (0,0) circle (1);
	\draw[thick] (-1,0) -- (-2,0);
	\node[above] at (-2,0) {$O_{h_{\alpha_1}}$};
	\node[above] at (0,1) {$p'$};
	\node[below] at (0,-1) {$q'$};
	\draw[thick] (1,0) -- (2,0);
	\node[above] at (2,0) {$O_{h_{\alpha_2}}$};
	\node[scale=0.75] at (0,0) {$\tau'$};
	\end{tikzpicture}
	}}\overline{\vcenter{\hbox{
	\begin{tikzpicture}[scale=0.5]
	\draw[thick] (0,0) circle (1);
	\draw[thick] (-1,0) -- (-2,0);
	\node[above] at (-2,0) {$O_{h_{\alpha_1}}$};
	\node[above] at (0,1) {$p'$};
	\node[below] at (0,-1) {$q'$};
	\draw[thick] (1,0) -- (2,0);
	\node[above] at (2,0) {$O_{h_{\alpha_2}}$};
	\node[scale=0.75] at (0,0) {$\tau'$};
	\end{tikzpicture}
	}}}~,
\end{split}
\ee
where the last line comes from the usual Virasoro conformal block expansion. 

In particular, through inserting a complete basis of Alekseev-Shatashvili states on each boundary of the annulus and taking trace, we reproduce the wormhole contribution that has a topology of torus$\times$interval with operator insertions. This comes from the idenfication of Hilbert space in Section \ref{sec:3d_Liouville_AS}, and identification of the two pairs of probe particle insertions across boundaries as four Liouville operators $\hat{V}_{\alpha_1} \hat{V}_{\alpha_1} \hat{V}_{\alpha_2} \hat{V}_{\alpha_2}$ in Section \ref{sec:correlation_function}. Following this identification, we compute the trace in the coupled Alekseev-Shatashvili theories/Liouville theory and obtain
\be
\Bigg{|}\int_0^\infty dP' dQ'   \hat{C}_{DOZZ}(P',P_{\alpha_1},Q') \hat{C}_{DOZZ}(P',P_{\alpha_2},Q') \vcenter{\hbox{
	\begin{tikzpicture}[scale=0.5]
	\draw[thick] (0,0) circle (1);
	\draw[thick] (-1,0) -- (-2,0);
	\node[above] at (-2,0) {$O_{h_{\alpha_1}}$};
	\node[above] at (0,1) {$P'$};
	\node[below] at (0,-1) {$Q'$};
	\draw[thick] (1,0) -- (2,0);
	\node[above] at (2,0) {$O_{h_{\alpha_2}}$};
	\node[scale=0.75] at (0,0) {$\tau$};
	\end{tikzpicture}
	}}\vcenter{\hbox{
	\begin{tikzpicture}[scale=0.5]
	\draw[thick] (0,0) circle (1);
	\draw[thick] (-1,0) -- (-2,0);
	\node[above] at (-2,0) {$O_{h_{\alpha_1}}$};
	\node[above] at (0,1) {$P'$};
	\node[below] at (0,-1) {$Q'$};
	\draw[thick] (1,0) -- (2,0);
	\node[above] at (2,0) {$O_{h_{\alpha_2}}$};
	\node[scale=0.75] at (0,0) {$\tau'$};
	\end{tikzpicture}
	}} \Bigg{|}^2~.
\ee

The above answer matches with what we expect from the CFT ensemble in \cite{Collier:2022bqq}, as the ensemble average sets $p=p',q=q'$, and
\be
\begin{aligned}
&\overline{G_{\alpha_1,\alpha_2}(\tau,\bar{\tau}) G_{\alpha_1,\alpha_2}(\tau',\bar{\tau}')} \\
&=4\sum_{p,q} \overline{c^2_{p 1 q}} \overline{ c^2_{p 2 q} } \Bigg{|}\vcenter{\hbox{
	\begin{tikzpicture}[scale=0.5]
	\draw[thick] (0,0) circle (1);
	\draw[thick] (-1,0) -- (-2,0);
	\node[above] at (-2,0) {$O_{h_{\alpha_1}}$};
	\node[above] at (0,1) {$p$};
	\node[below] at (0,-1) {$q$};
	\draw[thick] (1,0) -- (2,0);
	\node[above] at (2,0) {$O_{h_{\alpha_2}}$};
	\node[scale=0.75] at (0,0) {$\tau$};
	\end{tikzpicture}
	}}\vcenter{\hbox{
	\begin{tikzpicture}[scale=0.5]
	\draw[thick] (0,0) circle (1);
	\draw[thick] (-1,0) -- (-2,0);
	\node[above] at (-2,0) {$O_{h_{\alpha_1}}$};
	\node[above] at (0,1) {$p$};
	\node[below] at (0,-1) {$q$};
	\draw[thick] (1,0) -- (2,0);
	\node[above] at (2,0) {$O_{h_{\alpha_2}}$};
	\node[scale=0.75] at (0,0) {$\tau'$};
	\end{tikzpicture}
	}} \Bigg{|}^2\\
&\approx 4\Bigg{|}\int_0^\infty dP' dQ'  \rho_0(P') \rho_0(Q') C_0(P',P_{\alpha_1},Q') C_0(P',P_{\alpha_2},Q') \vcenter{\hbox{
	\begin{tikzpicture}[scale=0.5]
	\draw[thick] (0,0) circle (1);
	\draw[thick] (-1,0) -- (-2,0);
	\node[above] at (-2,0) {$O_{h_{\alpha_1}}$};
	\node[above] at (0,1) {$P'$};
	\node[below] at (0,-1) {$Q'$};
	\draw[thick] (1,0) -- (2,0);
	\node[above] at (2,0) {$O_{h_{\alpha_2}}$};
	\node[scale=0.75] at (0,0) {$\tau$};
	\end{tikzpicture}
	}}\vcenter{\hbox{
	\begin{tikzpicture}[scale=0.5]
	\draw[thick] (0,0) circle (1);
	\draw[thick] (-1,0) -- (-2,0);
	\node[above] at (-2,0) {$O_{h_{\alpha_1}}$};
	\node[above] at (0,1) {$P'$};
	\node[below] at (0,-1) {$Q'$};
	\draw[thick] (1,0) -- (2,0);
	\node[above] at (2,0) {$O_{h_{\alpha_2}}$};
	\node[scale=0.75] at (0,0) {$\tau'$};
	\end{tikzpicture}
	}} \Bigg{|}^2\\
&=4\Bigg{|}\int_0^\infty dP' dQ'   \hat{C}_{DOZZ}(P',P_{\alpha_1},Q') \hat{C}_{DOZZ}(P',P_{\alpha_2},Q') \vcenter{\hbox{
	\begin{tikzpicture}[scale=0.5]
	\draw[thick] (0,0) circle (1);
	\draw[thick] (-1,0) -- (-2,0);
	\node[above] at (-2,0) {$O_{h_{\alpha_1}}$};
	\node[above] at (0,1) {$P'$};
	\node[below] at (0,-1) {$Q'$};
	\draw[thick] (1,0) -- (2,0);
	\node[above] at (2,0) {$O_{h_{\alpha_2}}$};
	\node[scale=0.75] at (0,0) {$\tau$};
	\end{tikzpicture}
	}}\vcenter{\hbox{
	\begin{tikzpicture}[scale=0.5]
	\draw[thick] (0,0) circle (1);
	\draw[thick] (-1,0) -- (-2,0);
	\node[above] at (-2,0) {$O_{h_{\alpha_1}}$};
	\node[above] at (0,1) {$P'$};
	\node[below] at (0,-1) {$Q'$};
	\draw[thick] (1,0) -- (2,0);
	\node[above] at (2,0) {$O_{h_{\alpha_2}}$};
	\node[scale=0.75] at (0,0) {$\tau'$};
	\end{tikzpicture}
	}} \Bigg{|}^2~.
\end{aligned}
\ee
We like to emphasize that in the presence of two boundaries, the extra factors of $\rho_0$ coming from the difference of $C_0$ and $\hat{C}_{\text{DOZZ}}$ cancel with the $\rho_0$ from the Cardy spectrum, further leading to a Liouville result. In this paper, we hope that we have been clear on addressing that the relation between Liouville and 3d gravity with two boundaries is not simply a coincidence.

\begin{figure}
\begin{center}
\begin{overpic}[scale=0.5]{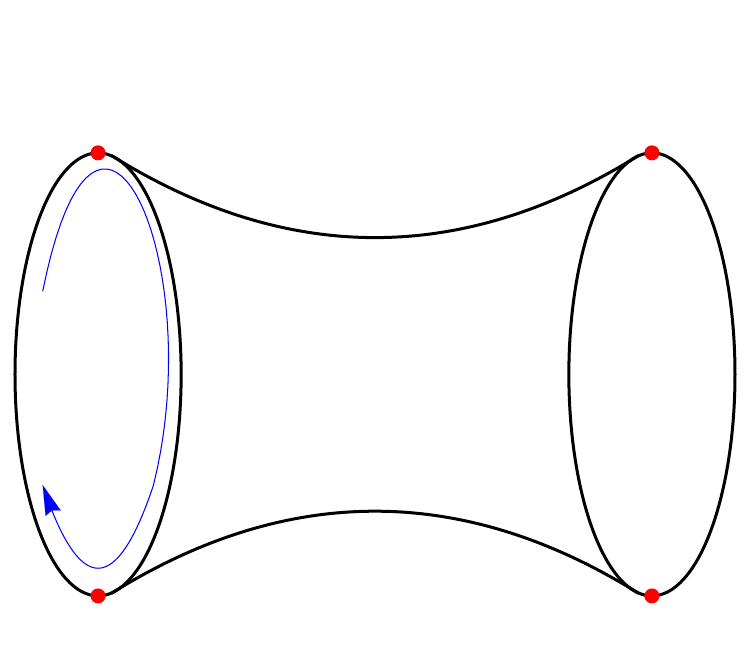}
\put(-5,72){\parbox{0.2\linewidth}{
		\begin{equation*}
			O_{h_{\alpha_1}}
\end{equation*}}}
\put(-5,4){\parbox{0.2\linewidth}{
		\begin{equation*}
			O_{h_{\alpha_2}}
\end{equation*}}}
\put(70,72){\parbox{0.2\linewidth}{
		\begin{equation*}
			O_{h_{\alpha_1}}
\end{equation*}}}
\put(70,4){\parbox{0.2\linewidth}{
		\begin{equation*}
			O_{h_{\alpha_2}}
\end{equation*}}}
\put(-10,30){\parbox{0.2\linewidth}{
		\begin{equation*}
			\color{blue}{\widetilde{\beta}}
\end{equation*}}}
\end{overpic}  
\end{center}
\caption{The two-boundary disk wormhole in JT gravity. There are two operator insertions on each boundary that are separated in Lorentzian time $t$ and Euclidean time $\widetilde{\beta}/2$. }\label{fig:stanfordWH}
\end{figure}
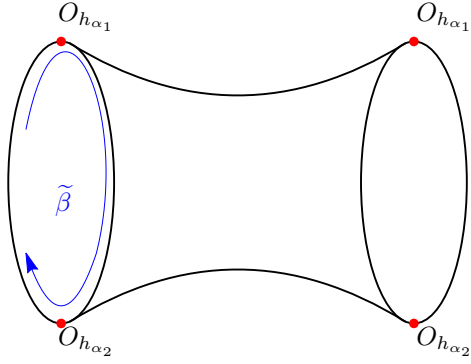

We also want to point out the connection to the wormholes that describes quantum noise in JT gravity \cite{Stanford:2020wkf}. We just need to take the Schwarzian limit of the correlated torus two-point function, which is given by $\beta_L = -2 \pi i \tau \rightarrow +\infty$ and $\beta_R = 2 \pi i \bar{\tau} \rightarrow 0$. In doing so, the contribution from the right-movers is given by the vacuum block and we choose not to keep track of it as it only contributes as an overall normalization factor to the Schwarzian correlators \cite{Ghosh:2019rcj}. In the Schwarzian limit, the OPE coefficents of descendants are suppressed as the black-hole states are given by $P' = b \sqrt{E_1}, Q' = b \sqrt{E_2}$ where $E_1,E_2$ are held fixed. For the external operators, $\alpha = b h \sim \mathcal{O}(b)$. We finally have
\begin{equation}
    \begin{split}
        \int_0^\infty dP' dQ'  \rho_0(P') \rho_0(Q') &\rightarrow \int_0^\infty dE_1 dE_2  \sinh(2 \pi \sqrt{E_1})\sinh(2 \pi \sqrt{E_2})~, \\
        C_0\left(b \sqrt{E_1},i\left(\frac{Q}{2} - b h_{\alpha_1} \right),b \sqrt{E_2}\right) &\rightarrow \frac{\prod_{\pm 1,2} \Gamma(h_{\alpha_1} \pm_1 i \sqrt{E_1} \pm_2 i \sqrt{E_2})}{\Gamma(2 h_{\alpha_1})}~, \\
    \vcenter{\hbox{
	\begin{tikzpicture}[scale=0.5]
	\draw[thick] (0,0) circle (1);
	\draw[thick] (-1,0) -- (-2,0);
	\node[above] at (-2,0) {$O_{h_{\alpha_1}}$};
	\node[above] at (0,1) {$P'$};
	\node[below] at (0,-1) {$Q'$};
	\draw[thick] (1,0) -- (2,0);
	\node[above] at (2,0) {$O_{h_{\alpha_2}}$};
	\node[scale=0.75] at (0,0) {$\tau$};
	\end{tikzpicture}
	}}\vcenter{\hbox{
	\begin{tikzpicture}[scale=0.5]
	\draw[thick] (0,0) circle (1);
	\draw[thick] (-1,0) -- (-2,0);
	\node[above] at (-2,0) {$O_{h_{\alpha_1}}$};
	\node[above] at (0,1) {$P'$};
	\node[below] at (0,-1) {$Q'$};
	\draw[thick] (1,0) -- (2,0);
	\node[above] at (2,0) {$O_{h_{\alpha_2}}$};
	\node[scale=0.75] at (0,0) {$\tau'$};
	\end{tikzpicture}
	}} &\rightarrow e^{-\tilde{\beta}(E_1+E_2)}~,
    \end{split}
\end{equation}
where we have taken $z_1,z_1' = \frac{i \tilde{\beta}}{2 b^2}$; $z_2,z_2' = 0$ and $ \tilde{\beta} = b^2 \beta_L= b^2 \beta_L'$ in reproducing the last relation. In principle, the moduli of the two boundaries are different but the identification between the prime and unprime parameters is for computing the variance of the wormhole contribution in JT gravity coupled to matter. We subsequently obtain 
\begin{equation}
\begin{split}
    G^L_{\alpha_1,\alpha_2}(\tau,\tau) &\rightarrow \Big{|} \text{Tr}\left[e^{-\tilde{\beta} \frac{\tilde{H}}{2}}O_{h_{\alpha_1}}(t)e^{-\tilde{\beta} \frac{\tilde{H}}{2}} O_{h_{\alpha_2}}(0) \right]\Big{|}^2~, \\
    &=  \int_0^\infty dE_1 dE_2  \sinh(2 \pi \sqrt{E_1})\sinh(2 \pi \sqrt{E_2})e^{-\tilde{\beta}(E_1+E_2)} \langle E_1| \hat{G}_{\alpha_1} | E_2 \rangle  \langle E_2| \hat{G}_{\alpha_2} | E_1 \rangle~,
\end{split}
\end{equation}
where
\begin{equation}\label{eq:JT_DOZZ}
    \langle E_1| \hat{G}_{\alpha} | E_2 \rangle = \frac{\prod_{\pm 1,2} \Gamma(h_{\alpha} \pm_1 i \sqrt{E_1} \pm_2 i \sqrt{E_2})}{\Gamma(2 h_{\alpha})}~.
\end{equation}
As shown in Figure \ref{fig:stanfordWH}, $e^{-\tilde{\beta} \frac{\tilde{H}}{2}}$ represents a Euclidean time evolution of $\tilde{\beta}/2$ on the disk and the two operators $O_{h_{\alpha_1}},O_{h_{\alpha_2}}$ are also separated in Lorentzian time $t$. $ \langle E_1 | \hat{G}_{\alpha} | E_2 \rangle$ is the matrix element of the operator insertion $\hat{G}_\alpha$ that acts on the Hilbert space of a disk$\times$interval in JT gravity. This is exactly the result obtained in \cite{Stanford:2020wkf}.

\section{Counterterms in renormalizing the action}\label{app:counterterm}
In this appendix, we like to derive the counterterms needed in normalizing the defect action. A detailed procedure in normalizing the wormhole action has been demonstrated in \cite{Collier:2022bqq}. The result is given by
\begin{equation}\label{eq:Tom_Sct}
    -S_{\text{ct}}^{\text{there}}(\eta_i) = \frac{c}{3 }\sum_i \left((1-2 \eta_i)\ln(1-2 \eta_i) +(1-2 \eta_i) \ln \epsilon - \ln R+2 \eta_i^2 \ln \epsilon_i\right) ~,
\end{equation}
where their Liouville field lives on a disk with cutoff radius $R$. We notice that \eqref{eq:Tom_Sct} is not local as it depends on $R$ due to the chosen normalization of the wormhole action. It was mentioned that the following GHY term at $|z| = R$
\begin{equation}
    \frac{1}{8 \pi G_N}\int \sqrt{h} K = \frac{1}{2 G_N}\left(2 \ln R + \ln\frac{2}{\epsilon}\right)~,
\end{equation}
was not included in the computation of the action. Had we included it in normalizing the defect action and adding the relevant terms such that $-S_{\text{ct}}(0) = \frac{c}{3}(\ln 2-1)$, the counterterm is given by
\begin{equation}
    -S_{\text{ct}}(\eta_i)= \frac{c}{3}\sum_i \left((1-2 \eta_i)\ln(1-2 \eta_i) -2 \eta_i\ln \epsilon+2 \eta_i^2\ln \epsilon_i+\ln 2 -1\right)~,
\end{equation}
which is what we have in \eqref{eq:npoint}. The reason of requiring $-S_{\text{ct}}(0) = \frac{c}{3}(\ln 2-1)$ is because each massless operator inserted in the gravitational action is related to the identity operator in Liouville by the following normalization factor 
\begin{equation}
   \frac{1}{\sqrt{S_L\left(P\right)S_{\mathds{1}P}}} \approx e^{\frac{c}{6}(\ln 2 -1)}~,
\end{equation}
where $P = \frac{i }{2}\left(b+\frac{1}{b}\right)$.
\section{ADM Decomposition}\label{eq:ADM_Decomp}
\begin{figure}
\begin{center}
\begin{overpic}[scale=0.8]{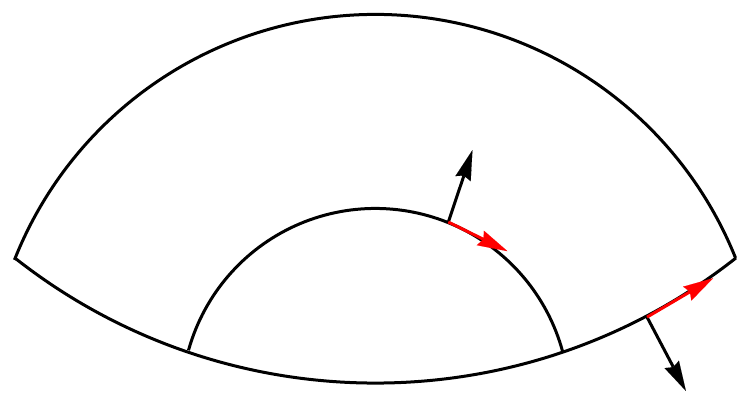}
\put(30,28){\parbox{0.2\linewidth}{
		\begin{equation*}
			\Sigma_{ZZ}'
\end{equation*}}}
\put(30,54){\parbox{0.2\linewidth}{
		\begin{equation*}
			\Sigma'
\end{equation*}}}
\put(30,38){\parbox{0.2\linewidth}{
		\begin{equation*}
			\mathcal{M}'
\end{equation*}}}
\put(30,12){\parbox{0.2\linewidth}{
		\begin{equation*}
		\mathcal{M}-	\mathcal{M}'
\end{equation*}}}
\put(-5,10){\parbox{0.2\linewidth}{
		\begin{equation*}
		\mathcal{B}''
\end{equation*}}}
\put(30,-2){\parbox{0.2\linewidth}{
		\begin{equation*}
		\mathcal{B}'-\mathcal{B}''
\end{equation*}}}
\put(85,20){\parbox{0.2\linewidth}{
		\begin{equation*}
		\Gamma
\end{equation*}}}
\put(-15,20){\parbox{0.2\linewidth}{
		\begin{equation*}
		\Gamma
\end{equation*}}}
\put(50,35){\parbox{0.2\linewidth}{
		\begin{equation*}
		u^\mu
\end{equation*}}}
\put(48,15){\parbox{0.2\linewidth}{
		\begin{equation*}
		\color{red}{n^\mu}
\end{equation*}}}
\put(73,0){\parbox{0.2\linewidth}{
		\begin{equation*}
		r^\mu
\end{equation*}}}
\put(80,10){\parbox{0.2\linewidth}{
		\begin{equation*}
		\color{red}{t^\mu}
\end{equation*}}}
\end{overpic}  
\end{center}
\caption{Figure shows the cross section of the spacetime geometry of the wavefunction, $\mathcal{M}$.}\label{fig:ADMdecomp}
\end{figure}
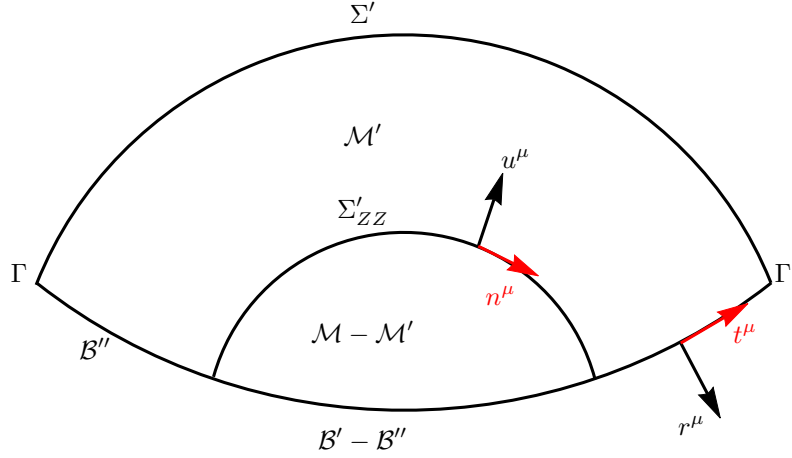

In this appendix, we like to compute the gravitational action in \eqref{eq:Sgrav_Phistates3} by performing an ADM split on the scalar curvature and extrinsic curvature components.  We find that this decomposition is convenient for the $(\Phi_0, J)$-basis Wheeler-DeWitt wavefunction computation in Section \ref{subsec:Quasi_Fuchsian} and here, we use the $\Phi_0$-states as an exercise on this machinery.

As shown in Figure \ref{fig:ADMdecomp}, with the outward pointing normal vector of $\mathcal{B}''$ to be $r_\mu$ and the forward pointing normal of $\Sigma',\Sigma'_{ZZ}$ surfaces, which we now denote as $\Sigma_\Phi$ surfaces, to be $u_\mu$, we have the following boost relations between vector fields
\begin{equation}\label{eq:boost_v}
    \begin{split}
        r_\mu &= \gamma_v n_\mu - \gamma_v v u_\mu = \cos \alpha n_\mu - \sin \alpha u_\mu~, \\
        t_\mu &= \gamma_v u_\mu + \gamma_v v n_\mu = \cos \alpha u_\mu + \sin \alpha n_\mu~,
    \end{split}
\end{equation}
where $r \cdot t = 0, n \cdot u = 0$, $\gamma_v = (1+v^2)^{-1/2}$ and $v = \tan \alpha$. 

The wormhole slicing of the hyperbolic metric is in ADM form
\begin{equation}
    ds^2 = \cosh^2 \rho e^\Phi dY^2 + d\rho^2 + \cosh^2 \rho e^\Phi dX^2~,
\end{equation}
where $X=\text{Re}(z)$ and $Y=\text{Im}(z)$ is the chosen direction for ``time'' evolution. The unit normal $u_\mu$ to $\Sigma_\Phi$ is given by $u_\mu = N \nabla_\mu Y$ where $N = \cosh \rho e^{\Phi/2}$ is the lapse function. With $n_\mu$ being the outward unit normal to the constant $\rho$ surface, we obtain the projection $\sigma_{\mu\nu}$ onto the intersection between the $\Sigma'_\Phi$ slices and $\rho=$constant slices
\begin{equation}
    \sigma_{\mu\nu} = g_{\mu\nu} - n_\mu n_\nu - u_\mu u_\nu = g_{\mu\nu} - r_\mu r_\nu - t_\mu t_\nu~.
\end{equation}
We are ready to compute the following gravitational action using ADM decomposition
\begin{equation}\label{eq:app_Sgrav_Phi0}
\begin{split}
    -S_{\text{grav}}(\Phi_0)  &= \frac{1}{16 \pi G_N}\int_{\mathcal{M}'} d^3 x\sqrt{g}\left(R+2\right)+ \frac{1}{8 \pi G_N}\int_{\mathcal{B}''} d^2 x\sqrt{\gamma}\left(\Theta-1\right)\\
        &+ \frac{1}{8 \pi G_N}\int_{\Sigma'_{ZZ}}^{\Sigma'} d^2 x\sqrt{h} K +\frac{1}{8 \pi G_N}\int_{\Gamma} dx \sqrt{\sigma_\Gamma} \left(\theta-\frac{\pi}{2}\right)~.
\end{split}
\end{equation}
Using the following decomposition of the Ricci scalar
\begin{equation}\label{eq:R_decomp}
    R^{(3)} = R^{(2)}+K^2 - K_{\mu\nu}K^{\mu\nu}-2 \nabla_\mu(K u^\mu - a^\mu)~,
\end{equation}
where $K_{\mu\nu} = h^\alpha_\mu \nabla_\alpha u_\nu$ and $a^\nu = u^\mu \nabla_\mu u^\nu$, we obtain 
\begin{equation}\label{eq:ADM_Fuchsian}
\begin{split}
     -S_{\text{grav}}(\Phi_0) &= \frac{1}{16 \pi G_N}\Big{(}\int_{\mathcal{M}'} d^3 x\sqrt{g}\left(R^{(2)}+K^2 - K_{\mu\nu}K^{\mu\nu}+2\right)+2\int_{\mathcal{B}''} d^2 x\sqrt{\gamma}\left(\gamma_v k-1 -  t^\mu \nabla_\mu \alpha\right) \\
     &+2 \int_{\Gamma} dx \sqrt{\sigma_\Gamma} \left(\theta-\frac{\pi}{2}\right)\Big{)}~.
\end{split}
\end{equation}
In deriving this result, we have used the following splittings of $\Theta$ and $K$
\begin{equation}\label{eq:K_splitting}
    \begin{split}
    \Theta &= \gamma_v k - \gamma_v v l - \gamma_v n\cdot a +\gamma_v v u\cdot b- t^\mu \nabla_\mu \alpha~, \\
    K &= l - u \cdot b~,
    \end{split}
\end{equation}
where
\begin{equation}
    \begin{split}
        k &= \sigma^{\mu\nu} \nabla_\mu n_\nu = \frac{1}{\sqrt{\sigma}}n^\mu \partial_\mu \sqrt{\sigma}~, \\
        l &= \sigma^{\mu\nu} \nabla_\mu u_\nu = \frac{1}{\sqrt{\sigma}}u^\mu \partial_\mu \sqrt{\sigma}~, \\
        b^\nu &= n^\mu \nabla_\mu n^\nu~, 
    \end{split}
\end{equation}
and the decomposition of the extrinsic curvature components is derived in \cite{Brown:2000dz}. There are sign differences between \cite{Brown:2000dz} and us due to analytic continuation from Lorentzian signature to Euclidean signature. To prevent possible confusion, we rederive the splittings here.

Given that the induced metric of the $\Sigma'_\Phi$ slices to be $h_{ij}$ and the forward pointing normal to be $u_\mu$, the extrinsic curvature is given by
\begin{equation}
\begin{split}
    K_{\mu\nu} &= h^\alpha_\mu \nabla_\alpha u_\nu~.
\end{split}
\end{equation}
The induced metric $h_{ij}$ can be expressed as
\begin{equation}
    h_{ij} = \sigma_{ij}+n_i n_j~,
\end{equation}
where $\sigma_{ij}$ is the projection onto a codimension-2 surface with outward normal $n_i$ that foliates $\Sigma'_\Phi$. This allows the following decomposition of the trace of extrinsic curvature $K$
\begin{equation}
\begin{split}
    K &= l -u \cdot b~.
\end{split}
\end{equation}
We have used Leibniz rule $n^\mu \nabla_\nu u_\mu = -u^\mu \nabla_\nu n_\mu$ in deriving this expression. 

Similarly, the extrinsic curvature of $\mathcal{B}''$ embedded in $\mathcal{M}$ is given by
\begin{equation}
\begin{split}
    \Theta_{\mu\nu} &=\gamma^\alpha_\mu \nabla_\alpha r_\nu~, 
\end{split}
\end{equation}
where $r_\mu$ being the outward unit normal and $\gamma_{ab}$ being the induced metric of $\mathcal{B}''$. To arrive at the splitting of $\Theta$ in \eqref{eq:K_splitting}, we use the identity
\begin{equation}
    \nabla_\mu u_\nu = K_{\mu\nu}+u_\mu a_\nu~,
\end{equation}
to obtain
\begin{equation}
    \nabla_\mu r_\nu = \gamma_v \nabla_\mu n_\nu - \gamma_v v K_{\mu\nu} - \gamma_v v u_\mu a_\nu - t_\nu \nabla_\mu \alpha~.
\end{equation}
Finally, using
\begin{equation}
    \sigma^{\mu\nu}\Theta_{\mu\nu} = \gamma_v k - \gamma_v v l~,
\end{equation}
and
\begin{equation}
\begin{split}
    t^\mu t^\nu \nabla_\mu r_\nu &= \gamma_v^2  t^\mu u^\nu \nabla_\mu n_\nu - \gamma_v^3 v^3 n^\mu n^\nu \nabla_{\mu} u_\nu - \gamma_v^3 v^2  (n \cdot a) - t^\mu \nabla_\mu \alpha ~, \\
    &= - \gamma_v n\cdot a +\gamma_v v u\cdot b- t^\mu \nabla_\mu \alpha~,
\end{split}
\end{equation}
we obtain \eqref{eq:K_splitting}.

We notice that the gravitational action in \eqref{eq:app_Sgrav_Phi0} simplifies as there is some total derivative terms from the bulk that cancels with the boundary terms of the $\Sigma'_\Phi$ slices, reducing to \eqref{eq:ADM_Fuchsian}. With $R^{(2)} = -2$, the bulk term in \eqref{eq:ADM_Fuchsian} vanishes and we automatically reproduce \eqref{eq:Sgrav_Phi0}
\begin{equation}
\begin{split}
      -S_{\text{grav}}(\Phi_0)  &= -\frac{1}{4 \pi G_N}\int_{B} dz d\bar{z} \left(\frac{1}{4}(\partial \Phi \bar{\partial} \Phi + e^\Phi)\right) + \frac{\pi}{G_N \beta \epsilon_y}~,
\end{split}
\end{equation}
where we consider the same Liouville solution in \eqref{eq:HH_Liouville_sol1}.
\section{Gluing of ZZ boundaries in gravity}
\label{app:ZZ_gluing}
\begin{figure}
\begin{center}
\begin{overpic}[scale=0.8]{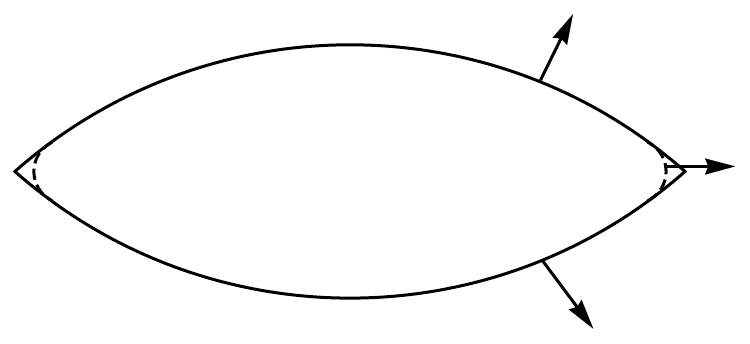}
\put(30,42){\parbox{0.2\linewidth}{
		\begin{equation*}
			\Sigma_{ZZ}'
\end{equation*}}}
\put(65,42){\parbox{0.2\linewidth}{
		\begin{equation*}
			u^\mu
\end{equation*}}}
\put(65,7){\parbox{0.2\linewidth}{
		\begin{equation*}
			r^\mu
\end{equation*}}}
\put(80,25){\parbox{0.2\linewidth}{
		\begin{equation*}
			\Bar{r}^\mu
\end{equation*}}}
\put(30,20){\parbox{0.2\linewidth}{
		\begin{equation*}
			\mathcal{M}-\mathcal{M}'
\end{equation*}}}
\put(90,17){\parbox{0.2\linewidth}{
		cap}}
\put(30,0){\parbox{0.2\linewidth}{
		\begin{equation*}
			\mathcal{B}'-\mathcal{B}''
\end{equation*}}}
\end{overpic}  
\end{center}
\caption{Figure shows the cross section of the small cylindrical region $\mathcal{M} - \mathcal{M}'$. We regulate the codimension two surface $\Sigma_{ZZ} \cap (\mathcal{B}' - \mathcal{B}'')$ with a spherical cap surface that has unit normal $\Bar{r}^\mu$. }\label{fig:ZZgluing}
\end{figure}
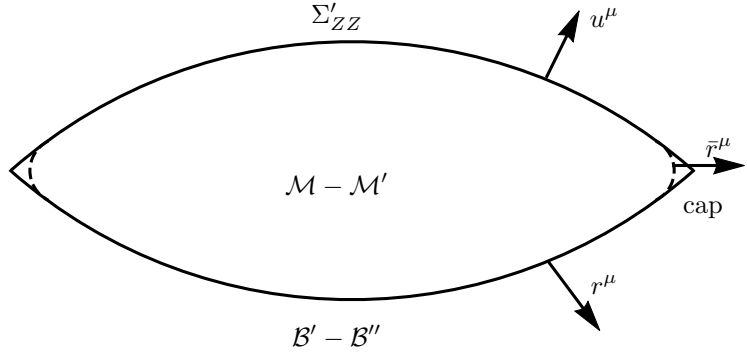
In this subsection, we show the bulk+AdS boundary contribution of the small cylindrical region $\mathcal{M} - \mathcal{M}'$ that is carved out due to the presence of the ZZ boundary conditions (as shown in Figure \ref{fig:HH_Wavefunctioncutoff} and \ref{fig:ZZ_term}) is equal to the extrinsic curvature contribution at the codimension-one carved out surface $\Sigma'_{ZZ}$. The contribution of the carved out surface to the on-shell action of the Hartle-Hawking wavefunction is given by
\begin{equation}\label{eq:Sgrav_carve}
    \begin{split}
        -S' &= \frac{1}{16 \pi G_N}\left(\int_{\mathcal{M} - \mathcal{M}'}d^3 x\sqrt{g}(R+2)+ 2\int_{\mathcal{B} - \mathcal{B}'}d^2 x \sqrt{\gamma}(\Theta-1) \right)~.
    \end{split}
\end{equation}
Using \eqref{eq:R_decomp} for the ADM splitting of the Ricci scalar
\begin{equation}
    R^{(3)} = R^{(2)}+K^2 - K_{\mu\nu}K^{\mu\nu}-2 \nabla_\mu(K u^\mu - a^\mu)~,
\end{equation}
and \eqref{eq:K_splitting} for the extrinsic curvature components
\begin{equation}
    \begin{split}
    \Theta &= \gamma_v k - \gamma_v v l - \gamma_v n\cdot a +\gamma_v v u\cdot b- t^\mu \nabla_\mu \alpha~, \\
    K &= l - u \cdot b~,
    \end{split}
\end{equation}
\eqref{eq:Sgrav_carve} becomes
\begin{equation}\label{eq:Sgrav_carve_decomp}
    \begin{split}
        -S' &= \frac{1}{16 \pi G_N}\left(\int_{\mathcal{M} - \mathcal{M}'}d^3 x\sqrt{g}(R+2)+ 2\int_{\mathcal{B} - \mathcal{B}'}d^2 x \sqrt{\gamma}(\Theta-1) \right)~, \\
        &=  \frac{1}{16 \pi G_N}\left(\int_{\mathcal{M} - \mathcal{M}'}d^3 x\sqrt{g}(R^{(2)}+K^2 - K_{\mu\nu}K^{\mu\nu}+2)+ 2\int_{\mathcal{B} - \mathcal{B}'}d^2 x \sqrt{\gamma}(\gamma_v k-1 - t^\mu \nabla_\mu \alpha) -2\int_{\Sigma'_{ZZ}}d^2 x \sqrt{h} K\right)~, \\
        &-\frac{1}{8 \pi G_N}\int_{\text{cap}}d^2 x \sqrt{\Bar{\gamma}}(K(\bar{r} \cdot u) - \bar{r} \cdot a)~, \\
    \end{split}
\end{equation}
where $\bar{r}$ is the normal vector to the spherical cap around the joint $(\mathcal{B}' - \mathcal{B}'')\cap \Sigma_{ZZ}$, as shown in Figure \ref{fig:ZZgluing}. In shrinking the cap to zero size, $\sqrt{\bar{\gamma}} \rightarrow 0$, and with the term $(K(\bar{r} \cdot u) - \bar{r} \cdot a)$ being regular around the cap, the last term in \eqref{eq:Sgrav_carve_decomp} vanishes. In the limit $\mathcal{M} - \mathcal{M}'$ becomes a zero size region, the bulk+AdS boundary term of \eqref{eq:Sgrav_carve_decomp} vanish both for the $\Phi_0$-basis and $(\Phi_0,J)$-basis Wheeler-DeWitt wavefunction. Hence, we have shown
\begin{equation}
    -S' = \frac{1}{16 \pi G_N}\left(\int_{\mathcal{M} - \mathcal{M}'}d^3 x\sqrt{g}(R+2)+ 2\int_{\mathcal{B} - \mathcal{B}'}d^2 x \sqrt{\gamma}(\Theta-1) \right) = -\frac{1}{8 \pi G_N}\int_{\Sigma'_{ZZ}}d^2 x \sqrt{h} K~.
\end{equation}

\section{Density of states for the ramp}\label{app:ramp}
In this appendix, we provide details in evaluating the following integral
\be\label{eq:ramp_app}
\rho_{\text{ramp}}(P',Q')=\int_0^\infty d\lambda \, \lambda \, {\cos(4\pi P' \lambda) \cos(4\pi Q' \lambda)}~.
\ee
We first notice that $\cos(4\pi P' \lambda)$ is the density of states for the trumpet partition function in JT gravity \cite{Penington:2023dql,Saad:2019pqd}
\be
Z_{\text{trumpet}}(\beta_1, \lambda)=\frac{\sqrt{\pi } e^{-\frac{4 \pi ^2 \lambda^2}{\beta_1 }}}{2 \sqrt{\beta_1 }}=\int_{0}^\infty dP' \cos(4\pi P' \lambda) e^{-\beta_1 P^{\prime 2}}~,
\ee
and the double-trumpet partition function is given by \cite{Penington:2023dql,Saad:2019pqd}
\be
Z(\beta_1,\beta_2)=\int_0^\infty d\lambda \, \lambda \, Z_{\text{trumpet}}(\beta_1, \lambda) Z_{\text{trumpet}}(\beta_2, \lambda)=\frac{\sqrt{\beta_1 \beta_2}}{32\pi (\beta_1+\beta_2)}~.
\ee
Assuming $E_1,E_2 < 0$, we compute the contribution to the correlator of resolvents \cite{Saad:2019lba}
\be\label{eq:resolvent}
R(E_1,E_2)=\int_0^\infty d\beta_1 d\beta_2 e^{\beta_1 E_1 +\beta_2 E_2} \frac{\sqrt{\beta_1 \beta_2}}{32\pi (\beta_1+\beta_2)}=\frac{1}{64} \frac{1}{\sqrt{-E_1}\sqrt{-E_2}\left(\sqrt{-E_1} + \sqrt{-E_2}\right)^2}~.
\ee

Next, we continue \eqref{eq:resolvent} to positive energy, thus obtaining the density of states for the ramp from the analytical continued  $R(\pm,\pm) = R(E_1\pm i \epsilon,E_2\pm i \epsilon)$ on the corresponding branch\footnote{The denominator comes from the Jacobian when we perform a change of variables.}
\be
\begin{aligned}
\rho_{\text{ramp}}(E_1,E_2)&=\int_0^\infty d\lambda \, \lambda \, \frac{{\cos(4\pi \sqrt{-E_1} \lambda) \cos(4\pi \sqrt{-E_2} \lambda)}}{4\sqrt{-E_1} \sqrt{-E_2}}\\
&= \frac{R(+,+)+R(-,-)-R(+,-)-R(-,+)}{(-2\pi i)^2}\\
&=-\frac{1}{64\pi^2} \frac{E_1+E_2}{\sqrt{E_1 E_2} (E_1-E_2)^2}~.
\end{aligned}
\ee
Using the relation $E_1=P'^2,E_2=Q'^2$ and taking into account the Jacobian, we obtain\footnote{In principle, we can apply Cauchy's integral theorem to deform the integration contour from real to imaginary axis on the four exponential oscillatory integrals of \eqref{eq:ramp_app}. After Wick rotation, the integral is exponentially damped and we obtain \eqref{eq:ramp_ans}. }
\be\label{eq:ramp_ans}
\rho_{\text{Ramp}}(P',Q')=-\frac{P'^2+Q'^2}{16 \pi ^2 \left(P'^2-Q'^2\right)^2}~,
\ee
and $\rho_{\text{Ramp}}(P',Q')$ is responsible for level repulsion in random matrix theory.

\nocite{*}

\bibliographystyle{ourbst}
\bibliography{gravity.bib}

\end{document}